\documentclass[letterpaper,11pt]{article}
\usepackage[utf8]{inputenc}

\pdfoutput=1 
\usepackage{jheppub}
\usepackage{graphicx,verbatim}
\usepackage{blindtext}
\usepackage[T1]{fontenc}
\usepackage{epsf}
\usepackage{mathtools}
\usepackage{lmodern}
\usepackage{dsfont}
\usepackage[mathscr]{euscript}

\usepackage{amsmath}
\usepackage{amsfonts}
\usepackage{amssymb}
\usepackage{mathrsfs}
\usepackage{slashed}
\usepackage{xcolor}
\usepackage{caption}
\usepackage{subcaption}
\usepackage{makecell}

\newcommand{\sssslash}{\mathbin{/\mkern-6mu/\mkern-6mu/\mkern-6mu/}}

\def\tU{\mathtt{U}}

\def\fg{\mathfrak{g}}
\def\qe{\mathfrak{q}}
\def\kq{\mathfrak{q}}
\def\fq{\mathfrak{q}}

\def\fR{\mathfrak{R}}
\def\fL{\mathfrak{L}}
\def\fsl{\mathfrak{sl}}
\def\fh{\mathfrak{h}}

\def\bz{\mathbf{z}}
\def\by{\mathbf{y}}
\def\bx{\mathbf{x}}
\def\ba{\mathbf{a}}

\def\bL{\mathbf{L}}
\def\bM{\mathbf{M}}

\def\BC{\mathbb{C}}
\def\BR{\mathbb{R}}
\def\BE{\mathbb{E}}
\def\BZ{\mathbb{Z}}

\def\BP{\mathbb{P}}
\def\BT{\mathbb{T}}
\def\BL{\mathbb{L}}
\def\BH{\mathbb{H}}

\def\bI{\mathbf{I}}
\def\bbE{\mathbf{E}}

\def\bbA{\mathbb{A}}

\def\CalB{\mathcal{B}}

\def\CalA{\mathcal{A}}

\def\CalP{\mathcal{P}}
\def\CalR{\mathcal{R}}
\def\CalZ{\mathcal{Z}}
\def\CalH{\mathcal{H}}

\def\CalO{\mathcal{O}}

\def\CalS{\mathcal{S}}

\def\qe{\mathfrak{q}}
\def\CalL{\mathcal{L}}

\def\Tr{{\rm Tr}}

\def\ve{{\varepsilon}}

\def\sK{\mathsf{K}}

\def\EQ{\EuScript{Q}}
\def\EW{\EuScript{W}}
\def\EN{\EuScript{N}}
\def\EY{\EuScript{Y}}
\def\EX{\EuScript{X}}

\def\EZ{\EuScript{Z}}
\def\EG{\EuScript{G}}
\def\EH{\EuScript{H}}

\def\EG{\EuScript{G}}
\def\EB{\EuScript{B}}
\def\ER{\EuScript{R}}
\def\ET{\EuScript{T}}
\def\EM{\EuScript{M}}
\def\EL{\EuScript{L}}
\def\EC{\EuScript{C}}
\def\EF{\EuScript{F}}
\def\ED{\EuScript{D}}

 \def\p{\partial}
 
 \def\a{\alpha}
 \def\b{\beta}
 \def\g{\gamma}
 \def\d{\delta}

 \def\th{\theta}
 
 \def\k{\kappa}
 \def\l{\lambda}
 \def\m{\mu}
 \def\n{\nu}
 
 \def\r{\rho}

 \def\s{\sigma}
 \def\t{\tau}
 \def\th{\theta}
 \def\z{\zeta }
 
 \def\G{\Gamma}
 \def\D{\Delta}

 \def\S{\Sigma}
 \def\L{\Lambda}
 \def\O{\Omega}
 \def\o{\omega }
 \def\U{\Upsilon}
 
 \def\bn{\mathbf{n}}
  \def\bU{\mathbf{U}}
 
 \def\bl{\boldsymbol{\lambda}}

 \makeatletter
\newsavebox{\@brx}
\newcommand{\llangle}[1][]{\savebox{\@brx}{\(\m@th{#1\langle}\)}%
  \mathopen{\copy\@brx\kern-0.5\wd\@brx\usebox{\@brx}}}
\newcommand{\rrangle}[1][]{\savebox{\@brx}{\(\m@th{#1\rangle}\)}%
  \mathclose{\copy\@brx\kern-0.5\wd\@brx\usebox{\@brx}}}
\makeatother

\def\beq{\begin{equation}}
\def\eeq{\end{equation}}

\newcommand{\ket}[1]{\left|{#1} \right\rangle}

\title{{Parallel surface defects, \\ Hecke operators, and quantum Hitchin system}}
\author[a]{Saebyeok Jeong}
\author[b]{, Norton Lee}
\author[c]{, and Nikita Nekrasov}

\affiliation[a]{Department of Theoretical Physics, CERN, \\ 1211 Geneva 23, Switzerland}
\affiliation[b]{Center for Geometry and Physics, Institute for Basic Science (IBS),\\Pohang 37673, Republic of Korea}
\affiliation[c]{Simons Center for Geometry and Physics, \\ Stony Brook University, Stony Brook, NY 11794-3636, USA}
\emailAdd{saebyeok.jeong@cern.ch}
\emailAdd{nortonxs@gmail.com}
\emailAdd{nnekrasov@scgp.stonybrook.edu}

\preprint{CERN-TH-2023-057, CGP23015}

\vspace{2cm}
\abstract{We examine two types of half-BPS surface defects $-$ regular monodromy surface defect and canonical surface defect $-$ in four-dimensional gauge theory with $\EN=2$ supersymmetry and $\Omega_{\varepsilon_1,\ve_2}$-background. Mathematically, we investigate integrals over the moduli spaces of parabolic framed sheaves over $\BP^2$. Using analytic methods of $\EN=2$ theories, we demonstrate that the former gives a twisted $\ED$-module on $\text{Bun}_{G_{\mathbb{C}}}$ while the latter acts as a Hecke operator. In the limit $\ve_2 \to 0$, the cluster decomposition implies the Hecke eigensheaf property for the regular monodromy surface defect. The eigenvalues are given by the opers associated to the canonical surface defect. We derive, in our $\EN=2$ gauge theoretical framework, that the twisted $\ED$-modules assigned to the opers in the geometric Langlands correspondence represent the spectral equations for quantum Hitchin integrable system. A duality to topologically twisted four-dimensional $\EN=4$ theory is discussed, in which the two surface defects are mapped to Dirichlet boundary and 't Hooft line defect. This is consistent with earlier works on the $\EN=4$ theory approach to the geometric Langlands correspondence.}

\begin{document}

\maketitle
\section{Introduction}
Geometric Langlands correspondence \cite{BD1,BD2,Drinfeld1987,Frenkel:2005ef,Gaitsgory:2002dt} relates geometric structures on the moduli spaces of Hitchin's equations on a Riemann surface $\EC$ \cite{hitchin1987} associated to Langlands dual groups. The correspondence states that for each ${}^L G_\BC$-local system on $\EC$, there is a twisted $\ED$-module on the moduli stack $\text{Bun}_{G_\BC} (\EC)$ of $G_\BC$-bundles over $\EC$ on which the Hecke operators act \textit{diagonally}. When the ${}^L G_\BC$-local system is chosen to be an oper, the associated twisted $\ED$-module can be characterized by a system of partial differential equations, which describe the spectral equations for the quantum Hitchin integrable system.

A gauge theoretical approach to the geometric Langlands correspondence was established based on a topologically twisted (GL-twisted) four-dimensional $\EN=4$ gauge theory on $\S \times \EC$ \cite{Kapustin:2006pk}, where $\S$ is a two-dimensional manifold possibly with boundaries and $\EC$ is the Riemann surface on which the geometric Langlands correspondence is studied. The main idea was to compactify the GL-twisted $\EN=4$ theory with the compact gauge group $G$ on the Riemann surface $\EC$ to obtain a topological sigma model of maps $\S \to \EuScript{M}_H (G,\EC)$, where the target $\EuScript{M}_H (G,\EC)$ is the moduli space of Hitchin's equations on $\EC$ defined by $G$. The geometric Langlands correspondence can then be viewed as a consequence of the S-duality of the GL-twisted $\EN=4$ theory, which exchanges the gauge group $G$ to its Langlands dual group ${}^L G$. The S-duality descends to the mirror symmetry of the topological sigma model, exchanging its target $ \EuScript{M}_H (G,\EC)$ to $\EuScript{M}_H ({}^L G,\EC)$ \cite{Strominger:1996it} and therefore providing the duality between the geometric structures therein. The $\EN=4$ theory formulation of the geometric Langlands correspondence was extended to include ramifications at marked points $S = \{p_1,p_2,\cdots, p_n\} \subset \EC$ by introducing half-BPS monodromy surface defects supported on $\S \times \{p_i \} \subset \S \times \EC$ \cite{gukwit2}. 

Meanwhile, a gauge theoretical approach to quantization of integrable systems was developed based on a topologically twisted (Donaldson-Witten twist \cite{cmp/1104161738}) four-dimensional $\EN=2$ gauge theory on $\BC^2$. There is a classical integrable system emergent on the Coulomb branch of the $\EN=2$ theory \cite{Seiberg:1994rs,Seiberg:1994aj,Gorsky:1995zq,Martinec:1995by,Seiberg:1996nz}. For the $\EN=2$ theories that descend from the six-dimensional $\EN=(0,2)$ theory with a partial topological twist compactified on $\EC$ (called class $\CalS$) \cite{gai1}, the integrable system is shown to be precisely the Hitchin integrable system defined on $\EC$ \cite{Gaiotto:2009hg}. The quantization was achieved by subjecting the $\EN=2$ gauge theory to the $\O_{\ve_1,\ve_2}$-background \cite{Nekrasov:2002qd} and taking the limit of turning off one of the $\O$-background parameters ($\ve_2 \to 0$), where the two-dimensional $\EN=(2,2)$ super-Poincar\'{e} symmetry is restored \cite{Nekrasov:2009rc}. The four-dimensional $\EN=2$ theory is then effectively described by a two-dimensional $\EN=(2,2)$ theory, where the twisted chiral ring represented on the discrete vacua can be viewed as giving rise to the desired quantization \cite{Nekrasov:2009uh,Nekrasov:2009ui}. It was then realized that the topological sigma model with the Hitchin moduli space target, which appeared in the $\EN=4$ theory approach to the geometric Langlands correspondence, can also be studied directly by compactifying the $\EN=2$ theory along a two-torus \cite{Nekrasov:2010ka}. In particular, such a $\EN=2$ theory approach was used to characterize the space of open string states between the two branes of the topological sigma model, created as a result of the torus compactification, as the space of conformal blocks of the vertex algebra at the junction of the two branes, explaining the result of \cite{Alday:2009aq}.

The reduction of the $\EN=2$ theory to the topological sigma model opened up an alternative gauge theoretical approach of studying the geometric Langlands correspondence. The mirror symmetry applied to the space of open string states stretched between these specific branes (a canonical coisotropic brane and a brane of opers in the case of \cite{Nekrasov:2010ka}) and its manifestation in the vertex algebra (a $\mathcal{W}$-algebra in the case of \cite{Nekrasov:2010ka}) is indeed an example of the geometric Langlands correspondence \cite{Feigin:1991wy,10.1155/S1073792891000119}. It is expected that the $\EN=2$ theory approach to the quantization of integrable systems and the consequences of \cite{BD1,BD2} on the quantization of the Hitchin system are connected in this way. However, this connection has not been fully established due to a lack of understanding on how to account for different types of twisted $\ED$-modules on $\text{Bun}_{G_\BC} (\EC;S)$ and the ${}^L G_\BC$-opers in the $\EN=2$ gauge theoretical framework.

The goal of the present work is to elucidate such an enhancement of the $\EN=2$ theory approach to the geometric Langlands correspondence can be accomplished by incorporating half-BPS surface defects. Specifically, we investigate two types of half-BPS surface defects: the regular monodromy surface defect, which is defined by assigning a singularity of the fields along a surface \cite{gukwit}, and the canonical surface defect, which is defined by coupling a two-dimensional $\EN=(2,2)$ sigma model to the four-dimensional theory \cite{aggtv,Gaiotto:2009fs}.

It was realized in \cite{Jeong:2018qpc} that the canonical surface defect (supported on, say, the $\BC _{\ve_1}$-plane) gives a $\EN=2$ gauge theoretical construction of the ${}^L G_\BC$-opers by the twisted chiral ring relation of the 2d/4d coupled system \cite{Gaiotto:2009fs,Gaiotto:2013sma} subject to the $\O$-background. In section \ref{sec:TQoper}, we will generalize this construction to arbitrary ranks (see \eqref{eq:oper}). The mentioned twisted chiral ring relation takes the form of an operator-valued differential equation,
\begin{align}
    0 = \left( \p_y ^N + t_2 (y) \p_y ^{N-2} + \cdots  +  t_N (y) \right) X(y),
\end{align}
where $y$, valued in an open patch in $\EC$, is the complexified FI parameter of the 2d $\EN=(2,2)$ gauge theory defining the canonical surface defect; $X(y)$ is its surface observable; and $t_i (y)$ are meromorphic functions whose coefficients are combinations of the local chiral observables of the 4d $\EN=2$ gauge theory (i.e., the trace invariants of the complex adjoint scalar). In the limit $\ve_2 \to 0$, the surface observable and the local observables can be arbitrarily separated on the topological $\BC_{\ve_2}$-plane, leading to the factorization of their correlation functions$-$computed in a vacuum specified by the Coulomb moduli $\ba$$-$into products of individual vacuum expectation values (vevs). Namely, we get
\begin{align}
    0 =  \left( \p_y ^N + t_2 (\ba;y) \p_y ^{N-2} + \cdots  +  t_N (\ba;y) \right) \chi(\ba;y) \equiv \r_\ba \chi(\ba;y),
\end{align}
where $t_i (\ba;y)$ and $\chi(\ba;y)$ denote the vevs of the local and surface observables at the vacuum $\ba$ in the limit $\ve_2\to 0$. We show that the differential operator $\r_\ba$ can be brought into an oper ${}^L G_{\BC}$-connection having regular singularities at marked points, i.e., $\r_\ba \in \text{Op}_{{}^L G_\BC} (\EC;S)$ (see \eqref{eq:operNN}). The Coulomb moduli $\ba$ provide holomorphic coordinates on the affine space of opers through the vevs of the local observables. As conjectured in \cite{Nekrasov:2011bc} and proven in \cite{Jeong:2018qpc}, this gauge theoretical construction of opers directly implies that the twisted superpotential, governing the effective dynamics of the $\EN=2$ gauge theory in the limit $\ve_2 \to 0$, coincides with the generating function of the oper Lagrangian submanifold in the moduli space of ${}^L G_{\BC}$-local systems. 

Meanwhile, it was shown in \cite{Alday:2010vg,Nikita:IV,Nikita:V,Nekrasov:2021tik} that the regular monodromy surface defect in the 4d $\EN=2$ gauge theory gives solutions to the Knizhnik-Zamolodchikov (KZ) equations by its vacuum expectation value $\boldsymbol\Psi(\ba;\boldsymbol\g)$. The solutions to the KZ equations are coinvariants composed of modules over the affine Kac-Moody algebra $\widehat{\mathfrak{g}}$, twisted by elements of $\text{Bun}_{G_\BC} (\EC;S)$. Varying the element $E \in \text{Bun}_{G_\BC} (\EC;S)$, the $E$-twisted coinvariants are organized into sections of a sheaf over $\text{Bun}_{G_\BC} (\EC;S)$, on which the Ward identities are realized by differential equations giving rise to the structure of twisted $\ED$-module \cite{Frenkel:2005pa}. This twisted $\ED$-module is precisely the sheaf of differential operators on $\text{Bun}_{G_\BC} (\EC;S)$, twisted by a line bundle $\mathfrak{L}\to \text{Bun}_{G_\BC} (\EC;S)$.\footnote{More precisely, a tensor product of complex powers of line bundles. See \eqref{eq:twist}.} Thus, in order to incorporate the results of \cite{Alday:2010vg,Nikita:IV,Nikita:V,Nekrasov:2021tik} within the framework of the geometric Langlands correspondence, one must establish the precise parametrization of a parabolic bundle $E \in \text{Bun}_{G_\BC} (\EC;S)$, for which the vev $\boldsymbol\Psi(\ba;\boldsymbol\g)$ of the monodromy surface defect is viewed as an $E$-twisted coinvariant, in terms of the monodromy parameters $\boldsymbol\g$ associated with the defect. Moreover, one must identify the data of the twisting $\mathfrak{L}$ in terms of the gauge theory parameters. In section \ref{sec:reg}, we will demonstrate that the monodromy parameters $\boldsymbol\g$ are identified with holomorphic coordinates on an open patch of $\text{Bun}_{G_\BC}(\EC;S)$, through the way that the KZ equation is shown to be satisfied by $\boldsymbol\Psi(\ba;\boldsymbol\g)$ (see \eqref{eq:bungpara} and \eqref{eq:kzgauge}). Moreover, we will also show that the associated twisting $\mathfrak{L}$ is determined by the mass parameters of the 4d $\EN=2$ gauge theory. It also precisely matches with the monodromy data of the ${}^L G_\BC$-oper $\r_\ba$ labeled by the same Coulomb moduli $\ba$, as anticipated in the geometric Langlands correspondence (see \eqref{eq:twgauge} and \eqref{eq:monoparameter2}).

Note that the vev $\boldsymbol\Psi(\ba;\boldsymbol\g)$ of the monodromy defect is evaluated at the vacuum specified by the Coulomb moduli $\ba $. It is important to split the Coulomb moduli as $\ba = [\ba]+\ve_1 \mathbf{n}$, into continuous parameters $[\ba]$ and discrete parameters $\mathbf{n}$ by their decomposition on the $\ve_1$-lattice. The continuous parameters $[\ba]$, along with the mass parameters, determine the $\widehat{\mathfrak{g}}$-modules themselves underlying the twisted coinvariants (see \eqref{eq:kzgauge}). Consequently, they are held fixed when working with given $\widehat{\fg}$-modules. The vev $\boldsymbol\Psi([\ba]+\ve_1 \mathbf{n};\boldsymbol\g)$ at each remaining discrete parameter $\mathbf{n}$ then provides a section of the sheaf of twisted coinvariants on $\text{Bun}_{G_\BC} (\EC;S)$. These vevs with varying $\mathbf{n}$ are distinguished elements in the space of twisted coinvariants that will provide sections of the Hecke eigensheaf corresponding to the oper $\r_\ba$.\footnote{The space of twisted coinvariants considered here is infinite-dimensional. In this situation, the notion of expanding a general twisted coinvariant into an infinite linear combination of distinguished elements requires specifying a topology and a corresponding completion in which such infinite sums converge. In the present paper, we do not endow the full space with such a topology. Accordingly, our results should be understood as constructing a family of KZ solutions and the subspace they span inside the algebraic solution space. Investigating appropriate topological completions is an interesting problem that we leave for future work.}

Then, in section \ref{sec:Hecke} we consider the configuration of parallel surface defects where the canonical surface defect is inserted on top of the regular monodromy surface defect. Since the parameter space of the canonical surface defect is given by the Riemann surface $y\in \EC$, it is naturally expected that the fusion of the two surface defects leads to a \textit{Hecke modification}, which replaces the parabolic $G_\BC$-bundle $E$ by a new bundle $E'$, isomorphic to $E$ on $\EC \setminus\{y\}$ but with a small neighborhood of $y \in \EC$ glued to the rest by a non-trivial transition function.\footnote{More precisely, the Hecke operators are labeled by the coweights of $G_\BC$ (see \eqref{eq:hecke}). We will see that the canonical surface defect gives the Hecke operator associated with the fundamental weight of ${}^L G_\BC = SL(N)$.} We show that the fusion indeed gives rise to the expected modification of the monodromy parameters $\boldsymbol\g$, by investigating the analytic constraints on the correlation functions of the two defects (see \eqref{eq:kznew}). In the vertex algebra approach of the geometric Langlands correspondence, it is known the Hecke modifications are implemented by an additional insertion of the \textit{twisted} vacuum module, which is the image of the vacuum $\widehat{\mathfrak{g}}$-module under its spectral flow \cite{Frenkel:2005pa,Teschner:2010je,Gaiotto:2021tsq}. We verify that the constraints on the correlation function of the parallel surface defects are identical to those for the twisted coinvariants with insertion of the twisted vacuum module (see \eqref{eq:constcoin}). 

In the limit $\ve_2 \to 0$, we define the Hecke operator by properly integrating the image of Hecke modification (see \eqref{eq:HeckeOp}). In our gauge theory realization, the space of Hecke modifications is parametrized by fugacities associated with local observables supported at the interface of the two defects. We then define the Hecke operator as a certain contour integral of the correlation function of the two surface defects and the local observables on this space of modifications. Namely, we have
\begin{align}
    H_y \boldsymbol\Psi (\ba) := \oint \lim_{\ve_2\to 0} \boldsymbol\U (\ba;y),\qquad y\in \EC,
\end{align}
where $\boldsymbol\U(\ba;y) = \left\langle X(y) \boldsymbol\Psi \right\rangle_\ba$ denotes the correlation function. We show that the action of the so-defined Hecke operator is diagonal on the twisted coinvariant $\boldsymbol\Psi(\ba)$ constructed from the monodromy surface defect, due to the fact that the two surface defects can be arbitrarily separated on the $\BC_{\ve_2}$-plane. Schematically, we have
\begin{align}
      H_{y} \boldsymbol\Psi(\mathbf{a}) = \lim_{\ve_2\to 0} \left\langle X(y) \boldsymbol\Psi \right\rangle_\ba = \chi(\mathbf{a};y) \boldsymbol\Psi(\mathbf{a}).
\end{align}
This factorization property is precisely the $\EN=2$ gauge theoretical account of the Hecke eigensheaf property of the vev of the regular monodromy surface defect. In other words, the distinguished elements of the space of twisted coinvariants constructed by the vevs $\boldsymbol\Psi(\mathbf{a})$ of the monodromy surface defect with varying discrete label $\mathbf{n}\in \BZ^{N-1}$ are shown to be the one which diagonalize the action of the Hecke operator. That is, for each $\mathbf{n}\in \BZ^{N-1}$ the vev $\boldsymbol\Psi(\mathbf{a})$ is the section of a Hecke eigensheaf. Its \textit{eigenvalue} is the oper solution $\chi(\ba;y)$ associated with the oper $\r_\ba$ that we constructed from the canonical surface defect (see \eqref{eq:factor}). 

Further, we uncover the constraints on the correlation function are organized into the \textit{operator-valued} oper equation, where the Laurent coefficients are replaced by the quantum Hamiltonians $(\hat{H}_k)_{k=2}^{N}$ of the quantum Hamiltonians of the Hitchin integrable system (see \eqref{eq:univM}). Combining it with the usual oper equation constructed from the canonical surface defect, we verify the vacuum expectation value of the regular monodromy surface defect provides common eigenfunctions of the mutually commuting quantum Hamiltonians $(\hat{H}_k)_{k=2}^{N}$ (see \eqref{eq:spectralfinal}). Namely,
\begin{align}
   0=\left( \hat{H}_k - E_k (\mathbf{a})\right) \boldsymbol\Psi(\mathbf{a};\boldsymbol\g), \quad\quad k=2,3,\cdots, N,
\end{align}
The eigenvalues $(E_k (\ba))_{k=2}^N$ are given by the vacuum expectation values of the local chiral observables, which span the space of opers by the result of \cite{Jeong:2018qpc} (and its higher-rank generalization that is established in the present work). This is the $\EN=2$ theoretical derivation of the statement that the Hecke eigensheaf assigned to a ${}^L G_\BC$-oper is the quotient of the sheaf of rings of twisted differential operators on $\text{Bun}_{G_\BC}$ by the ideal generated by the oper \cite{BD1}. Our derivation establishes a direct connection between the consequence of the geometric Langlands correspondence on the quantization of Hitchin system and the $\EN=2$ theoretical framework of quantizing integrable systems.

The paper is organized as follows. In section \ref{sec:hitchin}, we review the essential notions of the moduli space of Hitchin's equations with ramifications and geometric structures defined on it. We also recall the topological sigma model and the relation between the mirror symmetry and the geometric Langlands correspondence. In section \ref{sec:bbd}, we introduce a duality between the $\EN=2$ theory and the GL-twisted $\EN=4$ theory by embedding them to string theory, from which we motivate to study specific half-BPS surface defects in the $\EN=2$ theory. In section \ref{sec:sgo}, we present the construction of the $\EN=2$ gauge theory and the surface defects from the gauge origami. In section \ref{sec:TQoper}, we examine the analytic constraints obeyed by the vacuum expectation value of the canonical surface defect using the $qq$-characters. We generalize the construction of ${}^L G_{\BC}$-opers of \cite{Jeong:2018qpc}. In section \ref{sec:reg}, we revisit the KZ equations obeyed by the vacuum expectation value of the regular monodromy surface defect, showing it gives a distinguished family of twisted coinvariants enumerated by the Coulomb moduli. In section \ref{sec:Hecke}, we consider the configuration of parallel surface defects, matching the constraints on their correlation function with the constraints on the twisted coinvariants with insertion of a twisted vacuum module. We define a bi-infinite generalization of the twisted vacuum module to construct non-vanishing coinvariants when there are lowest-weight and highest-weight Verma modules with generic weights. We also show the insertion of the (bi-infinite generalization of) twisted vacuum module gives the action of Hecke modification on the twisted coinvariants. Then, we define the Hecke operator by taking a certain contour integral in the space of Hecke modifications. The factorization property obeyed by the outcome of the integral is shown to yield the Hecke eigensheaf property. In section \ref{sec:gaudin}, we construct the universal oper from a current algebra. By connecting it to the constraints on the correlation function of the parallel surface defects, we get concrete expressions for the elements in the Gaudin algebra represented on the given module in gauge theoretical terms. It was then verified that the sections of the twisted $\ED$-module that we constructed from the regular monodromy surface defect give common eigenfunctions of the quantum Hitchin Hamiltonians, with the eigenvalues given by the holomorphic coordinates on the space of opers. We conclude with discussions in section \ref{sec:dis}. The appendices contain some computational details.

\paragraph{Acknowledgment}
The authors thank Dylan Butson, Mykola Dedushenko, Pavel Etingof, Boris Feigin, Edward Frenkel, Alba Grassi, Nafiz Ishtiaque, David Kazhdan, Hee-Cheol Kim, Zohar Komargodski, Shota Komatsu, Kimyeong Lee, Miroslav Rap\v{c}\'{a}k, Alexander Tsymbaliuk, and Philsang Yoo for discussions on related subjects. The work of SJ is supported by CERN and CKC fellowship. The work of NL is supported by IBS project IBS-R003-D1.

\section{Higgs bundles, local systems, and integrable systems} \label{sec:hitchin}
We begin by reviewing essential concepts in the geometric Langlands correspondence. We also recall the relation between the mirror symmetry of the topological sigma model and the geometric Langlands correspondence.

\subsection{Moduli space of Hitchin's equations with ramifications}
Let $\EuScript{C}$ be a compact Riemann surface with $n$ distinct marked points $S = \{ p_1, p_2, \cdots, p_n \} \subset \EuScript{C}$. We sometimes denote the effective divisor of the marked points by the same letter, $S = \sum_{i=1} ^n p_i$. Let $G$ be a compact Lie group with the Lie algebra $\mathfrak{g} = \text{Lie}(G)$. We call the maximal torus $\mathbb{T} \subset G$ and the Cartan subalgebra $\mathfrak{t} = \text{Lie}(\mathbb{T}) \subset \mathfrak{g}$. We consider a smooth $G$-bundle $E \to \EuScript{C}$ and pairs $(A,\phi)$ of connection $A$ on $E$ and adjoint-valued $1$-form $\phi \in \Omega^1 (\EuScript{C},\text{ad}(E))$.

For each marked point $p_i$, $i=1,2,\cdots, n$, we pick a triple $(\a_i,\b_i,\g_i) \in \BT \times \mathfrak{t} \times \mathfrak{t}$ (modulo the action of the Weyl group) which commutes precisely with a Levi subgroup $\mathbb{L}_i \subset G$ (i.e., $\mathbb{L}_i$-regular). Define the space $\EuScript{W}\left((\a_i,\b_i,\g_i;p_i)_{i=1} ^n \right)$ as the space of pairs $(A,\phi)$ with the following singular behavior at each $p_i$:
\begin{align} \label{eq:singular}
\begin{split}
    &A= \a_i d\th + \cdots \\
    &\phi = \b_i \frac{dr}{r} - \g_i d\th+ \cdots,
\end{split}
\end{align}
where $(r,\th)$ are radial coordinates near $p_i$ and the ellipses indicate terms less singular than $1/r$. Let us call $\EuScript{G}_S$ the group of $G$-gauge transformations of the bundle $E$ which takes value in $\mathbb{L}_i$ at $p_i$. These gauge transformations preserve the singular behaviors \eqref{eq:singular} by construction. 

The space $\EuScript{W}\left((\a_i,\b_i,\g_i;p_i)_{i=1} ^n \right)$ is hyper-K\"{a}her with the K\"{a}hler forms given by
\begin{align}
\begin{split}
    &\o_I = -\frac{1}{4\pi} \int_{\EuScript{C}} \Tr\,(\d A \wedge \d A - \d\phi \wedge \d \phi),\\
    &\o_J = \frac{i}{2\pi} \int_{\EuScript{C}}  d^2 z \, \Tr\, (\d \phi_{\bar{z}} \wedge \d A_z + \d \phi_z \wedge \d A_{\bar{z}}), \\
    &\o_K = \frac{1}{2\pi} \int_{\EuScript{C}} \Tr\, \d\phi \wedge \d A,
\end{split}
\end{align}
which are $(1,1)$-forms with respect to the complex structures $I$, $J$, and $K$, respectively. The $\EuScript{G}_S$-action on the space $\EuScript{W}\left((\a_i,\b_i,\g_i;p_i)_{i=1} ^n \right)$ preserves the hyper-K\"{a}hler structure. The corresponding hyper-K\"{a}hler moment maps are
\begin{align}
\begin{split}
    & \mu_I = \frac{1}{2\pi} \int_{\EuScript{C}} \Tr\, \epsilon (F_A - \phi \wedge \phi), \\
    &\mu_J = -\frac{i}{2\pi} \int_{\EuScript{C}}  d^2 z \, \Tr\, \epsilon(D_z \phi_{\bar{z}} + D_{\bar{z}} \phi_z),  \\
    &\mu_K = \frac{1}{2\pi} \int_{\EuScript{C}} d^2 z  \, \Tr\, \epsilon (D_z \phi_{\bar{z}} - D_{\bar{z}} \phi_z),
\end{split}
\end{align}
 where $\epsilon \in \Omega^0 (\EuScript{C},\text{ad}(E))$ is an element of the Lie algebra of $\EuScript{G}_S$. Thus we can define the moduli space $\EuScript{M}_H \left( G,\EC;S; (\a_i,\b_i,\g_i)_{i=1} ^n \right)$ as the hyper-K\"{a}hler quotient $\EuScript{W}\left((\a_i,\b_i,\g_i;p_i)_{i=1} ^n \right) \sssslash \EuScript{G}_S$, namely, $\vec{\mu}^{-1} (0) / \EuScript{G}_S$. This is called the moduli space of Hitchin's equations with ramifications on $S$.

When it does not cause any confusion, we will abbreviate the notation for the Hitchin moduli space with ramifications on $S$ as $\EuScript{M}_H (G,\EC;S)$, understanding $S$ to denote both the marked points and the ramification data assigned there.

Since the moduli space $\EuScript{M}_H (G,\EC;S)$ is constructed by a hyper-K\"{a}hler quotient, it is also hyper-K\"{a}hler, admitting $\BP^1$-worth of complex structures. We may parameterize these complex structures by $w\in \BP^1$ as
\begin{align} \label{eq:compstr}
    I_w = \frac{1-w \bar{w}}{1+w \bar{w}} I + \frac{i(w-\bar{w})}{1+w\bar{w}}J + \frac{w+\bar{w}}{1+ w\bar{w}}K,
\end{align}
so that $w=0,\infty$ gives $\pm I$, $w =\mp i$ gives $\pm J$, and $w=\pm 1$ gives $\pm K$. The holomorphic variables in the complex structure $I_w$ are $A_z + w^{-1} \phi_z$ and $A_{\bar{z}} - w {\phi}_{\bar{z}}$. From below, we give descriptions of the moduli space $\EuScript{M}_H (G,\EC;S)$ as a complex manifold in a chosen complex structure $I_w$.

\subsubsection{Parabolic Higgs bundles}
Upon choosing a particular complex structure, $I$ for instance, the hyper-K\"{a}hler quotient can be studied from a geometric invariant theory quotient $\nu^{-1} _I (0)/\EuScript{G}_{S,\BC}$, where $\nu_I = \mu_J + i \mu_K$ is the complex moment map holomorphic in $I$ and $\EuScript{G}_{S,\BC}$ is the complexification of $\EuScript{G}_S$. The quotient $\nu^{-1} _I (0)/\EuScript{G}_{S,\BC}$ admits a holomorphic description as the moduli space of stable parabolic Higgs bundles \cite{Simp, Kon,Nak,Kapustin:2006pk,gukwit2}, as we briefly recall here.

For each $\mathbb{L}_i$-regular $\a_i$, we associate a parabolic subgroup $\mathcal{P}_i$ of $G_\BC$ in the following way. We regard $\a \in \mathfrak{t}$ as an element in the complexified Lie algebra $\mathfrak{t}_{\BC} \subset \mathfrak{g}_\BC$, and obtain a parabolic subalgbera $\mathfrak{p} \subset \mathfrak{g}_\BC$ spanned by elements $\psi \in \mathfrak{g}_{\BC}$ satisfying
\begin{align}
    [\a ,\psi] = i \l \psi,\quad \l \geq 0.
\end{align}
We associate a parabolic subgroup $\mathcal{P} \subset G_{\BC}$ as its Lie group. Note that the Levi subgroup $\mathbb{L} \subset G$ that preserves $\a$ is the maximal compact subgroup of $\mathcal{P}$, namely, $\mathbb{L} = G \cap \mathcal{P}$.

Any connection $A$ on a smooth $G$-bundle $E\to \EuScript{C}$ endows a holomorphic structure by the $\bar{\p}_A$ operator on the bundle $E$ (more precisely, its complexification; we still denote it by $E$ following the convention in \cite{Kapustin:2006pk,gukwit2}), being automatically integrable in complex dimension 1. Thus, along with the complexification of gauge transformations, $E$ becomes a holomorphic $G_\BC$-bundle over $\EuScript{C}$. The structure group reduces to the parabolic subgroup $\mathcal{P}_i \subset G_\BC$ at each marked point $p_i \in S$. Such a holomorphic $G_\BC$-bundle is called a parabolic $G_\BC$-bundle. Also, we write $\phi = \varphi + \bar{\varphi}$, splitting $\phi$ into the $(1,0)$ and $(0,1)$ parts. The Hitchin equations and the boundary condition \eqref{eq:singular} imply $\varphi$, called the Higgs field, is holomorphic away from $S$ and has a simple pole at each $p_i \in S$; namely, $\varphi \in H^0 (\EuScript{C},  \text{ad}(E) \otimes K_\EC (S))$. The residue of $\varphi$ at $p_i$ takes value in the parabolic subalgebra $\mathfrak{p}_i$; in particular, it is given by $\s_i = \frac{1}{2} (\b_i + i \g_i)$ modulo the nilpotent ideal $\mathfrak{n}_i \subset \mathfrak{p}_i$. A pair $(E,\varphi)$ of parabolic $G_\BC$-bundle $E$ and the Higgs field $\varphi$ satisfying the above condition is called a parabolic Higgs bundle (with fixed eigenvalues $\s_i$ of residues).

The stability of a parabolic Higgs bundle $(E,\varphi)$ is a condition on the parabolic degrees and the slopes of $E$ and its holomorphic subbundles preserved by $\varphi$. We will not present the detail of the stability condition. Instead, we only remark here that even though the stability depends on $\a_i$ in general, it is independent of $\a_i$ for generic $\s_i = \frac{1}{2} (\b_i + i \g_i)$. This genericity assumption is always made throughout our work. With this assumption, as a complex manifold in complex structure $I$, the space $\EuScript{M}_H \left(G,\EC;S \right)$ is isomorphic to the moduli space of stable parabolic Higgs bundles with fixed eigenvalues of residues. We denote this space by $\EuScript{M}_{\text{Higgs}} (G_\BC,\EC;S;(\s_i)_{i=1} ^n)$.

\subsubsection{Parabolic local systems}
We may also view $\EuScript{M}_H \left(G,\EC;S \right)$ as a complex manifold in complex structure $J$. Consider the complex moment map $\nu_J = \mu_K + i \mu_I$ holomorphic in $J$, which implies the flatness of the complex-valued connection $\CalA = A+i \phi$. Thus, the holomorphic description of the quotient $\nu^{-1} _J (0) /\EuScript{G}_{S,\BC}$ is given by the moduli space of stable parabolic local systems \cite{Nak,Kapustin:2006pk,gukwit2}.

A parabolic local system is a flat $G_\BC$-connection on $\EuScript{C} \setminus S$, with a constraint on the monodromy around each $p_i \in S$. If the subgroup of $G_\BC$ that commutes with $U_i = \exp (-2\pi(\a_i-i\g_i))$ is precisely the Levi subgroup $\mathbb{L}_i$ (that is, $U_i$ is $\mathbb{L}_i$-regular) for each marked point $p_i \in S$, the constraint is that the monodromy $M_i$ around $p_i$ is conjugate to $U_i$. We will always impose this regularity assumption.

The stability condition for a parabolic local system involves the parabolic degrees and the slopes of the underlying parabolic $G_\BC$-bundle and its holomorphic subbundles. We will not state the detail of the condition here. We only remark that the stability condition depends on $\b_i$ in general, but is independent of $\b_i$ if $\g_i + i\a_i$ is generic. This genericity assumption is always made throughout our work. As a complex manifold in complex structure $J$, the space $\EuScript{M}_H \left(G,\EC;S; (\a_i,\b_i,\g_i)_{i=1} ^n \right)$ is equal to the moduli space of stable parabolic local systems with conjugacy classes of monodromies around marked points fixed, which we denote by $\EuScript{M}_{\text{loc}} \left(G_\BC,\EC;S; (U_i)_{i=1} ^n \right)$.

In fact, in generic complex structure $I_w$, $w \neq 0,\infty$, the Hitchin moduli space with ramifications admits a description as the moduli space of parabolic local systems. The Hitchin equations imply the complex-valued connection $\CalA = (A_z + w^{-1} \phi_z )dz + (A_{\bar{z}} -w \phi_{\bar{z}})d\bar{z}$ is flat. Thus the hyper-K\"{a}hler quotient is almost isomorphic to the previous case ($w = -i$), leading to its description as the moduli space of stable parabolic local systems, except that the monodromy at each marked point $p_i \in S$ are now conjugate to $U_i = \exp\left( -2 \pi \left( \a_i + \left( \frac{\g(w-w^{-1})}{2} + \frac{i\b (w+ w^{-1})}{2} \right) \right) \right)$. With this subtlety understood, we will still denote this space by $\EuScript{M}_{\text{loc}} \left(G_\BC,\EC;S; (U_i)_{i=1} ^n \right)$.

\subsection{Hitchin fibrations, Parabolic bundles, and opers}
\subsubsection{Complete integrability and Hitchin fibration}
Let us restrict to the case $\mathfrak{g} = A_{N-1}$ and $G = PSU(N)$. As a complex manifold in $I$, we view the Hitchin moduli space with ramifications as the moduli space of parabolic Higgs bundles $\EuScript{M}_{\text{Higgs}} (PGL(N),\EC;S)$. The Hitchin fibration is the projection
\begin{align} \label{eq:hf}
\begin{split}
    p: \EuScript{M}_{\text{Higgs}} \left(PGL(N),\EC;S\right) &\longrightarrow \mathcal{B} = \bigoplus_{k=2} ^N H^0 (\EuScript{C} , K_{\EuScript{C}} ^{\otimes k} \otimes \CalO (k S) )\\
    (E,\varphi)\quad &\longmapsto \quad \left( \text{Tr}\,\varphi^k \right)_{k=2} ^N,
\end{split}
\end{align}
which is holomorphic Lagrangian in the complex structure $I$. More precisely, the coefficients of the $k$-th order poles of $\text{Tr}\, \varphi^k$ at the marked points are fully determined by the eigenvalues of the residues $\s_i$ of the Higgs field $\varphi$, so that they are just numbers fixed by the initial data $(\s_i)_{i=1} ^n$. Thus we regard the space $\CalB$ to be spanned by the remaining Laurent coefficients only. The space $\mathcal{B}$ is called the Hitchin base, and the holomorphic functions on $\mathcal{B}$ are called the \textit{classical} Hamiltonians. The fibers of the projection are abelian varieties at generic points on the Hitchin base. This is called the Hitchin fibration that endows the moduli space of parabolic Higgs bundles $\EuScript{M}_\text{Higgs} (G_\BC,\EC;S)$ with structure of an algebraic integrable system. This classical integrable system is called the Hitchin integrable system.

The spectral curve of the Hitchin integrable system is given by a degree $N$ polynomial equation valued in $T^* \EC$,
\begin{align}
    \S_{\mathbf{E}^{(0)}} : \quad \{(z,\l) \; \vert \; 0= \det(\l - \varphi (z))  \}  \subset T^* \EC,
\end{align}
defined by the characteristic polynomial of the Higgs field $\varphi$ evaluated at a point $\mathbf{E}^{(0)} \in \mathcal{B}$ on the Hitchin base. By the coefficients of $\l^{N-k}$, $k=2,3,\cdots, N$, the characteristic polynomial generates all the independent trace invariants $\text{Tr}\,\varphi^k$, namely, the classical Hamiltonians. We can think of fixing a point $\mathbf{E}^{(0)}$ on the Hitchin base as assigning \textit{energies} to the classical Hamiltonians. Conversely, the Hitchin base $\mathcal{B}$ can be thought of as the space of spectral curves. 

The fiber of the Hitchin fibration \eqref{eq:hf} is the Prym variety of the projection $\Sigma_{\mathbf{E} ^{(0)} } \to \EC$, namely, the kernel of the map $J(\Sigma_{\mathbf{E} ^{(0)} }) \to J(\EC)$. In the case we mainly consider $\EC=\BP^1$, $J(\EC)$ is trivial so that the Hitchin fiber is just the Jacobian of the spectral curve.

\subsubsection{Second fibration and parabolic bundles}
As the moduli space of parabolic Higgs bundles, the ramified Hitchin moduli space is endowed with another fibration called the Hitchin's second fibration. Consider the forgetful map holomorphic in $I$,
\begin{align} \label{eq:h0}
\begin{split}
    \pi_0:  \EuScript{M}_{\text{Higgs}} \left(G_\BC,\EC;S \right) &\longrightarrow \text{Bun}  _{G_\BC} (\EuScript{C};S) \\
    (E,\varphi) \quad&\longmapsto \quad E,
\end{split}
\end{align}
where $\text{Bun} _{G_\BC} (\EuScript{C};S)$ is the moduli space of stable parabolic $G_\BC$-bundles on $\EuScript{C}$ (it is defined on an open dense subset because $(E,\varphi)$ may be a stable Higgs pair even when $E$ is not a stable parabolic $G_\BC$-bundle). The fiber of this map is a linear space parametrized by the Higgs field $\varphi \in H^0 (\EC, \text{ad}(E) \otimes K_\EC (S) )$, but in generic cases we consider ($\b_i, \g_i \neq 0$ and generic for all $p_i \in S$) the polar part of $\varphi$ is not nilpotent and hence $\varphi$ does not represent a cotangent vector to $\text{Bun}_{G_\BC} (\EC;S)$. Instead, the difference between two Higgs fields does represent a cotangent vector. This implies the space $\EuScript{M}_{\text{Higgs}}$ contains an open dense subset that is an affine deformation of the cotangent bundle $T^* \text{Bun} _{G_\BC} (\EuScript{C};S)$.

In fact, in generic complex structure $I_w$, $w\neq 0, \infty$, the Hitchin moduli space with ramifications also admits a holomorphic Lagrangian projection from an open dense subset to the moduli space of stable parabolic $G_\BC$-bundles on $\EC$,
\begin{align} \label{eq:secfibgen}
\begin{split} 
    \pi_w : \EuScript{M}_{\text{loc}} (G_\BC,\EC;S) &\longrightarrow \text{Bun}  _{G_\BC} (\EC;S) \\
    \CalA ^{(w)} &\longmapsto \bar{\p} := \p_{\bar{z}}+ \CalA_{\bar{z}} ^{(w)},
\end{split}
\end{align}
where we endow the rank $N$ flat bundle with a holomorphic structure by the $\CalA_{\bar{z}} ^{(w)} = A_{\bar{z}} - w \phi_{\bar{z}}$ part of the flat connection. The fiber of this map at a parabolic $G_\BC$-bundle $E$ is the space of parabolic $\l$-connections on $E$ (with $\l=w$), which are the linear maps $\nabla  \in \text{End}(E) \otimes K_\EC (S)$ commuting with the $\bar{\p}$-operator of $E$, obeying $\nabla  (f\cdot s) = f \nabla  s + \l \,\p f \wedge s$ for any function $f$ and section $s$ of $E$, and preserving the parabolic structures of $E$ at all the marked points $S$ \cite{DA}. A parabolic $\l$-connection itself is not a cotangent vector at a given parabolic bundle, but the difference between two parabolic $\l$-connections is. Hence, it also follows that the open dense subset of $\EuScript{M}_{\text{loc}}$ is isomorphic to an affine deformation of the cotangent bundle $T^* \text{Bun}_{G_\BC} (\EC;S)$. Note that we recover the previous projection $\pi_0$ from parabolic Higgs bundles to parabolic $G_\BC$-bundles in the limit $w \to 0$, where a $\l$-connection is just a Higgs field $\varphi$.

On the \textit{automorphic} side of the geometric Langlands correspondence, we encounter twisted $\ED$-modules on $\text{Bun}_{G_\BC} (\EC;S)$. In this work, we will always consider the case $G_\BC = PGL(N)$. Since $G_\BC=PGL(N)$ is not simply-connected, the moduli space of stable parabolic $PGL(N)$-bundles is a disjoint union of connected components labelled by the characteristic class valued in $\pi_1(PGL(N))= \BZ_N$, measuring the obstruction to lifting to a parabolic $SL(N)$-bundle. For $d \in \BZ_N$, we denote the corresponding connected component by $\text{Bun}_{G_\BC} (\EC;S)_d$.

The twisted $\ED$-modules are sheaves of modules over the sheaf of rings of differential operators twisted by (complex powers of) line bundles over $\text{Bun}_{G_\BC} (\EC;S)$. Thus we need to know how to classify the line bundles over $\text{Bun}_{G_\BC} (\EC;S)$. Let us restrict to the connected component $\text{Bun}_{G_\BC} (\EC;S)_0$ where parabolic $PGL(N)$-bundles uplift to parabolic $SL(N)$-bundles. The theorem  \cite{LS} states that
\begin{align} \label{eq:pic}
    \text{Pic} \left( \text{Bun}_{G_\BC} (\EC;S)_0  \right)= \BZ  \oplus  \bigoplus_{i=1} ^n \L_{wt,\mathbb{L}_i},
\end{align}
where $\BZ$ is generated by the determinant line bundle $\EL$ and $\L_{wt,\mathbb{L}_i}$ is the sublattice of the weight lattice $\L_{wt}$ of $G$ that is invariant under the Weyl group of the $i$-th Levi subgroup $\mathbb{L}_i$, generated by the tautological line bundles over the flag variety $G_\BC/\CalP_i$.\footnote{To be precise, we need to consider the moduli \textit{stack} of parabolic $G_\BC$-bundles to correctly account for the determinant line bundle $\EuScript{L}$. We will not discuss the determinant line bundle part of the Picard group, and only consider the line bundles over the stable subset.}

\subsubsection{Opers}
We just have seen that a parabolic local system on $\EC$ can be viewed as a parabolic $G_\BC$-bundle $E \to \EC$ with a meromorphic flat connection $\nabla$. Let us first consider the case $G_\BC = PGL(N)$. We call a parabolic local system an \textit{oper} (with regular singularities at $S \subset\EC$) if $E$, viewed as the projectivization of a parabolic vector bundle of rank $N$, admits a filtration $0 = E_0 \subset E_1\subset  \cdots \subset E_N = E$ of subbundles satisfying \cite{BD2}
\begin{itemize}
    \item $\nabla E_i \subset E_{i+1} \otimes K_{\EC}(S)$ 
    \item $\nabla:E_i/E_{i-1} \to E_{i+1}/E_i \otimes K_\EC (S)$ is an isomorphism for each $i=1,2,\cdots, N-1$.
\end{itemize}
We denote the subspace of $\EuScript{M}_{\text{loc}} (G_\BC,\EC;S)$ spanned by the opers with regular singularities at $S$ by $\text{Op}_{G_\BC} (\EC;S)$. Since the holomorphic structure $\bar{\p} = \p_{\bar{z}} +\CalA_{\bar{z}} ^{(w)}$ is fixed up to gauge transformations, the complex symplectic structure
\begin{align}
    \O_{I_w} = \frac{i}{2\pi} \int_\EC  d^2 z  \; \text{Tr}\, \d\CalA_{\bar{z}} ^{(w)} \wedge \d \CalA_{{z}} ^{(w)}
\end{align}
vanishes on $\text{Op}_{G_\BC} (\EC;S)$. It also turns out that $\dim \text{Op}_{G_\BC} (\EC;S)=\frac{1}{2} \dim \EuScript{M}_{\text{loc}} (G_\BC,\EC;S)$. This implies the space $\text{Op}_{G_\BC} (\EC;S)$ of opers is a complex Lagrangian submanifold in  $\EuScript{M}_{\text{loc}} (G_\BC,\EC;S)$. The oper condition uniquely determines the underlying parabolic $G_\BC$-bundle $E$ up to its connected component. The space $\text{Op}_{G_\BC} (\EC;S)$ of opers is precisely the preimage of the projection \eqref{eq:secfibgen} at the oper bundle $E$.

On the \textit{Galois} side of the geometric Langlands correspondence, in fact, we will have the space of opers defined by the Langlands dual group ${}^L G_\BC = SL(N)$. Thus we consider the moduli space $\EuScript{M}_{\text{loc}} ({}^L  G_\BC,\EC;S)$ of stable parabolic local systems defined by flat $SL(N)$-bundles. This implies the local systems, including the opers $\text{Op}_{ {}^L G_{\BC}} (\EC;S) \subset \EuScript{M}_{\text{loc}} ({}^L  G_\BC,\EC;S)$, are rank $N$ parabolic vector bundles equipped with a trivialization of the determinant line bundle $\det E$.

\subsection{Quantization by branes and geometric Langlands correspondence} \label{subsec:sigmamodel}
The quantization of the ring of holomorphic functions on the Hitchin moduli space can be implemented by topological sigma model of maps from the two-dimensional worldsheet $\S$ to the Hitchin moduli space $\EuScript{M}_H$ \cite{Kapustin:2006pk,gukwit2,Gukov:2008ve}. Since the Hitchin moduli space is hyper-K\"{a}hler, a two-dimensional sigma model with target $\EuScript{M}_H$ has $\EN=(4,4)$ supersymmetry. A topological sigma model can be obtained by the standard twisting procedure \cite{Witten:1991zz} once a $\EN=(2,2)$ subalgebra is picked. Picking a $\EN=(2,2)$ subalgebra amounts to picking a pair $(w_+,w_-)$ out of $\BP^1$-worth of complex structures \eqref{eq:compstr} in $\EuScript{M}_H$, for which the corresponding topological sigma model is defined on the maps holomorphic in $I_{w_+}$ and antiholomorphic in $I_{w_-}$. Among these $\BP^1 \times \BP^1$ choices, a particular $\BP^1$ subset $(w_+,w_-)=(-t,t^{-1})$, $t\in \BP^1$, is distinguished because the corresponding topological sigma models descend from four-dimensional gauge theories with special twists: the GL-twisted $\EN=4$ theory \cite{Kapustin:2006pk} and the Donaldson-Witten twisted $\EN=2$ theory subject to the $\O$-background \cite{Nekrasov:2010ka}.\footnote{An alternative approach to the geometric Langlands correspondence is to view this $\BP^1$-family as deformations of the $B$-model in the symplectic structure $\o_I$, which descends from the holomorphic-topological twist \cite{Kapustin:2006hi} in the $\EN=4$ theory side and the Donaldson-Witten twist without the $\O$-background in the $\EN=2$ theory side. We thank Philsang Yoo for his explanation on this approach.}

For $t=\pm i$, we have $w_+ = w_- = \mp i$. The associated topological sigma model is the $B$-model in the complex structure $\pm J$. For all the other values $t \neq \pm i$, the topological sigma model is an $A$-model in some symplectic structure $\o_t$ with B-field $B_t$, which are given by 
some $t$-dependent linear combinations of $\o_I$ and $\o_K$. This $A$-model only depends on the cohomology class of $B_t + i \o_t$, which is $-[\o_I]$ times the \textit{canonical parameter} $\k \in \BP^1$ given by
\begin{align} \label{eq:tkrel}
    \k = \text{Re}\,\t + i\, \text{Im}\,\t \,\frac{t-t^{-1}}{t+t^{-1}} = -\frac{\ve_2}{\ve_1}. 
\end{align}
In the $\EN=4$ gauge theory perspective, $\t$ is the complexified gauge coupling \cite{Kapustin:2006pk}. In the $\EN=2$ gauge theory perspective, it is the modulus of the torus fibered over $\S$, where the total space of the fibration is the worldvolume of the four-dimensional gauge theory \cite{Nekrasov:2010ka}.\footnote{To be precise, the sigma model description of \cite{Nekrasov:2010ka} is obtained by setting $\k = \t$ (or $\bar{\t}$), which yields $t=\infty$ (or $0$) by \eqref{eq:tkrel}. This precisely gives the $A$-model in the symplectic structure $- \o_I$ (or $+ \o_I$). In this work, rather, we vary $\k$ with $\t$ fixed, to land on different topological sigma models (namely, different values of $t$) by \eqref{eq:tkrel}.} We mainly work in the $\EN=2$ gauge theory setup, where the canonical parameter $\k$ is determined by the ratio of the $\O$-background parameters $\ve_1$ and $\ve_2$ (see \eqref{eq:kzgauge}). We regard $\t$ as a fixed parameter so that the above equation determines $t$, or equivalently the symplectic structure and the B-field that the $A$-model is defined by, for a given $\k = -\frac{\ve_2}{\ve_1}$. In fact, we will always take $\text{Re}\, \t =0$, in which $\k = 0$ (i.e.  $\ve_2 = 0$) gives $t =\pm 1$, namely, the $A$-model in $(\text{Im}\,\t) \o_K$ with a B-field given only by the two-dimensional $\th$-angles $\eta_i \in {}^L \mathbb{T}$.

\subsubsection{Canonical coisotropic brane and brane of opers} \label{subsubsec:canoper}
The $A$-model on the Hitchin moduli space $\EuScript{M}_H$, viewed as a symplectic manifold in $\o_t$, may admit branes supported on coisotropic submanifolds \cite{Kapustin:2001ij}. A distinguished coisotropic brane is the canonical coisotropic brane $\EuScript{B}_{cc}$ supported on the whole moduli space $\EuScript{M}_H$ with a rank 1 unitary Chan-Paton bundle. Let $F_t$ be the curvature of the Chan-Paton line bundle. The condition for defining an $A$-brane is $(\o_t ^{-1} (F_t+B_t))^2=-1$, where $B_t$ is the B-field.

For any generic $t \neq \pm i$, the solution $F_t$ is always found. In fact, it can be shown that $\EB_{cc}$ is also a $B$-brane with respect to the complex structure in which $F_t+ B_t$ is an $(1,1)$-form, and finally an $A$-brane with respect to still another symplectic structure (associated to $w_t \equiv \frac{1-t}{1+t}$ \eqref{eq:compstr}) by the hyper-K\"{a}hlerity of $\EM_H$. With the relation \eqref{eq:tkrel} understood, we refer to such a brane as of $(A,B,A)_\k$ type. Thus, the canonical coisotropic brane $\EB_{cc}$ is a brane of $(A,B,A)_\k$ type. Note that when $\k = 0$, $\EB_{cc}$ becomes an $(A,B,A)$-brane compatible with the $A$-models in the symplectic structures $\o_I$ and $\o_K$ and the $B$-model in the complex structure $J$.

The canonical coisotropic brane $\EB_{cc}$ is a space-filling brane in $\EM_H$ which looks like an affine deformation of the cotangent bundle over $\text{Bun}_{G_\BC} (\EC;S)$ in the complex structure $I_w$. Hence, if we forget about the ring structure, the sheaf of $(\EB_{cc} ,\EB_{cc})$ strings is the sheaf of holomorphic functions on this affine bundle. The multiplication defined by joining strings promotes the sheaf of $(\EB_{cc} ,\EB_{cc})$ strings to the sheaf of rings of differential operators acting on sections of a line bundle $\mathfrak{L}_\k \to \text{Bun}_{G_\BC} (\EC;S)$ \cite{Kapustin:2006pk, gukwit2}, which we denote by $\ED_{\mathfrak{L}_\k}$. Note that such a sheaf is well-defined even when the twisting is a complex power of a line bundle. In fact, the sheaf $\ED_{\fL_\k}$ of differential operators is defined upon a twist of this kind, given by
\begin{align} \label{eq:twist}
    \mathfrak{L}_\k = \EuScript{L}^{ \k-h^\vee} \otimes \bigotimes _{i=1} ^n \CalL_i ^{ -(\eta_i + i (\text{Im}\,\t) \g_i ^* )},
\end{align}
where $\EuScript{L}$ is the determinant line bundle and $\CalL_i ^{g v_i} = (\CalL_i^{v_i})^g$ with $v_i \in \L_{wt,\mathbb{L}_i}$ and $g\in \BC$ denotes the $g$-power of the line bundle whose first Chern class is $v_i$. $h^\vee$ is the dual Coxeter number of $G$ ($h^\vee = N$ for $G=PSU(N)$). Note that when $\k=0$, the determinant line bundle part of the twisting becomes $\EuScript{L}^{-h^\vee} \simeq K_{\text{Bun} _{G_\BC} (\EC;S)} ^{\frac{1}{2}}$, the square-root of the canonical line bundle over $\text{Bun} _{G_\BC} (\EC;S)$. All the other parts of the twist are independent of $\k$. For simplicity, let us denote $\fL_{\k=0} = \fL$.

The quantization of the Hitchin integrable system can be understood as a consequence of the mirror symmetry applied to the topological sigma models with branes (see table \ref{tab:sigmabranes}). The mirror symmetry replaces the group $G$ to its Langlands dual group $G \to {}^L G$ and maps the canonical parameter by $\k \to {}^L\k = -\frac{1}{\k}$.\footnote{Here, we only consider simply-laced cases. For non-simply-laced cases, $\k \to -\frac{1}{n_{\mathfrak{g}} \k}$ with $n_{\mathfrak{g}} = 2\text{ or }3$. See \cite{Kapustin:2006pk,gukwit2,Frenkel:2018dej} for example.} In particular, the $A$-model with respect to the symplectic structure $\o_K$ defined on the target $\EM_H (G,\EC;S)$ ($\k=0$) is mirror dual to the $B$-model with respect to the complex structure $J$ defined on the target $\EM_H ({}^L G ,\EC;S)$ (${}^L \k =\infty$). The canonical coisotropic $(A,B,A)$-brane $\EB_{cc}$ in the former dualizes to the \textit{brane of opers} ${}^L \EB_{op}$, which is an $(A,B,A)$-brane supported on the oper submanifold $\text{Op}_{{}^L G_\BC} (\EC;S) \subset \EM_H ({}^L G ,\EC;S)$ with trivial Chan-Paton line bundle, in the latter \cite{BD1,BD2,Nekrasov:2010ka, Gaiotto:2011nm}. Thus, the mirror symmetry between the $(\EB_{cc},\EB_{cc})$ strings in the $A$-model and the $({}^L \EB_{op},{}^L \EB_{op})$ strings in the $B$-model yields the isomorphism between the sheaf $\ED_{\mathfrak{L}}$ of rings of twisted differential operators on $\text{Bun}_{G_\BC} (\EC;S)$ and the sheaf of holomorphic functions on $\text{Op}_{{}^L G_\BC} (\EC;S)$. Accordingly, we establish the equivalence between the global sections of the two sheaves:
\begin{align}
D_\fL \equiv \G(\text{Bun}_{G_\BC} (\EC;S),\ED_{\mathfrak{L}}) \simeq \text{Fun}\, \text{Op}_{{}^L G_\BC} (\EC;S).
\end{align}
In particular, the ring $D_\fL$ of global sections of $\ED_\fL$ is commutative. To be precise, the marked point data are acted on by the mirror symmetry, and thus mapped under this isomorphism in a nontrivial manner. It turns out that the map is given by $\eta_i = -{}^L \a_i$ and $(\text{Im}\,\t) \g_i ^* = {}^L \g_i$ \cite{Frenkel:2006nm, gukwit2}. Thus, the complex exponents $-(\eta_i +i (\text{Im}\,\t)\g_i ^*)$ of the line bundles in the twisting $\mathfrak{L}$ are identified with the eigenvalues ${}^L \a_i - i {}^L \g_i $ of the monodromies of the ${}^L G_{\BC}$-oper at the marked points $p_i \in S$.

An oper $\r \in \text{Op}_{{}^L G_\BC} (\EC;S)$ is by definition a homomorphism $\r:\text{Fun}\, \text{Op}_{{}^L G_\BC} (\EC;S) \to \BC$. In turn, the above isomorphism induces a homomorphism $\tilde\r : D_{\mathfrak{L}} \to \BC$. We assign a left $\ED_\fL$-module $\D_\r$ by the quotient
\begin{align} \label{eq:heceigsh}
\D_\r = \ED_{\fL} / \text{ker}\,\tilde\r \cdot \ED_\fL.
\end{align}
Note that $\dim D_\fL = \dim \text{Op}_{{}^L G_\BC} (\EC;S) = \frac{1}{2} \dim \EM_H (G,\EC;S) $. Let us denote $\frac{1}{2} \dim \EM_H (G,\EC;S)$ generators of $D_\fL$ by $\hat{H}_i$, and let us denote $\tilde{\r} (\hat{H}_i) = E_i$. The $\ED_\fL$-module $\D_\r$ represents a system of differential equations
\begin{align} \label{eq:schrodinger}
0 = \left(\hat{H}_k - E_k \right)\psi, \quad\quad k=1,2,\cdots, \frac{1}{2} \dim \EM_H (G,\EC;S),
\end{align}
which are the spectral equations for the quantum integrable system. Namely, the $\ED_\fL$-module $\D_\r$ provides the common eigenfunctions of the mutually commuting quantum Hamiltonians $\hat{H}_k$ with the eigenvalues $E_k = \tilde{\r} (\hat{H}_k)  $.

\subsubsection{Brane of $\l$-connections and $\ED$-module of $\d$-functions}
We have seen the $(\EB_{cc},\EB_{cc})$ strings in the $A$-model form the sheaf $\ED_{\mathfrak{L}_\k}$ of rings of twisted differential operators on $\text{Bun}_{G_\BC} (\EC;S)$. For any $A$-brane $\EB'$, the $(\EB_{cc} ,\EB')$ strings naturally provide a sheaf of (left) modules over $\ED_{\mathfrak{L}_\k}$, namely, a twisted $\ED$-module on $\text{Bun}_{G_\BC} (\EC;S)$, since the $(\EB_{cc},\EB_{cc})$ strings act from the left by joining of the strings. Accordingly, there is a correspondence between the $A$-branes and the twisted $\ED$-modules on $\text{Bun}_{G_\BC} (\EC;S)$. As an immediate example, take the canonical coisotropic $A$-brane $\EB' = \EB_{cc}$ itself. Then the sheaf $\ED_{\mathfrak{L}_\k}$ of rings is a sheaf of modules over itself, tautologically.

A distinguished class of $A$-branes is the one supported on the fiber of the projection $\pi_{w_t}: \EM_{\text{loc}} (G_\BC,\EC;S) \to \text{Bun}_{G_\BC} (\EC;S)$ \eqref{eq:secfibgen} at a given parabolic $G_\BC$-bundle $E \in \text{Bun}_{G_\BC} (\EC;S)$. We observed that the fiber at $E$ is the affine space of parabolic $\l$-connections on $E$. Note also that the fiber is complex Lagrangian with respect to $\O_{I_{w_t}}$. Hence it indeed supports an $(B,A,A)_\k$-brane with trivial Chan-Paton line bundle. We denote this $(B,A,A)_\k$-brane by $\EuScript{F}' _E$, and call it the \textit{brane of $\l$-connections}. Note that when $\k = 0$, $I_{w_t} = \pm I$ so that $\EuScript{F}'_E$ becomes a Lagrangian $(B,A,A)$-brane supported on the preimage of $E$ with respect to the $I$-holomorphic map $\pi_0 : \EM_{\text{Higgs}} (G_\BC,\EC;S) \to \text{Bun}_{G_\BC} (\EC;S)$ \eqref{eq:secfibgen} from the parabolic Higgs bundles to the parabolic $G_\BC$-bundles (see table \ref{tab:sigmabranes}).
 
It is expected that the twisted $\ED$-module corresponding to the brane of $\l$-connections $\EuScript{F}' _E$ (namely, $(\EB_{cc}, \EF' _E)$-strings) is the sheaf of $\d$-functions supported at $E$ \cite{Ben-Zvi:2004, Frenkel:2010, Frenkel:2018dej}. The sections of this sheaf are the $E$-twisted coinvariants of the $\widehat{\mathfrak{g}}$-modules associated to the parabolic structures at the marked points $S\subset \EC$ at level $k = \k - h^\vee$ (see section \ref{subsec:slN}). This is precisely the fiber of the sheaf $\ED_{\fL_\k}$ at $E$ \cite{Ben-Zvi:2004}.

At $\k=0$ (the \textit{critical} level $k = -h^\vee$), there is a large center $Z (\widehat{\mathfrak{g}}) \subset \tilde{U}_{-h^\vee} (\widehat{\mathfrak{g}})$ of the completed enveloping algebra of $\widehat{\mathfrak{g}}$. It can be shown that a ${}^L G_\BC$-oper $\r \in \text{Op}_{{}^L G_\BC} (\EC;S)$ restricted to the disk around each marked point $p \in S$ induces a homomorphism $Z(\widehat{\mathfrak{g}}) \to \BC$ \cite{Frenkel:2002fw,Frenkel:2004qy}. Take the quotient of the $\widehat{\mathfrak{g}}$-modules by these central characters. The space of $E$-twisted coinvariants of these quotient $\widehat{\mathfrak{g}}$-modules is identified with the fiber of the $\ED_{\fL}$-module $\D_\r$ \eqref{eq:heceigsh}, associated to the oper $\r \in \text{Op}_{{}^L G_\BC} (\EC;S)$, at $E \in \text{Bun}_{G_\BC} (\EC;S)$.

\subsubsection{Hecke operators and Hecke eigensheaves} \label{subsubsec:heckeop}
The Hecke correspondence $\EH$ is the space of quadruples $(E,E',y,\iota) $, where $E, E' \in \text{Bun}_{G_\BC} (\EC;S)$, $y\in \EC \setminus S$, and $\iota$ is an isomorphism between $E$ and $E'$ restricted to $\EC \setminus \{y\}$. We define the projections $h^\leftarrow : \EH \to \text{Bun}_{G_\BC} (\EC;S)$ and $h^\rightarrow : \EH \to \EC \times \text{Bun}_{G_\BC} (\EC;S)$ by $h^\leftarrow (E,E',y,\iota) = E$ and $h^\rightarrow (E,E',y,\iota) = (y,E')$.

The fiber of $h^\rightarrow$ over $(y,E')$ is the space of all possible pairs $(E,\iota)$, namely, the parabolic $G_\BC$-bundles $E$ that are isomorphic to $E'$ away from $y \in \EC$. Such a parabolic $G_\BC$-bundle is called a Hecke modification of $E'$ at $y \in \EC$. To classify Hecke modifications, we may cover $\EC$ by $\EC\setminus\{y\}$ and a small neighborhood of $y$ and assign a transition function on their intersection. With a local coordinate $t$ in the neighborhood of $y$, a $G_{\BC}$-valued holomorphic function on the intersection modulo $G_\BC$-valued gauge transformations (right action) on the neighborhood of $y$ defines a transition function for the image $E$ of a Hecke modification at $y$. Thus, the space of Hecke modifications is isomorphic to the affine Grassmannian $\text{Gr}_{G_\BC} = G_\BC (\!(t)\!) / G_\BC [[t]]$.

Let us be given with an irreducible representation $V_\l$ of ${}^L G$ (of ${}^L  G_\BC$ by complexification) associated to a dominant integral coweight $\l \in \text{Hom}(U(1),\mathbb{T})$ of $G$. Then $\l$ analytically continues to give an orbit of the left action of $G_\BC [[t]]$ in $\text{Gr}_{G_\BC}$. Let $\text{IC}_\l$ be the intersection cohomology of this $G_\BC[[t]]$-orbit, extended to the whole $\text{Gr}_{G_\BC}$ by zero away from itself.\footnote{It can be thought of as the constant sheaf on this $G_\BC[[t]]$-orbit if the orbit is compact. This is the case if and only if the coweight $\l$ is \textit{minuscule}. When the $G_\BC[[t]]$-orbit is non-compact, we need to consider its closure where additional complication is required to properly define $\text{IC}_\l$ (see \cite{Frenkel:2005pa, Kapustin:2006pk}). In this work, we will only consider minuscule representations.} The Hecke operator $H_\l$ associated to $V_\l$ is defined by
\begin{align} \label{eq:hecke}
H_{\l} (\EF) = h^\rightarrow _* \left( h^{\leftarrow*} (\EF) \otimes \text{IC}_\l \right),
\end{align}
for a (twisted) $\ED$-module $\EF$ on $\text{Bun}_{G_\BC} (\EC;S)$.

Let $E \in \EM_{\text{loc}} ({}^L G_\BC ,\EC;S)$ be a ${}^L G_\BC $-local system on $\EC$, realized by a flat ${}^L G_\BC$-bundle. Then the associated bundle $V_\l ^E \equiv E \times _{{}^L G_\BC} V_\l$ also defines a local system on $\EC$. A (twisted) $\ED$-module $\EuScript{F}$ is called a Hecke eigensheaf if there is an isomorphism 
\begin{align} \label{eq:heckeeigen}
H_{\l}(\EuScript{F}) \xrightarrow{\sim} V_\l ^E  
\boxtimes \EuScript{F},
\end{align}
for each dominant integral coweight $\l$ of $G$. In this sense, the local system $E$ is the \textit{eigenvalue} of the Hecke operator for the Hecke eigensheaf $\EF$.

In this work, we will study the case of $G_\BC = PGL(N)$ (i.e., ${}^L G _\BC = SL(N)$). Further, we mainly restrict our attention to the simplest dominant integral (minuscule) coweight of $G$ corresponding to the $N$-dimensional representation of ${}^L G_\BC = SL(N)$. The corresponding $G_\BC [[t]]$-orbit, namely, the space of Hecke modifications, is the projective space $\BP^{N-1}$. The associated Hecke operator is defined by an integral on this space.

The geometric Langlands correspondence states that for a given ${}^L G_\BC$-local system $E$ there is a Hecke eigensheaf on $\text{Bun}_{G_\BC} (\EC;S)$ with the eigenvalue $E$. When restricted to the local system represented by an oper $\r \in \text{Op}_{{}^L G_\BC} (\EC;S)$, the corresponding Hecke eigensheaf was conjectured to be the $\ED_\fL$-module $\D_\r$ \eqref{eq:heceigsh} obtained by the quotient by the ideal generated by the character of the oper \cite{BD1}.

\begin{table}[h!]
    \centering \scalebox{0.78}{
    \begin{tabular}{c|c|c}
        $B$-model in $J$  & $A$-model in $\o_K$ & 4d $\EN=2$ theory origin \\ 
         \hline   $\EM_H({}^L G,\EC;S)$ & $\EM_H(G,\EC;S)$ & \makecell{Coulomb moduli space\\after compactifying along $T^2$-fiber }  \\
         \hline  ${}^L \k = \infty$ & $\k=0$ & $\ve_2=0$ ($\k=-\frac{\ve_2}{\ve_1} =0$) \\
        \hline  Brane of ${}^L G_\BC$-opers ${}^L\EB_{op}$ & Canonical coisotropic brane $\EB_{cc}$ & New boundary at the origin of $\BR^2 _{\ve_1}$  \\ 
           Canonical coisotropic brane ${}^L\EB_{cc}$ & Brane of $G_\BC$-opers $\EB_{op}$ & New boundary at the origin of $\BR^2 _{\ve_2}$ \\
           - & Brane of $\l$-connections $\EuScript{F}_E '$ & Monodromy surface defect \\ 
         \makecell{$B$-brane $\mathbf{B}_{\r_\ba}$ supported at\\an oper $\r_\ba\in \text{Op}_{{}^L G_\BC}(\EC;S)$} & \makecell{Lagrangian $A$-brane $\mathbf{F}_\ba$\\supported on a Hitchin fiber\\(a Hecke eigensheaf on $\text{Bun}_{G_\BC} (\EC;S)$)} & \makecell{Boundary condition at infinity\\specified by Coulomb moduli $\ba$} \\
         \hline \makecell{Wilson line operator at $y \in \EC\setminus S$\\(carrying fundamental weight of ${}^L G_\BC$)} & \makecell{Hecke operator at $y\in \EC\setminus S$\\(for fundamental coweight of $G_\BC$)} & \makecell{Canonical surface defect\\(w/ cpx. FI parameter $y\in \EC\setminus S$)}
    \end{tabular}}
    \caption{Mirror-dual topological sigma models, their branes, and the Hecke operator relevant to the geometric Langlands correspondence. The Wilson line operator in the $B$-model is the line defect that acts by tensoring any brane with the representation bundle at the point $y\in \EC \setminus S$. The third column displays the 4d $\EN=2$ gauge theory origin of these branes and the Hecke operator, as we establish throughout the work. The \textit{new boundaries} indicate those emerging under the compacitifcation along the $T^2$-fiber of $\BR^4 = \Sigma \tilde{\times} T^2$, where $\S = \BR^+ \times \BR^+$ is the worldsheet of the resulting topological sigma model.}
    \label{tab:sigmabranes}
\end{table}

\subsection{Genus-0 with marked points and Gaudin model}
The main example that we consider in the present work is the Riemann sphere $\EC= \BP^1$ with marked points $S$. The associated Hitchin integrable system is well-known to be the Gaudin model.

\subsubsection{The case of four marked points} \label{subsubsec:g0four}
We explicitly present the moduli spaces associated with our main example, the Riemann sphere $\EuScript{C} = \mathbb{P}^1$ with four marked points $S=\{0,\qe,1,\infty\} \subset \mathbb{P}^1$. As before, we will also restrict to the case of $\mathfrak{g}=A_{N-1}$ with $G = PSU(N)$ and $G_\BC= PGL(N,\BC)$ (namely, ${}^L G = SU(N)$ and ${}^L G_\BC = SL(N,\BC)$). We note here that the parabolic structures at four marked points are chosen in such a way that the associated four-dimensional $\EN=2$ theory of class $\CalS$ will be the $SU(N)$ gauge theory with $N$ fundamental and $N$ antifundamental hypermultiplets, in an appropriate weak coupling regime.

The parabolic structures at the marked points are chosen by specifying the parabolic subgroups $\mathcal{P}_{i} \subset PGL(N,\BC)$ for each $i=0,\qe,1,\infty$. We choose
\begin{align}
\begin{split}
&\CalP_0 = \begin{pmatrix}
* & * & *&  \cdots  &* &* \\ 0 & * & * & \cdots &* & * \\ 0 & 0 & * & \cdots & * & *  \\ \cdots & \cdots & \cdots & \cdots & \cdots & \cdots \\ 0 & 0 & 0 & \cdots & 0 & * 
\end{pmatrix}, \quad \CalP_\infty = \begin{pmatrix}
* & * & *&  \cdots  &* &* \\ 0 & * & * & \cdots &* & * \\ 0 & 0 & * & \cdots & * & *  \\ \cdots & \cdots & \cdots & \cdots & \cdots & \cdots \\ 0 & 0 & 0 & \cdots & 0 & * 
\end{pmatrix}, \\ 
& \CalP_\qe = \begin{pmatrix}
* & * &   \cdots & * &* \\ * & * &  \cdots & * & * \\ * & * &\cdots & *& * \\ \cdots & \cdots &  \cdots & \cdots & \cdots \\ * & * & \cdots & * & * \\ 0 & 0 & \cdots & 0 & *   
\end{pmatrix}, \quad \CalP_1 =  \begin{pmatrix}
* & * &   \cdots & * &* \\ * & * &  \cdots & * & * \\ * & * &\cdots & *& * \\ \cdots & \cdots &  \cdots & \cdots & \cdots \\ * & * & \cdots & * & * \\ 0 & 0 & \cdots & 0 & *   
\end{pmatrix}.
\end{split}\end{align}
The marked points at $0$ and $\infty$ are called \textit{maximal} in the sense that $PGL(N)/\CalP_{0,\infty}$ are isomorphic to the space $F(N)$ of complete flags in $\BC^N$ with dimension $\dim F(N) = \frac{N(N-1)}{2}$, while the marked points at $\qe$ and $1$ are called \textit{minimal} since $PGL(N)/\CalP_{\qe,1}$ are isomorphic to the projective space $\BP^{N-1}$ of lines in $\BC^N$ with dimension $\dim \BP^{N-1} = N-1$, which is the minimal flag variety. In the simplest case of $G_\BC = PGL(2)$, the two parabolic subgroups are equivalent and there is no distinction between maximal and minimal marked points, but for general $N \geq 2$ considered in the present work this is not the case.

At this point, we emphasize that the geometric Langlands correspondence actually involves twisted $\ED$-modules on the moduli \textit{stack} of all parabolic $G_\BC$-bundles, instead of on its stable subset. In the genus-0 case $\EC=\BP^1$, in particular, it is crucial to consider the moduli stack to account for the the determinant line bundle part of the twisting \eqref{eq:twist}, $\EL^{\k -h^\vee}$. In this work, we will restrict to a stable subset with this subtlety understood.

The connected component of the moduli space of stable parabolic $PGL(N)$-bundles containing the trivial bundle is isomorphic to $\left(\bigtimes_{i=0,\qe,1,\infty} G_\BC / \CalP_i \right) / SL(N,\BC)$, where $SL(N,\BC)$ acts diagonally.\footnote{All the other connected components can also be reached by Hecke modifications at marked points. See \cite{Etingof:2021eub} for instance.} As just discussed, each quotient $G_\BC /\CalP_i$ is isomorphic to a flag variety, giving
\begin{align} \label{eq:BunGsim}
    \text{Bun} _{PGL(N)} (\BP^1;S)_0 = \left( F(N) \times \BP^{N-1} \times \BP^{N-1}  \times F(N) \right) / SL(N).
\end{align}
A simple dimension count shows $\dim \text{Bun}  _{PGL(N)} (\BP^1;S)_0 = N-1$, which is indeed the half of $\dim \EM_H (PGL(N),\EC;S) = 2(N-1)$.

Finally, let us describe the Picard group of this moduli space.  Note that $\BL_0 = \BL_\infty = \BT$, while $\BL_\qe = \BL_1 = S\left(U(N-1) \times U(1) \right)$. The Weyl group of the former is trivial, so that the invariant sublattice is just the whole weight lattice of $SU(N)$, $\L_{wt,\BL_{0,\infty}} = \L_{wt}$. On the other hand, the Weyl group of the latter equals to the Weyl group of $SU(N-1)$. The invariant sublattice is one-dimensional, $\L_{wt,\BL_{\qe,1}} \simeq \BZ$. The Picard group of the moduli space is therefore given by, as a special case of \eqref{eq:pic},
\begin{align}
    \text{Pic}\left( \text{Bun}_{PGL(N)} (\BP^1 ;S)_0 \right) = \L_{wt} \oplus \BZ\, \oplus \BZ\, \oplus \L_{wt}.
\end{align}
Here, we remind that we are missing the determinant line bundle $\EL$ part of the Picard group since we only consider a stable subset of the stack. Each lattice in the direct sum represents the pullbacks of the line bundles over each flag variety on the right hand side of \eqref{eq:BunGsim} under the natural projection. Namely, it is generated by the pullbacks of the tautological line bundles over the respective flag varieties (in our case, the space of complete flags $F(N)$ and the projective space $\BP^{N-1}$).

\section{From branes to boundaries and defects} \label{sec:bbd}
The topological sigma model with the Hitchin moduli space target reviewed in the previous section can originate from two different four-dimensional gauge theories: GL-twisted $\EN=4$ gauge theory \cite{Kapustin:2006pk, gukwit2} and $\O$-deformed Donaldson-Witten twisted $\EN=2$ gauge theory \cite{Nekrasov:2010ka}.

In this section, we illustrate a duality connecting these two frameworks by embedding them to string theory. We will recall the IIB brane setup for the gauge origami \cite{Nikita:I,Nikita:II,Nikita:III}, from which our $\EN=2$ gauge theory is constructed, and show how it can be dualized to the twisted M-theory \cite{Costello:2016nkh} and further to the $(p,q)$-web of fivebranes in IIB \cite{Leung:1997tw} to which the GL-twisted $\EN=4$ gauge theory is naturally embedded. The brane picture turns out to provide useful intuition on how different BPS objects on two sides $-$ as we will see shortly, surface defects in the $\EN=2$ theory and codimension-one boundaries and line defects in the $\EN=4$ theory $-$ are mapped to each other. Such an identification will be consistent with the exact results that we will derive on the $\EN=2$ theory side.

\begin{table}[h!] 
    \centering
    \begin{tabular}{ c||c|c|c|c|c|c|c|c|c|c } 
        \text{IIB Branes} & 0 & 1 & 2 & 3 & 4 & 5 & 6 & 7 & 8 & 9  \\ \hline\hline
         D3 & \rm{x} & \rm{x}& \rm{x}& \rm{x} &  & & & &  &  \\  KK5${}_m$ & \rm{x} & \rm{x} & \rm{x} & \rm{x} & &  &  & & \rm{x} & \rm{x}   \\ \hline KK5${}_l$ & \rm{x} & \rm{x} & &  &\rm{x}& \rm{x} & & & \rm{x} &  \rm{x} \\ D3 & \rm{x} & \rm{x} & & & \rm{x}& \rm{x}  &&&
    \end{tabular}
    \caption{IIB brane configuration for gauge origami}
    \label{table:IIBori}
\end{table}

\begin{table}[h!]
    \centering
    \begin{tabular}{c|c|c|c|c}
        $\BC_{\ve_1}$ & $\BC_{\ve_2}$ & $\BC_{\ve_3}$ & $\BC _{\ve_4}$ & $\mathbb{R}^2$  \\ \hline
        $x^0,x^1$ & $x^2,x^3$ & $x^4, x^5$ & $x^6,x^7$ & $x^8,x^9$ 
    \end{tabular}
    \caption{Spacetime of the IIB theory for gauge origami}
    \label{table:spacetimeori}
\end{table}

\subsection{Gauge origami and twisted M-theory} \label{subsec:oritwM}
In the present work, the $\EN=2$ gauge theory and its half-BPS defects are constructed from the \textit{gauge origami} \cite{Nikita:I,Nikita:II,Nikita:III}. The gauge origami is a configuration of intersecting stacks of D3-branes in the IIB theory on the ten-dimensional spacetime $X \times \BR^2$, where $X$ is a local Calabi-Yau four-fold. There are at most six stacks of D3-branes occupying all possible complex two-cycles in $X$ preserving $U(1)^3 \subset SU(4)$ of the isometry, located at certain positions on the transverse plane $\BR^2$ (see the tables \ref{table:IIBori}, \ref{table:spacetimeori}). On top of them, we may introduce D($-1$)-instantons which can be dissolved into the worldvolume of the D3-branes. They are called \textit{spiked instantons}. Upon a proper twist, there is a supercharge in which the $X$ is topological and the $\BR^2$ is holomorphic under its cohomology.

Since $X$ is topological, we may implement the $\O$-background associated to its $U(1)^3 \subset SU(4)$ isometry. It can be geometrically implemented by using the T-duality. First, compactify $\BR^2$ to a torus $T^2$ and T-dualize twice along the cycles of the torus. Then the gauge origami is dualized to intersecting stacks of D5-branes, which also wrap the dualized torus $\check{T}^2$, while the spiked instantons become D1-branes wrapping $\check{T}^2$. Now regard the ten-dimensional spacetime $X\times \check{T}^2$ as not just a product space but a fibration of $X$ over $\check{T}^2$, where $X$ is rotated by the $U(1)^3 \subset SU(4)$ isometry along the two cycles of the torus. Such a nontrivial fibration admits a preserved supercharge, yielding three independent $\O$-background parameters $\sum_{i=1} ^4 \ve_i =0$. Finally, we T-dualize twice along the cycles of $\check{T}^2$ back to $T^2$, and take the limit where the size of the dual torus $\check{T}^2$ shrinks to zero. Then the original torus $T^2$ decompactifies back to $\BR^2$, establishing the $\O$-background for the gauge origami.

To engineer the $\EN=2$ gauge theory, we start by choosing our local Calabi-Yau four-fold to be an $A_{m-1}$-type singularity, $X = \BC^2 _{12} \times \BC^2 _{34} /\BZ_m$ (see the second line of table \ref{table:IIBori}). Then we insert a single stack of $N$ D3-branes at the singularity, $\BC^2 _{12} \times \{0\} \subset X$. The effective field theory on the worldvolume of the D3-branes is the $\EN=2$ supersymmetric $\hat{A}_{m-1}$-quiver gauge theory \cite{Douglas:1996sw}. We may turn off gauge couplings for two consecutive gauge nodes to reduce further to the $A_{m-2}$ linear quiver gauge theory. The $A_{m-1}$ singularity is the limit of the $m$-centered Taub-NUT space where all the centers are brought to the origin, which in turn can be viewed as the transverse geometry of the Kaluza-Klein monopoles in IIB. Note that the D3-branes (the first line of table \ref{table:IIBori}) are located at the center of the Taub-NUT geometry. Note also that, when the Taub-NUT space is viewed as a circle bundle over $\BR^3$, the $U(1)^2$ rotation on $\BC_{\ve_3}\times \BC_{\ve_4}$ continuously deforms to the rotations on the base $\BR^2 \subset \BR^3$ and the circle fiber.

The gauge origami is a powerful framework since it is designed to produce partition functions expressed as finite-dimensional equivariant integrals, which further can be exactly computed by equivariant localization. This indeed will be our methodology of studying the geometric Langlands correspondence in the $\EN=2$ gauge theory framework from section \ref{sec:sgo}. Our aim here is, on the other hand, to show the gauge origami is in fact coherent with a seemingly distinct context $-$ the twisted M-theory \cite{Costello:2016nkh} $-$ in which the relation with the GL-twisted $\EN=4$ theory is more manifest. From now on, we will illustrate how the gauge origami and the twisted M-theory are connected to each other by a string duality.

Since the M-theory is a uplift of the IIA theory, it is tempting to apply a T-duality to our IIB setup for the gauge origami. An obstruction to the simplest application of T-duality is the $\O$-background associated to the isometry of $X$; If we would keep the mostly refined $\Omega$-background with three independent parameters, all the two-planes have to be viewed as a cigar, namely, a circle fibration over a semi-infinite line to keep the $U(1)^3$ isometry. Then T-dualization can only be performed along these circle fibers, where we must carefully take account for the effect of the fibration at the tip of the cigar.

To apply the T-duality in a simple manner, we will unrefine our $\O$-background so that the $\O$-background parameter associated with the rotation of the circle fiber of the Taub-NUT space is turned off, leaving only two independent parameters. T-dualizing along the circle fiber, the $m$-centered Taub-NUT space is dualized to $m$ NS5-branes which are separated along the circle $x^4$ \cite{Tong:2002rq,Witten:2009xu}, while the $N$ D3-branes become $N$ D4-branes stretched between those NS5-branes along the $x^4$-direction (see the first two rows in table \ref{table:twM} and table \ref{table:spacetime}). We recognize this is precisely the well-known IIA brane setup realizing the very same $\EN=2$ gauge theory that we obtained in the IIB gauge origami setting, on the non-compact part of the worldvolume of the D4-branes \cite{Witten1997}. The only modification made here is subjecting it to the $\O$-background associated with the $U(1)^2 \subset SU(3) $ isometry of $\BC_{\ve_1} \times \BC_{\ve_2} \times \BC_{\ve_4}$, which can now be treated as a local Calabi-Yau threefold.

This IIA brane setup naturally uplifts to the M-theory with the M-theory circle $x^{10}$, in which both D4-branes and NS5-branes become M5-branes. The $m$ M5-branes from NS5-branes are local on the torus of $x^4$ and $x^{10}$. The positions of these M5-branes on the torus determine the complexified gauge couplings ($\th$-angles and gauge couplings) in the point of view of the $\EN=2$ gauge theory. We will call the locations of the M5-branes on $\EC$ \textit{marked points}. In the cohomology of the preserved supercharge (the $\Omega$-deformed Donaldson-Witten supercharge \cite{Nekrasov:2002qd} in the $\EN=2$ theory point of view), the dependence on the complexified gauge couplings is only holomorphic. In this sense, the uplift from IIA to M-theory is \textit{holomorphic}, allowing us to view the torus of $x^4$ and $x^{10}$ as a Riemann surface \cite{Gaiotto:2009hg}. This is precisely the Riemann surface $\EC$ associated with our $\EN=2$ theory viewed as a theory of class $\CalS$ \cite{gai1}.

We may take the limit of ungauging two consecutive gauge nodes by \textit{removing} one of the M5-branes originated from NS5-branes, so that the M5-branes from D4-branes are now stretched to the infinity in both directions of $x^4$. In this way $x^4$ is decompactified, modifying the Riemann surface to an infinite cylinder. It is convenient to treat this $(x^4,x^{10})$-cylinder with two ends at infinity as a sphere $\EC = \BP^1$ with two additional marked points at $0$ and $\infty$. In total, we have $m+1$ marked points $S = \{ 0, p_2, p_3, \cdots, p_{m}, \infty\}  \subset \BP^1$; the two marked points are at the ends, while at each $p_i$ of the rest $m-1$ marked points a single M5-brane ends. Note that the holomorphic $(x^8,x^9)$-plane serves as the fiber of the trivial cotangent bundle over the $(x^4,x^{10})$-cylinder. Thus, we regard the complex manifold on $(x^8,x^8,x^4,x^{10})$ as $T^* \EC$.

Therefore, we are led to the M-theory defined on the 11-dimensional spacetime $\BC^3 \times \BR \times T^* \EC$, which is topological on $\BC^3 \times \BR$ and holomorphic on $T^*\EC$ under the twist. Moreover, the $\O$-background for the $U(1)^2 \subset SU(3)$ isometry of $\BC^3$ is turned on.\footnote{It is desirable to explicitly check our $\O$-background for the IIB gauge origami described above dualizes to the $\O$-background implemented in \cite{Costello:2016nkh} for the twisted M-theory, after the unrefinement. Conversely, it would be interesting to clarify the meaning of the refinement in the twisted M-theory, which would allow a deviation from the Calabi-Yau threefold condition $\ve_1+\ve_2 + \ve_4 =0$. This is beyond our scope and we leave it to a future work.} This is precisely the setup for the twisted M-theory which led to the non-commutative five-dimensional $\mathfrak{gl}(1)$ Chern-Simons theory on $T^*\EC \times \BR$, under different IIA reduction and localization \cite{Costello:2016nkh} (in the presence of the $A_{l-1}$ singularity introduced below, $\mathfrak{gl}(1)$ is replaced by $\mathfrak{gl}(l)$). For instance, the M5-branes which have appeared as a result of the dualization so far would be viewed as surface defects on holomorphic surfaces in this non-commutative five-dimensional Chern-Simons theory, as discussed in \cite{Costello:2016nkh}. See \cite{Haouzi:2024qyo,Ishtiaque:2024orn} for studies of intersecting line and surface defects in the 5d Chern-Simons theory relevant to our work.

\begin{table}[h!]
    \centering
    \begin{tabular}{ c||c|c|c|c|c|c|c|c|c|c||c } 
        \text{IIA/M branes} & 0 & 1 & 2 & 3 & 4 & 5 & 6 & 7 & 8 & 9 & 10 \\ \hline\hline
         D4/M5 & \rm{x} & \rm{x}& \rm{x}& \rm{x} & \rm{x} & & & & & & \rm{x} \\  NS5/M5 & \rm{x} & \rm{x} & \rm{x} & \rm{x} & & & & & \rm{x}  & \rm{x}   & \\ \hline KK5${}_l$/KK6${}_l$ & \rm{x} & \rm{x} & &  &\rm{x}& \rm{x}  & & & \rm{x} &  \rm{x} &\rm{x} \\ D2/M2 & \rm{x} & \rm{x} & & &  & \rm{x} && &&
    \end{tabular}
    \caption{Twisted M-theory with M-branes and $A_{l-1}$ singularity}
    \label{table:twM}
\end{table}

\begin{table}[h!]
    \centering
    \begin{tabular}{c|c|c|c|c}
        $\BC_{\ve_1}$ & $\BC_{\ve_2}$ & $\BC_{\ve_4}$ & $T^* \EuScript{C}$ & $\mathbb{R}$  \\ \hline
        $x^0,x^1$ & $x^2,x^3$ & $x^6, x^7$ & $x^8,x^9, x^4, x^{10}$ & $x^5$ 
    \end{tabular}
    \caption{Spacetime of the twisted M-theory}
    \label{table:spacetime}
\end{table}

\subsection{Reduction to GL-twisted $\EN=4$ theory and topological sigma model} \label{subsec:gln4}
The M-theory on a toric Calabi-Yau 3-fold is believed to be dual to a $(p,q)$-web of fivebranes in the IIB, where the web diagram is given by the toric diagram \cite{Leung:1997tw}. Applying this duality to our case, where the toric Calabi-Yau 3-fold is $\BC^3$, the $(p,q)$-web is simply composed of an NS5-brane, a D5-brane, and a (1,1) fivebrane, which are semi-infinite and join at a junction on a two-dimensional plane while the rest five dimensions of the worldvolumes are shared. This is precisely the IIB setup for the corner vertex algebra \cite{Gaiotto:2017euk}. See also \cite{Gaiotto:2019wcc,Gaiotto:2020dsq} for the duality between the twisted M-theory and IIB string theory.

The M5-branes in the twisted M-theory translate to D3-branes filling the faces of the web, ending on fivebranes. In our case, there are only D3-branes that fill in a single face, which is chosen to be the face $\S=\BR^+\times \BR^+$ between the NS5-brane and the D5-brane.\footnote{There is a $PSL(2,\BZ)$ duality from which we can choose which face to fill in \cite{Gaiotto:2017euk}. With the choice above, $\BC_{\ve_1}$ becomes a semi-infinite line in $\BR^2 _{01}$ supporting the D5-brane and $\BC_{\ve_2}$ becomes an orthogonal semi-infinite line supporting the NS5-brane. See the table \ref{table:IIBpq}, \ref{table:spacetimepq}.} There are two types of such D3-branes depending on where the transverse two dimensions of the worldvolume wrap. The first type wraps the Riemann surface $\EC$, while the second type wraps the fiber of the cotangent bundle $T^* \EC$ at each marked point $p_i \in S \subset \EC$.

On the worldvolume of the $N$ D3-branes of first type, the effective description is the $\EN=4$ $U(N)$ gauge theory on $\S \times \EC$, with the boundary conditions at two boundaries of $\S$ are prescribed by the NS5-brane and the D5-brane. These boundary conditions are known to be the (deformed) Neumann boundary condition and the (deformed) regular Nahm pole boundary condition \cite{Gaiotto:2008sa}, where the deformation $\k = - \frac{\ve_2}{\ve_1}$ determined by the $\O$-background parameters is compatible with one of the GL-twisted topological supercharge \cite{Gaiotto:2019wcc} (recall \eqref{eq:tkrel}). Now, the D3-branes of second type become half-BPS monodromy surface defects \cite{gukwit} located at $\S \times \{p_i\} \subset \S \times \EC$, at each marked point $p_i \in S$.\footnote{They should not be confused with the surface defects in the $\EN=2$ gauge theory that we introduce shortly. The monodromy surface defects described here are the ones in the $\EN=4$ gauge theory, which, in the point of view of the $\EN=2$ theory, determine the contents of the gauge theory itself. In other words, they correspond to the marked point data on the Riemann surface $\EC$ for the class $\CalS$ theory.} This is precisely the starting point of the GL-twisted $\EN=4$ gauge theory approach for the geometric Langlands correspondence on $\EC$ with ramification data on $S$ \cite{Kapustin:2006pk,gukwit2}.\footnote{To be precise, to study the geometric Langlands correspondence between the twisted $\EuScript{D}$-modules on $\text{Bun}_{PGL(N)} (\EC;S)$ and the $SL(N)$ local systems with parabolic structures at $S$, we would have to \textit{decouple} $U(1)$ part and consider the $G=PSU(N)$ gauge group (or ${}^L G=SU(N)$ in the S-dual frame). Here, we will not explain the details of the decoupling, but conduct the actual decoupling later after formulating the problem in our $\EN=2$ gauge theoretical setup.}

By compactifying the GL-twisted $\EN=4$ gauge theory on the Riemann surface  $\EC$, we are led to the topological sigma model of maps $\S \to \EuScript{M}_H $. The type of the topological sigma model is precisely determined by $t \in \BP^1$ parametrizing the topological supercharges in the GL-twist. As discussed earlier, we view that the canonical parameter $\k$ determines $t$ by the relation \eqref{eq:tkrel}, with the complexified gauge coupling $\t$ fixed (in fact, we take $\text{Re}\, \t = 0$). At generic $\k \neq  \infty$, this is an $A$-model in the symplectic structure $\o_t$ with a B-field $B_t$. Let us also recall that the topological sigma model becomes the $A$-model in $\o_K$ at $\k = 0$, and $B$-model in $J$ at $\k = \infty$.

At the boundaries of the worldsheet $\S$, the deformed Neumann and the deformed Nahm pole boundary conditions descend to the canonical coisotropic brane $\EuScript{B}_{cc}$ of $(A,B,A)_\k$ type and the brane of opers $\EuScript{B}_{op}$ of $(B,A,A)_\k$ type, respectively \cite{Nekrasov:2010ka, Gaiotto:2011nm, Frenkel:2018dej}. Thus we get the $(\EB_{cc},\EB_{op})$-strings in the $A$-model (in $\o_t$ with $B_t$). As the deformed Neumann and the deformed regular Nahm pole are S-dual to each other \cite{Gaiotto:2008ak}, the canonical coisotropic brane $\EuScript{B}_{cc}$ and the brane of opers $\EuScript{B}_{op}$ are mirror dual to each other. This precisely reflects the symmetry of exchanging the two complex planes of the four-dimensional worldvolume of the $\EN=2$ theory.

So far, we have only considered the basic gauge origami setup that engineers the $\EN=2$ gauge theory itself and how it is dualized to the GL-twisted $\EN=4$ theory and then reduced to the topological sigma model (the first two rows on the tables \ref{table:IIBori}, \ref{table:twM}). The purpose of this section is to motivate introducing additional constituents (the last two rows on the tables \ref{table:IIBori},  \ref{table:twM}) in the gauge origami by showing how they dualize to the GL-twisted $\EN=4$ theory and topological sigma model, explaining their roles in the geometric Langlands correspondence. In the twisted M-theory perspective, the two additional constituents are $A_{l-1}$ singularity and M2-branes. They will turn out to produce half-BPS surface defects on the $\EN=2$ gauge theory side.

\begin{table}[h!] 
    \centering
    \begin{tabular}{ c||c|c|c|c|c|c|c|c|c|c } 
        \text{IIB Branes} & 0 & 1 & 2 & 3 & 4 & 5 & 6 & 7 & 8 & 9  \\ \hline\hline  Fivebranes & \multicolumn{2}{|c|}{$\vdash$}  &  & &  & \rm{x} & \rm{x} & \rm{x} & \rm{x} & \rm{x}\\ 
        \hline
         D3 & $\vdash$ & $\vdash$& &  &  & & & &  \rm{x}& \rm{x}\\  D3 & $\vdash$ & $\vdash$ &  &  & &  & \rm{x} & \rm{x}&  &    \\ \hline  D1 & \multicolumn{2}{|c|}{$\vdash$}  & & & & \rm{x}  &&&
    \end{tabular}
    \caption{IIB brane configuration for $(p,q)$-web; the symbol $\vdash$ indicates semi-infinite filling}
    \label{table:IIBpq}
\end{table}

\begin{table}[h!]
    \centering
    \begin{tabular}{c|c|c|c}
        $\BR^2$ & $\BR^3$ & $\BR$ & $T^* \EuScript{C}$  \\ \hline
        $x^0,x^1$ & $x^2,x^3, x^4$ & $x^5$ & $x^6,x^7, x^8, x^{9}$ 
    \end{tabular}
    \caption{Spacetime of the IIB theory for $(p,q)$-web}
    \label{table:spacetimepq}
\end{table}

\paragraph{$A_{l-1}$-type singularity}
The twisted M-theory can be formulated not only on the simplest toric Calabi-Yau threefold $\BC^3$, but also on $\BC \times TN_{l-1}$ where $TN_{l-1}$ is the $l$-centered Taub-NUT space \cite{Costello:2016nkh}. When all the centers are brought to the origin, the Taub-NUT space has an $A_{l-1}$-type singularity $\BC^2/\BZ_{l}$ there. Dualized to the IIB setup of the gauge origami, this also amounts to adding Kaluza-Klein monopoles for which the transverse geometry can be viewed as the $A_{l-1}$ singularity. Since this transverse geometry intersects the worldvolume of the D3-branes realizing the $\EN=2$ gauge theory along the complex plane $\BC_{\ve_1}$, it gives rise to a half-BPS monodromy surface defect there \cite{gukwit,K-T} (see the third rows of table \ref{table:IIBori}, \ref{table:twM}).

Passing to the $(p,q)$-web of fivebranes in IIB, the $A_{l-1}$ singularity increases the number of D5-branes from $1$ to $l$, where the number of D3-branes ending on each D5-brane is determined by the number of the M5-branes carrying each representation of $\BZ_l$. In the effective $\EN=4$ gauge theory on the worldvolume of the D3-branes, it amounts to changing the corresponding boundary condition from the (deformed) regular Nahm pole to a (deformed) non-regular Nahm pole \cite{Gaiotto:2008sa}. We will give a description of the general $A_{l-1}$ singularity as a monodromy surface defect in the $\EN=2$ gauge theory. Later, we mainly focus on the case where $l=N$ and exactly one D3-brane ends on each of $N$ D5-branes, resulting in the deformed Dirichlet boundary condition that breaks the gauge group of the $\EN=4$ theory to the maximal torus. In the $\EN=2$ gauge theory side, the resulting surface defect also breaks the gauge group to the maximal torus. We will call such a surface defect to be \textit{regular} (see section \ref{subsec:mono}).\footnote{A caution in terminology: Somewhat confusingly, the regular monodromy surface defect in the $\EN=2$ theory corresponds to the extreme non-regular Nahm pole (Dirichlet) boundary in the GL-twisted $\EN=4$ theory. Conversely, the absence of monodromy surface defect (trivial defect) in the $\EN=2$ theory corresponds to the regular Nahm pole boundary in the GL-twisted $\EN=4$ theory.}

Upon reduction to the sigma model, the deformed Dirichlet boundary condition is expected to descend to the brane of $\l$-connections $\EuScript{F}' _{E}$ of $(B,A,A)_\k$ type, supported on the fiber of the $I_{w_t}$-holomorphic map $\pi_{w_t}$ \eqref{eq:secfibgen} from an open dense subset of the Hitchin moduli space with ramifications $\EuScript{M}_H$, viewed as the moduli space of stable parabolic local systems, to the moduli space of parabolic $G_{\BC}$-bundles, at $E \in \text{Bun}_{G_\BC} (\EC;S)$ \cite{Frenkel:2018dej}. Thus, we get the $(\EB_{cc},\EF' _{E})$-strings in the $A$-model (in $\o_t$ with $B_t$) which give the $E$-twisted coinvariants of the $\widehat{\fg}$-modules associated to the ramifications at $S$. 

It was indeed shown that the regular monodromy surface defect gives coinvariants of certain $\widehat{\mathfrak{g}}$-modules associated to the ramification data, by verifying the Knizhnik-Zamolodchikov equations are obeyed by its vacuum expectation value \cite{Nekrasov:2021tik} (see also \cite{Braverman:2004vv, Alday:2010vg}). It was, however, not clarified which parabolic $G_\BC$-bundle $E$ these coinvariants are twisted by. The regular monodromy surface defect carry the defect parameters $\mathbf{u}$ assigning the singularity of the gauge field and the magnetic fluxes along the surface \cite{gukwit}. By the duality that we described, it is naturally expected that these defect parameters provide holomorphic coordinates on the moduli space $\text{Bun}_{G_\BC} (\EC;S)$ of (stable) parabolic $G_\BC$-bundles on $\EC$. In section \ref{sec:reg}, we confirm that this is indeed the case. Moreover, we show the vacuum expectation value $\Psi(\mathbf{u})$ of the regular monodromy surface defect can be viewed as a (local) section of $\fL_\k \to \text{Bun}_{G_\BC} (\EC;S)$. Here, $\fL_\k$ is precisely the twisting \eqref{eq:twist} used for the sheaf $\ED_{\fL_\k}$ of differential operators on $\text{Bun}_{G_\BC} (\EC;S)$, associated to the $(\EB_{cc},\EB_{cc})$-strings.

\paragraph{M2-branes}
We recall Wilson and 't Hooft line defects play important roles in studying the geometric Langlands correspondence from the GL-twisted $\EN=4$ gauge theory \cite{Kapustin:2006pk,gukwit2}. When the canonical parameter $\k$ is generic, the line defects do not exist in the four-dimensional bulk, but they still can be supported on boundary with appropriate boundary condition \cite{Witten:2011zz}. 

In the twisted M-theory setup, the simplest line defects attached at boundaries originate from an M2-brane lying along the complex plane $\BC_{\ve_1}$ or $\BC_{\ve_2}$ (see the fourth rows of table \ref{table:IIBori}, \ref{table:twM}). In the IIB gauge origami picture, they are the D3-branes intersecting the worldvolume of the original stack of D3-branes along the complex plane $\BC_{\ve_1}$ or $\BC_{\ve_2}$, and therefore generate surface defects in the $\EN=2$ gauge theory. In the field theory limit, this surface defect arises as a 2d/4d coupled system. The two-dimensional $\EN=(2,2)$ gauged linear sigma model on the worldvolume of the M2-brane couples to the four-dimensional $\EN=2$ gauge theory, by gauging its flavor symmetry by the bulk gauge field restricted to the surface (see section \ref{subsec:Qcan}).

By dualizing to the $(p,q)$-web in IIB, the M2-brane wrapping $\BC_{\ve_2}$-plane becomes a fundamental string stretching from the junction of the fivebranes to the infinity along the NS5-brane; while the M2-brane wrapping $\BC_{\ve_1}$-plane becomes a D1-brane stretching from the junction to the infinity along the D5-brane(s). In the effective GL-twisted $\EN=4$ gauge theory, the former is the Wilson line defect placed at the deformed Neumann boundary, while the latter is the 't Hooft line defect placed at the deformed (non-)regular Nahm pole boundary, both of which are labelled by the $N$-dimensional representation of $U(N)$.

In the absence of the $A_{l-1}$ singularity, the Wilson line along the deformed Neumann boundary and the 't Hooft line along the deformed regular Nahm pole boundary are S-dual to each other, so that the two cases are symmetric. However, in the presence of the $A_{l-1}$ singularity, the two boundary conditions (Neumann and non-regular Nahm pole) are no longer S-dual to each other \cite{Gaiotto:2008ak}. Therefore, the two cases $-$ a Wilson line along the Neumann boundary and a 't Hooft line along the non-regular Nahm pole boundary $-$ have to be considered separately. In terms of the $\EN=2$ gauge theory, the former is the configuration of two surface defects \textit{intersecting} at the origin, while the latter is the configuration of two surface defects \textit{parallel} along the complex plane $\BC_{\ve_1}$; the two cases are obviously distinct.

Descending to the topological sigma model, a line defect attached at a brane defines a new brane (i.e., it is a functor acting on the category of branes) \cite{Kapustin:2006pk}. The Wilson line labelled by the representation $R$ of $G$ acts by modifying the Chan-Paton bundle of the brane to its tensor product with the bundle associated to the universal bundle in the representation $R$ restricted to the insertion point $y \in \EC$. In our case, it is the canonical coisotropic brane that is dressed with the Wilson line, and we may denote this new brane by $\EB_{cc} ^{y, R}$ \cite{Frenkel:2018dej}. It is expected that the $(\EB_{cc} ^{y,R}, \EF ' _E)$-strings give the $E$-twisted coinvariants of the tensor product of the $\widehat{\mathfrak{g}}$-modules associated to the ramifications at $S$ and the $\widehat{\mathfrak{g}}$-module induced from the representation $R$, viewed as a $\fg$-module. Indeed, the correlation functions of intersecting surface defects in the $\EN=2$ gauge theory setup were shown to be such $E$-twisted coinvariants since the former satisfy the Knizhnik-Zamolodchikov equations for the latter \cite{Jeong:2021bbh}. The configuration of intersecting surface defects was used, in particular, to study the isomonodromy problem appearing in the limit $\k \to \infty$ (namely, $\ve_1 \to 0$) \cite{Jeong:2020uxz}. 

In this work, we do not discuss insertion of Wilson lines but rather focus on a 't Hooft line attached to the Dirichlet boundary; namely, the configuration of parallel surface defects in the $\EN=2$ gauge theory. The 't Hooft line labelled by the dominant integral coweight $\l \in \text{Hom}(U(1),\BT)$ of $G$ descends to the Hecke operator labelled by the same $\l$ acting on $A$-branes \cite{Kapustin:2006pk}. We have reviewed in section \ref{subsubsec:heckeop} how it acts on the twisted $\ED$-modules associated to the $A$-branes. See also \cite{Balasubramanian:2017gxc} for a related realization of the Hecke operator in the vertex algebra perspective.

\subsection{Reduction to $\EN=2$ theory of class $\CalS$}
In the twisted M-theory setup, the low-energy effective theory on the worldvolume of the M5-branes wrapping $\EC$ is the six-dimensional $\EN=(0,2)$ theory of type $A_{N-1}$ on $\BC_{\ve_1} \times \BC_{\ve_2} \times \EC$, with codimension-two defects on $\BC_{\ve_1}\times \BC_{\ve_2} \times \{p_i\}$ for each marked point $p_i \in S \subset \EC$. By compactifying along the Riemann surface $\EC$, we obtain the four-dimensional $\EN=2$ theory of class $\CalS$ where the contents of the theory are determined by $\EC$ and the codimension-two defects \cite{gai1}. This gives a clear explanation of why the $\EN=2$ gauge theory constructed in the IIB gauge origami setup has to be of class $\CalS$.

The $A_{l-1}$ singularity gives rise to another type of codimension-two defect in the six-dimensional $\EN=(0,2)$ theory, supported on $\BC_{\ve_1} \times \{ 0 \} \times \EC$. Upon compactification along $\EC$, it descends to a half-BPS monodromy surface defect of the $\EN=2$ theory of class $\CalS$, defined by assigning monodromy of gauge fields along the circle linking the surface (the $\BC_{\ve_1}$-plane in this case) \cite{gukwit, K-T}. The Levi subgroup it preserves is determined by the $\BZ_l$-representations that the M5-branes carry. As mentioned above, we mainly consider the case where $l=N$ and the gauge group is broken to the maximal torus (see section \ref{subsec:mono} for the detail). This is the case that we refer to as the \textit{regular} monodromy surface defect, which corresponds to the deformed Dirichlet boundary condition in the GL-twisted $\EN=4$ theory side. We recall that the codimension-two defects of the six-dimensional $\EN=(0,2)$ theory are characterized by the global symmetry they carry \cite{Tachikawa:2011dz}. In the case where $l=N$ and the gauge symmetry of the class $\CalS$ theory is broken to the maximal torus, the global symmetry is $G=PSU(N)$. Thus when the six-dimensional theory is compactified along $\EC$, the parameter space of the regular monodromy surface defect is naturally given by $\text{Bun}_{G_\BC} (\EC;S)$ \cite{Gukov:2014gja}.

The M2-brane engineers a codimension-four defect of the six-dimensional $\EN=(0,2)$ theory, supported on $\BC_{\ve_1} \times \{0\} \times \{ y\}$ \cite{Gaiotto:2011tf}. It is local on $\EC$ and here we set $y \in \EC\setminus S$ to be its location. Under the reduction to the class $\CalS$ theory, it descends to the \textit{canonical} surface defect which can be thought of as coupling two-dimensional $\EN=(2,2)$ sigma model living on the non-compact part of the worldvolume of the M2-brane, by gauging the flavor symmetry with the bulk gauge field restricted to the surface (see section \ref{subsec:Qcan} for the detail). Note that the parameter space of the canonical surface defect is naturally given by the $y$-space $\EC\setminus S$ \cite{Gaiotto:2009fs}.

We remark here that there is an M-brane transition connecting these two types of surface defects \cite{Frenkel:2015rda}. We will study the manifestation of this brane transition at the level of the correlation functions of the surface defects in a separate work.

\section{Surface defects from gauge origami} \label{sec:sgo}
In the previous section, we have seen that half-BPS surface defects $-$ the regular monodromy surface defect and the canonical surface defect $-$ of the $\EN=2$ gauge theory 
are expected to descend to the brane of $\l$-connections and the Hecke operator acting on branes. We will confirm this within our 4d $\EN=2$ gauge theoretical setting in section \ref{sec:reg} and \ref{sec:Hecke}, using the analytic constraints on their vacuum expectation values and their correlation functions that will be established in section \ref{sec:TQoper}. In this section, we provide a detailed construction of these defects within the IIB gauge origami framework, thereby setting the stage for the subsequent sections.

We will first present the gauge origami configuration \cite{Nikita:I,Nikita:II,Nikita:III} of intersecting stacks of D3-branes on an orbifold in the IIB string theory.\footnote{To be precise, we will distinguish the terminology for the gauge origami and the spiked instantons in this work. The former refers to the configuration of intersecting D3-branes, while the latter refers to the D($-1$)-instantons dissolved into the worldvolume of such D3-brane configurations.} Remind that the role of intersecting D3-branes and orbifolds are to engineer 
\begin{enumerate}
    \item The four-dimensional $\EN=2$ supersymmetric gauge theory as the low-energy effective theory on the worldvolume of one of the stacks of D3-branes
    \item The half-BPS surface defects on this theory by an orbifold singularity or another stack of D3-branes which intersects the original stack on a complex plane
    \item The $qq$-character for the combined system by another intersecting stack of D3-branes.
\end{enumerate}
The correlation functions of the BPS defects (e.g., the ones in 2 and 3) are computed as the partition function of the gauge origami, that is, an equivariant integral of the equivariant Euler class of certain vector bundle over the associated moduli space of spiked instantons. We sometime call this spiked instanton partition function. 

Let us denote the ten-dimensional spacetime of the IIB theory as $X \times \BR^2$. Here $X$ is called the \textit{gauge origami worldvolume}. As we introduced earlier in table \ref{table:IIBori} and \ref{table:spacetimeori}, we will consider two types of orbifold $X = \BC^2 _{12} \times \BC^2 _{34} / \G_{34}$ and $X = \BC_1 \times \BC^3 _{234} /\G_{34}\times \G_{24}$, with cyclic groups $\G_{34} = \BZ_m$ and $\G_{24} =\BZ_l$. Here, $\G_{ab}$ acts on $\BC^2 _{ab}$ as
\begin{align} \label{eq:Gact}
    (z_a,z_b) \mapsto (\varpi z_a , \varpi^{-1} z_b),
\end{align}
where $\varpi$ is the generator of the cyclic group $\G_{ab}$ represented as ($m$-th or $l$-th) root of unity.

The gauge origami is intersecting stacks of D3-branes placed on non-compact complex 2-planes in the gauge origami worldvolume $X$, preserving its $SU(4)$-isometry, with generic positions on $\BR^2$ unless specified otherwise. There can be $\begin{pmatrix} 4 \\ 2 \end{pmatrix} = 6$ stacks at most. Let $\mathbf{6} = \{ab \, \vert \, a,b =1,2,3,4,\, a < b\}$ denote the set of these six choices of complex two-planes. Let the number of D3-branes inserted on $A \in \mathbf{6}$ be $N_{A}$. Let $a_{A,\a} \in \BR^2$, $A\in \mathbf{6}$, $\a=1,\cdots, N_A$ denote their locations on $\BR^2$ which are assumed to be generic. By slightly abusing the notation, we denote the Chan-Paton spaces of six stacks of D3-branes also by the same letters $N_A$. They carry representations of $\G$ and therefore can be decomposed as
\begin{align} \label{eq:Ndecom}
    N_A = \bigoplus_{\substack{i \in \{0,1,\cdots, m-1\} \\ \o \in \{0,1,\cdots, l-1\}}} N_{A,i,\o} \CalR_i \otimes \fR_\o,
\end{align}
where $\CalR_i$ (resp. $\fR_\o$) is the one-dimensional irreducible representation of $\G_{34}$ (resp. $\G_{24}$) of weight $i$ (resp. of weight $\o$). When $X = \BC^2 _{12} \times \BC^2 _{34} / \G_{34}$, we simply ignore the $\G_{24}$ part.

The low-energy effective theory on the gauge origami worldvolume is, therefore, several (at most 6) four-dimensional supersymmetric gauge theories living on the intersecting stacks interacting with each other through their interfaces. The global symmetry group is $\mathsf{H}=  \bigtimes_{A\in \mathbf{6}} \bigtimes_{{\substack{i \in \{0,1,\cdots, m-1\} \\ \o \in \{0,1,\cdots, l-1\}}}} U(N_{A,i,\o}) \times SU(4)$, where the first factor is the global gauge symmetry rotating the D3-branes carrying the same representation of $\G$ and the second factor is the isometry of $X$.

We may insert D$(-1)$-instantons dissolved into the worldvolume of the gauge origami, and the point-like BPS objects induced in this way in the intersecting gauge theories are called the \textit{spiked instantons}. To preserve the $SU(4)$-isometry they are located at the origin of $X$, while on $\BR^2$ they are attached to one of the D3-branes. Let $K$ be the Chan-Paton space of the D$(-1)$-instantons. It is decomposed according to which D3-brane they are placed on: $K = \bigoplus_{A \in \mathbf{6} } \bigoplus_{\a=1} ^{N_A} K_{A,\a}$. Let us denote $K_A :=  \bigoplus_{\a=1} ^{N_A} K_{A,\a}$ for each $A \in \mathbf{6}$.

The moduli space of spiked instantons admits an ADHM-like realization in terms of the linear maps between the vector spaces $N_A$ and $K_A$. The discrete fixed points of this moduli space under the action of the maximal torus $\mathsf{T}_\mathsf{H}$ of the symmetry group $\mathsf{H}$ are classified by a set of partitions $\boldsymbol\l = \left( \l^{(A,\a)} \right)_{A\in \mathbf{6},\, \a \in \{1,2,\cdots, N_A\}}$, where $\vert \boldsymbol\l \vert = \vert K \vert = \dim K$. 

Since $\mathsf{T}_\mathsf{H}$ is an abelian group, the spaces $N_A$ and $K_A$ are decomposed into its one-dimensional irreducible representations. Let us present this decomposition by writing the equivariant Chern characters of these spaces,
\begin{align} \label{eq:Echar}
\begin{split}
    &N_A = \sum_{\a=1} ^{N_A} e^{a_{A,\a}} \\
    &K_A = \sum_{\a=1} ^{N_A} \sum_{(i,j) \in \l^{(A,\a)}} e^{a_{A,\a} + (i-1)\ve_a + (j-1) \ve_b} ,
\end{split}
\end{align}
where $\left( a_{A,\a} \right)_{A\in \mathbf{6}, \a\in \{1,2,\cdots, N_A\}}$ and $\left( \ve_a \right)_{a=1,2,3,4}$ are equivariant parameters for the global gauge symmetry and the isometry of $X$, respectively. The former are called the Coulomb moduli and the latter are called the $\O$-background parameter. Note that there are only three independent $\O$-background parameters, $\sum_{a=1} ^4 \ve_a = 0$. For convenience, we will use the notation $q_a = e^{\ve_a}$, $P_a = 1-q_a$, $q_{ab}= q_a q_b$, and $P_{ab} = (1-q_a)(1-q_b)$. Note that each one-dimensional subspace represented by each term in the equivariant Euler character \eqref{eq:Echar} carries a representation of $\G$, according to the decomposition \eqref{eq:Ndecom} and the $\G$-action on $X$ \eqref{eq:Gact}.

By supersymmetric localization, the path integral for these intersecting gauge theories reduces to a sum of equivariant integrals of the Euler class of certain vector bundle over the moduli spaces of spiked instantons with varying instanton numbers, where the measure is given by the gauge couplings to the power of instanton numbers. Let us define the universal sheaf for the instantons on the $A$-th stack ($A \in \mathbf{6}$) as $S_A = N_A - P_A K_A$. The partition function is computed as \cite{Nikita:III}
\begin{align} \label{eq:spinstpart}
\begin{split}
\CalZ = \sum_{\boldsymbol\l} \boldsymbol\qe ^{\vert \boldsymbol\l \vert} &\BE \left[ - \sum_{A \in \mathbf{6}} \frac{{P}_{\text{min}\bar{A}} {S}_{A} {S}_{A} ^*}{{P}_{A} ^*} - \sum_{A=12,13,14} q_A ^{-1} {S}_{A} {S}_{\bar{A}}^* + \sum_{\substack{A <B \\ \vert A \cap B \vert = 1}} q_{\text{max}B} {P}_{\overline{A \cup B}} \frac{{S}_{A} {S}_{B} ^*}{{P}_{A \cap B} ^*}  \right]^\G,
\end{split}
\end{align}
where $\BE \left[ \cdots \right]$ is the symbol to take the product of the equivariant weights and the notation $[ \cdots ]^\G$ is to pick up the part invariant under the action of $\G$ (namely, $\G_{34}$ or $\G_{34} \times \G_{24}$). Also, we used the notation $\bar{A} = ab$ where $a,b \notin A$ and $a<b$. The summation is taken over all the partitions $\boldsymbol\l = \left( \l^{(A,\a)} \right)_{A\in \mathbf{6},\, \a \in \{1,2,\cdots, N_A\}}$. The gauge couplings are collectively denoted by $\boldsymbol\qe = \left( \qe_{i,\o} \right)_{\substack{i\in \{0,1,\cdots, m-1\} \\ \o \in \{0,1,\cdots, l-1\}}}$. We also used the notation $\boldsymbol\qe ^{\vert \boldsymbol\l \vert} = 
\prod_{\substack{i\in \{0,1,\cdots, m-1\} \\ \o \in \{0,1,\cdots, l-1\}}} \qe_{i,\o} ^{\vert K_{i,\o} \vert}$ where $ K_{i,\o} $ is the subspace of $K$ carrying the representation $\CalR_i \otimes \fR_\o$ of $\G$.

\subsection{$\EN=2$ supersymmetric gauge theory}
We first consider the case $X = \BC^2 _{12} \times \BC^2 _{34} / \G_{34}$ with the cyclic group $\G_{34} = \BZ_m$. To engineer four-dimensional gauge theory, we insert a stack of D3-branes at the orbifold singularity, $\BC^2_{12} \times \{0\} \subset X$. The action of $\G_{34}$ breaks half of the supersymmetry, leaving a $\EN=2$ supersymmetric theory on the worldvolume of the D3-branes. This worldvolume theory is shown to be the $\widehat{A}_{m-1}$-quiver gauge theory by studying the spectrum of open string ending on the D3-branes \cite{Douglas:1996sw}. By ungauging two consecutive gauge nodes, i.e., turning off their gauge couplings, we can also get the linear $A_{m-2}$-quiver gauge theory.

This can be immediately shown at the level of partition functions as follows. We will restrict to the case $m=3$ from now on. Let there be $N$ D3-branes carrying the $\BZ_3$-representation $\CalR_i$ for each $i=0,1,2$. The gauge origami is represented by the equivariant Chern character of the Chan-Paton space, which is now a representation of $\BZ_3$. It is simply
\begin{align}
     N_{12} & = \sum_{\alpha =1} ^N e^{a_\alpha} \cdot \CalR_0 + \sum_{\alpha=1 } ^N e^{m_\alpha^+ - \ve_4} \cdot \CalR_1 + \sum_{\alpha=1} ^N e^{m_\alpha^- + \ve_4} \cdot \CalR_{2}
\end{align}
The spiked instanton partition function reads
\begin{align}
\begin{split}
\CalZ_S &= \sum_{\boldsymbol\l} \prod_{i=0,1,2} \qe_i ^{\vert \boldsymbol\l_i \vert} \BE \left[ -\frac{ P_3 S_{12} S_{12}^*}{P_{12} ^*} \right]^{\BZ_3} \\
&= \sum_{\boldsymbol\l} \prod_{i=0,1,2} \qe_i ^{\vert \boldsymbol\l_i \vert} E \left[ - \frac{ S_0 S_0 ^* + S_1 S_1 ^* + S_2 S_2 ^* - q_{12} ^{-1} S_0 S_1 ^* -  q_3  q_4 ^{-2} S_1 S_2 ^* - q_{12}^{-1} S_2 S_0 ^* }{P_{12} ^*} \right],
\end{split}
\end{align}
where we defined the universal sheaf $S_i \equiv N_i -P_{12} K_i$ for the instantons associated to the $i$-th node. This is precisely the partition function of the $\hat{A}_2$-quiver $U(N)$ gauge theory \cite{Nekrasov:2002qd,Nekrasov:2003rj}. We may turn off the gauge couplings $\qe_1$ and $\qe_2$, sending them to zero. Then some of the universal bundles become flavor bundles, $S_1 = M^+$ and $S_2 = M^-$, so that the partition function reads
\begin{align} \label{eq:a1part}
    \CalZ = \sum_{\boldsymbol\l} \qe ^{\vert\boldsymbol\l\vert} \BE \left[ - \frac{(S-M^+ - q_{12} ^{-1} M^-) S^*  }{P_{12} ^*} \right].
\end{align}
Namely, the global gauge symmetry for the ungauged nodes becomes the flavor symmetry, converting the relevant Coulomb moduli to the mass parameters for hypermultiplets. The resulting expression is the partition function of the $U(N)$ gauge theory with $N$ fundamental and $N$ anti-fundamental hypermultiplets, i.e., the $A_1$-quiver $U(N)$ gauge theory. This is our main example of the $\EN=2$ gauge theory throughout the work.

So far, all the equivariant parameters $-$ the Coulomb moduli $(a_\a)_{\a=1} ^N$, the hypermultiplet masses $(m^\pm _\a)_{\a=1} ^N$, and the $\O$-background parameters $\ve_1$ and $\ve_2$ $-$ have been taken to be generic. We consider a limit that plays an important role throughout the work: $\ve_2 \to 0$. This is the limit where the two-dimensional $\EN=(2,2)$ super-Poincar\'{e} symmetry on $\BC_2$ is restored. The four-dimensional $\EN=2$ theory is effectively described by a two-dimensional $\EN=(2,2)$ theory, governed by a twisted superpotential $\widetilde{\EW}$. This twisted superpotential is obtained from the partition function in the limit, \cite{Nekrasov:2009rc}\footnote{This limiting behavior of the partition function can also be obtained from the limit shape of the partitions in the ensemble \eqref{eq:a1part} \cite{Nekrasov:2003rj, Poghossian:2010pn}.}
\begin{align}
    \lim_{\ve_2\to 0} \CalZ(\mathbf{a},\mathbf{m},\boldsymbol\ve;\qe) = e^{\frac{\widetilde{\EW} (\mathbf{a},\mathbf{m},\ve_1;\qe) }{\ve_2}}.
\end{align}

When local or non-local observables are inserted, the path integral computes the vacuum expectation values or the correlation functions of them. As an equivariant integral, it localizes to the ensemble average over the same fixed points $\{\boldsymbol\l\}$. Thus, given a $\EN=2$ theory observable $\CalO$, there is corresponding observable $\CalO[\boldsymbol\l]$ defined on the ensemble of partitions so that the vacuum expectation value is written as
\begin{align}
    \langle \CalO \rangle_{\mathbf{a}} = \sum_{\boldsymbol\l} \qe^{\vert \boldsymbol\l \vert} \, \CalO[\boldsymbol\l]\, \BE \left[ - \frac{(S-M^+ - q_{12} ^{-1} M^-) S^*  }{P_{12} ^*} \right].
\end{align}
We explicitly write the Coulomb moduli $\mathbf{a}$ among other parameters (the masses $(m_\a ^\pm)_{\a=1} ^N$ for the hypermultiplets, the $\O$-background parameters $\ve_1$ and $\ve_2$, and the gauge coupling $\qe$) when we emphasize the dependence of vacuum expectation values (or correlation functions) on them.

\subsection{$Q$-observable and canonical surface defect} \label{subsec:Qcan}
We turn to the BPS defects in the $\EN=2$ supersymmetric gauge theory built from additional constituents in the gauge origami setup. Here, we recall the \textit{$Q$-observable}, which descends from single D3-brane placed on $\BC^2 _{13} \subset X$ intersecting the worldvolume of the gauge theory $\BC^2 _{12}$ along the complex plane $\BC_1$ (the fourth row of the table \ref{table:IIBori}), and the \textit{canonical surface defect} obtained from the $Q$-observable by a certain transition.

\subsubsection{$Q$-observable} \label{subsubsec:qobs}
We present the explicit form of the $Q$-observable for the case of $X = \BC^2 _{12} \times \BC^2 _{34} /\G_{34}$ with $\G_{34} = \BZ_3$. As we described above, the underlying gauge theory is the $A_1$-quiver gauge theory after turning off some of the gauge couplings, and the $Q$-observable will be a half-BPS surface observable defined on this theory. The reason for its nomenclature is explained later in section \ref{sec:TQoper} (see \eqref{eq:qTQ}).

Thus we add a single D3-brane wrapping $\BC^2 _{13} \subset X$, located at $x \in \BR^2$. There is a choice of the representation that the Chan-Paton space of the new D3-brane carries under the action of $\G_{34} = \BZ_3$. Here, we choose $\CalR_1$.\footnote{Another choice $\CalR_0$, which leads to the \text{dual} $Q$-observable, will be discussed somewhere else. Still another choice $\CalR_2$ yields a decoupled surface defect after the ungauging from $\hat{A}_2$-quiver to $A_1$-quiver. This is not very interesting for our purpose and we neglect the case of $\CalR_2$.} Then the gauge origami is represented by the following equivariant Chern characters of the Chan-Paton spaces:
\begin{align}
\begin{split}
    N_{12} & = \sum_{\alpha=1}^N e^{a_\a}  \cdot \CalR_0 + \sum_{\alpha=1} ^N e^{m_\alpha^+ - \ve_4} \cdot \CalR_1 + \sum_{\alpha=1} ^N e^{m_\alpha^- + \ve_4} \cdot \CalR_{2} \\
    N_{13} & = e^{x+\ve_1+\ve_3} \cdot \CalR_1.
\end{split}
\end{align}
By applying \eqref{eq:spinstpart} to this case, the spiked instanton partition function is given by
\begin{align}
    \CalZ = \sum_{\boldsymbol\l} \prod_{i=0,1,2} \qe_i ^{\vert \boldsymbol\l_{12,i} \vert + \vert \boldsymbol\l_{13,i} \vert} \BE \left[ - \frac{P_3 S_{12} S_{12} ^*}{P_{12} ^*} - \frac{P_2 S_{13} S_{13} ^*}{P_{13} ^*} + q_3 P_4 \frac{S_{12} S_{13} ^*}{P_{1}^ *} \right]^{\BZ_3}.
\end{align}
We take the decoupling limit $\qe_1 = \qe_2 = 0$ which yields the $A_1$-quiver gauge theory. The instantons with nonzero $\BZ_3$-charges are forbidden, and the universal sheaf for the instantons on the $\BC^2 _{13}$ becomes trivial: $S_{13} = N_{13}$. Thus the second term only contributes an overall multiplicative factor which can be omitted. The third term yields the $Q$-observable that we desired. Explicitly, the expression is
\begin{align}
    \CalZ = \sum_{\boldsymbol\l} \qe ^{\vert \boldsymbol\l \vert} \BE \left[ -\frac{(S- M^+ -q_{12} ^{-1} M^-) S^* }{P_{12} ^*} \right] \BE \left[ - \frac{e^x (S^*- M^{+*})}{P_1 ^*} \right].
\end{align}
Thus, the spiked instanton partition function gives the vacuum expectation value of the $Q$-observable, whose expression at the partition $\boldsymbol\l$ is given by
\begin{align} \label{def:Q}
    Q(x)[\boldsymbol\l] = \frac{1}{  \prod_{\a=1} ^N \ve_1 ^{\frac{x-m_\a ^+}{\ve_1}} \G \left( 1+ \frac{x-m^+ _\a}{\ve_1} \right) } \BE \left[- \frac{e^x S^*[\bl]}{P_1 ^*} \right],
\end{align}
where the 1-loop contribution from the $M^+$ term is regularized to the $\G$-function. An immediate computation reveals the dependence of the $Q$-observable on the partitions $\boldsymbol\l$ more clearly,
\begin{align} \label{eq:Qdetail}
    Q(x) [\boldsymbol\l] = \prod_{\a=1} ^N \frac{\ve_1 ^\frac{m_\a ^+ -a_\a}{\ve_1}}{\G\left( 1+ \frac{x-m^+ _\a}{\ve_1} \right) \G\left( l(\l^{(\a)}) - \frac{x-a_\a}{\ve_1} \right)} \prod_{i=1} ^{l(\l^{(\a)}) } \left( i-1 - \frac{x-a_\a-\l_i ^{(\a)} \ve_2}{\ve_1} \right),
\end{align}
where $l(\l^{(\a)})$ is the length of the partition $\l^{(\a)} = \left(\l^{(\a)} _1,\cdots, \l^{(\a)} _{l (\l^{(\a)})} \right)$. It will be important that there are $2N$ semi-infinite rays of zeros. $N$ of them are from the $\G$-functions $\G\left( 1+\frac{x-m^+_\a}{\ve_1} \right)$, while the other $N$ originate from the $\G$-functions $\G\left( l(\l^{(\a)}) - \frac{x-a_\a}{\ve_1} \right)$, in the denominator. These two groups of zeros have the positions growing in the opposite directions, with the $\ve_1$-spacing.

In the limit $\ve_2 \to 0$, the four-dimensional $\EN=2$ theory is effectively described by the two-dimensional $\EN=(2,2)$ theory on the $\BC_2$-plane. Since the $Q$-observable is a surface defect supported on the $\BC_1$-plane, it becomes a local observable in the limit and does not alter the effective theory itself. Thus, the vacuum expectation value becomes
\begin{align} \label{eq:Qobsns}
    \lim_{\ve_2\to 0}\Big\langle Q(x) \Big\rangle_{\mathbf{a}} = e^{\frac{\widetilde{\EW} (\mathbf{a})}{\ve_2}} Q(\mathbf{a};x),
\end{align}
where $\widetilde{\EW}(\mathbf{a})$ is the twisted superpotential for the effective $\EN=(2,2)$ theory. We have abused the notation a bit and denote the normalized vacuum expectation value in the limit $\ve_2 \to 0$ by the same letter $Q$.

\begin{figure}[h!]
    \centering
    \includegraphics[width=0.5\textwidth]{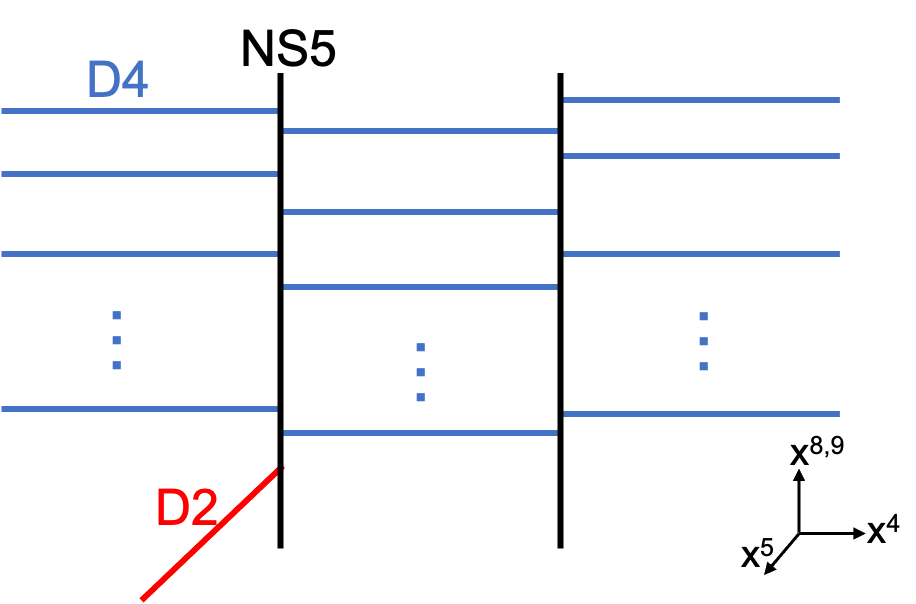}
    \caption{IIA brane picture for $Q$-observable}
    \label{fig:ns5d4d2-2}
\end{figure}

Recall that the above IIB gauge origami configuration of intersecting D3-branes is T-dualized to the IIA branes, where the D3-brane on $\BC^2 _{13}$ becomes a D2-brane (see Figure \ref{fig:ns5d4d2-2}). The D2-brane end on one of the NS5-branes according to the $\BZ_m$-representation that the D3-brane carries. In our case, we took $m=3$ so that there are three NS5-branes, and two of them remain after the ungauging that decompactifies $x^4$. The D2-branes ending on these two NS5-branes correspond to choosing the $\BZ_3$-representation carried by the Chan-Paton space of the D3-brane on $\BC^2 _{13}$ as $\CalR_1$ and $\CalR_0$ in \eqref{eq:gosetup}, for which we call the corresponding surface defects in the $\EN=2$ gauge theory the $Q$-observable and the \textit{dual} $Q$-observable, respectively.\footnote{More generally, if we had $m \geq 3$ then the $\EN=2$ gauge theory is the linear $A_{m-2}$-quiver gauge theory, realized by D4-branes stretched between $m-1$ NS5-branes and infinity. Hence, the number of inequivalent $Q$-observables is $m-1$ in total. We will present the detail of inequivalent $Q$-observables in a separate work.}

In this work, we will not discuss much about the dual $Q$-observable, but focus on the $Q$-observable originating from the D2-brane ending on the NS5-brane on the left (by choosing $\CalR_1$ as in \eqref{eq:gosetup}). In the field theory limit of the IIA setup, the effective theory on the worldvolume of the D2-brane is easily read off to be the $\EN=(2,2)$ gauged linear sigma model with
\begin{itemize}
    \item $U(1)$ gauge group
    \item $N$ chiral multiplets of charge $+1$ and $N$ chiral multiplets of charge $-1$.
\end{itemize}
The flavor symmetry is $U(N)_+ \times U(N)_-$. The $U(N)_+$ subgroup rotating only the chiral multiplets of charge $+1$ descends from a subgroup of the bulk flavor symmetry, identifying the twisted masses for the $U(N)_+$ subgroup with half of the hypermultiplet masses $\left( m^+ _\a \right)_{\a=1} ^N$. The other $U(N)_-$ subgroup is gauged by the bulk gauge field. If there were no bulk gauge coupling ($\qe \to 0$), it is simply to turn on the twisted masses for $U(N)_-$ given by the (four-dimensional) Coulomb moduli $\left( a_\a \right)_{\a=1} ^N$, but in general there is a nontrivial dynamics due to the coupling to the bulk theory in four-dimensions. In the expression \eqref{eq:Qdetail} of the $Q$-observable $Q(x)[\boldsymbol\l]$, this coupling is reflected on its dependence on $\boldsymbol\l$. 

The location of the D2-brane on the $(x^8, x^9)$-plane is precisely given by $x$, which becomes the vacuum expectation value of the complex scalar in the $\EN=(2,2)$ vector multiplet. The complexified FI parameter of the gauged linear sigma model is turned off so that this non-compact Coulomb branch can open up only then.

\subsubsection{Canonical surface defect} \label{subsubsec:canonsurf}
So far, we have argued that the $Q$-observable $Q(x)$ can be viewed as a surface defect generated by coupling a two-dimensional gauged theory in the Coulomb phase. By tuning the vacuum expectation value $x$ to specific values, the complexified FI parameter can be turned on where the gauged linear sigma model is brought to nonlinear sigma model phase. We refer to the induced surface defect as the \textit{canonical} surface defect.

\begin{figure}[h!]
    \centering
    \begin{subfigure}[b]{0.45\textwidth}
       \centering
       \includegraphics[width=\textwidth]{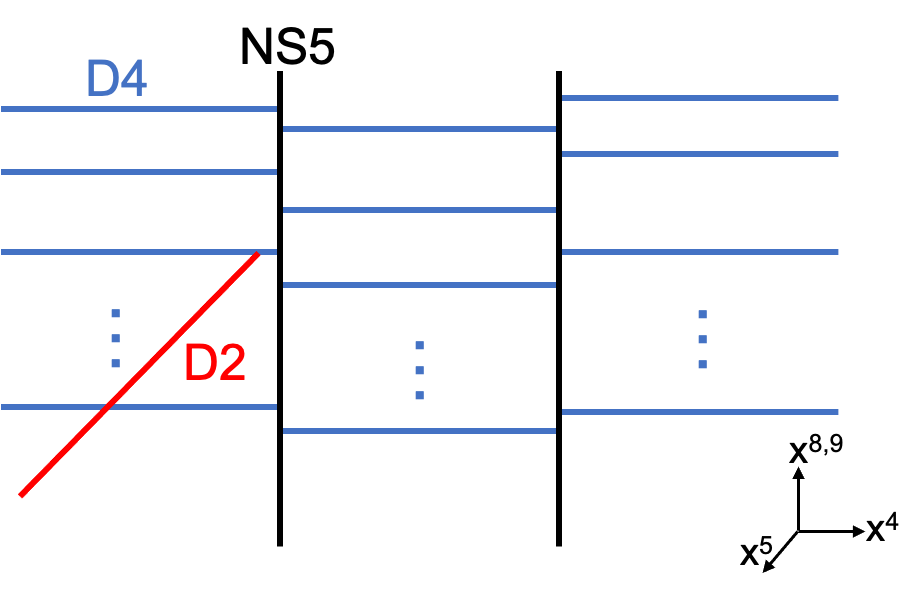}
       \caption{}
       \end{subfigure}
       \quad\quad
\begin{subfigure}[b]{0.45\textwidth}
   \centering
    \includegraphics[width=\textwidth]{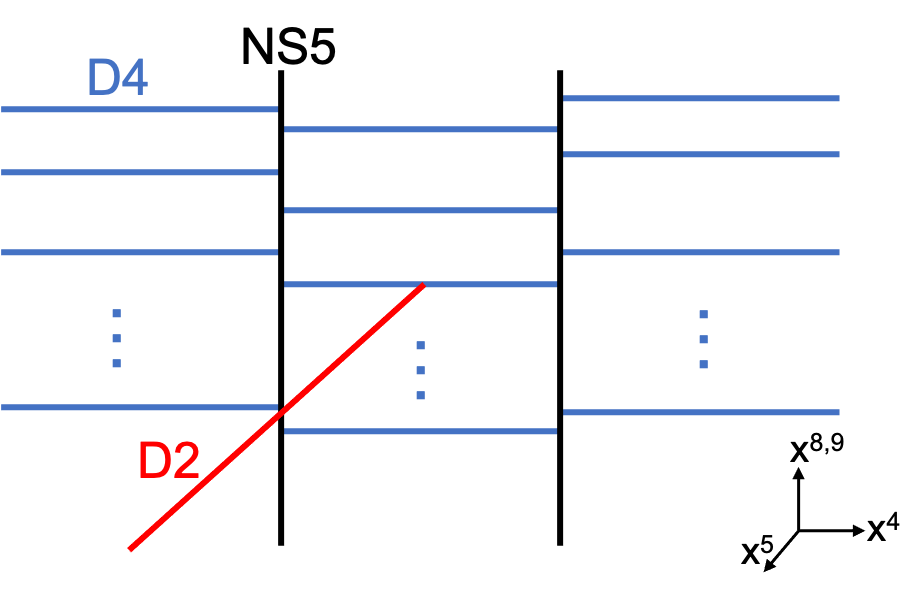}
    \caption{}
\end{subfigure}
         \caption{IIA brane picture for canonical surface defect with (a) positive real FI parameter (b) negative real FI parameter}
    \label{fig:ns5d4d2-1}
\end{figure}

In the IIA brane picture, we bring the D2-brane to one of the $2N$ D4-branes ($N$ from left and $N$ from right) on the $(x^8,x^9)$-plane on top of each other (see Figure \ref{fig:ns5d4d2-1}). This allows to deviate the complexified FI parameter, which we denote by $\log y$, from the locus of the non-compact Coulomb branch by moving the D2-brane along the $x^4$-direction of the D4-brane. The real FI parameter becomes positive or negative depending on whether the D2-brane moves toward left or right, for which there are $N$ choices respectively. In either case, the two-dimensional theory is then described by the nonlinear sigma model with the target being the total space of $\CalO(-1) \otimes \BC^N \to \BP^{N-1}$, where the FI parameter becomes the K\"{a}hler parameter. The $N$ choices of D4-branes that the D2-brane ends on precisely correspond to the choice of the vacuum at infinity among $N$ discrete vacua. Away from the singularity of the non-compact Coulomb branch, these two nonlinear sigma model descriptions (corresponding to positive or negative real FI parameter) are smoothly connected to each other, undergoing the flop transition of the target space across the zero K\"{a}hler parameter.

Integrating out the degrees of freedom residing in the two-dimensional theory produces a surface observable for the four-dimensional theory. In our case, the two-dimensional theory is $A$-twisted and subject to the $\O$-background on $\BC_{\ve_1}$. When the FI parameter is positive, the expression for the surface observable at the fixed point $\boldsymbol\l$ of the instanton moduli space is found as \cite{Jeong:2018qpc}
\begin{align} \label{eq:canon}
    X_\a (y)[\bl] \equiv y^{\frac{\bar{m}^- -\ve_2}{\ve_1} +\frac{N-5}{2}} (y-\qe)^{-\frac{\bar{m}^- -\bar{a} -\ve}{\ve_1}} (y-1) ^{\frac{\bar{m}^+ -\bar{a}}{\ve_1} +1} \sum_{x \in L_\a} y  ^{-\frac{x}{\ve_1}} Q (x)[\bl],
\end{align}
where $L_\a  = m^+ _\a +  \ve_1 \BZ$ is the $\ve_1$-lattice centered at $m^+ _\a$. The complicated perturbative prefactor does not play an important role in this section, but its meaning will be clearer in section \ref{subsec:oper}. The choice of $\a \in \{1,2,\cdots,N\}$ corresponds to the choice of the vacuum at infinity. Because of the $\G$-function in the denominator of the $Q$-observable \eqref{def:Q}, the summation gets non-zero contributions only from non-negative integers. Thus, the vacuum expectation value of the new surface defect observable $X_\a (y)$ is convergent and well-defined in the domain $0<\vert  \qe \vert <1 < \vert y \vert $. Note that the singular locus where the Coulomb branch opens up is exactly $y=1$.

The smooth transition of the nonlinear sigma model descriptions across (but away from) the singular locus $y=1$ is manifest at the level of the vacuum expectation value of this surface observable, using its contour integral representation of Mellin-Barnes type \cite{Jeong:2018qpc}.\footnote{For the partition functions of $\EN=(2,2)$ gauged linear sigma models (without coupling to the four-dimensional theory) expressed as contour integrals of Mellin-Barnes type, see \cite{Hori:2013ika} for instance.} Without describing the detail, we only note that the observable \eqref{eq:canon} can be made convergent also if we alternatively choose the $\ve_1$-lattices to be centered at the Coulomb moduli, $L_\a = a_\a + \ve_1 \BZ$, $\a=1,2,\cdots, N$, due to the other $\G$-function in the denominator of the $Q$-observable \eqref{def:Q}. It is immediate to see the series converge in the domain $0<\vert \qe \vert< \vert y \vert<1$, and they define surface observables there. The Mellin-Barnes integral representation provides the analytic continuation formula between the vacuum expectation values of these surface observables in the two domains. 

Moreover, the surface defect in the domain $0<\vert \qe \vert < \vert y \vert <1$ admits a dual description as coupling the same two-dimensional sigma model in a different way, where the twisted masses for the $U(N)$ subgroup of the flavor symmetry are identified with $m^- _\a$ instead of $m^+ _\a$. In this way, the same surface observable is obtained as a $\ve_1$-lattice sum of the dual $Q$-observables. Then, we can further analytically continue the vacuum expectation value of the surface defect to the remaining domain $0< \vert y \vert<\vert \qe \vert < 1$ after the flop transition \cite{Jeong:2018qpc}. Thus, the parameter $y$-space of the surface defect covers the whole Riemann surface $\EC\setminus S = \BP^1 \setminus \{0,\qe,1,\infty\}$ associated to the class $\CalS$ theory that the two-dimensional theory couples to. The so-obtained surface defect will be called the canonical surface defect, adopting the terminology of \cite{Gaiotto:2011tf}. See also \cite{Dimofte:2010tz} for its relation to the refined BPS invariants in the dual side of the geometric engineering \cite{Katz:1996fh}.

To sum up, the $Q$-observable and the canonical surface defect both can be viewed as coupling the same two-dimensional gauged linear sigma model to the four-dimensional $\EN=2$ theory, but the two-dimensional theory is in the Coulomb phase in the former while it is in the nonlinear sigma model phase in the latter. There are $N$ choices for the vacuum at infinity of the canonical surface defect. At each convergence domain in the parameter $y$-space $\EC$ and at each choice $\a\in \{1,2,\cdots, N\}$ of the vacuum at infinity, the canonical surface defect $X_\a (y)$ is expressed as a $\ve_1$-lattice sum of the $Q$-observables (or dual $Q$-observables in some open patches of $\EC$).

For a chosen convergence domain of $y$ and a chosen vacuum $\a \in  \{1,2,\cdots, N\}$, we can compute the vacuum expectation value of the canonical surface defect. In the limit of turning off the $\O$-background parameter for the $\BC_2$-plane, $\ve_2 \to 0$, the canonical surface defect becomes a local defect in the effective two-dimensional theory on $\BC_2$. Thus, the vacuum expectation value becomes
\begin{align} \label{eq:cansurfns}
    \lim_{\ve_2 \to 0} \Big\langle X_\a(y) \Big\rangle_{\mathbf{a} } = e^{\frac{\widetilde{\EW} (\mathbf{a})}{\ve_2}} \chi_\a (\mathbf{a};y),
\end{align}
where $\widetilde{\EW} (\mathbf{a})$ is again the effective twisted superpotential which is not affected by the presence of the canonical surface defect, and $\chi_\a (\mathbf{a};y)$ is the normalized vacuum expectation value of the canonical surface defect at the chosen vacuum $\a$ in the limit $\ve_2 \to 0$.

\subsection{Monodromy surface defect} \label{subsec:mono}
Next, we turn to another kind of half-BPS surface defects defined by assigning singularity of the gauge field along a surface. The singularity of the gauge field can be modelled by placing the gauge theory on an orbifold \cite{K-T}. Consider the following $\BZ_l$-action on $\BC^2$,
\begin{align}
(z_1,z_2) \mapsto (z_1 ,\zeta z_2),\quad\quad \zeta= \exp \frac{2\pi i}{l}.
\end{align}
Then the quotient space $\BC\times \BC/\BZ_l$ has a singularity along $\BC \times \{0\}$.

In the gauge origami point of view, we are replacing the worldvolume to be $X= \BC_1 \times \BC^3 _{234} / \G_{34} \times \G_{24}$, introducing an additional orbifold singularity from the action of $\G_{24} = \BZ_l$ on the spacetime (the third row of the table \ref{table:IIBori}). Then the Chan-Paton spaces of the D3-branes are representations of $\G_{24}$ as well as $\G_{34}$. We keep our choice for the $\G_{34}$-representations intact, while newly assigning representations of $\G_{24}$. The choice will be packaged into the coloring functions,
\begin{align}
\begin{split}
    &c_{A,i}:N_{A,i} \longrightarrow \{0,1,\cdots, l-1\}, \quad\quad A\in \mathbf{6},\;\; i=0,1,\cdots, m-1.
\end{split}
\end{align}
Once the coloring functions are chosen, the Chan-Paton spaces are decomposed accordingly,
\begin{align}
\begin{split}
    &N_A = \bigoplus_{i=0} ^{m-1} \bigoplus_{\o=0} ^{l-1} N_{A,i,\o} \CalR_i \otimes \fR_\o,\quad\quad A\in \mathbf{6}.
\end{split}
\end{align}
In explicit terms, the $l$-th root of unity $\zeta = \exp \frac{2\pi i}{l}$, viewed as the generator of $\G_{24}=\BZ_l$, is represented as $\O_{N_{A}} = \bigoplus_{i=0} ^{m-1}\bigoplus_{\o=0} ^{l-1}  \eta^\o \mathds{1}_{N_{A,i,\o}}   \in \text{End}(N_{A})$.

The spiked instantons are D$(-1)$-instantons dissolved into the gauge origami worldvolume. The Chan-Paton space of the D$(-1)$-instantons is also a representation of $\G_{34} \times \G_{24}$,
\begin{align}
    K = \bigoplus_{i=0} ^{m-1} \bigoplus_{\o=0} ^{l-1} K_{i,\o} \CalR_i \otimes \fR_\o.
\end{align}
In explicit terms, $\zeta = \exp \frac{2\pi i}{l}$ is represented as $\O_{K} = \bigoplus_{i=0} ^{m-1}\bigoplus_{\o=0} ^{l-1}\eta^\o \mathds{1}_{K_{i,\o}} \in \text{End}(K)$.

The spiked instanton partition function is still given by the expression \eqref{eq:spinstpart}, with the projection to the part invariant under $\G=\G_{34} \times \G_{24}$. Let us consider the case in which we have only one stack of D3-branes for $A=12 \in \mathbf{6}$. The effective worldvolume theory in the presence of the orbifold singularity can be mapped to the $\EN=2$ theory with a prescribed singularity of the gauge field along a surface, defining a monodromy surface defect. From now on, we explain how this mapping is implemented at the level of the moduli space of instantons.

We will consider our main example of the $A_1$-quiver $\EN=2$ gauge theory. For this, we started with $m=3$ and $\dim N_{12,i} = N$ for $i=0,1,2$ and all the other $N_{A,i} = \varnothing$. We turned off the gauge couplings $\qe_1=\qe_2 = 0$ which further restricts to $K_1 = K_2 = \varnothing$. The generalized ADHM construction for the moduli space of spiked instantons just reduces to the usual ADHM construction for the moduli space of instantons, as we have illustrated. In the presence of the additional orbifold singularity, we impose the equivariance condition
\begin{align}
    B_1 = \O_K ^{-1} B_1 \O_K,\quad \zeta B_2 = \O_K ^{-1} B_2 \O_K,\quad   I = \O_K ^{-1} I \O_N,\quad \zeta J = \O_N ^{-1} J \O_K.
\end{align}
It leads to the decomposition of these ADHM data according to the assigned coloring functions $c_{12,i} : N_{12,i} \to \{0,1,\cdots, l-1\}$, $i\in\{0,1,2\}$. Thus we have
\begin{align}
\begin{split}
    &B_{1,\o} : K_\o \to K_\o, \quad 
    B_{2,\o} : K_\o \to K_{\o-1}\\
    &I_\o : N_\o \to K_\o,\quad  J_\o: K_\o \to N_{\o-1}.
\end{split}
\end{align}
The ADHM equation also fractionalizes into $N$ equations
\begin{align}
    0=B_{1,\o-1} B_{2,\o} - B_{2,\o} B_{1,\o} + I_{\o-1} J_\o : K_\o \to K_{\o-1},\quad\quad \o = 0,1,\cdots, N-1.
\end{align}
The moduli space $\mathfrak{M}^{\G_{24}} _{\mathbf{k}}$ of instantons on the orbifold is the locus of these equations modulo the $\bigtimes_{\o = 0} ^{l-1} GL(K_\o)$ action.

As noted above, the $\EN=2$ gauge theory placed on the orbifold singularity can be mapped to the $\EN=2$ gauge theory with a monodromy surface defect. This mapping is realized on their moduli spaces of instantons, in the following way. Let us choose $\t \in \{0,1,\cdots, l-1\}$ and define
\begin{align}
\begin{split}
    &B_1 \equiv B_{1,\t} : K_\t \to K_\t \\
    &B_2 \equiv B_{2,\t+1} B_{2,\t+2} \cdots B_{2,N-1} B_{2,0} \cdots B_{2,\t} : K_\t \to K_\t \\
    &I \equiv \sum_{\o =0} ^{l-1} B_{2,\t+1} \cdots B_{2,\o-1} I_{\o-1} : N \to K_\t \\
    &J \equiv \sum_{\o=0} ^{l-1} J_{\o} B_{2,\o+1} \cdots B_{2,\t} : K_\t \to N.
\end{split}
\end{align}
Then it is straightforward to check that they satisfy the usual ADHM equation, 
\begin{align}
    0 = [B_1, B_2]+IJ : K_\t \to K_\t.
\end{align}
Thus we find a projection map $\r_\t :\mathfrak{M}^{\G_{24}} _{\mathbf{k}} \to \mathfrak{M} _{k_\t}$ for each chosen $\t\in \{0,1,\cdots, l-1\}$. When the equivariant integral over $\mathfrak{M}^{\G_{24}} _{\mathbf{k}}$ is performed, we can first integrate over each fiber of the projection $\r_\t$ and then integrate over the base $\mathfrak{M}_{k_\t}$. Therefore, the first step of this procedure defines an observable of the $\EN=2$ gauge theory without the orbifold singularity, and the partition function is nothing but the vacuum expectation value of this observable. We interpret this observable as the surface observable for the monodromy defect, created after the mapping.

The fixed points in the moduli space of instantons on the orbifold are classified by the colored partitions $\{\hat{\boldsymbol\l}\}$. The projection map $\r_\t$ restricts to the map between the fixed points of the moduli spaces under the symmetry group action, $\r_\t : \{\hat{\boldsymbol\l}\} \to \{\boldsymbol\l\}$, whose image $\boldsymbol\l$ is the partition composed by taking only the $\mathfrak{R}_\t$ part of the colored partition $\hat{\boldsymbol\l}$. In explicit terms, it is given by
\begin{align}
   \boldsymbol\l = \left(\l ^{(\a)}\right)_{\a=1} ^N,\quad\quad  \l^{(\a)} = \left( \l_i ^{(\a)} \right)_{i=1} ^{l\left(\hat{\l} ^{(\a)} \right)}, \quad\quad \l_{i} ^{(\a)} = \left\lfloor \frac{\hat{\l}_i ^{(\a)} +c(\a) -\t +l -1}{l} \right\rfloor,
\end{align}
where $\lfloor \cdots \rfloor$ is the floor function.

Thus, we realize the mapping from the $\EN=2$ gauge theory on the orbifold singularity to the $\EN=2$ gauge theory with a prescribed singularity of the gauge field as the projection map $\r$ of the fixed points of the moduli space of instantons. The gauge origami partition function is, as a sum over these fixed points, decomposed into a summation over the image $\{\boldsymbol\l\}$ of $\r$ and a summation over the fibers $\r^{-1} (\boldsymbol\l)$ of $\r$ at each element $\boldsymbol\l$ in the image. The former gives the usual measure of the $\EN=2$ gauge theory partition function \eqref{eq:a1part}, while the latter precisely gives the surface observable $\Psi_{c} (\mathbf{u}) [\boldsymbol\l]$ at the fixed point $\boldsymbol\l$ that we desired. By construction, the surface observable is labelled by the coloring function $c = (c_{i})_{i=0,1,2}$ and the counting parameters $\mathbf{u} = (u_\o)_{\o=0} ^{l-1}$ of the $\BZ_l$-charges which determine the singular behavior of the gauge field after the mapping.

Let us explicitly construct the surface observable starting from the gauge origami on the orbifold created by the action of $\G_{34} \times \G_{24}$. We will restrict to the case where $l=N$ and the coloring functions are one-to-one, $c_i (\o) = \o$, $\o=0,1,\cdots, N-1$ for all $i=0,1,2$. The choice of $\t \in \{0,1,\cdots, N-1\}$ is only conventional and we set $\t = N-1$. The resulting monodromy surface defect has a singularity of the gauge field which breaks the global gauge symmetry to the maximal torus. We refer such a monodromy surface defect to be \textit{regular}. As noted earlier, the Chan-Paton space of the D3-branes carries a representation of $\G_{24}$ as well as $\G_{34}$. Thus, we write its equivariant Chern character as (here we use hats to distinguish the presence of $\G_{24}$ from its absence after the mapping $\r$)
\begin{align}
\begin{split}
    \hat{N}_{12}  & = \sum_{\o=0}^{N-1} \left( e^{a_{\o}} \CalR_0 + e^{m^+_{\o}+\ve_1+\ve_3} \CalR_1 + e^{m^-_{\o}-\ve_1-\ve_3} \CalR_{2}  \right) \otimes \hat{q}_2^{\o} \fR_{\o}.
\end{split}
\end{align}
The universal sheaf for the instantons on the orbifold, carrying $\BZ_N$-representations, can be written as
\begin{align}
 \hat{S}_{12} = \sum_{\o=0}^{N-1} S_\o \hat{q}_2^{\o} \CalR_0 \otimes \fR_{\o} + q_{13} M_\o^+ \hat{q}_2^{\o} \CalR_1 \otimes \fR_{{\o}} + q_{13}^{-1} M_\o^- \hat{q}_2^{\o} \CalR_{-1} \otimes \fR_{{\o}} ,
\end{align}
where we defined $S_\o \equiv N_\o - P_1 K_\o + q_2 ^{\d_{\o,N-1}}P_1 K_{\o-1}$, $M^\pm _\o \equiv e^{m^\pm _\o}$. Then we get the universal sheaf for the instantons in the absence of the orbifold and the flavor bundle by
\begin{align}
    S = \sum_{\o=0}^{N-1} S_\o = N_{12} - P_{12} K_{12,N-1}, \quad M^\pm = \sum_{\o=0}^{N-1} M_\o^\pm.
\end{align}
Thus, $S_\o$ should be thought of as giving a filtration of the universal sheaf for the instantons with certain constraints on the consecutive quotients (see section \ref{sec:Hecke}).

The gauge origami partition function is still given by \eqref{eq:spinstpart}. Using the projection $\r$, it can be organized into the vacuum expectation value of an observable,
\begin{align} \label{eq:vevmono}
\begin{split}
    \CalZ &= \sum_{\hat{\boldsymbol\l}} \prod_{\o=0} ^{N-1} \hat{\qe}_\o ^{\vert K_\o \vert} E \left[ - \frac{\hat{P}_3 \hat{S}_{12} \hat{S}_{12}^* }{ \hat{P}_{12}^*} \right]^{\BZ_3 \times \BZ_N} \\
    &= \sum_{\boldsymbol\l} \qe^{k} E \left[ \frac{-SS^* + M^+S^* + S(M^-)^*}{P_{12}^*} \right] \times \\  &\quad\quad \times \sum_{\hat{\boldsymbol\l} \in \r^{-1} (\boldsymbol\l)} \prod_{\o=0} ^{N-1} u_\o ^{k_{\o-1} - k_\o} \BE \left[ \sum_{\o_1<\o_2} \frac{S_{\o_1}S_{\o_2}^*}{P_1^*} - \sum_{\o_1\leq \o_2} \frac{ M_{\o_1}^+ S_{\o_2}^* + S_{\o_1}(M_{\o_2}^-)^*}{P_1^*} \right],
\end{split}
\end{align}
where we have redefined the instanton counting parameters by $\hat{\qe}_\o = \frac{u_{\o+1}}{u_\o}$ ($u_{\o+N} = \qe u_\o$), so that $\qe$ is the complexified gauge coupling after the mapping and $(u_\o)_{\o=0} ^{N-1}$ are the monodromy defect parameters. The defect parameters $(u_\o)_{\o=0} ^{N-1}$ are defined up to an overall scaling leaving us only $N-1$ independent parameters. The observable $\Psi(\mathbf{u})$ is only dependent on their ratios and therefore well-defined. As apparent in the expression in the second line, the partition function is the vacuum expectation value of the surface observable $\Psi(\mathbf{u})$, obtained as
\begin{align}
 \Psi(\mathbf{u})[\boldsymbol\l] = \sum_{\hat{\boldsymbol\l} \in \r^{-1} (\boldsymbol\l)} \prod_{\o=0} ^{N-1} u_\o ^{k_{\o-1} - k_\o} \BE \left[ \sum_{\o_1<\o_2} \frac{S_{\o_1}S_{\o_2}^*}{P_1^*} - \sum_{\o_1\leq \o_2} \frac{ M_{\o_1}^+ S_{\o_2}^* + S_{\o_1}(M_{\o_2}^-)^*}{P_1^*} \right].
\end{align}
Let us incorporate the perturbative contribution, writing the full vacuum expectation value as
\begin{align} \label{eq:vevregmono}
\begin{split}
    &\Psi(\mathbf{a};\mathbf{u};\qe) = \Psi^{\text{pert}} (\mathbf{a};\mathbf{u};\qe) \sum_{\{ \boldsymbol\l \} } \qe^{\vert \boldsymbol\l \vert } \, \Psi(\mathbf{u})[\boldsymbol\l] \, E \left[ \frac{-SS^* + M^+S^* + S(M^-)^*}{P_{12}^*} \right], \\
    & \Psi^{\text{pert}} =\qe^{-\frac{\sum_{\o=0} ^{N-1} a_\o ^2 }{2\ve_1\ve_2} +\frac{N\ve_1}{\ve_2} \left( \frac{\bar{m}^--\bar{a}}{\ve_1}-1
    \right)\left( - \frac{N(\bar{m}^- -\bar{a})}{\ve_1} -\frac{\bar{a}}{\ve_1} + \frac{N-1}{2} \right)} (1-\qe)^{ N \left( \frac{\bar{m}^-- \bar{a}}{\ve_1}-1
    \right)\left( \frac{\bar{a}-\bar{m}^+ -\ve_1}{\ve_2} +1
    \right)} \times \\
    &\quad\quad\quad\quad \times \prod_{\o=0}^{N-1} \left( \frac{u_{\o} + u_{\o+1} + \cdots + u_{\o+N-1}}{u_{\o}}\right) ^{-\frac{m^- _{\o} -a_{\o}}{\ve_1} +1}
\end{split}
\end{align}
where we abuse the notation a bit and use the same letter $\Psi$ to denote the vacuum expectation value of the regular monodromy surface defect.

For notational convenience, we will use the double bracket $\llangle \cdots \rrangle$ to indicate the expectation value on the ensemble over the colored partitions $\{\hat{\boldsymbol\l}\}$. Namely, the vacuum expectation value of the regular monodromy surface defect can be written as
\begin{align}
    \Psi(\mathbf{a};\mathbf{u};\qe) = \llangle 1 \rrangle_\mathbf{a} = \left\langle \Psi(\mathbf{u}) \right\rangle_\mathbf{a}.
\end{align}

If we turn off the $\O$-background parameter for the $\BC_2$-plane, $\ve_2 \to 0$, the effective description of the gauge theory is the two-dimensional $\EN=(2,2)$ gauge theory on $\BC_2$. Since the monodromy surface defect is supported on the $\BC_1$-plane, it is a local defect in the effective two-dimensional theory. At the level of the vacuum expectation value, we get
\begin{align} \label{eq:regdefns}
        \lim_{\ve_2 \to 0} \Psi (\mathbf{a};\mathbf{u};\qe) = e^{\frac{\widetilde{\EW} (\mathbf{a};\qe)}{\ve_2}} \psi(\mathbf{a};\mathbf{u};\qe),
\end{align}
where $\widetilde{\EW} (\mathbf{a};\qe)$ is the effective twisted superpotential for the $\EN=(2,2)$ theory, and $\psi(\mathbf{a};\mathbf{u};\qe)$ is the normalized vacuum expectation value of the regular monodromy surface defect in the limit $\ve_2 \to 0$. It is important to note that the surface defect does not affect the effective theory itself, implying $\widetilde{\EW} (\mathbf{a};\qe)$ only gets contributions from the bulk and does not depend on the defect parameters $\mathbf{u}$, since the surface defect becomes a local observable in the effective two-dimensional theory.

\subsection{Parallel surface defects and fractional $Q$-observables} \label{subsub:parallel}
We have constructed half-BPS surface defects of the $\EN=2$ gauge theory with two distinct origins: coupling to two-dimensional sigma model (the $Q$-observable and the canonical surface defect) and assigning singularity of the gauge field (the monodromy surface defect). We may lay these two kinds of surface defects on the same surface, say, the complex plane $\BC_1$. On the transverse plane $\BC_2$, they must be put at the origin on top of each other to preserve the isometry. We will refer to such a configuration of surface defects as being \textit{parallel}.

Let us recall that the $\EN=2$ gauge theory partition function reduces to an ensemble average over partitions $\{\boldsymbol\l\}$, enumerating the fixed points of the moduli space of instantons under the symmetry group action, with certain measure. The $Q$-observable at a given partition $\boldsymbol\l$ was
\begin{align}
    Q(x)[\boldsymbol\l] =  \BE \left[ - \frac{e^x (S^* [\bl] - M^{+*} )}{P_1 ^*} \right].
\end{align}
Now, we place the gauge theory on the orbifold that was used to set up the monodromy defect. In the gauge origami point of view, we replace the gauge origami worldvolume by $X = \BC_1 \times \BC^3 _{234} / (\G_{34} \times \G_{24})$. We noted that the equivariant Chern character of the universal bundle becomes a sum of fractionalized ones, $S = \sum_{\o=0} ^{N-1} S_\o$. It follows that the $Q$-observable factorizes as
\begin{align} \label{eq:bulkQfrac}
    Q(x) [\boldsymbol\l] = \BE \left[ - \frac{e^x \sum_{\o=0} ^{N-1} (S_\o ^* [\hat{\boldsymbol\l}]  -M^{+*} _\o )}{P_1 ^*} \right]  = \prod_{\o=0} ^{N-1} Q_\o (x)[\hat{\boldsymbol\l}],\quad \quad \hat{\boldsymbol\l}\in \r^{-1} (\boldsymbol\l).
\end{align}
Note that even though the individual piece in the product is well-defined only as an observable at a colored partition $\hat{\boldsymbol\l}$, the product does not depend on which $\hat{\boldsymbol\l}$ we take as long as its image under the projection is the given partition, $\r(\hat{\boldsymbol\l}) = \boldsymbol\l$. Each piece $Q_\o (x)$, $\o=0,1,\cdots, N-1$, in the product will be called \textit{fractional $Q$-observable}. 

The fractional $Q$-observable $Q_\o (x)$ is engineered by a fractional D3-brane on $\BC_{13} ^2$ carrying the representation $\mathfrak{R}_\o$ of $\G_{24} = \BZ_N$. The above observation for the $Q$-observable inspires us to consider a stack of $N$ fractional D3-branes on $\BC_{13} ^2$, in which exactly one of them carries the representation $\mathfrak{R}_\o$ of $\G_{24}= \BZ_N$ for each $\o \in \{0,1,\cdots, N-1\}$. Accordingly, the equivariant Chern characters of the Chan-Paton spaces are now
\begin{subequations}
\begin{align}
    \hat{N}_{12} &  = \sum_{\omega=0}^{N-1} \left( e^{a_{\omega}} \hat{q}_2^{\omega} \CalR_0 \otimes \fR_{\omega} + e^{m^+_{\omega}+\ve_1+\ve_3} \hat{q}_2 ^{\o} \CalR_{1} \otimes \fR_{\omega} + e^{m^-_{\omega}-\ve_1-\ve_3} \hat{q}_2^{\omega} \CalR_2 \otimes \fR_{\omega} \right) \nonumber\\
    \hat{N}_{13} & = \sum_{\omega=0}^{N-1} e^{x_\o + \ve_1+\ve_3} \hat{q}_2^{\omega} \CalR_1 \otimes \fR_{\omega}.
\end{align}
\end{subequations}
Just as in the case without $\G_{24}$, the gauge origami gives gauge theories interacting through the two-dimensional intersection along $\BC_1$. By integrating out degrees of freedom on $\BC^2 _{13}$, we get a surface observable of the gauge theory on the orbifold by $\G_{24}$. At the same time, we have seen that the orbifold of $\G_{24}$ can be compensated by the monodromy surface defect $\Psi(\mathbf{u})$. Therefore, the gauge origami partition function translates to the correlation function of two different types of surface defects, possibly with local observables lying on their interface as we will see below.

By a direct computation of the gauge origami partition function, we obtain the vacuum expectation value of what we call the \textit{generalized $Q$-observable}, defined by
\begin{align}\label{def:ER-n}
    \EQ(\mathbf{x}) [\hat{\boldsymbol\l}] \equiv \prod_{\o=0}^{N-1} {Q}_{\o}(x_\o) [ \hat{\boldsymbol\l}]
\end{align}
Recall that $x_\o \in \BR^2$ is the position of the fractional D3-brane wrapping $\BC^2 _{13}$, carrying the $\G_{24}$-representation $\mathfrak{R}_\o$, on the $\BR^2$ transversal to $X$. Instead of the most generic positions $x_\o$, we sometimes set the reference value $x \in \BR^2$ and consider relative $\ve_1$-integral shifts. Namely, we choose $x_\o = x - n_\o \ve_1$ with $n_{N-1} = 0$ and $n_\o \in \BZ$ for $\o=0,1,\cdots, N-2$. When making this choice, we will write
\begin{align}
      \EQ_\bn (x) [\hat{\boldsymbol\l}] \equiv \prod_{\o=0}^{N-1} {Q}_{\o}(x-n_\o \ve_1) [ \hat{\boldsymbol\l}].
\end{align}
Note that we recover the original $Q$-observable if we set $\mathbf{n} = 0$, i.e., $\EQ_{(0,\cdots,0)} (x)[\hat{\boldsymbol\l}] = Q(x)[\boldsymbol\l]$ by the observation \eqref{eq:bulkQfrac}. Thus, the vacuum expectation value of $\EQ_{(0,\cdots,0)} (x)$ in the ensemble over the colored partitions is the correlation function of the monodromy surface defect and the $Q$-observable, $\llangle \EQ_{(0,\cdots,0)} (x) \rrangle =  \left\langle 
Q(x) \Psi(\mathbf{u}) \right\rangle$.

For general $\bn \in \BZ^{N-1}$, the generalized $Q$-observable $\EQ_\mathbf{n} (x)$ can be broken into the $Q$-observable and the local observables located at the origin,
\begin{align}
    \EQ_\bn (x) [\hat{\boldsymbol\l}] = Q(x)[\boldsymbol\l]\, \BE \left[ - \sum_{\o=0} ^{N-1} \sum_{i=0} ^{n_\o-1} e^{x- i \ve_1} S_\o ^* [\hat{\boldsymbol\l}] \right].
\end{align}
We interpret the latter part as the contribution from a $0$-observable at the interface of the two surface defects. A crucial difference between the absence and the presence of the additional $0$-observable is that, only in the former case, the two surface defects can be arbitrarily separated on the topological $\BC_2$-plane in the $\ve_2 \to 0$ limit. In other words, the correlation function completely factorizes in this limit,
\begin{align} \label{eq:factor}
    \lim_{\ve_2 \to 0} \llangle \EQ_{(0,\cdots, 0)} (x) \rrangle = \lim_{\ve_2 \to 0}  \left\langle 
Q(x) \Psi(\mathbf{u}) \right\rangle = e^{\frac{\widetilde{\EW} (\mathbf{a})}{\ve_2}} Q(\mathbf{a};x) \psi (\mathbf{a};\mathbf{u}),
\end{align}
where we abuse the notation on the right hand side a bit, and use the same letter $Q$ to indicate the normalized vacuum expectation value of the $Q$-observable in the limit $\ve_2 \to 0$ \eqref{eq:Qobsns}, and $\psi(\mathbf{a};\mathbf{u})$ is the normalized vacuum expectation value of the regular monodromy surface defect in the limit $\ve_2 \to 0$ \eqref{eq:regdefns}. The complete factorization will be an important feature of the parallel surface defect configuration leading to the Hecke eigensheaf property, as we explain later in section \ref{subsubsec:hecke}.

Similar to the construction of the canonical surface defect from the $Q$-observable, we consider the Fourier transform of the generalized $Q$-observable,
\begin{align}
     \sum_{\mathbf{x} \in \mathbf{L}} \prod_{\o=0} ^{N-1} y_\o  ^{-\frac{x_\o}{\ve_1}} \llangle \EQ (\mathbf{x}) \rrangle,
\end{align}
where we introduced the fugacities $y_\o \in \BC$. Here, we are summing $\mathbf{x}$ over an $N$-dimensional $\ve_1$-lattice $\mathbf{L}$. When we specialize to $\EQ_\bn (x)$ where all the $N$ $\ve_1$-lattices are chosen to be the same $L$, the Fourier transform of the vacuum expectation value is written as
\begin{align} \label{eq:G-F-N}
\begin{split}   \U(\mathbf{a};\mathbf{u},\boldsymbol{\m};\qe,y) = \U^{\text{pert}} (\mathbf{a};\mathbf{u};\boldsymbol\m;\qe,y) \sum_{x \in L} y^{-\frac{x}{\ve_1}}   \sum_{\bn \in \BZ^{N-1}} \prod_{\o=0}^{N-1} \m_\o ^{ n_{\overline{\o-1}} -n_\o} \llangle  \EQ_{\bn}(x) \rrangle,
\end{split}
\end{align}
where
\begin{align} \label{eq:correlpert}
\begin{split}
         \U^{\text{pert}}  &=  \kq^{ -\frac{\sum_{\o=0}^{N-1} a_\o ^2 }{2\ve_1 \ve_2}+ \frac{N\ve_1}{\ve_2}  \left(-\frac{N(\bar{m}^--\bar{a})}{\ve_1} -\frac{\bar{a}}{\ve_1} +\frac{N-1}{2}\right) \left( \frac{\bar{m}^--\bar{a}}{\ve_1}-1 \right)}  (1-\qe)^{ N \left( \frac{\bar{m}^--\bar{a}}{\ve_1}-1 \right) \left( \frac{\bar{a}-\bar{m}^+ -\ve_1}{\ve_2} +\frac{1}{N} +1 \right)} \times \\
    & \quad \times  y^{\frac{\bar{m}^-}{\ve_1} -\frac{N+3}{2}  } (y-\kq)^{-\frac{ \bar{m}^- -\bar{a} }{\ve_1} +1}  (y-1)^{\frac{\bar{m}^+-\bar{a} - \frac{\ve_2}{N} }{\ve_1} +1  } \times \\
     &  \quad \times \prod_{\o=0} ^{N-1} \left( \frac{u_{\o}+u_{\o+1}+\cdots +u_{\o+N-1}}{u_{\o}} \right) ^{-\frac{m^- _\o -a_{\o}}{\ve_1}+1} \frac{\m_{N-1}}{\m_\o} \times \left( \frac{\m_{N-1}}{\m_0} \right)^{\frac{\ve_2}{\ve_1}},
    \end{split}
\end{align}
where we made a reparametrization of fugacities by
\begin{align} \label{eq:changevar}
    y_\o = \frac{\mu_{\o+1}}{\mu_\o}, \quad \mu_{\o+N} = y \mu_\o, \quad \o=0,1,\cdots, N-1.
\end{align}
The new fugacities $(\m_\o)_{\o=0} ^{N-1}$ are ambiguous by an overall scaling, but the vacuum expectation value $\U(\mathbf{a};\mathbf{u},\boldsymbol{\m};\qe,y)$ only depends on their ratios and thus admits a well-defined expression in $y$ and $\boldsymbol\m$. Note that the 0-th Laurent coefficient of the fugacities $(y_\o)_{\o=0} ^{N-2}$ gives precisely the canonical surface defect $X(y)$ as a result of the summation over $x \in L$, upon a proper choice of the $\ve_1$-lattice $L$. Namely, the 0-th Laurent coefficient is the correlation function of the regular monodromy surface defect $\Psi(\mathbf{u})$ and the canonical surface defect $X(y)$, i.e., $\langle X(y) \Psi(\mathbf{u}) \rangle$. This $0$-th Laurent coefficient is distinguished because in the limit $\ve_2 \to 0$, the correlation function completely factorizes just as \eqref{eq:factor}:
\begin{align} \label{eq:factor2}
    \lim_{\ve_2 \to 0} \Big\langle X_\a (y) \Psi(\mathbf{u}) \Big\rangle_{\mathbf{a}} = e^{\frac{\widetilde{\EW}(\mathbf{a})}{\ve_2}} \chi_\a (\mathbf{a};y) \psi(\mathbf{a};\mathbf{u}),
\end{align}
where $\chi_\a (\mathbf{a};y)$ is the normalized vacuum expectation value of the canonical surface defect at the vacuum $\a\in \{1,2,\cdots,N\}$ in the limit $\ve_2 \to 0$ \eqref{eq:cansurfns}, and $\psi(\mathbf{a};\mathbf{u})$ is the normalized vacuum expectation value of the regular monodromy surface defect in the limit $\ve_2 \to 0$ \eqref{eq:regdefns}. Again, this property will be important to derive the Hecke eigensheaf property from the $\EN=2$ theory framework.

\section{TQ equations and opers} \label{sec:TQoper}
In this section, we derive a difference equation satisfied by the vacuum expectation value of the $Q$-observable (introduced in section \ref{subsubsec:qobs}), and a differential equation satisfied by the vacuum expectation value of the canonical surface defect (introduced in section \ref{subsubsec:canonsurf}). These equations are called the TQ equation and the oper differential equation, respectively. In particular, our construction leads to the ${}^L G_\BC$-opers, as differential operators defined on $\EC$, and their solutions expressed in $\EN=2$ gauge theoretical terms. This generalizes the construction of \cite{Jeong:2018qpc} to arbitrary $N$.

Recall that the gauge origami construction of the defects involves two stacks of D3-branes in the IIB string theory on $X \times \BR^2$, where $X = \BC^2 _{12} \times \BC^2 _{34}/\G_{34}$ or $X = \BC_1 \times \BC^3 _{234} /\G_{34}\times \G_{24}$. To derive the difference/differential equations satisfied by their vacuum expectation values, we introduce an additional D3-brane supported on $\BC^2_{34}$. Since its worldvolume $\BC^2 _{34}$ is transverse to the worldvolume of the gauge theory $\BC^2 _{12}$ in $X$, they give rise to a BPS local observable. These local observables are called $qq$-characters \cite{Nikita:I}. Note that the $qq$-characters are labeled by their positions $x$ in the $\BR^2$-plane transverse to the gauge origami worldvolume $X$.

The $qq$-characters encode the analytic properties of the gauge theory partition functions by their regularity property \cite{Nikita:II,NaveenNikita}. Namely, the gauge theory correlation functions involving the $qq$-characters are regular as functions of $x$. It follows that, when the $qq$-characters are expanded at large $x$, the coefficients of negative powers of $x$ have vanishing vacuum expectation values. These coefficients are combinations of local chiral observables in the gauge theory in the presence of surface defects \cite{Jeong:2019fgx}. In turn, the analytic properties of the $qq$-characters encode the chiral ring relations of the gauge theory with surface defects, expressed as constraints on their vacuum expectation values \cite{Nikita:IV,Jeong:2017pai,Jeong:2017mfh}.

\subsection{Quantum TQ equations} \label{eq:subsubop}
Let us begin with the $qq$-character in the presence of the $Q$-observable. In the gauge origami setup, we insert an additional D3-brane on $\BC^2 _{34}$ carrying the representation $\CalR_0$ of $\G_{34} = \BZ_3$. We will turn off the gauge couplings $\qe_1 = \qe_2 = 0$ in the end, so other choices of representation would not allow the instantons to dissolve into the worldvolume of the new D3-brane, giving rather trivial results. For our choice, the gauge origami is represented by the following equivariant Chern characters of the Chan-Paton spaces:
\begin{align}\label{eq:gosetup}
\begin{split}
    N_{12} & = \sum_{\alpha=1} ^N e^{a_\alpha} \cdot \CalR_0 + \sum_{\alpha=1} ^N e^{m_\alpha^+ - \ve_4} \cdot \CalR_1 + \sum_{\alpha=1}^N e^{m_\alpha^- + \ve_4} \cdot \CalR_{2} \\
    N_{13} & = e^{x'+\ve_1+\ve_3} \cdot \CalR_1 \\
    N_{34} & = e^x \cdot \CalR_0.
\end{split}
\end{align}
By computing the partition function by applying \eqref{eq:spinstpart}, we get a correlation function of the $qq$-character and the $Q$-observable which is regular in $x$ by the compactness of the moduli space of spiked instantons. More explicitly, we get
\begin{align}
    \Big\langle \ET(x,x') Q(x') \Big\rangle =\left\langle(x'-x) Q(x') {\EY}(x + \ve)+ \fq (x'-x-\ve_2) Q(x') \frac{P^+ (x) P^- (x+\ve)}{{\EY}(x)}  \right\rangle,
\end{align}
where $\ET(x,x')$ is a polynomial in $x$ of degree $N+1$ obtained straightforwardly by expanding the right hand side in $x \to \infty$, up to the 0-th order in $x$. 

Here, the degree $N$ polynomials $P^\pm (x) = \prod_{\a=1}^N (x-m^\pm _\a)$ encode the mass parameters of the hypermultiplets and the $\EY$-observable generates the local chiral observables $\text{Tr}\, \phi^k$, where $\phi$ is the complex adjoint scalar in the $\EN=2$ vector multiplet, by its Laurent coefficients,
\begin{align}
    \EY(x) [\boldsymbol\l] \equiv x^N \exp \left[ - \sum_{k=1} ^N \frac{1}{k x^k} \text{Tr}\, \phi^k [\boldsymbol\l] \right].
\end{align}
Hence, by construction, the coefficients of the polynomial $\ET(x,x')$ are given by the local chiral observables $\Tr\,\phi^k$ and the mass parameters.

By setting $x=x'$ and $x=x'-\ve_2$, we obtain respectively
\begin{align}
\begin{split}
    &\Big\langle \ET(x,x) Q(x) \Big\rangle = -\ve_2 \qe \left\langle Q(x) \frac{P^+ (x) P^- (x+\ve)}{{\EY}(x)} \right\rangle = -\ve_2 \qe P^- (x+\ve) \Big\langle Q(x-\ve_1) \Big\rangle\\
    & \Big\langle \ET(x-\ve_2,x) Q(x)  \Big\rangle  = \ve_2 \Big\langle Q(x) \EY (x+\ve_1) \Big\rangle = \ve_2 P^+ (x+\ve_1)
 \Big\langle Q(x+\ve_1) \Big\rangle.
\end{split}
\end{align}
By subtracting the first equation from the second, we get
\begin{align} \label{eq:qTQ}
     \Big\langle P^+ (x+\ve_1)   Q(x+\ve_1)  + \qe P^- (x+\ve)  Q(x-\ve_1) -  T(x) Q(x) \Big\rangle = 0,
\end{align}
where we defined the polynomial $T(x)$ of degree $N$ by
\begin{align}
\begin{split}
    T(x) &= \frac{\ET(x-\ve_2,x) - \ET(x,x)}{\ve_2} \\&= (1+\qe) x^N + N\left(\ve_1 -\bar{a} +\qe(\bar{a}-\bar{m}^+ - \bar{m}^- +\ve) \right) x^{N-1} + \cdots.
\end{split}
\end{align}
We call this equation \eqref{eq:qTQ} the \textit{quantum TQ equation}.

By construction, the coefficients of the polynomial $T(x)$ are the local observables $\Tr\, \phi^k$ (and the mass parameters) of the gauge theory. Even though these operators are well-defined BPS objects for all the non-negative integers $k \in \BZ_{\geq 0}$, only $N-1$ of them are mutually independent in the case of the $SU(N)$ gauge theory since they are subject to chiral ring relations.\footnote{The chiral ring relations relating the observables $\text{Tr}\, \phi^k$ with $k>N$ to the ones with $k \leq N$ can be derived by using the $qq$-characters \cite{Jeong:2019fgx}.} $T(x)$ can be regarded as the generating function of $N-1$ independent combinations of the observables $\Tr\, \phi^k$, $k=2,3,\cdots, N$. 

Recall that, due to the $\Omega$-background, any local operator has to be located at the origin of the worldvolume $\BC^2_{12}$ not to ruin the spacetime isometry. By the same token, the surface defect $Q(x)$, which wraps the first complex plane $\BC_1$, is located at the origin of the second complex plane $\BC_2$. Therefore, when both the $Q$-observable and the local observable $T(x)$ are inserted, $T(x)$ lies on top of $Q(x)$ becoming a defect local observable. The above equation \eqref{eq:qTQ} implies this defect local operator is given by a certain shift operator acting on the Coulomb parameter $x$ of the $Q$-observable, which is the vacuum expectation value of the complex scalar of the $\EN=(2,2)$ vector multiplet. Differently put, the local observable $T(x)$ lying on top of the surface defect $Q(x)$ initiates a defect fusion, resulting in two $Q$-observables with the Coulomb parameters shifted by $\ve_1$.

If we turn off the $\Omega$-background parameter $\ve_2 \to 0$ for the complex plane $\BC_2$ transverse to the plane $\BC_1$ of the surface defect $Q(x)$, the observables $T(x)$ and $Q(x)$ can be arbitrarily separated on $\BC_2$. The cluster decomposition implies the correlation function factorizes without any contact term. Thus, the equation \eqref{eq:qTQ} leads to
\begin{align} \label{eq:baxTQ}
    P^+ (x+\ve_1) Q(\mathbf{a};x+\ve_1) +\qe P^- (x+\ve_1) Q(\mathbf{a};x-\ve_1) = T(\mathbf{a};x) Q(\mathbf{a};x),
\end{align}
where we abuse the notation a bit and use the same letter $T$ and $Q$ for the respective normalized vacuum expectation values in the limit $\ve_2 \to 0$. In a separate work, we will give a proper interpretation of this equation as the Baxter TQ equation for the quasi-periodic XXX spin chain, evaluated at an eigenstate.\footnote{Along the line of \cite{Lee:2020hfu, Jeong:2021bbh}, we will present a $\EN=2$ gauge theoretical account of the bispectral duality between the Gaudin model and the XXX spin chain with bi-infinite modules. See also \cite{Gorsky:1997jq, Mironov:2012ba, Gaiotto:2013bwa} for other studies on classical and quantum duality between the two integrable systems.} Namely, we would like to view $T(x)$ and $Q(x)$ in \eqref{eq:baxTQ} as \textit{eigenvalues} of the monodromy operator and the $Q$-operator for a given common eigenstate. In this sense, the quantum TQ equation \eqref{eq:qTQ} is the $\ve_2 \neq 0$ uplift of the Baxter TQ equation, explaining its nomenclature.

\subsection{Canonical surface defect and opers} \label{subsec:oper}
To make a direct contact with the geometric Langlands correspondence and the quantization of Hitchin integrable system that we introduced, let us take a $\ve_1$-lattice sum of the quantum TQ equation \eqref{eq:qTQ}. Then the quantum TQ equation becomes an operator-valued $N$-th order differential equation in $y$
\begin{align} \label{eq:FTTQ}
    0 = \left[y P^+ (-\ve_1 y \p_y )  + \frac{\qe}{y} P^- (-\ve_1 y \p_y + 2\ve_1+\ve_2) - T(-\ve_1 y \p_y) \right] \sum_{x \in L} y^{-\frac{x}{\ve_1}} Q(x).
\end{align}
Here, $y\p_y$ evidently commutes with itself so that the polynomials in $y\p_y$ are well-defined. The coefficient of the highest-order differential $\p_y ^{N}$ is not an operator but a simple function in $y$. Thus, after multiplying the perturbative prefactor for the canonical surface defect \eqref{eq:canon} further, the equation is brought into a more canonical form as
\begin{align} \label{eq:Qoper}
\begin{split}
    0&= \left[\p_y ^{N}  + \sum_{k=2} ^{N} t_k (y) \p_y ^{N-k} \right] {X}(y) \\
    &= \left[ \p_y ^N + t_2 (y) \p_y ^{N-2} +  \cdots + t_{N-1} (y) \p_y + t_N (y) \right]{X}(y).
\end{split}
\end{align}
Note that the perturbative part of the canonical surface observable precisely makes the coefficient of the next-to-highest $(N-1)$ order differential to vanish. We also point out that this equation holds as an operator in the $\EN=2$ gauge theory, so that it does not know about the choice of the vacuum at infinity. This includes the vacuum for the two-dimensional $\EN=(2,2)$ theory on the canonical surface defect, determined by the choice of the lattice $L$, as well as the Coulomb moduli $\mathbf{a}$ for the four-dimensional $\EN=2$ theory. Indeed, the operator-valued equation \eqref{eq:Qoper} does not depend on either of them (this is why the subscript of $X(y)$ was omitted). Rather, at each choice of the vacuum, the vacuum expectation value of \eqref{eq:Qoper} can be taken to yield nontrivial relations of individual correlation functions of the observables involved in the equation.
 
It is straightforward to see that the only singularities of the operator-valued meromorphic functions $t_k (y)$, $k=2,3,\cdots, N$ are at $S=\{0,\qe,1,\infty\}$, which we call the \textit{marked points} as before. The highest-order Laurent coefficients of $t_k (y)$ are determined solely by the multiplication of the perturbative prefactor in a simple manner; in particular, the order is given by $\text{ord}_{p} (t_k (y)) = -k$ at each marked point $p\in S$. Similar to the quantum TQ equation, this equation shows how the local observables $t_k (y)$ translate to defect local observables on the canonical surface defect ${X}(y)$. We call this equation the \textit{quantum oper} ($\ve_2 \neq 0$ uplift of the oper), for the reason to be explained from below.\footnote{The quantum opers defined here are \textit{not} the $q$-opers, which are another deformation of opers to difference equations. See \cite{Aganagic:2017smx, Koroteev:2018jht} for more details about the $q$-opers and the associated (quantum) $q$-Langlands correspondence. The $q$-opers and the (quantum) $q$-Langlands correspondence are supposed to correspond to the uplift of our four-dimensional $\EN=2$ theory framework to the five-dimensional $\EN=1$ theory compactified on a circle.}\footnote{We remark that the 2d/4d coupled system giving the canonical surface defect can alternatively be constructed by partially Higgsing $\EN=2$ gauge theory with larger gauge group \cite{Hanany:2004ea}. In the point of view of the vertex algebra at the junction (recall section \ref{sec:bbd}), imposing the Higgsing condition precisely corresponds to constraining one of the vertex operators to be degenerate \cite{Nikita:IV,Jeong:2018qpc}. See \cite{Teschner:2010je,Poghosyan:2016mkh,Nikita:IV,Jeong:2017mfh,Jeong:2017pai} for other works studying the quantum oper equation \eqref{eq:Qoper} as the null-vector decoupling equation \cite{Belavin:1984vu}.}

When we turn off the $\Omega$-background parameter $\ve_2 \to 0$ for $\BC_2$, the local observables $t_k (y)$ and the canonical surface observable ${X}(y)$, which is local on $\BC_2$, can be arbitrarily separated on the topological plane $\BC_2$. Therefore, in the quantum oper equation \eqref{eq:Qoper}, their correlation function factorizes into the simple product of the vacuum expectation values of each,
\begin{align}
    \lim_{\ve_2 \to 0} \left\langle t_k (y) \p_y ^{N-k} {X} _\a (y) \right\rangle_{\mathbf{a}} = e^{\frac{\widetilde{\EuScript{W}}(\mathbf{a})}{\ve_2}} \, t_k (\mathbf{a};y) \, \p_y ^{N-k} \chi_\a (\mathbf{a};y),
\end{align}
where we slightly abuse the notation on the right hand side and use the same letter $t_k$ to indicate their normalized vacuum expectation values in the limit $\ve_2 \to 0$, and $\chi_\a (\mathbf{a};y)$ is the normalized vacuum expectation value of the canonical surface defect in the limit $\ve_2 \to 0$, at the vacuum $\a \in \{1,2,\cdots, N\}$. $\widetilde{\EuScript{W}}(\mathbf{a})$ is the twisted superpotential for the effective two-dimensional theory on $\BC_2$, which does not get contribution from the canonical surface defect and does not depend on the defect parameter $y$. Thus, the $y$-derivative does not act so that it simply factors out from the equation. In turn, the equation in this limit becomes
\begin{align} \label{eq:oper}
    0 = \left[ \p_y ^N + t_2 (\mathbf{a};y) \p_y ^{N-2} + \cdots + t_{N-1} (\mathbf{a};y) \p_y + t_N(\mathbf{a};y) \right] \chi(\mathbf{a};y) \equiv \r_\mathbf{a} \chi(\mathbf{a};y).
\end{align}
Namely, $t_k (\mathbf{a};y)$'s are now truly scalar-valued meromorphic functions in $y$ whose Laurent coefficients are given by the mass parameters $(m^\pm _\a)_{a=1}^N$ of the hypermultiplets and the normalized vacuum expectation values of the local observables $\Tr\, \phi^k$, $k=2,3,\cdots, N$, in the limit $\ve_2 \to 0$, $\lim_{\ve_2 \to 0} \left\langle\text{Tr}\,\phi^k \right\rangle_{\mathbf{a}}$. We denote these $N-1$ independent combinations of $\lim_{\ve_2 \to 0} \left\langle\text{Tr}\,\phi^k \right\rangle_{\mathbf{a}}$ that appear as the Laurent coefficients by ${E}_k (\mathbf{a})$, $k=2,3,\cdots, N$. Thus, the meromorphic functions $t_k (\mathbf{a};y)$ simply multiply $\chi(\mathbf{a};y)$ or its $y$-derivatives from the left. The above equation implies the normalized vacuum expectation value $\chi(\mathbf{a};y)$ of the canonical surface defect in the limit $\ve_2 \to 0$ is annihilated by the differential operator $\r_\mathbf{a}$ in the square bracket, which we call the \textit{oper}. We also sometimes call $\chi(\mathbf{a};y)$ the \textit{oper solution}. Since there are $N$ discrete vacua of the surface defect theory, we actually obtain $N$ independent oper solutions $\chi_\a (\mathbf{a};y)$, $\a=1,2,\cdots, N$, which span the solution space of the $N$-th order differential equation. This is the generalization of the construction in \cite{Jeong:2018qpc} to arbitrarily higher ranks.

These opers which arise from our gauge theoretical construction are precisely the $SL(N)$-opers developed in \cite{DS,Ben-Zvi:1999snd,BD1,BD2}, defined on the genus-0 curve $\BP^1$ with regular singularities at marked points $S =\{0,\qe,1,\infty\}$. More specifically, the monodromies at the regular singularities at $S$ are fixed, obtained straightforwardly by solving \eqref{eq:FTTQ} locally near each marked point. These monodromies are semisimple, given by
\begin{align} \label{eq:mono}
\begin{split}
    &M_0 = \text{diag}\left(  (-)^{N-1} e^{-2\pi i \frac{m ^- _\a -\bar{m}^-}{\ve_1}} \right)_{\a=1} ^N \quad\quad M_\infty = \text{diag}\left(  (-)^{N-1} e^{2\pi i \frac{m^+ _\a - \bar{m}^+}{\ve_1}} \right)_{\a=1} ^N \\
    &M_\qe = \text{diag}\left( e^{-2\pi i \frac{\bar{m}^- -\bar{a}}{\ve_1}} ,\cdots,   e^{-2\pi i \frac{\bar{m}^- -\bar{a}}{\ve_1}} ,  e^{2\pi i (N-1) \frac{\bar{m}^- -\bar{a}}{\ve_1}} \right) \\
    &M_1 = \text{diag}\left( e^{2\pi i \frac{\bar{m}^+ -\bar{a}}{\ve_1}} ,\cdots,   e^{2\pi i \frac{\bar{m}^+ -\bar{a}}{\ve_1}} ,  e^{-2\pi i (N-1) \frac{\bar{m}^+ -\bar{a}}{\ve_1}} \right),
\end{split}
\end{align}
where $M_p$ is the monodromy at the marked point $p \in S$ valued in the conjugacy class in $SL(N)$. Namely, the monodromies at $0$ and $\infty$ are maximal, while the monodromies at $\qe$ and $1$ are minimal. As apparent from the expression, there are $2N$ degrees of freedom for such monodromy data, and they are fixed precisely by the $2N$ mass parameters of the hypermultiplets $\left( m^ \pm _\a \right)_{\a=1} ^N$. Just as we fix the mass parameters in the $\EN=2$ gauge theory once and for all, we will always regard these monodromy data to be fixed and being included in the choice of the marked points, (somewhat loosely) denoting them altogether by $S$. We denote the space of such opers by $\text{Op}_{SL(N)} (\BP^1;S)$, so that $\r_{\mathbf{a}} \in \text{Op} _{SL(N)} (\BP^1;S)$. 

Indeed, we can bring \eqref{eq:oper} into a meromorphic flat connection on a rank $N$ parabolic vector bundle over $\BP^1$, whose horizontal sections are $(N-1)$-jets of the oper solution $\chi(y)$,
\begin{align} \label{eq:operNN}
    0 = \left[ \p_y + \begin{pmatrix} 0 & t_2 & t_3 & \cdots &\cdots & t_N \\ -1 & 0 & 0 & \cdots & \cdots & 0 \\ 0 & -1 & 0 & \cdots &\cdots  & 0 \\ \cdots & \cdots & \cdots &  \cdots &\cdots & \cdots \\ 0 & 0 & 0 & \cdots & -1 & 0  \end{pmatrix} \right]\begin{pmatrix} \p_y ^{N-1} \chi \\ \p_{y} ^{N-2} \chi \\ \vdots \\ \p_y \chi \\ \chi \end{pmatrix}.
\end{align}
We can expand the meromorphic functions $t_k (y)$ in Laurent polynomials at each marked point $z_i \in S$. Then, by making a upper triangular gauge transformation, we may bring the meromorphic connection into the form of
\begin{align}
   \p_y + \frac{1}{y-z_i}  \begin{pmatrix} * & * & * & \cdots &\cdots & *\\ -1 & * & * & \cdots & \cdots & * \\ 0 & -1 & * & \cdots &\cdots  & * \\ \cdots & \cdots & \cdots &  \cdots &\cdots & \cdots \\ 0 & 0 & 0 & \cdots & -1 & *  \end{pmatrix},
\end{align}
where $*$ is holomorphic in the neighborhood of $z_i$. This is the $SL(N)$-oper with regular singularities at $S$ (and the prescribed monodromies \eqref{eq:mono} there), in the form appearing in \cite{Frenkel:2005pa}. Taking account for holomorphic change of variables, we also find that the oper solution is valued in $\left(-\frac{N-1}{2}\right)$-differentials, $\chi \in K_\EC ^{-\frac{N-1}{2}}$, and thus $\r _{\mathbf{a}}: K_\EC ^{-\frac{N-1}{2}} \to K_\EC ^{\frac{N+1}{2}} \otimes \CalO (N S)$.

Recall that the space of opers $\text{Op}_{SL(N)} (\BP^1;S) \subset \EM_{\text{loc}} (SL(N),\BP^1;S) $ is a complex Lagrangian submanifold in the moduli space of parabolic local systems. In particular, in our case of the sphere with four marked points, $\dim \text{Op}_{SL(N)} (\BP^1;S) = \frac{1}{2} \dim \EM_{\text{loc}} (SL(N),\BP^1;S)  =N-1$. By construction, the opers $\r_\mathbf{a}$ \eqref{eq:oper} constructed from the canonical surface defect of the $\EN=2$ gauge theory indeed form a $(N-1)$-dimensional affine space spanned by the normalized vacuum expectation values of the chiral observables in the limit $\ve_2\to 0$, $\lim_{\ve_2 \to 0} \left\langle \text{Tr}\,\phi^k \right\rangle _{\mathbf{a}}$, $k=2,3,\cdots, N$. In turn, the Coulomb moduli $\mathbf{a}$ (more precisely, the dimensionless parameters $\frac{\mathbf{a}}{\ve_1}$) of the $\EN=2$ gauge theory provide holomorphic coordinates on $\text{Op}_{SL(N)} (\BP^1;S)$.

\begin{figure}
    \centering
    \includegraphics[width=0.7\textwidth]{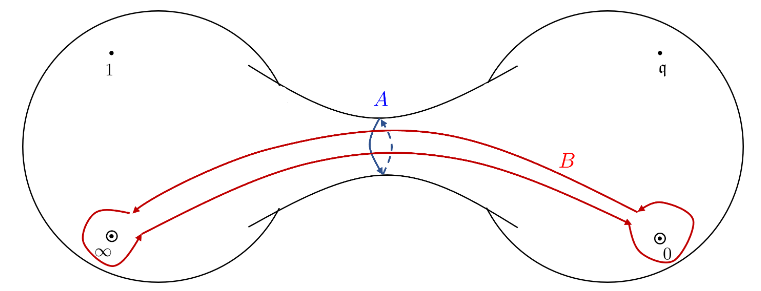}
    \caption{$\BP^1$ with four marked points $S=\{0,\qe,1\,\infty\}$. The double circles represent the maximal marked points at $0$ and $\infty$, while the dots denote the minimal marked points at $\qe$ and $1$. The $A$-cycle is drawn in blue and the $B$-cycle is drawn in red.}
    \label{fig:sphere}
\end{figure}

To see more clearly how the Coulomb moduli parameterize the space of opers, we introduce a Darboux coordinate system on the moduli space of parabolic local systems $\EM_{\text{loc}} (SL(N),\BP^1;S) $. We use the higher-rank generalization \cite{Jeong:2018qpc} of the Nekrasov-Rosly-Shatashvili coordinates \cite{Nekrasov:2011bc} (higher-rank generalization of the complexified Fenchel-Nielsen coordinates \cite{FenchelNielsen+2002}) on $\EM_{\text{loc}} (SL(N),\BP^1;S)$.\footnote{See also \cite{Hollands:2017ahy} for the spectral network construction \cite{Gaiotto:2012rg, Hollands:2013qza} of the higher-rank Fenchel-Nielsen coordinates.} Note that the monodromy defines an isomorphism
\begin{align}
    \EM_{\text{loc}} (SL(N),\BP^1;S) \xrightarrow{\sim} \text{Hom}\left(\pi_1 (\BP^1\setminus S) ,SL(N)\right)/SL(N),
\end{align}
where the monodromies at the marked points are fixed to be \eqref{eq:mono}. Other than these four loops enclosing only single marked point, there are two independent loops in $\pi_1 (\BP^1\setminus S)$; the $A$-cycle enclosing two marked points $0$ and $\qe$, and the $B$-cycle enclosing two marked points $0$ and $\infty$ (see Figure \ref{fig:sphere}). Now, half of the Darboux coordinates $(\boldsymbol\a_\a)_{\a=1} ^{N-1}$ are defined by the eigenvalues of the monodromy along the $A$-cycle. Their Darboux pairs $(\boldsymbol\b_\a)_{\a=1} ^{N-1}$ can be properly defined from the trace invariants of the monodromy along the $B$-cycle, for which we do not present the detail here (the exact expressions can be found in \cite{Jeong:2018qpc}). 

In our $\EN=2$ theoretical construction, the oper solutions $\chi_\a (\mathbf{a};y)$ in the domain $0< \vert \qe \vert <\vert y \vert <1$ are obtained by the normalized vacuum expectation values of the canonical surface defect \eqref{eq:canon} in the limit $\ve_2 \to 0$, with the $\ve_1$-lattices chosen to be centered at the Coulomb moduli; $L_\a = a_\a + \ve_1 \BZ$, $\a=1,2,\cdots,N$. Then, it is straightforward to compute the monodromy along the $A$-cycle simply by reading off the critical exponents of the so-obtained oper solutions as
\begin{align}
    M_A = \text{diag}\left((-)^{N-1} e^{-2\pi i \frac{a_\a -\bar{a}}{\ve_1}} \right)_{\a=1} ^N.
\end{align}
Therefore, the Coulomb moduli are identified with half of the Darboux coordinates $\boldsymbol\a_\a = \frac{a_\a-\bar{a}}{\ve_1}$, $\a=1,2,\cdots, N-1$, at the oper $\r_{\mathbf{a}}$ constructed from the canonical surface defect.

Since the space of opers is a complex Lagrangian submanifold, when restricted to the space of opers the other half $(\boldsymbol\b_\a)_{\a=1} ^{N-1}$ of the Darboux coordinates are determined once $(\boldsymbol\a)_{\a=1} ^{N-1}$ are chosen. In particular, there is a generating function $S$ that satisfies locally
\begin{align}
    \boldsymbol\b_\a = \frac{\p S(\boldsymbol\a)}{\p \boldsymbol\a_\a},\quad\quad \a=1,2,\cdots, N-1.
\end{align}
It was conjectured in \cite{Nekrasov:2011bc} that this generating function $S$ is identical to the effective twisted superpotential $\widetilde{\EW}(\mathbf{a})$ (more precisely, the dimensionless function $S(\boldsymbol\a) = \frac{1}{\ve_1} \widetilde{\EW}(\mathbf{a})$), which governs the effective two-dimensional $\EN=(2,2)$ theory of the $\EN=2$ theory in the limit $\ve_2 \to 0$. The statement was proven explicitly for our case of the sphere with four marked points in \cite{Jeong:2018qpc}. All in all, the oper $\r_\mathbf{a} \in \text{Op}_{SL(N)} (\BP^1;S)$ that we constructed from the canonical surface defect is specified by the Coulomb moduli through simple relation $\boldsymbol\a_\a = \frac{a_\a -\bar{a}}{\ve_1}$, $\a=1,2,\cdots, N-1$.

\subsection{Parallel surface defects and fractional quantum TQ equation}
Having constructed the opers from the canonical surface defect, we now consider the configuration in which the canonical surface defect is inserted on top of the monodromy surface defect. Using the $qq$-characters, we will derive analytic constraints on the correlation function of the two defects. 

Later (see Section \ref{sec:Hecke}), these constraints will be applied in the $\EN=2$ gauge-theoretic realization of the geometric Langlands correspondence to show that inserting a canonical surface defect realizes the action of a Hecke operator in the limit $\ve_2\to 0$. This action is diagonal, as a consequence of the cluster decomposition of the two defects (see \eqref{eq:factor2}). This derivation of the Hecke eigensheaf property further establishes that the vacuum expectation value of the monodromy surface defect corresponds to a section of a Hecke eigensheaf.

Since the canonical surface defect was realized as a $\ve_1$-lattice summation of the $Q$-observables, we will first start from the configuration of the $Q$-observable on top of the regular monodromy surface defect, and then perform the proper $\ve_1$-lattice summation to convert the $Q$-observable to the canonical surface defect.

Thus, we consider the gauge origami setup involving gauge theories in three complex planes: ${\BC}^{2}_{12}$, where the original physical theory lives, ${\BC}^{2}_{13}$ producing the surface defect along the ${\BC}_{1}$ complex line, and ${\BC}^{2}_{34}$, generating a point-like observable, the $qq$-character. The difference from the previous section \ref{eq:subsubop} is that the gauge origami worldvolume includes an additional orbifold, $X = \BC_1 \times \BC_{234} ^3 /\G_{34}\times \G_{24}$ giving rise to the monodromy surface defect. The data of gauge theories is encoded in the equivariant Chern characters of the respective Chan-Paton spaces:
\begin{align}
\begin{split}
    \hat{N}_{12} & = \sum_{\omega'=0}^{N-1} \left( e^{a_{\omega'}} \hat{q}_2^{\omega'} \CalR_0 \otimes \fR_{\omega'} + e^{m^+_{\omega'}+\ve_1+\ve_3} \hat{q}_2 ^{\o'} \CalR_{1} \otimes \fR_{\omega'} + e^{m^-_{\omega'}-\ve_1-\ve_3} \hat{q}_2^{\omega'} \CalR_2 \otimes \fR_{\omega'} \right) \\
    \hat{N}_{13} & = \sum_{\omega'=0}^{N-1} e^{x'_{\o'}+\ve_1+\ve_3} \hat{q}_2^{\omega'} \CalR_1 \otimes \fR_{\omega'} \\
    \hat{N}_{34} & = e^x \hat{q}_2^{\omega} \CalR_0 \otimes \fR_\omega
\end{split}
\end{align}
Here, $a_{\omega}$ stand for the Coulomb moduli of the ${\BC}^{2}_{12}$ theory, $m^{\pm}_{\omega}$ translate to the hypermultiplet masses, while $x'_{\omega}$ and $x$ are the parameters of the observables. The gauge origami partition function produces an observable whose vacuum expectation value is regular $x$. Namely, the partition function can be written as
\begin{align} \label{eq:fracqq}
\begin{split}
&\llangle[\Big] \ET_{\o}(x,\bx') \EQ(\bx') \rrangle[\Big] \\
&= \llangle[\Big] (x-x'_\o) \EY_{\overline{\o+1}}(x+\ve_1+\delta_{\o,N-1}\ve_2) \EQ(\bx') + \kq_\o (x-x'_{\overline{\o+1}}+\delta_{\o,N-1}\ve_2) \frac{P_{\o+1} ^+ (x) P_\o ^- (x+\ve_1)}{\EY_\o(x)} \EQ(\bx') \rrangle[\Big],
\end{split}
\end{align}
where $\ET_\o(x,\bx')$ is a polynomial in $x$ of degree 2. It can be computed by the large $x$ expansion of the right hand side as
\begin{align}\label{eq:TwnN-2}
\begin{split}
    \ET_{\o}(x,\bx') 
    =& (x-x'_\o) (x-a_{\overline{\o+1}}+\ve_1 + \ve_2\delta_{\o,N-1}+\ve_1\nu_\o) \\
    & + \kq_\o (x-x'_{\overline{\o+1}}+\ve_2\delta_{\o,N-1})(x-m_{\o+1}^+-m_{{\o}}^-+a_\o-\ve_1\nu_{\o-1}+\ve_1) \\
    & +\ve_1 D^{(1)} _\o - {\qe}_\o \ve_1 D^{(1)} _{\o-1} +\frac{\ve_1 ^2}{2} \n_\o ^2 - \ve_1 a_{\overline{\o+1}} \n_\o  \\
    & +{\qe}_\o \left( \frac{\ve_1^2}{2} \n_{\o-1}^2 +(m_{\o+1}^+ - m_{{\o}}^- -a_\o - \ve_1 )\ve_1 \n_{\o-1} +P_{\o+1}^+ (a_\o) P_{{\o}}^-(a_\o+\ve_1) \right),
\end{split}
\end{align}
where we defined $\n \equiv k_\o - k_{\o+1}$ and
\begin{align}
\begin{split}
    & D_{\omega}^{(1)} \equiv \ve_2 k_{\omega} + \sum_{\Box\in {\sK}_{\omega}} \hat{c}_\Box - \sum_{\Box\in {\sK}_{\omega+1}} \hat{c}_\Box \equiv \ve_2 k_\omega + \hat{c}_\omega - \hat{c}_{\omega+1}, \\
    & \hat{c}_\Box = \hat{a}_\alpha + (i-1)\ve_1 + (j-1)\hat\ve_2 - \o\hat\ve_2, \quad \text{for} \quad \Box = (i,j) \in \l^{(\a)}.
\end{split}
\end{align}
In the equation \eqref{eq:fracqq}, let us set $x$ to be two special values to get
\begin{subequations}
\begin{align}
    \ET_\o (x=x'_{\overline{\o+1}} - \delta_{\o,N-1}\ve_2,\bx') \EQ(\bx') 
    & = (x_{\overline{\o+1}}'-\delta_{\o,N-1}\ve_2-x_\o') \EY_{\overline{\o+1}}(x_{\overline{\o+1}}'+\ve_1) \EQ(\bx') \nonumber\\
    & = (x_{\overline{\o+1}}'-\delta_{\o,N-1}\ve_2-x_\o') P_{\overline{\o+2}}^+(x_{\overline{\o+1}}+\ve_1) \EQ(\bx'+\ve_1 e_{\overline{\o+1}}) ,\\
    \ET_\o (x=x'_\o) \hat\ER(\bx') 
    & = \kq_\o (x_\o'-x_{\overline{\o+1}}'+\delta_{\o,N-1}\ve_2) \frac{P_{\o+1}^+(x'_\o)P_{{\o}}^-(x'_\o+\ve_1) }{\EY_\o(x'_\o)} \EQ(\bx') \nonumber\\
    & = \kq_\o P_{{\o}}^-(x'_\o+\ve_1) (x_\o'-x_{\overline{\o+1}}'+\delta_{\o,N-1}\ve_2) \EQ (\bx'-\ve_1e_{\o}),
\end{align}
\end{subequations}
where we used the notation $e_\o \equiv (\d_{\o,\o'})_{\o'=0} ^{N-1}$. Taking the difference of the two equations, we obtain
\begin{align} \label{eq:fracTQ}
     \llangle[\Big] T_\o(\bx) \EQ(\bx)  \rrangle[\Big]= P_{\overline{\o+2}}^+(x_{\overline{\o+1}}+\ve_1)   \llangle[\Big]\EQ(\bx+\ve_1e_{\overline{\o+1}}) \rrangle[\Big] + \kq_\o P_{{\o}}^-(x_\o+\ve_1)   \llangle[\Big]\EQ(\bx-\ve_1 e_\o ) \rrangle[\Big],
\end{align}
where we defined the degree 1 polynomial $T_\o (\bx) \equiv \frac{T_\o (x_\o,\bx) - T_\o (x_{\overline{\o+1}} - \delta_{\o,N-1}\ve_2,\bx)}{x_\o-x_{\overline{\o+1}}+\delta_{\o,N-1}\ve_2}$, which is directly computed as
\begin{align}
\begin{split}
    T_\o(\bx) &= x_{\overline{\o+1}} - a_{\overline{\o+1}} + \ve_1 + \ve_1\nabla^u_{\o+1} + \kq_\o (x_\o - m_{\o+1}^+ - m_{{\o}}^- + a_\o + \ve_1 - \ve_1\nabla^u_\o).
\end{split}
\end{align}
The above identities \eqref{eq:fracTQ} for different values of $\o=0,1,\cdots, N-1$ form a set of $N$ difference equations satisfied by the vacuum expectation value of the generalized $Q$-observable $\EQ(\bx)$. We call these equations the \textit{fractional quantum TQ equations}. In section \ref{sec:Hecke}, we will investigate the implications of the fractional quantum TQ equation on our $\EN=2$ theory approach to the geometric Langlands correspondence.

\section{Regular monodromy surface defect and coinvariants} \label{sec:reg}
The vertex algebra at the junction of the Neumann and the Dirichlet boundary conditions is known to be the affine vertex algebra, i.e., the VOA whose underlying vector space is the vacuum module\footnote{The highest-weight $\widehat{\fsl}(N)$-module induced from $\fsl(N)[[t]] \oplus \BC K$ where $\fsl(N)[[t]]$ acts trivially and $K$ acts by $k \in \BC$.} of the affine Kac-Moody algebra $\widehat{\fsl}(N)$ \cite{Frenkel:2018dej}. The M5-branes wrapping the fiber of the $T^* \EC \to \EC$ in the twisted M-theory become D3-branes that also end on the NS5-brane and D5-brane but wraps the fiber of $T^* \EC \to \EC$, being located at the marked points $S \subset \EC$. In the effective $\EN=4$ gauge theory, they are the half-BPS monodromy surface defects in the $\EN=4$ gauge theory \cite{gukwit2}. In the point of view of the vertex algebra at the junction $\EC$, they become vertex operators (modules over the vertex algebra) which are non-maximally-degenerate \cite{Prochazka:2018tlo}. Therefore, the vacuum expectation value of the regular monodromy surface defect is expected to be identified with a conformal block (or a coinvariant) formed by these vertex operators.\footnote{See also \cite{Bourgine:2019phm} for a realization of the monodromy surface defect (to be precise, the 5d $\EN=1$ uplift) in the representation theory of quantum toroidal algebra.}

It was verified that the vacuum expectation value of the regular monodromy surface defect in $\EN=2$ gauge theory gives a coinvariant for $\widehat{\fsl}(N)$ by showing the Knizhnik-Zamolodchikov equation is obeyed by the former, for the case of the $SU(N)$ gauge theory with $2N$ hypermultiplets corresponding to the 4-point genus-0 coinvariants for $\widehat{\fsl}(N)$ \cite{Nekrasov:2021tik}. In particular, the $\widehat{\fsl}(N)$-modules assigned to the four marked points are shown to be the modules induced from Verma modules and Heisenberg-Weyl (bi-infinite) modules over $\fsl(N)$. In this section, we clarify that the vacuum expectation value of the regular monodromy surface defect can be viewed as a section of a sheaf over $\text{Bun}_{G_\BC} (\EC;S)$. In this paper we shall not be too specific about the precise nature of $\text{Bun}_{G_\BC} (\EC;S)$. It appears one should be working over de Rham stack of bundles, rather than just a moduli space of stable bundles. We shall be working over an open subset.

\subsection{Genus-0 coinvariants for $\widehat{\fsl}(N)$} \label{subsec:slN}
Let us first briefly recall some relevant facts about the affine Kac-Moody algebra and coinvariants. See \cite{Frenkel:2005pa}, for instance, for more complete review. Recall that the (untwisted) affine Kac-Moody algebra $\widehat{\mathfrak{sl}}(N)$ is the central extension of the loop algebra of the Lie algebra ${\mathfrak{sl}}(N)$, namely, $\widehat{\fsl}(N) = \fsl(N) \otimes \BC (\!(t)\!) \oplus \BC K$. The generators of $\widehat{\fsl}(N)$ are
\begin{align}
\begin{split}
    &\big( J_a ^b \big)_n \quad\quad\quad a,b=1,2,\cdots, N, \quad a\neq b,\quad n\in \BZ \\
    & \big(h_i \big) _n \equiv \big( J^i _i \big)_n - \big( J^{i+1} _{i+1} \big)_n \quad\quad\quad i=1,2,\cdots,N-1,\quad n\in \BZ \\
    &K
\end{split}
\end{align}
which satisfy the commutation relations
\begin{align}
\begin{split}
&\Big[\big( J_a ^b \big)_n , \big( J_c ^d \big)_m \Big] = \d_c ^b \big( J_a ^d \big)_{n+m} - \d_a ^d \big( J_c ^b \big)_{n+m} + K n \d_a ^d \d_c ^b \d_{n+m,0} \\
&[\big(J_a ^b \big)_n, K] = 0
\end{split}, \quad\quad   n,m \in \BZ.
\end{align}
Equivalently, we can define the current $J_a ^b (z) \equiv \sum_{n\in \BZ} \big( J_a ^b \big)_n z^{-n-1}$, and cast it into the form of current algebra with the OPE
\begin{align}
    J_a ^b (z) J_c ^d (w) \sim \frac{K \d_a ^d \d_c ^b}{(z-w)^2} + \frac{\d_c ^b J_a ^d (w) - \d_a ^d J_c ^b (w)}{z-w}.
\end{align}

Let $V$ be a module over $\fsl(N)$. Then it is automatically a module over $\mathfrak{sl}(N) \otimes \BC[[t]] \oplus \BC K$, by declaring $\mathfrak{sl}(N) \otimes t \BC[[t]]$ acts trivially and $K$ acts as multiplication by $k \in \BC$. Thus we denote the induced $\widehat{\fsl}(N)$-module by $\mathbb{V}^k = \text{Ind}^{\widehat{\mathfrak{sl}}(N)} _{\mathfrak{sl}(N) \otimes \BC[[t]] \oplus \BC K} V$. $k$ is called the level of the module $\mathbb{V}^k$. Note that $\mathbb{V}^k \simeq U(\fsl(N) \otimes t^{-1} \BC[t^{-1}]) \otimes _\BC V$ as a vector space, so that $V$ is embedded in $\mathbb{V}^k$ as the subspace of $\fsl(N) \otimes t \BC[[t]]$-invariants. 

Let us prepare $n$ modules $\left( V_i\right)_{i=1} ^n$ over $\fsl(N)$. Consider $\BP^1$ with the global coordinate $z$ and $n$ distinct points $S =\{z_1,\cdots, z_n \} \subset \BP^1$. Suppose we are given with a parabolic $PGL(N)$-bundle $E \in \text{Bun}_{PGL(N)} (\BP^1;S)$. Let us define $\fsl(N)_{ \bz} ^E$ to be the Lie algebra of the global sections of the associated vector bundle, $\fsl(N) _\bz ^E = \G ( \BP^1 \setminus S, E\times_{PGL(N)} \fsl(N)) $. Any element of $\fsl(N)_{\bz} ^E$ can be expanded in Laurent series at each $z_i \in S$ and therefore defines an element of $\bigoplus_{i=1} ^n \fsl(N) \otimes (\!(z-z_i)\!)$. It can be shown that the central extension on the image of $\mathfrak{sl}(N)_{\bz} ^E$ is trivial, and thus $\mathfrak{sl}(N)_{\bz} ^E$ naturally acts on $\bigotimes_{i=1} ^n \mathbb{V}^k _{i}$. We define the space $C ^E _{\fsl(N)} \left( (V_i)_{i=1} ^n \right)$ of $n$-point genus-0 $E$-twisted conformal blocks as the space of linear functionals on $\bigotimes_{i=1} ^n \mathbb{V}^k _{i}$ which are invariant under the action of $\mathfrak{sl}(N)_{\bz} ^E$. That is, $\varphi : \bigotimes_{i=1} ^n \mathbb{V}^k _{i} \to \BC$ such that
\begin{align}
    \varphi (g \cdot v) = 0, \quad\quad g \in \fsl(N)_{\bz} ^E,\quad v \in \bigotimes_{i=1} ^n \mathbb{V}^k _{i}.
\end{align}
Its dual space
\begin{align}
    H^E _{\fsl(N)} \left( (V_i)_{i=1} ^n \right) = \bigotimes_{i=1} ^n \mathbb{V}_i ^k \left/ \fsl(N) _\bz ^E \cdot \bigotimes_{i=1} ^n \mathbb{V}_i ^k \right.
\end{align}
is called the space of $E$-twisted coinvariants. It can be shown that the restriction to $\bigotimes_{i=1} ^n V_i$ in fact yields an isomorphism, so that $H ^E _{\fsl(N)} \left( (V_i)_{i=1} ^n \right)  \simeq \left( \bigotimes_{i=1} ^n V_i\right)^{\fsl(N)}$ \cite{Feigin:1994in}.\footnote{In this work, we allow modules that are infinite-dimensional and not necessarily of highest‑weight type. We use the vertex algebra definition of coinvariants (in the sense of \cite{Ben-Zvi:2004}). The resulting sheaves of coinvariants are quasi-coherent and carry a natural $\EuScript{D}$‑module structure; we do not assume local freeness or finite rank. We use the term \emph{conformal blocks} for the restricted/continuous duals of these coinvariants. With this convention, the Sugawara construction yields a projectively flat connection on conformal blocks, and analytic KZ solutions are sections of this dual local system. This is the sense in which our formulation yields the generically non‑polynomial KZ solutions.}

As we vary $E \in \text{Bun}_{PGL(N)} (\BP^1;S)$, the spaces $H^E _{\fsl(N)} \left( (V_i)_{i=1} ^n \right) $ of coinvariants are organized into a sheaf on $\text{Bun}_{PGL(N)} (\BP^1;S)$. The Ward identities are realized by differential equations on $\text{Bun}_{PGL(N)} (\BP^1;S)$, so that the space $H ^E _{\fsl(N)} \left( (V_i)_{i=1} ^n \right)$ of coinvariants give solutions to these differential equations locally around $E \in  \text{Bun}_{PGL(N)} (\BP^1;S)$. In other words, the sheaf of twisted coinvariants carries the structure of twisted $\ED$-module on $\text{Bun}_{PGL(N)} (\BP^1;S)$ (see \cite{Ben-Zvi:2004} for instance). This twisted $\ED$-module turns out to be the sheaf $\ED_{\fL_k}$ of rings of twisted differential operators itself.

The Sugawara construction of the stress tensor identifies the Virasoro generators $L_n$, $n\in \BZ$, in terms of the generators of the Kac-Moody algebra $\widehat{\fsl}(N)$. From the expression of $L_{-1}$, in particular, we derive the Knizhnik-Zamolodchikov equations obeyed by the section $\Psi$ of the sheaf of coinvariants,
\begin{align} \label{eq:KZ}
    0=\left[(k+N)\frac{\p}{\p z_i} - \sum_{j\neq i} \frac{ \sum_{a,b=1}^N \bar{T}_a ^b \vert_i \otimes  \bar{T}_b ^a \vert _j  }{z_i -z_j} \right]\Psi(\bz),\quad\quad i=1,2,\cdots, n,
\end{align}
where $\bar{T}_a ^b = T_a ^b - \d_a ^b \frac{1}{N} \sum_{c=1} ^N T^c _c$ are the generators $\fsl(N)$ in the standard basis, and the notation $\vert_i$ means it is represented on the $i$-th module in the tensor product.

\subsubsection{Modules over $\fsl(N)$}
We will review the construction of the relevant modules over the Lie algebra $\mathfrak{sl}(N)$. The $\widehat{\fsl}(N)$-modules which are used to construct the coinvariants in our interest are the ones induced from the $\fsl(N)$-modules introduced here. We first find maps from the universal enveloping algebra $U(\fsl(N))$ to the algebra of global sections of the sheaf of differential operators on flag varieties. Generically, the sheaf of modules over this algebra may not admit a global section, but we will fix an open patch where we define modules over the differential operators. We believe our construction is a manifestation of the Beilinson-Bernstein localization \cite{BB} (and its generalizations), which realizes modules over $\fsl(N)$ as twisted $\EuScript{D}$-modules on flag varieties, although we do not try to articulate in this direction.\footnote{We thank Dylan Butson for his explanation on the Beilinson-Bernstein localization.} For other works using Beilinson-Bernstein localization to geometric Langlands correspondence with ramifications, see \cite{Frenkel95} for instance.

We present Verma modules, which are lowest-weight $V_{\boldsymbol\z}$ or highest-weight $\tilde{V}_{\tilde{\boldsymbol\z}}$, and Heisenberg-Weyl modules $H^{\boldsymbol\t} _\s$ (HW modules), which are \textit{bi-infinite} in the sense that they are neither height-weight nor lowest-weight. We realize these modules by sections of complex powers of line bundles over appropriate flag varieties, which will be used to identify the coinvariants as a section of the tensor product of complex powers of line bundles over $\text{Bun}_{PGL(N)} (\EC;S)$.

\paragraph{Verma modules}
Consider the sequence of complex vector spaces $\left( F_i \right)_{i=1} ^{N}$, with $\dim F_i = i$. Consider the vector space of linear maps $\bU = (\tU_i )_{i=1} ^{N-1} \in \bigoplus_{i=1} ^{N-1} \text{Hom}(F_i , F_{i+1}) =: \EuScript{U}$. The gauge transformations $\EuScript{G} := \bigtimes_{i=1} ^{N-1} GL(F_i) $ acts on $\EuScript{U}$ as
\begin{align}
\tU_i \mapsto g_{i+1} \tU_i g_i ^{-1},\quad\quad g_i \in GL(F_i), \quad i =1,2,\cdots, N-1.
\end{align}
Let us restrict to the stable subset $\EuScript{U}^s \subset \EuScript{U}$, where all the maps $\tU_i$ are injective. Then the action of $\EuScript{G}$ on $\EuScript{U}^s$ is free. The complete flag variety is defined by the quotient $F(N) := \EuScript{U}^s / \EuScript{G}$.

There is a natural action of $h\in GL(F_N) = GL(N)$ on $F_N = \BC^N$, which acts on $\bU \in \EuScript{U}^s$ by $\tU_{N-1} \mapsto h\tU_{N-1}$. It descends to the action of $GL(N)$ to the quotient $F(N)$. Hence the Lie algebra $\mathfrak{gl}(N)$ induces vector fields on $F(N)$. Let us choose a basis $\left( e_a \right)_{a=1} ^N$ of $F_N$, whose dual basis is $\left( \tilde{e}^a \right)_{a=1}^N$, $\tilde{e}^b (e_a) = \d^b _a$. Denote the standard basis of $\mathfrak{gl}(N)$ by $T_a ^b = \tilde{e}^b \otimes e_a$, $a,b=1,2,\cdots, N$. They are represented as vector fields on $F(N)$ by
\begin{align} \label{eq:repflag}
    T_a ^b \vert_{V_{\boldsymbol\z}} = - \sum_{m=1} ^{N-1} \left(\tU_{N-1}\right)_m ^b \frac{\p}{\p \left(\tU_{N-1}\right) _m ^a},
\end{align}
where $\left(\tU_{N-1} \right)^a _m$ is the matrix elements of $\tU_{N-1}$ with respect to some chosen basis $\left(e^{(N-1)} _a\right)_{a=1} ^{N-1}$ of $F_{N-1}$ and $\left( e_a \right)_{a=1} ^N$. For notational convenience, we sometimes also write $J_a ^b = T_a ^b \vert_{V_{\boldsymbol\z}}$. Note that the vector fields \eqref{eq:repflag} are invariant under the gauge transformation of $GL(F_{N-1}) $ and therefore well-defined on $F(N)$.

We will realize a lowest-weight Verma $\mathfrak{gl}(N)$-module by sections of a line bundle over $F(N)$. Let us consider the rank $i$ (tautological) vector bundles over $F(N)$ whose fiber at $\bU$ is $(\tU_{N-1}\tU_{N-2} \cdots \tU_i )(F_i) $, for $i=1,2,\cdots, N-1$. By abusing the notation, we denote these tautological vector bundles by $F_i \to F(N)$. By taking the $i$-polyvector $\pi_i = \wedge^i (\tU_{N-1}\tU_{N-2} \cdots \tU_i )(F_i)$ of the image of $F_i$ in $F_N$, we also define line bundles $\det F_i$ over $F(N)$. 

Consider the $i$-form $\tilde\pi_0 ^i = \tilde{e}^1 \wedge\cdots \wedge\tilde{e}^i \in \wedge^i F_N ^*$ on $F_N$. Note that $\tilde\pi _0 ^i (\pi_i)$ gives a coordinate on the fiber of $\det F_i$, and thus it is a section of $\left( \det F_i \right)^{-1}$. Now, for given $N-1$ complex numbers $\boldsymbol\z \in \BC^{N-1}$, let us consider $ \O_{\boldsymbol\z} := \prod_{i=1}^{N-1} \left(\tilde{\pi}^i _0 (\pi_i)\right)^{\z_i}$ viewed as a section of the tensor product of \textit{complex $\z_i$-powers} of line bundles $\left(\det F_i\right)^{-1}$, namely, $\bigotimes_{i=1} ^{N-1} \left( \det F_i \right)^{-\z_i}$. Then we construct the differential operators twisted by this \textit{line bundle}, $\O_{\boldsymbol\z} ^{-1} T_a ^b \vert_{V_{\boldsymbol\z}} \O_{\boldsymbol\z}$. Note that even though $\bigotimes_{i=1} ^{N-1} (\det F_i) ^{-\z_i}$ is not well-defined as a line bundle, the corresponding twisted differential operators are well-defined. 

In the open subset $F(N)^\circ \subset F(N)$ where $\tilde{\pi}_0 ^i (\pi_i) \neq 0$ for all $i=1,2,\cdots,N-1$, consider the $\EG$-invariant functions $u^{(i)} _k : = \frac{\tilde{\pi}_0 ^i (e_k \wedge \tilde{e}^{i+1} \pi_i)}{\tilde{\pi}_0 ^i (\pi_i)}$, $1\leq k \leq i \leq N-1$. It can be shown \cite{Nekrasov:2021tik} that the space of polynomials $\BC \left[ u^{(i)} _k \right]$ forms the lowest-weight Verma module, under the $\mathfrak{sl}(N)$-action of the twisted differential operators $\O_{\boldsymbol\z} ^{-1} T_a ^b \vert_{V_{\boldsymbol\z}} \O_{\boldsymbol\z}$, with the weights of $\mathfrak{h}_i \in \mathfrak{t}$ on the lowest-weight state $1$ given by $ -\z_i$. We denote this lowest-weight Verma module by $V_{\boldsymbol\z}$.

To construct the highest-weight Verma modules, we consider the space of linear maps $\tilde\bU = ({\tilde\tU}_i)_{i=1} ^{N-1} \in \bigoplus_{i=1} ^{N-1} \text{Hom}(F_{i+1}, {F}_{i}  ) := \tilde{\EuScript{U}} $. We restrict to the open subset $\tilde{\EuScript{U}}^s \subset \tilde{\EuScript{U}}$ where all the maps $\tilde\tU _i$ are surjective. The action of the gauge transformations $\tilde{\EG} = \bigotimes_{i=1} ^{N-1} GL({F}_i)$ is free on $\tilde{\EuScript{U}} ^s$. The quotient is again the complete flag variety $F(N)$.

The natural action of $T_a ^b \in \mathfrak{gl}(N)$ on $F(N)$ is given by the vector fields
\begin{align}
    T_a ^b \vert_{\tilde{V}_{\tilde{\boldsymbol\z}}}  =\sum_{m=1} ^{N-1} (\tilde{\tU}_{N-1})^m _a \frac{\p}{(\p\tilde{\tU}_{N-1})^m _b},
\end{align}
where $(\tilde\tU_{N-1})^m _a$ is the matrix elements of $\tilde\tU$ with respect to the basis $\left( {e}_a \right)_{a=1} ^N$ of $F_N $ and some chosen basis $\left({e} ^{(i-1) } _a \right)_{a=1 }^{N-1}$ on $F_{N-1}$. Note that these vector fields do not depend on the choice of the basis on $F_{N-1} $, and therefore are well-defined in the quotient $F(N)$.

Consider the $i$-form $\tilde{\pi}^i \in \wedge^i F_N ^* $ which is a pullback of $\wedge^i F_i ^*$ under ${\tilde\tU}_{N-1} {\tilde\tU}_{N-2} \cdots {\tilde\tU}_{i+1} : F_N \to F_i $. Also take the $i$-polyvector $\pi_i ^0 = e_1 \wedge \cdots \wedge e_i \in \wedge^i F_N$. For given $N-1$ complex numbers $\tilde{\boldsymbol\z} \in \BC^{N-1}$, take $\tilde\O_{\tilde{\boldsymbol\z}} := \prod_{i=1 }^{N-1} \left( \tilde\pi^i (\pi_i ^0) \right)^{\tilde\z_i} $ which would have been a section of the line bundle $\bigotimes_{i=1} ^{N-1} \left(\det F_i  \right) ^{\tilde\z_i}$ over $F(N)$ if we had taken $\tilde\z_i \in \BZ$. We take all $\tilde\z_i$ to be generic, but still the twisted differential operators $\tilde\O_{\tilde{\boldsymbol\z}} ^{-1} T_a ^b \vert_{\tilde{V}_{\tilde{\boldsymbol\z}}} \tilde\O_{\tilde{\boldsymbol\z}}$ are well-defined. Then, in the open subset where $\tilde\pi^i (\pi_i ^0) \neq 0$ for all $i=1,2,\cdots, N-1$, we consider the space $\BC \left[ \tilde{u}^k _{(i)} \right]$ of polynomials in $\tilde{u}^k _{(i)} : = \frac{ (\tilde{e}^k \wedge \iota_{e_{i+1}} \tilde\pi^i) (\pi_i ^0)}{\tilde\pi^i (\pi_i ^0)}$, $1\leq k\leq i \leq N-1$. Under the $\fsl(N)$-action of the twisted differential operators, it is equipped with the highest-weight Verma module with the weights of $\mathfrak{h}_i \in \mathfrak{t}$ on the highest-weight state $1$ given by $\tilde{\z}_i$. We denote this highest-weight Verma module by $\tilde{V}_{\tilde{\boldsymbol\z}}$

\paragraph{Bi-infinite modules}
Next, consider the space of maps $ \text{Hom}(F_1, F_N)$. Restricting to the open subset where $X \in  \text{Hom}(F_1, F_N)$ is injective, the action of $GL(F_1) = \BC^\times$ is free. The quotient space is the projective space $\BP^{N-1}$. Note that $X$ gives homogeneous coordinates on $\BP^{N-1}$, which are sections of the hyperplane bundle $\CalO(1)$ over $\BP^{N-1}$.

The action of $GL(F_N) = GL(N)$ on $\text{Hom}(F_1, F_N)$ descends to the quotient $\BP^{N-1}$. The action of the Lie algebra $\mathfrak{gl}(N)$ induces vector fields on $\BP^{N-1}$,
\begin{align}
    T_a ^b \vert_{H _{\boldsymbol\t,\s}} = -X^b \frac{\p}{\p X^a}, \quad\quad a,b = 1,2,\cdots, N,
\end{align}
where $T_a ^b = \tilde{e}^b \otimes e_a$ is again the standard basis for $\mathfrak{gl}(N)$.

Each $X^a$, $a=1,2,\cdots, N$ gives a section of the hyperplane bundle $\CalO(1)$ over $\BP^{N-1}$. Let us given with $N$ complex numbers $h_a \in \BC$, $a=1,2,\cdots, N$. We consider $\boldsymbol\o_{\boldsymbol\t,\s} := \prod_{a=1} ^N (X^a) ^{h_a}$, where we defined $\t_i := h_i - h_{i+1}$, $i=1,2,\cdots, N-1$, and $\s:= \sum_{a=1} ^N h_a$. It would have been a section of the line bundle $\CalO(1) ^\s$, if we had $h_a \in \BZ$ which implies $\s \in \BZ$. We take all $h_a$ to be generic so that the section $\boldsymbol\o_{\boldsymbol\t,\s}$ is not well-defined nor is the \textit{line bundle} $\CalO(1)^\s$ itself. However, the twisted differential operators $\boldsymbol\o_{\boldsymbol\t,\s} ^{-1} T_a ^b \vert_{H_{\boldsymbol\t,\s}} \boldsymbol\o_{\boldsymbol\t,\s}$ are well-defined and this is all that we use.

In the open subset of $\BP^{N-1}$ where $X^a \neq 0$ for all $a=1,2,\cdots, N$, we consider the space of degree-0 Laurent polynomials $\BC ( X^1, \cdots, X^N )^{\BC^\times}$. It is equipped with a bi-infinite (namely, neither highest-weight nor lowest-weight) $\fsl(N)$-module structure under the $\fsl(N)$-action of the twisted differential operators, with the weights of $\mathfrak{h}_i \in \mathfrak{t}$ on $1$ given by $-\t_i$. We call this module the \textit{Heisenberg-Weyl} module, and denote it by $H_{\boldsymbol\t,\s}$.

\paragraph{Reducible bi-infinite modules and finite dimensional submodules} For generic $\boldsymbol\t$ and $\s$, the HW module $H_{\boldsymbol\t,\s}$ is irreducible. However, it becomes reducible when the parameters are tuned to special values. For example, consider the case $\boldsymbol\t = (h,0,\cdots, 0)$ and $\s = h$, where $h\in \BC$ is a generic complex number. Then the HW module $H_{(h,0,\cdots,0),h}$ contains the highest-weight Verma module $\tilde{V}_{(-h,0,\cdots, 0)}$, realized by the space of polynomials $\BC \left[ \frac{X^2}{X^1}, \cdots, \frac{X^N}{X^1} \right]$, as a proper submodule. Similarly, the HW module $H_{(0,\cdots, 0,-h),h}$ contains the lowest-weight Verma module ${V}_{(0,\cdots, 0,-h)}$ realized as the space of polynomials $\BC\left[\frac{X^1}{X^N},\cdots, \frac{X^{N-1}}{X^N} \right]$ as a proper submodule. 

For special values of $h \in \BC$, the $\widehat{\fsl}(N)$-modules induced from them, $\mathbb{V}^k _{(0,\cdots, 0,-h)} =\text{Ind}^{\widehat{\fsl}(N)} _{\fsl(N)\otimes \BC[[t]] \oplus \BC K} {V}_{(0,\cdots, 0,-h)}$ may not be irreducible even though the Verma module ${V}_{(0,\cdots, 0,-h)}$ itself is irreducible. This is indeed the case for the twisted vacuum module (namely, the spectral flow image of the vacuum $\widehat{\fsl}(N)$-module), for which $h=k$. We will study in detail in section \ref{sec:Hecke} as the realization of the Hecke operator in the affine Kac-Moody vertex algebra for $\widehat{\fsl}(N)$ \cite{Frenkel:2005pa}. 

It also happens that, at other special values of $h \in \BC$, the Verma $\fsl(N)$-modules $V_{(0,\cdots, 0,-h)}$ and $\tilde{V}_{(-h,0,\cdots,0)}$ become reducible. For example, take $h=1$ then the finite $N$-dimensional representation spanned by the monomials $\bigoplus_{a=1} ^N \BC \frac{X^a}{X^N}$ (resp. $\bigoplus_{a=1} ^N \BC \frac{X^a}{X^1}$) is contained in the Verma module $V_{(0,\cdots, 0,-1)}$ (resp. in $\tilde{V}_{(-1,0,\cdots,0)}$). Note that this is precisely the global sections of the hyperplane bundle $\CalO(1)$ over $\BP^{N-1}$ with homogeneous coordinates $(X^a)_{a=1}^N$, in the open patch $X^N \neq 0$ and $X^1 \neq 0$ respectively; a manifestation of the Borel–Weil–Bott theorem. The insertion of vertex operators associated to finite dimensional modules corresponds to the insertion of (coincident) canonical surface defects on the plane transverse to the plane of the regular monodromy surface defect, forming the configuration of intersecting surface defects in the $\EN=2$ gauge theory, as extensively studied in \cite{Jeong:2021bbh}. Indeed, this configuration of intersecting surface defects is dual to a Wilson line (labelled by a dominant integral weight of $G$) attached to the Neumann boundary in the GL-twisted $\EN=4$ theory side, as we discussed in section \ref{subsec:gln4}. We will not investigate the intersecting case in the present work, but focus on the configuration of parallel surface defects.

\subsubsection{4-point genus-0 Knizhnik-Zamolodchikov equation}
Let us consider the sphere $\BP^1$ with four marked points $S = \{z_1,z_2,z_3,z_4\} \subset \BP^1$. We assign a lowest-weight Verma module $V_{\boldsymbol\z}$ at $z_1$, Heisenberg Weyl modules $H _{\boldsymbol\t - \boldsymbol\z,\s}$ and $H _{\tilde{\boldsymbol\z} - \boldsymbol\t,\tilde{\s}}$ at $z_2$ and $z_3$ respectively, and finally a highest-weight Verma module $\tilde{V}_{\tilde{\boldsymbol\z}}$ at $z_4$. We construct the $E$-twisted coinvariants of the tensor products of the $\widehat{\fsl}(N)$-modules induced from these $\fsl(N)$-modules,
\begin{align}
     \boldsymbol\Psi \in \left( V_{\boldsymbol\z} \otimes H _{\boldsymbol\t -\boldsymbol\z,\s} \otimes  H _{\tilde{\boldsymbol\z} - \boldsymbol\t,\tilde{\s}} \otimes \tilde{V}_{\tilde{\boldsymbol\z}} \right)^{\fsl(N)},
\end{align}
from the geometric realization of the $\fsl(N)$-modules introduced above. 

Practically, we construct the solution to the Ward identities and the 4-point Knizhnik-Zamolodchikov equations. The Ward identities read
\begin{align}
    0=\sum_{i=1} ^4 L_i ^{0,\pm} \boldsymbol\Psi (\mathbf{z}), \quad\quad 0 = \sum_{i=1}^4 \big( \bar{T}_a ^b\big)\vert_i  \boldsymbol\Psi (\mathbf{z})
\end{align}
where $L_i ^a = z_i ^a (z_i \p_{z_i} -2 (a+1) \D_i )$, $a=0,\pm 1$, are the Virasoro generators and $\big( \bar{T}_a ^b\big)\vert_i$ are the $\fsl(N)$ generators represented on the $i$-th module. Here, $\D_i$ is the conformal weight of the $i$-th vertex operator, given by the quadratic Casimir of the associated module $\D_i = \frac{\sum_{a,b=1} ^N \big( \bar{T}_a ^b \big)\vert_i \big( \bar{T}_b ^a \big)\vert_{i}}{2(k+N)}$ due to the Sugawara construction of the stress tensor. We can solve the Ward identities by making an ansatz,
\begin{align} \label{eq:4ptansatz}
\begin{split}
&\boldsymbol\Psi(\mathbf{U},\tilde{\mathbf{U}},\mathbf{X};\mathbf{z})  = \Psi_0 (\bU,\tilde{\bU} ,\mathbf{X};\bz) \Psi (\boldsymbol\g;\qe) \\
&\Psi_0 =  \left( \frac{z_{2 1} ^2 z_{43}}{z_{41} z_{31} } \right)^{-\D_2} \frac{\prod_{a=1}^N \left( \tilde\pi ^a \left( X_2 \wedge \pi_{a-1} \right) \right)^{\b  _{a}} \left( \tilde\pi ^a \left( X_3 \wedge \pi_{a-1} \right) \right)^{\tilde{\b}  _{a}} \prod_{i=1} ^{N-1} \left( \tilde\pi ^i (\pi_i) \right)^{\a_i} }{z_{14} ^{\D_1 +\D_4 -\D_3} z_{13} ^{\D_1 +\D_3 -\D_4} z_{43} ^{\D_3 + \D_4 -\D_1}} \prod_{i=1} ^4 dz_i ^{\D_i}
\end{split}
\end{align}
where $\Psi(\boldsymbol\g;\qe)$ only depends on the invariants
\begin{align} \label{eq:bungpara}
\begin{split}
        &\g_a = \frac{\tilde{\pi}^a (X_2 \wedge \pi_{a-1})}{\tilde{\pi}^a (X_3 \wedge \pi_{a-1})}, \quad a=1,2,\cdots,N, \\
    &\qe = \frac{z_{21} z_{43}}{z_{24} z_{13}}.
\end{split}
\end{align}
Note that $(\g_a )_{a=1} ^N$ suffer from a remnant $\BC^\times$-redundancy, but the ratios of them are well-defined. Thus we require $\Psi(\boldsymbol\g;\qe)$ to be degree-0 in $\boldsymbol\g$; namely, $\sum_{a=1}^N \g_a \p_{\g_a} \Psi (\boldsymbol\g;\qe) = 0$.

The ansatz \eqref{eq:4ptansatz} gives a geometric realization of the $E$-twisted coinvariants. Let us remind the connected component of the moduli space of stable parabolic $PGL(N)$-bundles containing the trivial bundle, with the choice of the parabolic structures presented in section \ref{subsubsec:g0four}, is 
\begin{align}
    \text{Bun}_{PGL(N)} (\BP^1;S)_0 = \left( F(N) \times \BP^{N-1} \times \BP^{N-1} \times F(N)\right)/SL(N).
\end{align}
The ratios of $\boldsymbol\g$, which are invariants of the maps $(\bU, \tilde{\bU}, \mathbf{X})$, provide holomorphic coordinates on $\text{Bun}_{PGL(N)}(\BP^1;S)_0$. Note there are indeed $N-1$ independent ratios of $\boldsymbol\g$ and $\dim \text{Bun}_{PGL(N)} (\BP^1;S)_0  = N-1$. Thus, $\boldsymbol\Psi(\boldsymbol\g;\qe)$ is a coinvariant twisted by $E_{\boldsymbol\g} \in \text{Bun}_{PGL(N)} (\BP^1 ;S)_0$, the parabolic $PGL(N)$-bundle specified by the coordinates $\boldsymbol\g$. 

The $\fsl(N)$-modules are constructed by the sections of the line bundles over the flag varieties in the product. By taking the tensor product of the pullbacks of these line bundles under the natural projections, we identify the $E$-twisted coinvariant $\boldsymbol\Psi$ as a section of the line bundle $\mathfrak{L}$ over $ \text{Bun}_{PGL(N)} (\BP^1;S)_0$,
\begin{align} \label{eq:twmod}
    \mathfrak{L} = \bigotimes_{i=1} ^{N-1} \left( \det F_i \right)^{-\z_i} \boxtimes  \CalO (1) ^{\s}  \boxtimes  \CalO (1)^{\tilde{\s}}  \boxtimes \bigotimes_{i=1} ^{N-1} \left( \det {F}_i \right)^{\tilde{\z}_i},
\end{align}
where the complex powers, which are determined by the weights of the modules, fix the parameters $(\boldsymbol\b_{2},\tilde{\boldsymbol\b}, \boldsymbol\a)$ in the ansatz. Indeed, in the prefactor of the ansatz \eqref{eq:4ptansatz}, note that $\tilde{\pi}^b (X_2 \wedge \pi_{b-1})$ (resp. $\tilde{\pi}^b (X_3 \wedge \pi_{b-1})$) is a section of the line bundle $\left(\det F_{b-1} \right)^{-1} \boxtimes \CalO(1) \boxtimes \CalO \boxtimes \det F_b$ (resp. $\left(\det F_{b-1} \right)^{-1} \boxtimes \CalO \boxtimes \CalO(1) \boxtimes \det F_b$), while $\tilde{\pi}^i (\pi_i)$ is a section of the line bundle $\left(\det F_i\right) ^{-1} \boxtimes \CalO \boxtimes \CalO \boxtimes \det F_i$. By simple matching of the powers, we get the following relations that fully determine $3N-1$ parameters $(\boldsymbol\b,\tilde{\boldsymbol\b},\boldsymbol\a)$ in terms of the $3N-1$ module parameters $(\boldsymbol\z,\tilde{\boldsymbol\z},\boldsymbol\t,\s,\tilde{\s})$:

\begin{align} \label{eq:weights}
\begin{split}
\begin{split}
    &\z_i =  \b_{i+1} + \tilde\b_{i+1} +\a_i \\
    &\tilde{\z}_i = \b_{i} +\tilde\b_i +\a_i \\
    &\t_i -\z_i = \b_{i} -\b_{i+1} \\
    &\tilde{\z}_i -\t_i = \tilde\b _{i} - \tilde\b_{i+1}
\end{split} \quad\quad\;\; i=1,2,\cdots, N-1, \\
    &\s  = \sum_{b=1} ^N \b _{b}, \quad\quad\tilde\s  = \sum_{b=1} ^N \tilde\b _{b}
\end{split}
\end{align}

Strictly speaking, $\fL$ \eqref{eq:twmod} is not well-defined as a line bundle due to the complex powers in each factor. However, we can still define the differential operators on $ \text{Bun}_{PGL(N)} (\BP^1;S)_0$ twisted by $\fL$. The prefactor $\Psi_0$ in the ansatz \eqref{eq:4ptansatz} should really be understood as performing this twisting. The so-obtained twisted differential operators act on $\Psi(\boldsymbol\g;\qe)$, which is well-defined in an open patch of $\text{Bun}_{PGL(N)} (\BP^1;S)_0$.

To complete the construction of the coinvariants, we need to solve the 4-point KZ equation. This is achieved by simply requiring $\Psi(\boldsymbol\g;\qe)$ to solve the following reduced 4-point KZ equation,
\begin{align} \label{eq:redKZ4}
    0 = \left[-(k+N)\frac{\p}{\p \qe} + \frac{\hat{\CalH}_0}{\qe} + \frac{\hat{\CalH}_1}{\qe-1} \right] \Psi (\boldsymbol\g;\qe),
\end{align}
where $\hat{\CalH}_0$ and $\hat{\CalH}_1$ are twisted differential operators on $\text{Bun}_{PGL(N)} (\BP^1;S)$ expressed in the holomorphic coordinates $\boldsymbol\g$. The exact expressions of $\hat{\CalH}_0$ and $\hat{\CalH}_1$ are immediate to derive by twisting the differential operators $\sum_{a,b=1} ^N \big(\bar{T}_a ^b\big)\vert_i \big(\bar{T}_b ^a\big)\vert_j$ by $\fL$, computed as
\begin{align}\label{eq:H-4p}
\begin{split}
    \hat{\CalH}_0 = & -  \sum_{b>a} \frac{\g_b}{\g_a} \left( \g_a \p_{\g_a} + \beta_{a}  \right)  \left( (\g_a-\g_b)\p_{\g_b} + \frac{\g_a}{\g_b}\beta_{b} + \tilde\beta_{b} \right) + \sum_{a=1}^N \left( \g_a \p_{\g_a} + \beta_{a}  \right)   \left( \sum_{c=a+1}^N \beta_{c} + \tilde\beta_{c} + \sum_{i=a} ^{N-1} \alpha_i \right)  \\
    & - \frac{1}{N} \left( \sum_{a=1}^N \beta_{a} \right) \sum_{b=1}^N \left( \sum_{c=b+1}^N \beta_{c} + \tilde\beta_{c} + \sum_{i=b}^{N-1} \alpha_i \right) +\frac{(N-1)\s(\s+N)}{N} , \\
    \hat{\CalH}_1 =& \sum_{a,b=1}^N \frac{\g_b}{\g_a} \left( \g_a \p_{\g_a} + \beta_{a} \right) \left( - \g_b \p_{\g_b} + \tilde\beta_{b} \right) + \frac{1}{N} \sum_{a=1}^N\beta_{a}  \sum_{b=1}^N\tilde\beta_{b}.
\end{split}
\end{align}
We remind that the $3N-1$ parameters $(\boldsymbol\b,\tilde{\boldsymbol\b},\boldsymbol\a)$ are determined by the weights of the $\fsl(N)$-modules $(\boldsymbol\z,\tilde{\boldsymbol\z},\boldsymbol\t,\s,\tilde{\s})$ through the relations \eqref{eq:weights}. We present the detailed steps of the derivation in appendix \ref{app:4-point}.

\subsection{Regular monodromy surface defect and $\EuScript{D}$-module of coinvariants} \label{subsec:regdefcoin}
The vacuum expectation value $\Psi(\mathbf{a};\mathbf{u})$ \eqref{eq:vevregmono} of the regular monodromy surface defect is identified with a 4-point genus-0 $E$-twisted coinvariant by showing the former obeys the KZ equation for the latter \eqref{eq:redKZ4} \cite{Nekrasov:2021tik}. In the previous section, we have given geometric constructions to the modules over $\fsl(N)$ involved in the 4-point genus-0 $E$-twisted coinvariants, expressing it as a section of $\mathfrak{L}$, the tensor product of complex powers of line bundles over the moduli space of parabolic bundles $\text{Bun}_{PGL(N)} (\BP^1;S)$. Thus, the identification of the vacuum expectation value $\Psi(\mathbf{a};\mathbf{u})$ with a 4-point genus-0 $E$-twisted coinvariant conveys the very geometric meaning to the former. In particular, the vacuum expectation value $\Psi(\mathbf{a};\mathbf{u})$ forms a distinguished family, enumerated by the Coulomb moduli $\mathbf{a}$, of sections of the sheaf of twisted coinvariants on $\text{Bun}_{PGL(N)} (\BP^1;S)$.

Even though the presentation of the 4-point genus-0 $E$-twisted coinvariant was slightly different in \cite{Nekrasov:2021tik}, the identification can be achieved in a similar way and we will only present the result here. See appendix \ref{app:4-point} for explicit steps of the matching. The vacuum expectation value $\Psi(\mathbf{a};\mathbf{u};\qe)$ \eqref{eq:vevregmono} of the regular monodromy surface defect is shown to satisfy
\begin{align} \label{eq:kzgauge}
    0 = \left[ \frac{\ve_2}{\ve_1} \frac{\p}{\p \qe} + \frac{\hat{\CalH}_0}{\qe} + \frac{\hat{\CalH}_1}{\qe-1} \right]\Psi(\mathbf{a};\mathbf{u};\qe).
\end{align}
This is precisely the reduced 4-point KZ equation \eqref{eq:redKZ4} obeyed by the $E$-twisted coinvariants, with the following identification of the parameters:
\begin{itemize}
    \item Level: $k+N = -\frac{\ve_2}{\ve_1}$ 
    \item Coordinates on $\text{Bun}_{PGL(N)} (\BP^1;S)_0$: $\g_{\o} = \frac{u_{\o}+u_{\o+1}+ \cdots+u_{\o+N-1}}{\kq-1} $, \quad\quad $\o=0,1,\cdots, N-1$
    \item Weights of the modules: \begin{align} &\zeta_{i} =  \frac{m_i ^- - m_{i-1}^-   }{\ve_1} - 1 , \quad \tilde\zeta_i =   \frac{m^+_{i}-m_{i-1}^+  }{\ve_1} -1, \quad \tau_i =\frac{a_{i} - a_{i-1} }{\ve_1} - 1, \quad\quad \ i=1,\dots,N-1 , \nonumber \\ 
     &\sigma = N \frac{\bar{m}^- -\bar{a} -\ve_1}{\ve_1}, \quad  \tilde{\sigma} =-N \frac{\bar{m}^+ -\bar{a} +\ve_1}{\ve_1}  \nonumber \end{align}
\end{itemize}
Here, we are shifting the indices for $\g_a$ in \eqref{eq:H-4p} by 1, so that they now run from $0$ to $N-1$ instead of $1$ to $N$. Note that the weight parameters do not depend on $\ve_2$. We also remind only the ratios of $\boldsymbol\g$ are well-defined, parametrizing the $(N-1)$-dimensional space $\text{Bun}_{PGL(N)} (\BP^1;S)_0$.

Let us recall that inserting the regular monodromy surface defect in the $\EN=2$ gauge theory is dual to changing the (deformed) regular Nahm pole boundary to the (deformed) Dirichlet boundary in the GL-twisted $\EN=4$ theory (see section \ref{subsec:gln4}). Compactifying to the topological sigma model, the (deformed) Dirichlet boundary descends to the brane of $\l$-connections $\EF' _E$ at the parabolic $PGL(N)$-bundles $E$. The duality works precisely in the way that $E$ is the parabolic $PGL(N)$-bundle fixed by the monodromy defect parameters $\boldsymbol\g$, since the latter provide holomorphic coordinates on $\text{Bun}_{PGL(N)} (\BP^1;S)_0$. 

The twisted $\ED$-module associated to the brane of $\l$-connections $\EF' _E$ is the sheaf of $\d$-functions at $E$, whose global sections are the $E$-twisted coinvariants. However, we study the holomorphic dependence of the vacuum expectation value $\Psi(\mathbf{a};\boldsymbol\g)$ on the defect parameters $\boldsymbol\g$, rather than fixing $\boldsymbol\g$ at a particular value. Putting into the topological sigma model context, this is to study the spaces of $(\EB_{cc}, \EF' _E)$-strings with varying $E \in \text{Bun}_{PGL(N)} (\BP^1;S)$. In turn, we should view the vacuum expectation value $\Psi(\mathbf{a};\boldsymbol\g)$ as a section of the \textit{sheaf} of coinvariants \cite{Ben-Zvi:2004,Frenkel:2005pa} on $\text{Bun}_{PGL(N)} (\BP^1;S)$.

It is important to note that, even though $\Psi(\mathbf{a};\boldsymbol\g)$ is labeled by the continuous Coulomb moduli $\ba$, the $\fsl(N)$-modules underlying the twisted coinvariants only depend on their values $[\ba]$ on the $\ve_1$-lattice, $\ba := [\ba] + \ve_1 \mathbf{n}$ with $\mathbf{n} \in \BZ^N$. Thus, the continuous parameters $[\ba]$ are kept fixed when the $\fsl(N)$-modules are specified. The vevs $\left\{\Psi([\ba] + \mathbf{n} \ve_1;\boldsymbol\g) \;\vert\; \mathbf{n} \in \BZ^{N} \right\}$ of the monodromy defect labeled by the remaining discrete parameter $\mathbf{n}$ provide the mentioned sections of the sheaf of coinvariants. 

Recall that the Picard group of $\text{Bun}_{PGL(N)} (\BP^1;S)_0$ is given by
\begin{align}
    \text{Pic}(\text{Bun}_{PGL(N)} (\BP^1;S)_0) = \L_{wt} \oplus \L_{wt,\BL} \oplus \L_{wt,\BL} \oplus \L_{wt},
\end{align}
where the determinant line bundle $\EL$ is missing because we are restricting to the stable subset. Expressing the module parameters in terms of the gauge theory parameters, we get the twist \eqref{eq:twmod} given by
\begin{align} \label{eq:twgauge}
    \mathfrak{L} = \bigotimes_{i=1} ^{N-1} \left( \det F_i \right)^{ \frac{m_{i-1}^- - m_{{i}}^-  }{\ve_1} + 1} \boxtimes  \CalO (1) ^{N \frac{\bar{m}^- -\bar{a}-\ve_1}{\ve_1}}  \boxtimes  \CalO (1)^{-N \frac{\bar{m}^+ -\bar{a}+\ve_1}{\ve_1}}  \boxtimes \bigotimes_{i=1} ^{N-1} \left( \det {F}_i \right)^{- \frac{m^+_{i-1}-m_{i}^+  }{\ve_1} -1}.
\end{align}
Note that this is indeed independent of the level $k+N=-\frac{\ve_2}{\ve_1}$. Here, the determinant line bundles $(\det F_i)^{-1}$ of the tautological bundles over $F(N)$ correspond to the $N-1$ fundamental weights generating the weight lattice $\L_{wt}$ of $SL(N)$. The tautological bundle $\CalO(-1)= \CalO(1)^{-1}$ over $\BP^1$ corresponds to the single fundamental weight invariant under the Weyl group of $\BL$, generating $\L_{wt,\BL}$.

At this point, let us remind that the vacuum expectation value $\Psi(\boldsymbol\g;\qe)$ is defined by imposing a further boundary condition at infinity, the complex adjoint scalar approaching to the Coulomb moduli $\phi \to \mathbf{a}$. We will argue later that imposing this boundary condition translates into taking the quotient by the ideal corresponding to the $SL(N)$-oper fixed by certain coordinates $\mathbf{a}$, leading to the Hecke eigensheaf. For now, let us note that there is a precise matching between the twisting $\fL$ \eqref{eq:twgauge} for the sheaf $\ED_{\fL}$ on $\text{Bun}_{PGL(N)} (\BP^1;S)$ and the monodromy \eqref{eq:mono} of the $SL(N)$-oper from their gauge theoretical constructions, confirming the expectation for the geometric Langlands correspondence with ramifications (see section \ref{subsubsec:canoper}) \cite{Frenkel:2006nm, gukwit2}. Namely, the monodromies of the $SL(N)$-oper, constructed from the canonical surface defect in the $\EN=2$ theory, at the marked points $S=\{0,\qe,1,\infty\}$ are expressed in gauge theory parameters as \eqref{eq:mono}, which can be written in terms of the twisting parameters in \eqref{eq:twgauge} as
\begin{align} \label{eq:monoparameter2}
\begin{split}
    &M_0 = \exp \left(-2\pi i \sum_{i=1} ^{N-1}  \frac{m^- _{i-1} - m^- _{i}}{\ve_1} \fh_i \right) \\
    &M_\qe = \exp  \left(-2\pi i  N\frac{\bar{m}^- -\bar{a}}{\ve_1} \fh_{N-1} \right) \\
    &M_1 =\exp  \left(2\pi i N \frac{\bar{m}^+ -\bar{a}}{\ve_1}  \fh_{N-1} \right) \\
    &M_\infty = \exp \left(2\pi i \sum_{i=1} ^{N-1}  \frac{m^+ _{i-1} - m^+ _{i}}{\ve_1} \fh_i \right),
\end{split}
\end{align}
where $(\fh_i)_{i=1}^{N-1}$ are the Cartan elements corresponding to the fundamental weights appearing in the twisting $\fL$ \eqref{eq:twgauge} for $\ED_{\fL}$. Among them, $\fh_{N-1}$ is the one corresponding to the generator of $\L_{wt,\BL}$.

\section{Hecke operator and parallel surface defects} \label{sec:Hecke}
So far, we have studied insertion of a single half-BPS surface defect $-$ engineered by either coupling to a two-dimensional theory or assigning singularity of the gauge field $-$ in the $\EN=2$ gauge theory, and interpreted their vacuum expectation values in the context of the geometric Langlands correspondence. Recall that these defects are dualized respectively to the 't Hooft line defect and the deformed Dirichlet boundary condition in the GL-twisted $\EN=4$ gauge theory. In this section, we will study the configuration of two surface defects parallel on the same complex plane $\BC_{\ve_1}$, 
whose dual configuration is therefore the 't Hooft line attached to the deformed Dirichlet boundary. Compactified to the topological sigma model, this is the $(B,A,A)_\k$-brane of $\l$-connections $\EuScript{F}' _{\mathbf{u}}$ dressed with a Hecke operator \cite{Kapustin:2006pk,Frenkel:2018dej}.

Without going into details, let us mention the mathematical meaning of the regular monodromy and canonical surface defects. In supersymmetric gauge theory with $\EN =2$ supersymmetry, the topologically twisted path integral localizes onto the moduli space of four dimensional gauge instantons, or some modified version thereof. The moduli space is non-compact even for compact Euclidean spacetimes, due to the point-like instantons. On Euclidean spacetime admitting a complex structure (e.g. on ${\BC\BP}^2$ but not on $S^4$), for gauge group $SU(n)$ or a product of unitary groups, it is customary to partly compactify the moduli space by first identifying the instanton connections with rank $n$ holomorphic vector bundles, which then are generalized to rank $n$ torsion free sheaves $\EuScript{E}$. The regular surface defect, located at the vanishing locus $z_2 = 0$ of a holomorphic function $z_2$, evaluated at the bulk instanton configuration represented by the sheaf ${\EuScript{E}}$, is the integral over the moduli space of parabolic sheaves, which are the collections of sheaves ${\EuScript{E}}_{1}, {\EuScript{E}}_{2}, \ldots , {\EuScript{E}}_{N-1}$, obeying:
\[ z_{2} {\EuScript{E}} \subset {\EuScript{E}}_{1} \subset {\EuScript{E}}_{2} \subset \ldots \subset {\EuScript{E}}_{N-1} \subset {\EuScript{E}} \]
with certain constraints on the quotients ${\EuScript{E}}_{i}/{\EuScript{E}}_{i-1}$. By contrast, the $Q$-observable is the regularized Chern polynomial of the coherent sheaf ${\EuScript{E}}_{z_{2}=0}$ which is defined as cohomology of a complex $z_{2} {\EuScript{E}} \to {\EuScript{E}}$ of sheaves. In our configuration of parallel surface defects, we study similar complexes $z_{2} {\EuScript{E}}_{j} \to {\EuScript{E}}_{i}$ for various $i$ and $j$.

In the vertex algebra at the corner $-$ the affine Kac-Moody vertex algebra for $\widehat{\fsl}(N)$ at the junction of the deformed Neumann boundary and the deformed Dirichlet boundary in our case $-$ the Hecke operator corresponds to inserting the spectral flow image of the vacuum module (called the twisted vacuum module)  \cite{Teschner:2010je, Gaiotto:2021tsq}. In this section, we will establish a correspondence between the insertion of the twisted vacuum module for the $\widehat{\fsl}(N)$ and the gauge theory configuration of parallel surface defects. In fact, we discover that it is necessary to include the quotient of the bi-infinite $\widehat{\fsl}(N)$-modules $\mathbb{H}_{\boldsymbol\t,k} ^k$ where $\boldsymbol\t$ is generic. When inserted into the coinvariants, the vertex operator for this quotient module constrains the coinvariants in exactly the same way as the twisted vacuum module does. 

\subsection{Twisted vacuum module and bi-infinite generalization}
We first introduce the twisted vacuum module defined by the spectral flow image of the vacuum module. Then, we study the property of the twisted coinvariants when the twisted vacuum module is inserted. Finally, we present the bi-infinite generalization of the twisted vacuum module, in the sense that it constrains the coinvariants in the same way but it is neither highest-weight nor lowest-weight.

\subsubsection{Spectral flow and twisted vacuum module}
The affine Dynkin diagram for $\widehat{\fsl}(N)$ possesses the symmetry group $D_N$, the dihedral group of $N$ objects. Since the Chevally-Serre basis of $\widehat{\fsl}(N)$ is directly constructed from the affine Dynkin diagram, the symmetry group $D_N$ induces the automorphism group of $\widehat{\fsl}(N)$ \cite{FGV}. Among these $D_N$ automorphisms, we restrict our attention to the maximal abelian normal subgroup $\BZ_N$, induced by the rotation symmetry of $N$ object, and call them spectral flows. The $\widehat{\fsl}(N)$-modules are mapped in a non-trivial manner by a spectral flow.

Using the same notations as in section \ref{subsec:slN}, the $\BZ_N$ spectral flows are generated by
\begin{align}
\begin{split}
&J_a ^b (z) \to z^{-\o_a + \o_b} J_a ^b (z) - \frac{k \o_a \d^a _b}{z}, \\
&\left( \fh_i (z) \equiv J^i _i (z) - J^{i+1} _{i+1} (z) \to \fh_i (z) - \frac{k(\o_i -\o_{i+1})}{z} \right)
\end{split}
\end{align}
where we choose the fundamental weight $\o = \left(-\frac{1}{N},-\frac{1}{N}, \cdots, -\frac{1}{N} ,\frac{N-1}{N} \right)$.\footnote{Other fundamental weights for $\o$ differ from the one above by a root of $\mathfrak{sl}(N)$. The spectral flows generated by roots are inner automorphisms so that all those choices are equivalent as outer automorphisms. We will explain the meaning of the inner automorphisms in the $\EN=2$ gauge theory shortly. Note also that $N\o = (-1,-1,\cdots, N-1) \sim (0,0,\cdots, 0) = \text{id}$ as an outer automorphism; namely, $\o$ is the generator of the $\BZ_N$ outer automorphism group.}
The twisted vacuum module is the spectral flow image of the vacuum module, built from a state $\vert \S \rangle$ satisfying
\begin{align} \label{eq:sfmodinv}
\begin{split}
    &\big( J_N ^a\big)_{n-1} \vert \S \rangle = \big( J_a ^N\big)_{n+1} \vert \S \rangle= 0 , \quad\quad a\neq N,\; n\geq 0  \\
    &\big( J_a ^b\big)_{n} \vert \S \rangle = \big( J_b ^a\big)_{n} \vert \S \rangle = 0 , \quad\quad 1\leq a<b \leq N-1, \;  n\geq 0 \\
    &\big( \fh_i \big)_n \vert \S \rangle = 0 , \quad\quad i\neq N-1, \; n \geq 0 \\
    &\big( \fh_{N-1} \big)_n \vert \S \rangle = 0, \quad\quad n>0 \\ 
    &\big(\fh_{N-1} \big)_0 \vert \S \rangle = -k \vert \S \rangle.
\end{split}
\end{align}
Let $\S(w)$ be the operator corresponding to the state $\vert \S \rangle$ and define the \textit{spectral flow operator} by $\Sigma (\mathbf{x},w) \equiv \left( e^{\sum_{a=1} ^{N-1} x^a \left( J_a ^N \right)_0 } \Sigma \right) (w)$ (these are higher-rank generalizations of the ones defined in \cite{Gaiotto:2021tsq}). Note that $\left[\left(J_a ^N \right)_0 , \big(J_b ^N \big)_0\right] = 0$ for $a,b \neq N$ so that $e^{\sum_{a=1} ^{N-1} x^a \left(J_a ^N \right)_0 } = \prod_{a=1} ^{N-1} e^{x^a \left(J_a ^N \right)_0}$. We compute the OPEs of this operator with the current as
\begin{align} \label{eq:opespecsln-2}
\begin{split}
    &J_a ^b (z) \S (\mathbf{x},w) \sim \frac{x^b \p_{x^a} \S(\bx,w) }{z-w} ,\quad\quad  a,b\neq N\\
    &J_N ^b (z) \S (\mathbf{x},w) \sim  \frac{x^b\left(k-\sum_{c=1} ^{N-1} x^c \p_{x^c} \right) \S(\bx,w)}{z-w},\quad\quad b\neq N \\
    &J_a ^N (z) \S(\mathbf{x},w) \sim \frac{\p_{x^a} \S(\bx,w)}{z-w} ,\quad\quad a\neq N  \\
    &\fh_i (z) \S(\mathbf{x},w) \sim \frac{\left( x^i \p_{x^i} -x^{i+1}\p_{x^{i+1}} \right)\S(\bx,w)}{z-w},\quad\quad i\neq N-1\\
    &\fh_{N-1}(z) \S(\mathbf{x},w) \sim \frac{\left(x^{N-1} \p_{x^{N-1}} +\sum_{c=1} ^{N-1} x^c \p_{x^c} -k \right) \S(\bx,w)}{z-w} .
\end{split}
\end{align}
From these OPEs, we recognize the twisted vacuum module is in fact induced from the lowest-weight Verma module $V_{(0,\cdots, 0, -k)} \subset H_{(0,\cdots,0,-k),k}$ realized by the space of polynomials $\BC \left[ \frac{X^1}{X^N} ,\cdots, \frac{X^{N-1}}{X^N} \right] = \BC[x^1, \cdots, x^{N-1}]$, on which the $\fsl(N)$ generators act as the differential operators
\begin{align}
    \big( T^b _a \big) \vert _0 =  -X^b \frac{\p}{\p X^a} ,
\end{align}
twisted by $\left(X^N \right) ^{k}$. Indeed, we see the numerators of the OPEs \eqref{eq:opespecsln-2} are precisely these twisted differential operators.

However, the twisted vacuum module is not the full induced $\widehat{\fsl}(N)$-module $\text{Ind}_{\fsl(N)\otimes \BC[[t]] \otimes \BC K} ^{\widehat{\fsl}(N)} V_{(0,\cdots, 0, -k)}$ itself, but instead its quotient by a proper submodule. Note that there is a proper submodule generated by the states $\big(J_N ^a \big)_{-1} \vert \S \rangle$, $a=1,2,\cdots,N-1$. Following from the definition \eqref{eq:sfmodinv} of the twisted vacuum module, there are null relations $\big(J_N ^a \big)_{-1} \vert \S \rangle = 0$, $a =1,2,\cdots, N-1$, so that the twisted vacuum module is the quotient of the induced module $\text{Ind}_{\fsl(N)\otimes \BC[[t]] \otimes \BC K} ^{\widehat{\fsl}(N)} V_{(0,\cdots, 0, -k)}$ by this proper submodule. The null relations lead to
\begin{align}
\begin{split}
&0 = \left[ \big( J_N ^a \big)_{-1} -  \d_b ^a  \Big( x^b \big(  J^N _N \big)_{-1} + (x^b) ^2 \big( J_b ^N \big)_{-1} \Big) + x^b \big( J_b ^a \big)_{-1} \right] e^{x^b \left( J_b ^N \right)_0} \vert \S \rangle, \quad\quad a,b \neq N, 
\end{split}
\end{align}
which give
\begin{align} \label{eq:nullst}
            &0 = \left[ \big( J_N ^a \big)_{-1} - x^a \sum_{b=1} ^N x^b \big(J_b ^N \big)_{-1}  +\sum_{b=1} ^{N-1}  x^b \big( J_b ^a \big)_{-1} \right] e^{ \sum_{b=1 }^{N-1}  x^b \left( J_b ^N \right)_0} \vert \S \rangle , \quad\quad a =1,\cdots, N-1.
\end{align}
where we are using the notation $x^N  = 1$.

\subsubsection{Coinvariants with twisted vacuum module}
Let us consider coinvariants $\boldsymbol\U$ of $n+1$ $\widehat{\fsl}(N)$-modules, where one of them is the twisted vacuum module and the rest $n$ of them are the irreducible modules induced from certain $\fsl(N)$-modules. The coinvariants satisfy the Ward identities, the Knizhnik-Zamolodchikov equations, and finally the constraints associated to the null states \eqref{eq:nullst}.

\paragraph{The conformal Ward identities} The coinvariants satisfy the conformal Ward identities, which read
\begin{align} \label{eq:glconf}
    0 = \sum_{i=0} ^{n} L^{\pm,0} _i \boldsymbol\U (\bz),
\end{align}
where we wrote the Virasoro generators as
\begin{align}
    L^+_i = z_i ^2 \p_{z_i} + 2 \D_i z_i,\quad\quad L^-_i = \p_{z_i} , \quad\quad L^0 _i = z_i\p_{z_i} + \D_i,
\end{align}
with the conformal weights given by the quadratic Casimir, $\D_i = \frac{\sum_{a,b=1} ^N \big(\bar{T}^a _b\big)\vert_i \big(\bar{T}^b _a\big)\vert_i}{2(k+N)}$. For the twisted vacuum module, it is straightforward to compute the conformal weight from \eqref{eq:opespecsln-2} as $\D_0 = \frac{(N-1)k}{2N}$.

\paragraph{The current algebra Ward identities} Similar to the conformal Ward identities, we have the current algebra Ward identities realized as differential equations satisfied by the coinvariant $\boldsymbol\U (\bU,\tilde\bU,\mathbf{X};\bz)$. It is nothing but the global $\fsl(N)$-invariance constraint:
\begin{align} \label{eq:gslnconst}
0= \sum_{i=0} ^n \big( \bar{T}^b _a \big)\vert_i \boldsymbol\U (\bU,\tilde\bU,\mathbf{X};\bz), \quad a,b=1,2,\cdots, N,
\end{align}
where we remind that $\big( \bar{T}^b _a \big)\vert_i$, $a,b=1,2,\cdots, N$ are the $\fsl(N)$ generators represented on the $i$-th module.

\paragraph{The Knizhnik-Zamolodchikov equations} We also have the KZ equations associated to the $n+1$ vertex operators. The Sugawara construction\footnote{Let $K(J_0 ^\a , J^\b _0 ) = g^{\a\b}$. The Sugawara construction gives $T(z) = \frac{1}{2(k+N)} \sum_{\a} (J_\a J^\a) (z)$ where $J_\a$ is dual to $J^\a$ with respect to the Killing form $K$, i.e., $K(J_{\a,0} , J^\b _0) = \d_\a ^\b$. We write out $J_\a = g_{\a\b} J^\b$ by using $g_{\a\b} g^{\b\g} = \d_\a ^\g$.} of the stress tensor yields
\begin{align} \label{eq:sugsln}
\begin{split}
    L_{-1} &= \frac{1}{2(k+N)} \sum_{n\in \BZ} \sum_{\a,\b} \left[ K(J^\a _0 , J^\b_0) \right]^{-1} : J^\a _n J^\b _{-1-n} : \\
    &= \frac{1}{2(k+N)} \sum_{ n \in \BZ} \left[ \sum_{r,s=1} ^{N-1} \left( \text{min}\{r,s\} - \frac{r s}{N} \right) :\big( h_r \big)_n  \big( h_s \big)_{-1-n}: + \sum_{a\neq b}  : \big( J^a _b \big)_n    \big( J^b _a \big)_{-1-n} : \right] \\
    &= \frac{1}{k+N} \left[ \sum_{r,s=1} ^{N-1} \left( \text{min}\{r,s\} - \frac{rs}{N} \right) \big( h_r \big)_{-1}  \big( h_s \big)_{0} +  \sum_{a \neq b} \big( J^a _b \big)_{-1} \big( J^b _a \big)_0  + \cdots \right].
\end{split}
\end{align}
Applying this identity to the $i$-th primary state $\vert V_i\rangle$ for each $i=1,2,\cdots, n$ ($\vert V_0 \rangle = \vert \S (\bx) \rangle$ will be treated separately below), we get $n$ null-states. Note that the terms in the ellipses do not contribute since $\vert V_i \rangle$ is a primary state. Then the KZ equations follow from inserting these null-states into the coinvariants. The equations read
\begin{align} \label{eq:slNkz}
    0=\left[(k+N) \frac{\p}{\p z_i} - \sum_{\substack{j=1\\j\neq i}} ^n \frac{\big( \bar{T}_a ^b \big) \vert_i \otimes \big( \bar{T}_b ^a \big) \vert_j}{z_i -z_j} - \frac{\big( \bar{T}_a ^b \big) \vert_i \otimes \big( \bar{T}_b ^a \big) \vert_0}{z_i -z_0} \right] \boldsymbol\U (\bU,\tilde\bU,\mathbf{X};\bz),\quad i=1,2,\cdots, n.
\end{align}

\paragraph{Constraints from twisted vacuum module} So far, the equations we derived did not rely on the property of the twisted vacuum module. Due to the quotient in defining the twisted vacuum module, there were null-state \eqref{eq:nullst} which leads to the following constraint when inserted in the coinvariants
\begin{align} \label{eq:constsln}
 0 = \sum_{i=1} ^n \frac{1}{z_0-z_i} \left[ \big(T_N ^a\big)\vert_i - x^a \sum_{c=1} ^N x^c \big( T_c ^N \big)\vert_i + \sum_{b=1} ^{N-1} x^b \big(T_b ^a\big)\vert_i \right] \boldsymbol\U (\bU,\tilde\bU,\mathbf{X};\bz), \quad\quad a=1,2,\cdots, N-1.
\end{align}

Meanwhile, applying the defining relation \eqref{eq:sugsln} of the Sugawara tensor to the state $e^{ \sum_{a=1} ^{N-1} x^a \left( J_a  ^N \right)_0} \vert \S \rangle $, we obtain another null-state. Note that the terms in the ellipsis in \eqref{eq:sugsln} do not contribute due to the commutation relations and the definition of $\vert \S \rangle$ \eqref{eq:sfmodinv}. The operator corresponding to this null-state can be inserted in the coinvariants to yield
\begin{align} \label{eq:twkz}
    0= \left[ \frac{\p}{\p z_0} - \frac{1}{k+N} \sum_{i=1}^n \frac{1}{z_0-z_i} \left\{ \sum_{r,s=1} ^{N-1} \left(\text{min}\{r,s\} -\frac{r s}{N} \right) \big(\fh_r \big)\vert_i \big( \fh_s\big) \vert_0 + \sum_{a\neq b} \big( T^b _a \big)\vert _i \big( T^a _b \big) \vert_0  \right\} \right]\boldsymbol\U .
\end{align}

Using the $x^a$-derivatives of the constraint \eqref{eq:constsln}, all the $x^a$-derivatives of $\boldsymbol\U$ in \eqref{eq:twkz} can be cancelled, simplifying the equation into
\begin{align} \label{eq:slnlax1}
    0 = \left[\frac{\p}{\p z_0}  + \sum_{i=1} ^n \frac{1}{z_0-z_i} \left( \sum_{a=1} ^{N-1} \big(T_a ^N \big)\vert_i x^a+ \big(T^N _N \big)\vert_i - \frac{1}{N} \sum_{a=1} ^N \big( T^a _a \big)\vert_m  \right) \right]\boldsymbol\U .
\end{align}
Note that the explicit dependence on the level $k$ cancels nontrivially. 

Multiplying $x^b$ ($1\leq b < N$) to this equation and using \eqref{eq:constsln} again, we also get
\begin{align}\label{eq:slnlax2}
    0 = \left[\frac{\p}{\p z_0}  x^b + \sum_{i=1} ^n \frac{1}{z_0 -z_i} \left\{ \left( \big( T^b _b\big)\vert_i- \frac{1}{N} \sum_{a=1} ^N \big( T^a _a\big) \vert_i \right) x^b +\big( T_1 ^b \big)\vert_i + \sum_{c\neq b,N} \big( T_c ^b \big)\vert_i x^c \right\} \right]\boldsymbol\U(\bx,\bz).
\end{align}
Thus, the set of $N$ constraints \eqref{eq:constsln} and \eqref{eq:twkz} is equivalent to the new set of $N$ constraints \eqref{eq:slnlax1} and \eqref{eq:slnlax2}. The latter can be organized into a matrix form as follows,
\begin{align} \label{eq:slNlax}
\begin{split}
    0 &= \left[\frac{\p}{\p z_0} +\sum_{i=1}^n \frac{1}{z_0-z_i} \begin{pmatrix}
    \big( \bar{T}^1 _1 \big)\vert_i & \big( T_2 ^1 \big)\vert_i & \big( T_3 ^1 \big)\vert_i & \cdots & \big( T_N ^1 \big)\vert_i \\ \big( T_1 ^2 \big)\vert_i  & \big( \bar{T}_2 ^2 \big)\vert_i & \big( T_3 ^2 \big)\vert_i & \cdots & \big( T_N ^2 \big)\vert_i  \\ \vdots & \vdots & \vdots & \cdots & \vdots &  \\ \big( T_1 ^N \big)\vert_i & \big( T_2 ^N \big)\vert_i & \cdots & \cdots & \big( \bar{T}^N _N \big)\vert_i \end{pmatrix} \right] \begin{pmatrix}
    x^1 \boldsymbol\U(\bx,\bz) \\ x^2 \boldsymbol\U(\bx,\bz) \\ \vdots \\ x^{N-1} \boldsymbol\U(\bx,\bz)  \\  \boldsymbol\U(\bx,\bz)  \end{pmatrix} \\
    &= \left[ \frac{\p}{\p z_0} + \sum_{i=1} ^n \frac{\sum_{a,b=1}^N \big( \bar{T} ^b _a \big)\vert_i \otimes E^a _b}{z_0 -z_i}  \right] \bx \boldsymbol\U(\bx,\bz).
\end{split}
\end{align}
where we have defined $\big( \bar{T}_a ^b \big)\vert_i \equiv \big( T_a ^b \big)\vert_i - \d_a ^b \frac{1}{N} \sum_{c=1} ^N \big( T^c_c \big)\vert_i$ and $E_a ^b = e_a \otimes \tilde{e}^b$ is the standard basis of $\text{End}(\BC^N)$. Note that each $\big(\bar{T}^a _a \big)\vert_i$, $a=1,2,\cdots, N$, is a linear combination of $\big( \fh_r \big)\vert_i = \big( T^r _r \big)\vert_i - \big( T^{r+1} _{r+1} \big)\vert_i$, $r=1,2,\cdots,N-1$. Namely, each residue at $z=z_i$ is an $N\times N$ matrix whose entries are the $\fsl(N)$ generators represented on the $i$-th module. As they are $\fsl(N)$ generators, the trace vanishes; $\sum_{a=1} ^N \big(\bar{T}^a _a\big)\vert_i = 0$. We also used the notation $\bx = (x^1 ,x^2 , \cdots, x^{N-1}, 1)$, where we set $x^N = 1$ by convention.

It will turn out to be useful to put the equations in this particular matrix form, to make an explicit construction of the universal opers for the Gaudin model. The absence of the explicit $k$ dependence, as opposed to the usual Knizhnik-Zamolodchikov equations \eqref{eq:KZ}, will be a crucial property for this aspect. We elaborate on this construction later in section \ref{sec:gaudin}. For now, \eqref{eq:slNlax} is just a convenient way of organizing the constraints on the coinvariants with the twisted vacuum module inserted.

Note, however, that there is an obstruction if we would simply add the twisted vacuum module to the set of $n$ $\widehat{\fsl}(N)$-modules that was used to construct the $n$-point coinvariants to form $(n+1)$-point coinvariants. The problem is that, if we had taken the weights of the modules generic, the $n$-point coinvariants can be built only by two Verma modules (one lowest-weight and one highest-weight) and $n-2$ bi-infinite modules. Then it is not possible to form a coinvariant when the twisted vacuum module is added, because it is another lowest-weight module, unless the weight parameters are tuned in special ways. Rather, to form a $(n+1)$-point coinvariant the $n$ modules other than the twisted vacuum module should consist of one highest-weight module and $n-1$ bi-infinite modules.

To keep the two Verma modules of highest-weight and lowest-weight, we need a bi-infinite module that replace the twisted vacuum module but constrain the coinvariants in the same way as the twisted vacuum module does. In this sense, we extend the set of vertex operators corresponding to the Hecke modification.

\subsubsection{Reducible bi-infinite modules}
Let us consider the bi-infinite module $H_{\boldsymbol\t,\s}$, where $\t_r = h_r -h_{r+1}$, $r=1,2,\cdots, N-1$, and $\s = \sum_{a=1} ^N h_a$. We explicitly write it as a vector space $H_{\boldsymbol\t,\s}= \bigoplus_{\substack{\mathbf{n} \in \BZ^N \\ \sum_{a=1} ^N n_a = 0}} \BC \vert \mathbf{n} \rangle$, on which the $\fsl(N)$ generators act as
\begin{align}
\big(J_a ^b \big)_0 \vert \mathbf{n}\rangle = (h_b+n_b) \vert \mathbf{n} +\d_a -\d_b\rangle,
\end{align}
where we use the notation $\d_a = \left( \d_{a,b} \right)_{b=1,2,\cdots,N}$. We define the series that generates all the states as
\begin{align}
\begin{split}
    \vert \S (\bx) \rangle &:= \sum_{\substack{\mathbf{n} \in \BZ^N \\ \sum_{c=1} ^N n_c =0 }}  \prod_{a=1} ^N\frac{\G(n_a+1)}{ \G\left( h_a+n_a+1 \right)}  (X^a)^{h_a+n_a} \vert \mathbf{n} \rangle, \\ &=(X^N)^\s\sum_{\mathbf{n} \in \BZ^{N-1} }  \frac{\G\left( \s +1 -\sum_{b=1} ^{N-1} h_b \right) }{ \G\left( \s+1-\sum_{b=1} ^{N-1} (h_b+n_b) \right) } \prod_{a=1} ^{N-1}\frac{\G\left( h_a+1 \right)}{\G\left( h_a+n_a+1 \right)}  (x^a)^{h_a+n_a} \vert \mathbf{n} \rangle,
\end{split}
\end{align}
with formal parameters $x^a = \frac{X^a}{X^N}$, $a=1,2,\cdots, N-1$. Note that the $\fsl(N)$ generators now act on $\vert \S (\bx)\rangle$ as the twisted differential operators,
\begin{align}
\begin{split}
    &\big(J_a ^b \big)_0 \vert \S (\bx)\rangle = x^b \p_{x^a} \vert \S (\bx)\rangle, \quad\quad a\neq N,\\
    &\big(J_N ^b \big)_0 \vert \S (\bx)\rangle = x^b \left(\s - \sum_{c=1} ^{N-1} x^c \p_{x^c}  \right) \vert \S (\bx)\rangle,
\end{split}
\end{align}
where we are using the notation $x^N =1$. Note also that, if we set $h_a = 0$ for $a=1,2,\cdots, N-1$, the summation truncates to non-negative integers due to the $\G$-functions in the denominator. In turn, we get
\begin{align}
\begin{split}
    \vert \S (\bx) \rangle &= (X^N)^\s  \sum_{\mathbf{n} \in \left(\BZ_{\geq 0} \right)^{N-1}}  \frac{\G\left( \s+1 \right)}{\G \left( \s+1 -\sum_{b=1} ^{N-1} n_b\right)} \prod_{a=1} ^{N-1}  \frac{1}{n_a !} (x^a)^{n_a} \vert \mathbf{n}\rangle \\
    &=  (X^N)^\s   \sum_{\mathbf{n} \in \left(\BZ_{\geq 0} \right)^{N-1}} \prod_{a=1} ^{N-1} \frac{1}{n_a !} (x^a)^{n_a} \big( J_a ^N\big)_0 \vert 0 \rangle \\
    &= (X^N )^\s e^{\sum_{a=1} ^{N-1} x^a \big(J_a ^N\big) _0 } \vert 0\rangle,
\end{split}
\end{align}
which is precisely the state associated to the spectral flow operator for the twisted vacuum module, if we had set $\s=k$. Namely, if we set $\s = k$, there is a proper submodule of $\mathbb{H}^{k} _{(0,\cdots, 0,-k),k} : = \text{Ind}^{\widehat{\fsl}(N)} _{\fsl(N) \otimes \BC[[t]] \oplus \BC K} H _{(0,\cdots, 0,-k),k}$, and the quotient is nothing but the twisted vacuum module that we have studied.

Now, we will show that even at generic $\boldsymbol\t$, the induced module $\mathbb{H}^{k} _{\boldsymbol\t,k} : = \text{Ind}^{\widehat{\fsl}(N)} _{\fsl(N) \otimes \BC[[t]] \oplus \BC K} H _{\boldsymbol\t,k}$ is reducible, so that the quotient by the proper submodule generalizes the twisted vacuum module to generic values of the weight parameter $\boldsymbol\t$. It is a bi-infinite generalization of the twisted vacuum module, in the sense that the weights of the states grow in both directions. This is precisely the property that we need to construct non-vanishing twisted coinvariants when composed with a lowest-weight and a highest-weight Verma modules carrying generic weights.

The induced $\widehat{\fsl}(N)$-module $\mathbb{H}^{k} _{\boldsymbol\t,\s} $ is obtained by declaring the action of $K$ is the scalar multiplication by $k \in \BC$ and also acting the negative modes of the current $\big( J_a ^b \big)_{m}$, $m <0$, to the states in $H _{\boldsymbol\t,\s}$. Thus, the induced module is spanned by $\big(J_{a_1} ^{b_1} \big)_{m_1}\cdots \big(J_{a_p} ^{b_p} \big)_{m_p} \vert \mathbf{n}\rangle$, where $m_1 \leq m_2 \leq \cdots \leq m_p <0$. Let us call $\sum_{s=1} ^p m_s$ the degree of the state.

At degree $1$, there are $N^2$ states $\big( J_a ^b \big)_{-1} \vert \mathbf{n} -\d_a + \d_b \rangle$, $a,b=1,2,\cdots, N$, having the same weights $h_c + n_c$ under $\big( J_c ^c\big)_0$, $c=1,2,\cdots, N$. We find linear combinations $v[1]$ of these $N^2$ states annihilated by all the degree $1$ generators,
\begin{align} \label{def:v1}
    \big (J_a ^b \big)_1 v[1] = 0, \quad\quad a,b=1,2,\cdots, N.
\end{align}
As explicitly shown in appendix \ref{app:gentwvac}, non-zero solutions to the these conditions exist only when $\s = \sum_{a=1} ^N h_a = k$. In this case, there are $N-1$ solutions labelled by $a=1,2,\cdots, N-1$,
\begin{align}
\begin{split}
    v_{\mathbf{n}} ^a [1] &:= \left(k+1 - \sum_{c=1} ^{N-1} (h_c+ n_c) \right)\left(k+2 - \sum_{c=1} ^{N-1} (h_c+ n_c) \right) \big( J^a _N \big)_{-1} \vert \mathbf{n} \rangle \\
    & \quad  - (h_a+n_a)\left(k+2 - \sum_{c=1} ^{N-1} (h_c+ n_c) \right) \big( J^N _N \big)_{-1} \vert \mathbf{n} -\d_a \rangle \\
    &+\sum_{b=1} ^{N-1} (h_b+n_b) \left( \left(k+2 - \sum_{c=1} ^{N-1} (h_c+ n_c) \right) \big( J^a _b \big)_{-1} \vert \mathbf{n} - \d_b \rangle - (h_a +n_a -\d_a ^b) \big( J^N _b \big)_{-1} \vert \mathbf{n} -\d_a  - \d_b \rangle  \right).
\end{split}
\end{align}
These states generate a submodule of the induced $\widehat{\fsl}(N)$-module $\mathbb{H}^{k} _{\boldsymbol\t,k}$, which is proper by definition \eqref{def:v1}. We take the quotient by this proper $\widehat{\fsl}(N)$-module by setting the generating state to be a null-state, namely,
\begin{align}
    0 =  \left[ \big( J_N ^a \big)_{-1} - x^a \sum_{b=1} ^N x^b \big(J_b ^N \big)_{-1}  +\sum_{b=1} ^{N-1}  x^b \big( J_b ^a \big)_{-1} \right] \vert \S (\mathbf{x})\rangle ,\quad\quad a=1,2,\cdots, N-1.
\end{align}
Note that these constraints are identical to \eqref{eq:nullst} for the generating state of the twisted vacuum module. In this sense, the quotient module we just constructed is a bi-infinite generalization of the twisted vacuum module. When inserted, it constrains the twisted coinvariants in exactly the same way as the twisted vacuum module does.

\subsection{Hecke operator and parallel surface defects}
We will show the insertion of the (bi-infinite generalization of) twisted vacuum module gives the action of Hecke modification on the twisted coinvariants by solving the constraints. The Hecke operator is defined by an integral of Hecke modifications. We also verify these constraints are obeyed by the correlation function of parallel surface defects. We layout the consequences of such $\EN=2$ gauge theoretical construction.

\subsubsection{Action of Hecke modifications on coinvariants} \label{subsubsec:heckemod}
Inserting a twisted vacuum module (or its bi-infinite generalization) defines an action on the twisted coinvariants. Here, we show explicitly this action is a Hecke modification by solving the constraints on the coinvariants.

To illustrate, let us consider the rank $1$ case, $G_\BC = PGL(2)$. The two constraints from the (bi-infinite generalization of) twisted vacuum module can be written as a special case of \eqref{eq:slNlax} at $N=2$,
\begin{align}
    0 = \left[ \frac{\p}{\p z_0} + \sum_{i=1} ^n \frac{1}{z_0 - z_i} \begin{pmatrix}
        T^0 _i & T^- _i \\ T^+ _i & - T^0 _i
    \end{pmatrix} \right] \boldsymbol\U_{n+1},
\end{align}
where $T^a _i$, $a=\pm, 0$, are $\fsl(2)$ generators represented on the $i$-th module,
\begin{align}
    T^+_ i = -x_i ^2 \p_{x_i} +2 s_i x_i ,\quad \quad T^- _i = \p_{x_i}, \quad\quad T^0 _i = x_i \p_{x_i} -s_i.
\end{align}
The constraints are explicitly solved by making the following ansatz:
\beq \label{eq:anshecke1}
{\bf\U}_{n+1} = \prod_{i=1}^{n} \left( \frac{(x_{i}-x_{0})^{2}}{z_{i}-z_{0}} \right)^{s_{i}} \, {\boldsymbol\Psi}_{n}
\left( {\xi}_{1}, \ldots , {\xi}_{n} ; z_{1}, \ldots , z_{n} \right)
\eeq
where we defined
\beq \label{eq:xi}
{\xi}_{i} =-\frac{z_{i}-z_{0}}{x_{i}-x_{0}},\quad\quad i=1,2,\cdots, n.
\eeq
Then we further impose the usual $(n+1)$-point KZ equations \eqref{eq:slNkz} associated to the other $n$ modules, as well as the Ward identities. An explicit computation shows that the $(n+1)$-point KZ equations become exactly the $n$-point KZ equations satisfied by $\boldsymbol\Psi_n$, where $(x_i)_{i=1} ^n$ get replaced by the new variables $(\xi_i)_{i=1} ^n$,
\begin{align} \label{eq:kznew}
\begin{split}
   & 0 = \left[(k+2)\frac{\p}{\p z_i} - \sum_{\substack{j=1 \\ j\neq i}} ^n \frac{T' _{a,i} \otimes T'^a _j}{z_i -z_j} \right] \boldsymbol\Psi_n, \\
    &T'^+ _i = - \xi_i ^2 \p_{\xi_i} + 2 s_i \xi_i ,\quad\quad T'^- _i = \p_{\xi_i} ,\quad\quad T'^0 _i = \xi_i \p_{\xi_i} -s_i.
\end{split}
\end{align}
The difference compared to the usual KZ equations is that, the function $\boldsymbol\Psi_n$ scales under the following $6$-dimensional gauge symmetry
\begin{align} \label{eq:newtran}
\begin{split}
&\left( {\xi}_{i} , z_{i} \right) \mapsto \left( a {\xi}_{i} + b z_{i} + c , z_{i} \right) \\
&\left( {\xi}_{i} , z_{i} \right) \mapsto \left( \frac{{\xi}_{i}}{Cz_{i}+D} , \frac{A z_i +B}{C z_{i} +D} \right)
\end{split}
\end{align}
with $A,B,C,D, b, c \in \mathbb{C}$, $AD - BC = 1$, and $a \in \mathbb{C}^{\times}$, as
\begin{align} \label{eq:newinv}
\begin{split}
    & \boldsymbol\Psi_n (a\xi_i +b z_i +c ,z_i) =  a^{ \sum_{i=1} ^n s_i -\frac{k}{2}}\boldsymbol\Psi_n (\xi_i,z_i) \\
    & \boldsymbol\Psi_n \left( \frac{\xi_i}{Cz_i +D} , \frac{Az_i +B}{Cz_i+D}  \right) = \prod_{i=1}^{n} (Cz_i+D) ^{\frac{s_i (k-2s_i)}{k+2}} \boldsymbol\Psi_n (\xi_i,z_i).
\end{split}
\end{align}

Recall that the variables $(x_i)_{i=1} ^n$ subject to the $SL(2)$ gauge symmetry parametrize the connected component $\text{Bun}_{PGL(2)} (\BP^1;S)_0 $ corresponding to the trivial bundle over $\BP^1$. Also, the $\xi$-variables subject to the gauge symmetry of \eqref{eq:newtran} give birational coordinates on the other connected component $\text{Bun}_{PGL(2)} (\BP^1;S)_1 $ corresponding to $\CalO \oplus \CalO(1)$ \cite{Etingof:2021eeu}. Then, the relation \eqref{eq:xi} defines a map
\begin{align}
\begin{split}
    HM_{z_0,x_0}:\text{Bun}_{PGL(2)} (\BP^1;S)_0 &\to \text{Bun}_{PGL(2)} (\BP^1;S)_1 \\
    x_i \quad &\mapsto \quad \xi_i =- \frac{z_i-z_0}{x_i-x_0},
\end{split}
\end{align}
which was shown in \cite{Etingof:2021eeu} to be the action of the Hecke modification at $z_0 \in \BP^1$, where $x_0 \in \BP^1$ gives a choice of Hecke modification (namely, it parametrizes the space of Hecke modifications $\BP^1$).\footnote{The two connected components are in fact identified by another Hecke modification at a marked point \cite{Etingof:2021eeu}.}

Just as the usual KZ equations are reduced when written in terms of the $SL(2)$-invariants of $x_i$ and $z_i$, the KZ equations \eqref{eq:kznew} for $\boldsymbol\Psi_n$ can be solved in terms of the invariants of $(\xi_i,z_i)_{i=1} ^n$ under the action of \eqref{eq:newtran}. The invariants are the cross-ratios of $(z_i)_{i=1} ^n$,
\beq
{w}_{a} = \frac{z_{a+1,1} z_{n-1, n}}{z_{n-1,1} z_{a+1,n}},\quad\quad a=1,2,\cdots, n-3,
\eeq
and 
\beq
\eta_{a} =  \frac{[a+1,n,1]}{[n-1,n,1]} \frac{z_{n-1,n}}{z_{a+1,n}}, \quad\quad a=1,2,\cdots, n-3,
\eeq
where we defined
\beq
[a,b,c] = {\xi}_{a} z_{b,c} + {\xi}_{b} z_{c,a} + {\xi}_{c} z_{a,b}.
\eeq
To solve the KZ equations \eqref{eq:kznew} and the invariance condition \eqref{eq:newinv} for $\boldsymbol\Psi_n(\boldsymbol\xi,\mathbf{z})$, we take an ansatz
\begin{align}
\begin{split}
&\boldsymbol\Psi_n (\boldsymbol\xi,\mathbf{z}) \\
&=\prod_{j=1}^{n-1} \left( \frac{z_{j,n}^{2} z_{1,n-1}}{z_{1,n}z_{n-1,n}} \right)^{\frac{s_j\left(\frac{k}{2} -s_j \right)}{k+2}} \left( \frac{z_{1,n} z_{n,n-1}}{z_{1,n-1}} \right)^{\frac{s_n\left(\frac{k}{2} -s_n \right)}{k+2}}  \left( \frac{[n-1,n,1]^2}{z_{n,1}z_{1,n-1}z_{n-1,n}} \right)^{-\frac{k}{4}+\frac{1}{2} \sum_{i=1} ^n s_i }  {\Psi}_n\left( {\boldsymbol\eta}, {\mathbf{w}} \right)
\end{split}
\end{align}
The KZ equations reduce to differential equations satisfied by $\Psi_n (\boldsymbol\eta,\mathbf{w})$. To illustrate, let us take the simplest non-trivial case $n=4$ as an example. Let $u = \eta_{1}$, ${\qe} = {w}_{1}$, and also let us decompose $\Psi_4 (u,\qe)$ by perturbative and non-perturbative parts as
\begin{align}
& {\Psi}_4(u, {\qe}) = {\qe}^{A_{0}} (1-{\qe})^{A_{1}} u^{-{\nu}_{0}} (u-{\qe})^{-{\nu}_{\qe}} (u-1)^{-{\nu}_{1}}\, {\psi}_4 (u,{\qe}).
\end{align}
Then, the KZ equation becomes
\begin{align} \label{eq:anotherred}
\begin{split}
&  0 = \left[- (k+2) \frac{\partial }{\partial\qe} + \frac{u(u-1)(u-\qe)}{\qe(1-\qe)} \left( - \frac{\partial^{2} }{\partial u^{2}}
+ U (u, {\qe}) \right) \right]\psi_4 (u,\qe),  \\
& U (u, {\qe}) = \frac{{\nu}_{0}({\nu}_{0}-1)}{u^2} + \frac{{\nu}_{1}({\nu}_{1}-1)}{(u-1)^2}+\frac{{\nu}_{\qe}({\nu}_{\qe}+k+1)}{(u-{\qe})^2} + \\
& \qquad\qquad \frac{{\nu}_{\infty}({\nu}_{\infty}-1)-{\nu}_{0}({\nu}_{0}-1)-{\nu}_{1}({\nu}_{1}-1)-{\nu}_{\qe}({\nu}_{\qe} +k+1)}{u(u-{\qe})},
\end{split}
\end{align}
where
\begin{align} \label{eq:spinhec}
\begin{split}
& 2{\nu}_{0} = \frac{k+2}{2} - s_{1}  - s_{2} + s_{3} - s_{4} \, , \\
& 2{\nu}_{\qe} = -\frac{k+2}{2} - s_{1} - s_{2} - s_{3} +s_{4} \, , \\
& 2{\nu}_{1} = \frac{k+2}{2} + s_{1} - s_{2} - s_{3} - s_{4}\, , \\
& 2{\nu}_{\infty} =  \frac{k+2}{2} - s_{1} + s_{2} - s_{3} - s_{4} \, , \\
& \\
& A_{0} = -2 {\Delta}_{2} + \frac{s_{2} -s_{1}}{k+2} \left( \frac{k+2}{2} + s_{1}+s_{2}-s_{3}-s_{4} \right)\\
& 2A_{1} = -\frac{k+2}{4} - s_{2}-s_{3} + \frac{(s_{1}-s_{4})^{2} - (s_{2}-s_{3})^{2}}{k+2}.
\end{split}
\end{align}
This is precisely the reduced 4-point KZ equation. Thus, the insertion of the (bi-infinite generalization of) twisted vacuum module acts as a Hecke modification on the coinvariants. Given a $n$-point coinvariant ${\boldsymbol\Psi}_{n}( \bx; \bz)$, the image of the Hecke modification at the point $z_0$ with parameter $x_0$ is $\boldsymbol\Psi_n (\boldsymbol\xi;\bz)$ with the help of \eqref{eq:anshecke1}. We will define Hecke operator by integrating over $x_0$ in section \ref{subsubsec:hecke}. Note that there is a spin flip $s_4 \mapsto \frac{k+2}2 - s_4$ in \eqref{eq:spinhec} compared to the coinvariant before the Hecke modification. In turn, the reduced KZ equation \eqref{eq:anotherred} manifests the birational isomorphism of the two connected components of $\text{Bun}_{PGL(2)} (\BP^1;S)$ defined by another Hecke modification at a marked point.

\subsubsection{Geometric realization of coinvariants with twisted vacuum module}
Now, we realize the coinvariants in the presence of the bi-infinite generalization of the twisted vacuum module in a geometric way, for the simplest nontrivial case $n=4$; namely, we have the five-point coinvariant with two Verma modules, two bi-infinite modules, and the quotient of the bi-infinite module. We solve the Ward identities \eqref{eq:glconf} and \eqref{eq:gslnconst} by making an ansatz
\begin{align} \label{eq:twist5}
\begin{split}
&{\boldsymbol\U} (\bU,\mathbf{\tilde{U}},\mathbf{X};\bz) = \U_0 (\bU,\tilde{\bU},\mathbf{X};\bz) \Upsilon (\boldsymbol\g, \boldsymbol\m;\qe,y) \\
&\U_0 = \prod_{p=0,2} \left( \frac{z_{p 1} ^2 z_{43}}{z_{41} z_{31} } \right)^{-\D_p} \frac{\prod_{a=1}^N \prod_{p=0,2,3} \left( \tilde\pi ^a \left( X_p \wedge \pi_{a-1} \right) \right)^{\b  _{p,a}} \prod_{i=1} ^{N-1} \left( \tilde\pi ^i (\pi_i) \right)^{\a_i} }{z_{14} ^{\D_1 +\D_4 -\D_3} z_{13} ^{\D_1 +\D_3 -\D_4} z_{43} ^{\D_3 + \D_4 -\D_1}}\prod_{i=0} ^4 dz_i ^{\D_i},
\end{split}
\end{align}
where $\Upsilon (\boldsymbol\g, \boldsymbol\m;\qe,y)$ only depends on
\begin{align}
\begin{split}
    &\g_a = \frac{\tilde{\pi}^a (X_2 \wedge \pi_{a-1})}{\tilde{\pi}^a (X_3 \wedge \pi_{a-1})}, \quad \m_a  =  \frac{\tilde{\pi}^a (X_0 \wedge \pi_{a-1})}{\tilde{\pi}^a (X_3 \wedge \pi_{a-1})}, \quad a=1,2,\cdots,N, \\
    &\qe = \frac{z_{21} z_{43}}{z_{24} z_{13}} ,\quad y = \frac{z_{01} z_{43}}{z_{04} z_{13}}.
\end{split}
\end{align}
Note that $\g_a$ and $\m_a$ suffer from remnant $\BC^\times$ ambiguities, but the ratios among themselves are well-defined. Thus we take $\Upsilon(\boldsymbol\g,\boldsymbol\mu;\qe,y)$ to be a degree 0 meromorphic function of $\g_a$ and $\m_a$; namely, $\sum_{a=1} ^N \g_a \p_{\g_a} \Upsilon = \sum_{a=1} ^N \m_a \p_{\m_a} \Upsilon = 0$. As shown in section \ref{subsubsec:heckemod}, the ratios of $(\g_a)_{a=1} ^N$ are holomorphic coordinates on $\text{Bun}_{PGL(N)} (\BP^1;S)_0$, while $(\m_a)_{a=1}^N$ are homogeneous coordinates on the space of Hecke modifications $\BP^{N-1}$.

The weights of the Verma modules and the HW modules are determined by
\begin{align}
\begin{split}
\begin{split}
&\z_i = \sum_{p=0,2,3} \b_{p,i+1}  + \a_i \\
&\tilde{\z}_i = \sum_{p=0,2,3} \b _{p,i}  + \a_i \\
& \t_i -\z_i = \b_{2,i} - \b_{2,i+1}  \\
& \tilde{\z}_i - \t_i = \b _{3,i} - \b_{3,i+1}  +\b_{0,i}-\b_{0,i+1}
\end{split} ,\quad\quad\quad i=1,2,\cdots, N-1,
\\
\begin{split}
& \s_p  = \sum_{b=1} ^N \b _{p,b},
\end{split}\quad\quad\quad\quad\quad\quad\quad\quad\quad\quad\quad\quad\quad\quad\quad\quad p=2,3,
\end{split}
\end{align}
The weights of the bi-infinite generalization of twisted vacuum module is given by $(\b_{0,a})_{a=1} ^N$ so that $\s_0 = \sum_{a=1}^N \b_{0,a} =k$. In fact, the weights $\b_{0,a}$ are not arbitrary but are determined by the constraints \eqref{eq:constsln}. Even though the twisted coinvariant \eqref{eq:twist5} is defined on the open subset where we have bi-infinite generalization of the twisted vacuum module, it can be continued to another open subset where we have the twisted vacuum module instead by compensating the lowest-weight Verma module to a bi-infinite module. In this case, we have simply $\b_{0,a} = k \d_{a,N}$.

Now the equations \eqref{eq:slNlax} and \eqref{eq:slNkz} satisfied by the coinvariant $\boldsymbol\U$ get converted into differential equations for $\U$. Namely,
\begin{subequations} \label{eq:constcoin}
\begin{align}
    &0 = \left[ \frac{\p}{\p y} + \frac{\hat{\CalA}_0}{y} + \frac{\hat{\CalA}_\qe}{y-\qe} +\frac{\hat{\CalA}_1}{y-1} \right] \begin{pmatrix}
    \m_1 \Upsilon \\{\m_2} \Upsilon\\ \vdots \\ {\m_N} \Upsilon
    \end{pmatrix} \\
    & 0 = \left[-(k+N)\frac{\p}{\p \qe} + \frac{\hat{\CalH}_0}{\qe} +\frac{\hat{\CalH}_y}{\qe-y} + \frac{\hat{\CalH}_1}{\qe-1} \right]\Upsilon.
\end{align}
\end{subequations}
It is straightforward to compute the twisted differential operators $\hat{\CalA}_{0,\qe,1}$ and $\hat{\CalH}_{0,y,1}$. The result is (see appendix \ref{app:5point} for the detail of the derivation)
\begin{align}
\begin{split}
        \left( \hat{\CalA}_0 \right)_{ab} & = \th_{b>a} \left( \b _{2,b} \frac{\g_a}{\g_b} + \b _{3,b}+1 +(\g_a -\g_b)\p_{\g_b} - \m_b \p_{\m_b} \right) \\
    & \quad + \d^b _a \left( -k  -\sum_{c=a+1} ^N   \sum_{p=2,3} \b _{p,c}  -\sum_{i=a} ^{N-1} \a_i +\sum_{c=a+1} ^N \m_c \p_{\m_c} \right)  \\
    &  \quad +  \frac{1}{N} \d_a ^b \left\{ k+ \sum_{c=1} ^N \left( \sum_{p=0,2,3} \sum_{d=c+1} ^N \b _{p,d} + \sum_{i=c} ^{N-1} \a_i \right) \right\}\\
        \left( \hat{\CalA}_\qe \right)_{ab} &= - \g_a \p_{\g_b} - \b_{2,b} \frac{\g_a}{\g_b} + \frac{1}{N} \d^a _b \sum_{c=1} ^N \b _{2,c} \\
    \left( \hat{\CalA}_1 \right)_{ab} &= \g_b \p_{\g_b}+ \m_b \p_{\m_b}  -\b_{3,b} -1 + \frac{1}{N} \d^a _b \sum_{c=1}^N \b _{3,c} ,
\end{split}
\end{align}
and
\begin{align}
\begin{split}
    &\hat{\CalH}_0 =  - \sum_{a>b}  \left( \g_a \p_{\g_b} + \b _{2,b} \frac{\g_a}{\g_b}  \right)  \left( \b _{0,a} \frac{\m_b}{\m_a} + \b _{2,a} \frac{\g_b}{\g_a} + \b _{3,a} +  (\g_b-\g_a) \p_{\g_a} + (\m_b -\m_a)\p_{\m_a}  \right) +\frac{(N-1)\s(\s+N)}{N}   \\
    &\quad\quad+ \sum_{a=1}^N \left( \g_a \p_{\g_a} + \beta_{2,a} \right) \left( \sum_{c=a+1}^N \sum_{p=0,2,3} \beta_{p,c} + \sum_{i=a}^{N-1} \alpha_i \right) 
    -\frac{1}{N} \left( \sum_{a=1}^N \b _{2,a} \right) \left( \sum_{b=1}^N \left( \sum_{p=2,3} \sum_{c=b+1} ^N \b _{p,c} + \sum_{i=b} ^{N-1} \a_i \right) \right) \\
    & \hat{\CalH}_y =\sum_{a,b=1} ^N \left( \g_b \p_{\g_a} + \b _{2,a} \frac{\g_b}{\g_a} \right) \left( \m_a \p_{\m_b} +\b_{0,b} \frac{\m_a}{\m_b}  \right) - \frac{k}{N} \sum_{a=1} ^N \b _{2,a}  \\
    &\hat{\CalH}_1=\sum_{a,b=1} ^N \frac{\g_b}{\g_a} \left(\g_a \p_{\g_a} + \b _{2,a} \right)\left(-\g_b \p_{\g_b} - \m_b \p_{\m_b} +\b_{3,b} \right) - \frac{1}{N} \sum_{a=1} ^N \b _{2,a} \sum_{b=1} ^N \b_{3,b},
\end{split}
\end{align}
where we are using the notation
\begin{align}
    \th_P = \begin{cases} 1 \quad\quad \text{if $P$ is true} \\ 0 \quad\quad \text{if $P$ is false} \end{cases}.
\end{align}

\subsubsection{Parallel surface defects and twisted coinvariants}
We will verify the correlation function of parallel surface defects gives a twisted coinvariant with an insertion of the (bi-infinite generalization of) twisted vacuum module by showing the former satisfies the constraints on the latter, which we have derived just so far.

The starting point is the fractional quantum TQ equations \eqref{eq:fracTQ}. We take summations over $\ve_1$-lattices, converting the generalized $Q$-observables into the correlation function $\U(\mathbf{a};\mathbf{u},\boldsymbol\m;\qe,y)$ \eqref{eq:G-F-N} (after multiplying the perturbative prefactor $\U^{\text{pert}} $). Then the fractional quantum TQ equations become differential equations satisfied by $\U(\mathbf{a};\mathbf{u},\boldsymbol\m;\qe,y)$, which can be organized into a matrix equation as (see appendix \ref{app:5point} for the detail of the computation)
\begin{subequations} \label{eq:constalign}
\begin{align}
    &0 = \left[ \frac{\p}{\p y} + \frac{\hat{\CalA}_0}{y} + \frac{\hat{\CalA}_\qe}{y-\qe} +\frac{\hat{\CalA}_1}{y-1} \right] \begin{pmatrix}
    \m_0 \Upsilon \\{\m_1} \Upsilon\\ \vdots \\ {\m_{N-1}} \Upsilon
    \end{pmatrix}.
\end{align}
\end{subequations}
The fractional quantum TQ equations do not exhaust all the non-trivial constraints imposed by the relations \eqref{eq:fracqq} from the $qq$-characters. Note that $\ET_\o (x,\bx')$ \eqref{eq:TwnN-2} contains $\ve_1 (D_\o ^{(1)} - \qe_\o D_{\o-1} ^{(1)}) = \ve_1 u_{\o+1} (\p_{u_{\o+1}} -  \p_{u_\o} )$. If we multiply $\frac{u_{\o+1} +u_{\o+2} + 
\cdots + u_{\o+N}}{u_{\o+1}}$ and sum over all $\o=0,1,\cdots, N-1$, we get $\ve_1 (1-\qe) \qe \frac{\p}{\p \qe}$, a $\qe$-derivative. Using this fact, we can derive (see appendix \ref{app:5point} for detail of the derivation)
\begin{align} \label{eq:qeq}
        & 0 = \left[\frac{\ve_2}{\ve_1}\frac{\p}{\p \qe} + \frac{\hat{\CalH}_0}{\qe} +\frac{\hat{\CalH}_y}{\qe-y} + \frac{\hat{\CalH}_1}{\qe-1} \right]\Upsilon.
\end{align}
The equations \eqref{eq:constalign} and \eqref{eq:qeq} obeyed by the correlation function $\U(\mathbf{a};\mathbf{u},\boldsymbol\m;\qe,y)$ of the parallel surface defects are identical to the equations \eqref{eq:constcoin} for the coinvariants. The matching of the equations is under the following mapping of the parameters,
\begin{itemize}
    \item Level: $k+N = -\frac{\ve_2}{\ve_1}$ 
    \item Coordinates on $\text{Bun}_{PGL(N)} (\BP^1;S)_0$: $\g_{\o} = \frac{u_{\o}+u_{\o+1}+ \cdots+u_{\o+N-1}}{\kq-1} $, \quad\quad $\o=0,1,\cdots, N-1$
    \item Homogeneous coordinates on the space of Hecke modifications $\BP^{N-1}$ : $ (\m_\o)_{\o=0} ^{N-1}$
    \item Weights of the modules:  \begin{align} \zeta_{i} & = \frac{m^-_{{i}}-m_{{i-1}}^- }{\ve_1} - 1, \quad     \tilde\zeta_i  = \frac{m^+_{i} - m_{{i-1}}^+ }{\ve_1} -1 , \quad  \tau_{i}  = \frac{a_{{i}}-a_{{i-1}} }{\ve_1} - 1, \quad i=1,\dots,N-1 \nonumber  \\
    \s_0 &= k, \quad \sigma_2  = N \frac{\bar{m}^-- \bar{a}-\ve_1}{\ve_1},\quad
    \sigma_3  = -\frac{N(\bar{m}^+ -\bar{a} +\ve_1) -\ve_2}{\ve_1} . \nonumber \end{align}
\end{itemize}
Here, we are shifting the subscripts of $\g_\o$'s and $\m_\o$'s by 1, so that $\o$ runs from $0$ to $N-1$ instead of $1$ to $N$. Note that the weight parameters for the Verma modules and the bi-infinite modules without quotient match with the 4-point case in the limit $\ve_2 \to 0$. This is precisely what we will need when we study the Hecke eigensheaf property below.

\subsubsection{Hecke operator and eigensheaf property} \label{subsubsec:hecke}
The Hecke operator is defined by an integral of Hecke modifications (see section \ref{subsubsec:heckeop}). To see how the integral naturally appears in the $(n+1)$-point twisted coinvariants with the insertion of the twisted vacuum module, note that $\boldsymbol\U$ defines a section of $\CalO(1) ^{k}$ over the space of Hecke modifications $\BP^{N-1}$. More precisely, the twisting part can be written as
\begin{align}
    \frac{\U_0}{\Psi_0} = \left( \frac{z_{41}z_{31}}{z_{01}^2 z_{43}} dz_0 \right)^{\frac{k(N-1)}{2N}}\left(\bigwedge_{a=1} ^{N-1} dx_{0} ^a \right)^{-\frac{k}{N}} = (y^{-2} dy)^{\frac{k(N-1)}{2N}} \left(\bigwedge_{a=1} ^{N-1} dx_{0} ^a \right)^{-\frac{k}{N}} .
\end{align}
Thus at the critical level $k=-N$, it becomes a section of the canonical bundle $\CalO(-N) \simeq K_{\BP^{N-1}}$, and can be integrated along any $(N-1)$-cycle on $\BP^{N-1}$.\footnote{Since we only have the holomorphic volume form, we can only integrate along real $(N-1)$-dimensional cycles. } Thus, we define the Hecke operator $H_{z_0}$ on the $n$-point twisted coinvariants $\boldsymbol\Psi$ at the critical level as the integral of its Hecke modifications at $z_0 \in \BP^1 \setminus S$, namely, the $(n+1)$-point twisted coinvariants $\boldsymbol\U$ with the insertion of the twisted vacuum module at the critical level, given by
\begin{align} \label{eq:HeckeOp}
\begin{split}
    H_{z_0} \boldsymbol\Psi &:= \oint_C  \lim_{k \to -N}  \boldsymbol\U (z_0)   \\
    & = y^{N-1} dy^{-\frac{N-1}{2}} \Psi_0 \oint_C \bigwedge_{a=1} ^{N-1} dx_{0} ^a \lim_{k\to-N} \U(y),
\end{split}
\end{align}
for a given $(N-1)$-cycle $C$. We will specify which cycle to use shortly. We will also denote $H_y \Psi(\boldsymbol\g;\qe) := y^{N-1} dy^{-\frac{N-1}{2}} \oint_C \bigwedge_{a=1} ^{N-1} dx_{0} ^a \lim_{k\to-N} \U(y)$ when the twisting $\Psi_0$ is factored out.

Let us first examine some useful properties of the image of the Hecke operator. Note that the $\fsl(N)$ generators represented on the quotient of the bi-infinite module are total derivatives at the critical level $k=-N$,
\begin{align}
\begin{split}
    &\U_0 ^{-1} \big( T_a ^b \big) \vert_0 \U_0 = - \p_{x^a _0} x^b _0 ,\quad\quad a\neq b,\, a,b\neq N \\
    & \U_0 ^{-1} \big( T^b _N \big) \vert_0 \U_0 =  \sum_{c=1}^{N-1} \p_{x^c} x_0 ^b x^c _0 -(k+N) x_0 ^b ,\quad\quad b\neq N \\
    &\U_0 ^{-1} \big( T_a ^N \big) \vert_0 \U_0= - \p_{x_0 ^a} ,\quad\quad a \neq N \\
    & \U_0 ^{-1} \big( \mathfrak{h}_i \big) \vert_0 \U_0 = \p_{x^{i+1} _0} x _0 ^{i+1} - \p_{x _0 ^i} x_0 ^i,\quad\quad i \neq N-1 \\
    & \U_0 ^{-1} \big( \mathfrak{h}_{N-1} \big)\vert_0 \U_0 = -\p_{x^{N-1} _0} x _0 ^{N-1} - \sum_{c=1} ^{N-1} \p_{x _0 ^c} x _0 ^c +(k+N).
\end{split}
\end{align}
Thus the image of the Hecke operator $H_{z_0} \boldsymbol\Psi$ is again invariant under the $n$-point global $\fsl(N)$ transformations due to the integration. It follows that the image of the Hecke operator also satisfies the $n$-point KZ equations, since the additional term originating from the presence of the $(n+1)$-th module $-$ the (bi-infinite generalization of) twisted vacuum module $-$ vanishes after the integration. Also, if we factor out the following simple $z_0$-part in $\U_0$ which is a $\left(- \frac{N-1}{2} \right)$-differential on $\BP^1 $,
\begin{align} \label{eq:pertadd}
    \lim_{k\to -N} dz_0 ^{\D_0} \left( \frac{z_{01} ^2 z_{43}}{z_{41} z_{31}} \right)^{-\D_0}  = y^{N-1} dy^{-\frac{N-1}{2}},
\end{align}
then the remaining part of the image of the Hecke operator obeys the $n$-point conformal Ward identities. All in all, it implies that the Hecke operator maps the space of $n$-point twisted coinvariants to itself, with the coefficient valued in $K_{\BP^1} ^{-\frac{N-1}{2}}$. 

As explained in section \ref{subsec:regdefcoin}, the $\EN=2$ gauge theory in the presence of the regular monodromy surface defect provides a distinguished family $\Psi(\mathbf{a})$ of twisted coinvariants, enumerated by the Coulomb moduli $\mathbf{a}$. Our claim is that these elements diagonalize the action of the Hecke operator. For this, we use the equivalence of the Hecke modification, constructed as the twisted coinvariant with a further insertion of the special bi-infinite module, and the parallel surface defects in the $\EN=2$ gauge theory. In this identification, the reduced twisted coinvariant $\U(y)$ appearing in \eqref{eq:HeckeOp} is given by \eqref{eq:G-F-N}. If we gather the $(\m_\o)_{\o=0} ^{N-1}$ part (we shift the subscripts so that $\o$ runs from $0$ to $N-1$),
\begin{align}
\begin{split}
    &\oint_{C} \bigwedge_{\o=0} ^{N-2} d \left(\frac{\m_\o}{\m_{N-1}} \right) \prod_{\o'=0 } ^{N-1} \frac{\m_{N-1}}{\m_{\o'}} \sum_{\bn \in \BZ^{N-1}} \prod_{i=0} ^{N-2} y_i ^{n_i} (\cdots) \\
    &\quad = (-1)^{N-1} \oint_{C} \bigwedge_{\o=0} ^{N-2} \frac{dy_\o}{y_\o} \sum_{\bn \in \BZ^{N-1}} \prod_{i=0} ^{N-2} y_i ^{n_i} (\cdots),
\end{split}
\end{align}
where we changed the integration variables to $(y_\o)_{\o=0} ^{N-2}$ by using \eqref{eq:changevar}. At this point, we choose the $(N-1)$-dimensional cycle $C$ to be the product of small loops of $y_\o$ enclosing $0$. Then, the integral simply picks up the $0$-th Laurent coefficient of the integrand.

As we noted earlier (see section \ref{subsub:parallel}), the $0$-th Laurent coefficient is nothing but the correlation function of the regular monodromy surface defect and the canonical surface defect without any 0-observable on their interface. This correlation function factorizes into the product of the vacuum expectation values of each, since the limit $\ve_2 \to 0$ was taken where the two surface defects can be arbitrarily separated on the topological plane $\BC_2$ (i.e., by cluster decomposition). Explicitly, the correlation function has the following asymptotics in the limit $\ve_2 \to 0$,
\begin{align} \label{eq:factor}
    \lim_{\ve_2 \to 0 }\Big\langle X(\mathbf{a};y) \Psi(\mathbf{a};\mathbf{u}) \Big\rangle_{\mathbf{a}} = e^{\frac{\widetilde{\EuScript{W}}(\mathbf{a})}{\ve_2}} \chi(\mathbf{a};y) \psi(\mathbf{a};\mathbf{u}),
\end{align}
where the singular part $\widetilde{\EW}(\mathbf{a})$ is the twisted superpotential for the effective two-dimensional $\EN=(2,2)$ theory, and the regular part $\chi(\mathbf{a};y)$ (resp. $\psi(\mathbf{a};\mathbf{u})$) is the normalized vacuum expectation value of the canonical surface defect (resp. the regular monodromy surface defect) in the limit $\ve_2 \to 0$. In particular, $\chi(\mathbf{a};y)$ does not carry any dependence on $\mathbf{u}$, nor does $\psi(\mathbf{a};\mathbf{u})$ depend on $y$. In turn, all the $z_0$-dependent parts $-$ namely, the $\left(- \frac{N-1}{2} \right)$-differential $\chi(\mathbf{a}) := \chi(\mathbf{a};y) dy^{-\frac{N-1}{2}} $  $-$ completely factors out in the image of the Hecke operator, while the rest $-$ namely, $\Psi(\mathbf{a};\mathbf{u})= \exp \left(\frac{\widetilde{\EW}(\mathbf{a})}{\ve_2} \right) \psi(\mathbf{a};\mathbf{u})$ multiplied by the twisting factor $\Psi_0$ $-$ recovers the original $n$-point twisted coinvariant $\boldsymbol\Psi(\mathbf{a})$ at the critical level. That is, we get
\begin{align} \label{eq:heceigensf}
    H_{z_0} \boldsymbol\Psi(\mathbf{a}) = \chi(\mathbf{a};y) \boldsymbol\Psi(\mathbf{a}),
\end{align}
so that the distinguished twisted coinvariants $\boldsymbol\Psi(\mathbf{a})$ diagonalize the action of the Hecke operator. It should be noted that the perturbative prefactor \eqref{eq:correlpert} of the correlation function, combined with the contribution from \eqref{eq:pertadd}, leads to the correct perturbative prefactors of the vacuum expectation values of the canonical surface defect \eqref{eq:canon} and the regular monodromy surface defect \eqref{eq:vevregmono} in the limit $\ve_2 \to 0$, making the factorization property \eqref{eq:heceigensf} valid.

Moreover, note here the \textit{eigenvalue} $\chi(\mathbf{a}) \in K_{\BP^1} ^{-\frac{N-1}{2}}$ is identified with the normalized vacuum expectation value of the canonical surface defect in the limit $\ve_2 \to 0$. In section \ref{subsec:oper}, we reviewed \cite{Jeong:2018qpc} that this normalized vacuum expectation value is a solution to the $SL(N)$-oper $\r_{\mathbf{a}} \in \text{Op}_{SL(N)} (\BP^1;S)$ understood as a $N$-th order differential operator on $\BP^1 \setminus \{0,\qe,1,\infty\}$; $\r_\mathbf{a} \chi(\mathbf{a};y) = 0$. For a given position $y \in \BP^1 \setminus \{0,\qe,1,\infty\}$ of the Hecke operator, the oper solution $\chi(\mathbf{a};y)$ is expressed as a series which is convergent only in a particular domain. We showed in section \ref{subsec:oper} that the solution in the domains $0<\vert \qe \vert < 1 < \vert y \vert$ and $0<\vert \qe \vert <  \vert y \vert < 1$ can be obtained from $\ve_1$-lattice summations of $Q$-observables \cite{Jeong:2018qpc}.\footnote{The solutions in the domain $0<\vert \qe \vert <  \vert y \vert < 1$ admit another presentation in terms of the dual $Q$-observable. The solutions in the remaining domain $0<  \vert y \vert  <\vert \qe \vert < 1$ can also be obtained by using the dual $Q$-observable. See \cite{Jeong:2018qpc}.} In each domain, there are natural $N$ independent basis elements of the space of oper solutions acquired by choosing $N$ different $\ve_1$-lattices that make the series converge. For the domain $0<\vert \qe \vert < 1 < \vert y \vert$, these are the $\ve_1$-lattices centered at $N$ mass parameters for the fundamental hypermultiplets, $L_\a = m^+ _\a + \ve_1 \BZ$, $\a=1,2,\cdots, N$; for the domain $0<\vert \qe \vert <  \vert y \vert<1 $, the $\ve_1$-lattices centered at $N$ Coulomb moduli, $L_\a = a_\a + \ve_1 \BZ$, $\a=1,2,\cdots, N$. In the $\EN=2$ theory point of view, these $N$ choices come from which one among $N$ discrete vacua of the two-dimensional $\EN=(2,2)$ theory that defines the canonical surface defect to be assigned at the infinity of the $\BC_1$-plane, the non-compact support of the canonical surface defect. Thus, the correlation function of the parallel surface defects should also be labelled by these $N$ choices, which arise from making a choice of the $\ve_1$-lattice $L_\a$, $\a=1,2,\cdots, N$, in its definition \eqref{eq:G-F-N}: $\U_\a (\mathbf{a};y;\mathbf{u})$, $\a=1,2,\cdots,N$.

As reviewed in section \ref{subsubsec:heckeop}, Hecke operators are normally labelled by the dominant integral coweights of $G_\BC = PGL(N)$, as are the 't Hooft line defects in the GL-twisted $\EN=4$ theory (with the gauge group $G=PSU(N)$) that they descend from. However, in our case the 't Hooft line is semi-infinite, stretched from the junction of the two boundaries to the infinity along the Dirichlet boundary. The Weyl symmetry is broken at the infinity because of the boundary condition there; in turn, the 't Hooft lines, and therefore the corresponding Hecke operators, are labelled by the coweights of $G_\BC = PGL(N)$, not by their Weyl orbits. The Hecke operator \eqref{eq:HeckeOp} defined from the simplest non-trivial spectral flow carries a minuscule coweight of the $N$-dimensional representation of ${}^L G_\BC = SL(N)$. Thus, there are precisely $N$ Hecke operators defined as the integral \eqref{eq:HeckeOp} of $\U_\a$, $\a=1,2,\cdots, N$. This is also consistent with the $\EN=2$ theory point of view on the origin of the $N$ discrete choices, since this 't Hooft line associated to the simplest minuscule coweights in the GL-twisted $\EN=4$ theory is dual to the canonical surface defect in the $\EN=2$ theory (see section \ref{subsec:gln4}).

If we compose all the \textit{eigenvalues} $\chi_\a (\mathbf{a};y)$, $\a=1,2,\cdots,N$, of the $N$ distinct Hecke operators into a vector, we reconstruct the local system associated to the oper in the $N$-dimensional representation. This is precisely the manifestation of the defining relation \eqref{eq:heckeeigen} of the Hecke eigensheaf. Therefore, we conclude that the $n$-point twisted coinvariant $\boldsymbol\Psi(\mathbf{a})$ constructed from the regular monodromy surface defect is a section of the Hecke eigensheaf $\D_{\r_{\mathbf{a}}}$ on $\text{Bun}_{PGL(N)} (\BP^1;S)$ corresponding to the $SL(N)$-oper $\r_\mathbf{a}$ constructed from the canonical surface defect.

\subsubsection{Brane at infinity}
Let us try to rephrase what has been achieved so far in the topological sigma model description ($A$-model in $\o_t$ with B-field $B_t$, see section \ref{subsec:sigmamodel}). Recall that the insertion of the regular monodromy surface defect in the $\EN=2$ theory corresponds to the brane of $\l$-connections $\EF' _E$ of $(B,A,A)_\k$-type, while the insertion of the canonical surface defect on top of it corresponds to the Hecke operator attached to $\EF' _E$. The other boundary of the corner $\S$ is always associated to the canonical coisotropic brane $\EB_{cc}$ of $(A,B,A)_\k$-type.

We have seen that that the regular monodromy surface defect in the $\EN=2$ theory gives a family of twisted coinvariants enumerated by the Coulomb moduli $\mathbf{a}$. Since the Coulomb moduli specify the boundary condition of the complex scalar in the vector multiplet of the $\EN=2$ theory at infinity, they are mapped under the duality to the boundary condition for the $\EC$ components of the 1-form in the GL-twisted $\EN=4$ theory on $\Sigma \times \EC$, at infinity of the corner $\Sigma$. Thus, it descends to the boundary condition for the map $\Phi:\Sigma \to \EM_H$ of the topological sigma model under the compactification along $\EC$. Namely, fixing the Coulomb moduli $\mathbf{a}$ in the $\EN=2$ theory assigns a brane at infinity of $\Sigma$ in the topological sigma model. 

Let us restrict to the limit $\ve_2 \to 0$ first. This corresponds to $\k = 0$ according to \eqref{eq:tkrel}, in which the topological sigma model is the $A$-model in $\o_K$. In this case, the canonical coisotropic brane $\EB_{cc}$ is of $(A,B,A)$-type, while the brane of $\l$-connections $\EF' _E$ is of $(B,A,A)$-type. The mirror dual is the $B$-model in $J$, where $\EB_{cc}$ is dualized to the brane of opers ${}^L \EB_{op}$ of $(A,B,A)$-type. The mirror dual of the brane of $\l$-connections has not been well studied, and we do not discuss about it for the current purpose. To be consistent with the $\EN=2$ theory construction of the opers (see section \ref{subsubsec:canoper}), the brane at infinity is expected to pick a ${}^L G_\BC$-oper. Thus it is natural to associate the zero-brane $\mathbf{B}_{\r_{\mathbf{a}}}$ of $(B,B,B)$-type supported at the oper $\r_{\mathbf{a}} \in \text{Op}_{{}^L G_\BC} (\EC;S)$, in the $B$-model in $J$. Recall that the zero-brane $\mathbf{B}_{\r_{\mathbf{a}}}$ is an electric eigenbrane, on which the action of the Wilson lines is \textit{diagonal} (since we are in the mirror dual frame, i.e., the $B$-model in $J$, these Wilson lines are the one in the S-dual $\EN=4$ theory with the gauge group ${}^L G$) \cite{Kapustin:2006pk}.

Applying the mirror symmetry back to the original $A$-model in $\o_K$, the zero-brane $\mathbf{B}_{\r_{\mathbf{a}}}$ is dualized to the brane $\mathbf{F} _{\mathbf{a}}$ of $(B,A,A)$-type supported on a Hitchin fiber \eqref{eq:hf}, with a flat unitary Chan-Paton bundle \cite{Kapustin:2006pk}. Since it is the mirror dual of the electric eigenbrane, it is a magnetic eigenbrane; the action of the 't Hooft lines is diagonal on this brane \cite{Kapustin:2006pk}. Moreover, at $\k=0$ 't Hooft lines can be supported in the bulk of the GL-twisted $\EN=4$ theory, so that the 't Hooft line attached to the Dirichlet boundary can be detached to the bulk. In particular, 
we can deviate the finite end of the 't Hooft line from the junction of the two boundaries to infinity along the Neumann boundary. The 't Hooft line then acts diagonally on the brane at infinity $\mathbf{F} _\mathbf{a}$, giving the local system associated to the oper $\r_\mathbf{a}$ as the eigenvalue. This provides an explanation of the factorization \eqref{eq:factor} leading to the Hecke eigensheaf property \eqref{eq:heceigensf}. With the inclusion of this brane at infinity, the vacuum expectation value $\Psi(\mathbf{a};\boldsymbol\g)$ of the regular monodromy surface defect of the $\EN=2$ theory in the limit $\ve_2 \to 0$ is actually interpreted as the disk partition function of the $A$-model in $\o_K$ with three branes ($\EB_{cc}$, $\EF' _{\boldsymbol\g}$, and $\mathbf{F} _{\mathbf{a}}$) at the boundary making three junctions. A similar configuration was also considered in \cite{Balasubramanian:2017gxc}.

However, it is not so clear how this sigma model configuration deviates to $\k \neq 0$ (namely, $\ve_2 \neq 0$). The magnetic eigenbrane $\mathbf{F}_{\mathbf{a}}$ only exists at $\k=0$ and does not deform away from it \cite{Kapustin:2006pk,Frenkel:2018dej}. Nevertheless, the configuration of parallel surface defects in the $\EN=2$ theory is apparently defined for generic values of the $\O$-background parameters, although the factorization property \eqref{eq:factor} is lost. The canonical coisotropic brane of $(A,B,A)_\k$-type is mirror dual to the brane of opers of $(B,A,A)_{-\frac{1}{\k}}$-type. It seems to be natural to assign another Lagrangian brane of $(B,A,A)_{-\frac{1}{\k}}$-type supported on the Lagrangian submanifold on which the half of the Darboux coordinates $\boldsymbol\a$ are constant, transversally intersecting the brane of opers of $(B,A,A)_{-\frac{1}{\k}}$-type. Still, it is not understood what the image of this brane under the mirror symmetry is, and how in the limit $\k \to 0$ it (resp. its mirror) is related to the electric eigenbrane $\mathbf{B}_{\r_{\mathbf{a}}}$ of $(B,B,B)$-type (resp. the magnetic eigenbrane $\mathbf{F}_\mathbf{a}$ of $(B,A,A)$-type). Precisely identifying the brane at infinity and resolving the subtleties outlined above would require a direct construction of the sigma model branes starting from the $\EN=2$ theory, along the line of \cite{Nekrasov:2010ka}. We leave this to a future work.

\section{Gaudin algebra and universal opers from parallel surface defects} \label{sec:gaudin}
We have established the $\EN=2$ gauge theoretical construction of the opers and the corresponding Hecke eigensheaves. We remind that the Hecke eigensheaf $\D_{\r_{\mathbf{a}}}$ is expected to be the quotient of the sheaf $\ED_{\fL}$ of rings itself by the ideal generated by the oper $\r_{\mathbf{a}}$, as we have explained in \eqref{eq:heceigsh}. Here, we verify this by showing that the section $\boldsymbol\Psi(\mathbf{a})$ of the Hecke eigensheaf that we constructed from the regular monodromy surface defect in the $\EN=2$ gauge theory represents the system of differential equations \eqref{eq:schrodinger}, which are the spectral equations for the quantum Hitchin integrable system. In particular, we will explicitly construct the quantum Hamiltonians as mutually commuting twisted differential operators, at the example of sphere $\BP^1$ with marked points $S$. 

\subsection{Gaudin algebra}
The Hitchin integrable system associated to the sphere with $n$ marked points is known to be the Gaudin model, where the representations of $\fsl(N)$ giving the space of states $\bigotimes_{i=1}^n M_i$ are determined by the marked point data.

The quantum Hamiltonians are mutually commuting operators on $\bigotimes_{i=1}^n M_i$. This commutative algebra can be constructed from a maximal commutative subalgebra of $U(\fsl(N))^{\otimes n}$, called the \textit{Gaudin algebra}, represented on $\bigotimes_{i=1}^n M_i$. Furthermore, the Gaudin algebra admits an algebraic construction as the image of the \textit{universal Gaudin algebra}, a maximal commutative subalgebra of the universal enveloping algebra of a current algebra, under an evaluation map. We review this construction here first.

\subsubsection{Universal Gaudin algebra}
Let $\mathfrak{g}$ be a finite-dimensional simple Lie algebra over $\BC$. Define the current algebras $\widehat{\mathfrak{g}}_+ = \mathfrak{g}\otimes \BC[[t]]$ and $\widehat{\mathfrak{g}}_- = \mathfrak{g} \otimes t^{-1} \BC[[t^{-1}]]$, which are Lie subalgebras of the affine Lie algebra $\widehat{\mathfrak{g}} = \mathfrak{g}\otimes \BC((t)) \oplus \BC K$, i.e., the central extension of the loop algebra of $\mathfrak{g}$. For $k\in \BC$, let us define the vacuum module $V_k (\widehat{\mathfrak{g}}) = \text{Ind}^{\widehat{\mathfrak{g}}} _{\widehat{\mathfrak{g}}_+ \oplus \BC K} \BC v_k $ over $\widehat{\mathfrak{g}}$ at level $k$ as the $\widehat{\mathfrak{g}}$-module induced by the one-dimensional trivial $\mathfrak{g}$-module, namely, the quotient of $U(\widehat{\mathfrak{g}})$ modulo the left ideal generated by $\widehat{\mathfrak{g}}_+$ and $K-k$. Let $h^\vee$ be the dual Coxeter number of $\mathfrak{g}$, and construct the subspace of $\widehat{\mathfrak{g}}_+$-invariant vectors $\mathfrak{z}(\widehat{\mathfrak{g}}) = V_{-h^\vee} (\widehat{\mathfrak{g}}) ^{\widehat{\mathfrak{g}}_+}$. As vector spaces, $V_{-h^\vee} (\widehat{\mathfrak{g}})$ is isomorphic to $U(\widehat{\mathfrak{g}}_-)$. This induces an algebra structure on $\mathfrak{z}(\widehat{\mathfrak{g}})$ and identifies it as a subalgebra in $U(\widehat{\mathfrak{g}}_-)$. The subalgebra $\mathfrak{z}(\widehat{\mathfrak{g}}) \subset U ( \widehat{\mathfrak{g}}_-)$ is maximally commutative, called the universal Gaudin algebra \cite{Feigin:1991wy, Feigin:1994in}.\footnote{It is also called the Bethe subalgebra in some literature.}

Now we restrict to $\mathfrak{g}= \fsl(N)$. In this case, the universal Gaudin algebra admits another presentation \cite{Tal:2004,Chervov:2006xk, MTV:2006, Chervov:2009ck} that we introduce now. It was shown that the two constructions yield the identical maximal commutative subalgebra $\mathfrak{z}(\widehat{\fsl} (N)) \subset U(\widehat{\fsl}(N) _-)$ \cite{R:2006}. We will employ the latter presentation, since the universal oper which naturally arises in this construction will be used to connect to the $\EN=2$ gauge theoretical framework.

Write the generators of $\widehat{\mathfrak{sl}}(N) _{-} $ as $e_{ab} [-s] = e_{ab} \otimes t^{-s}$, $a,b=1,\cdots, N$ and $s \in \BZ_{> 0}$, with $\sum_{a=1} ^N e_{aa} = 0$. The defining commutation relations are given by $[e_{ab} [r] ,e_{cd} [s] ] = \d_{bc} e_{ad} [r+s] -\d_{ad} e_{cb} [r+s]$. 

Consider its universal enveloping algebra $U(\widehat{\mathfrak{sl}} (N)_{-})$. We compose the generating series by
\begin{align}
    L_{ab} (z) = \sum_{s=1} ^\infty e_{ba} [-s] z^{s-1} \in U(\widehat{\mathfrak{sl}}(N) _{-})[[z]], \quad a,b=1,\cdots, N,
\end{align}
and combine these series into the \textit{Lax matrix} $L(z) := \sum_{a,b=1} ^N E_{ab} \otimes L_{ab} (z) \in \text{End}(\BC^N) \otimes U(\widehat{\mathfrak{sl}} (N)_{-})[[z]]$, where $E_{ab}$ is the standard basis of $\text{End}(\BC^N)$. The commutation relations of $U(\widehat{\mathfrak{sl}} (N)_{-})$ can be written as
\begin{align} \label{eq:univcom}
    (z-z') [L_{ab} (z) , L_{cd} (z')] = \d_{ad} \left( L_{cb} (z) - L_{cb} (z') \right)- \d_{bc} \left(L_{ad}(z)-L_{ad}(z')\right),
\end{align}
or equivalently,
\begin{align}
     [L^{(13)} (z), L^{(23)}(z')] = \left[\frac{P^{(12)}}{z-z'} , L^{(13)} (z) + L^{(23)} (z')\right],
\end{align}
as elements in $\text{End}(\BC^N)\otimes \text{End}(\BC^N) \otimes U(\widehat{\mathfrak{sl}}_{N,-})[[z,z']]$, where $P \in \text{End}(\BC^N) \otimes \text{End}(\BC^N)$ is the exchange operator. Here, the superscripts indicate where the Lax matrices and the exchange operator are valued among the three pieces in the tensor product. Note that taking $z' \to z$, we also get
\begin{align} \label{eq:commder}
    [L_{ab} (z) ,L_{cd} (z)] = \d_{ad} \p_z L_{cb} (z) -\d_{bc} \p_z L_{ad} (z).
\end{align}
Also note that the two $\fsl(N)$-actions on the Lax matrix $L(z)$ cancel each other:
\begin{align}\label{eq:slN}
    [E_{ab} \otimes 1 + 1 \otimes e_{ab} ,L(z) ] = 0. 
\end{align}

It can be shown that for a given Gaudin Lax matrix $L(z)$, $\p_z - L(z)$ is a so-called \textit{Manin matrix} \cite{Chervov:2007bb}. A Manin matrix is a matrix with non-commuting entries and, nevertheless, a well-defined determinant in the column expansion.\footnote{The Manin matrix can alternatively be defined in a way that the determinant is well-defined in the row expansion. Here, we choose the convention that the column expansion is well-defined.} The determinant, which can be expanded as a $N$-th order differential operator as
\begin{align} \label{eq:univoper}
\det(\p_z - L(z))= \p_z ^N + \mathtt{t}_2 (z) \p_z ^{N-2} +\cdots + \mathtt{t}_{N-1} (z) \p_z + \mathtt{t}_{N} (z),
\end{align}
is called the \textit{universal} oper. The coefficients $\mathtt{t}_\a (z) \in U(\widehat{\fsl}(N)_-)[[z]]$, $\a=2,3,\cdots,N$, are called the universal Gaudin transfer matrices. The maximal commutative subalgebra generated by $\mathtt{t}_\a$ is precisely the universal Gaudin algebra $\mathfrak{z}(\widehat{\mathfrak{sl}}(N) ) \subset U(\widehat{\mathfrak{sl}}(N)_{-})$.

Now, we introduce a neat procedure \cite{Chervov:2009ck} to derive the universal Gaudin transfer matrices that will be useful in making a connection to the main theme of this work. Define the \textit{quantum} powers of the Lax matrix $L(z)$ by the following recursive relations:
\begin{align}
\begin{split}
    &L^{[0]} (z) := \mathds{1}_N, \\
    &L^{[m]} (z) := L^{[m-1]} L + \p_z L^{[m-1]} \in \text{End}(\BC^N) \otimes U(\widehat{\mathfrak{sl}}(N)_{-})[[z]], \quad m \in \BZ_{>0}.
\end{split}
\end{align}
Choose any $v\in \BC^N $. Define $C(z) \in \text{End}(\BC^N) \otimes U(\widehat{\mathfrak{sl}}(N)_{-})[[z]]$ by $C_{ab} = \sum_{c=1} ^N v_c L^{[N-a]} _{cb}$. Then, using the commutation relations \eqref{eq:commder}, one can show that
\begin{align} \label{eq:commu}
    C(\p_z - L) = \left[\p_z + \begin{pmatrix} 0 &  \mathtt{t}_2 & \cdots & \cdots & \mathtt{t}_N \\ -1 & 0 & \cdots& \cdots & 0 \\ 0 & -1 & \cdots & \cdots & 0 \\ \cdots & \cdots & \cdots & \cdots & \cdots \\ 0 & \cdots & 0 & -1 & 0  \end{pmatrix} \right] C,
\end{align}
where $\mathtt{t}_k (z) \in U(\widehat{\mathfrak{sl}}(N)_{-}) [[z]]$, $k=2,3,\cdots, N$, are exactly the universal Gaudin transfer matrices, generating the universal Gaudin algebra. Note that, if we assume there is an element $\U = (\U_1,\cdots, \U_N) \in \BC^N \otimes M$, where $M$ is a $\widehat{\fsl}(N)_-$-module, which annihilates the Manin matrix, $0=(\p_z -L)\vert_M\U$, then upon choosing $v=(0,\cdots, 0,1)$ it is straightforward to see that \eqref{eq:commu} implies
\begin{align} \label{eq:qoper}
    0=\left.\left[ \p_z ^N +  \mathtt{t}_2 (z) \p_z ^{N-2} + \cdots +  \mathtt{t}_{N-1} (z) \p_z +  \mathtt{t}_N (z) \right]\right\vert_M \U_{N}.
\end{align}
Namely, the $N$-th order differential operator appearing in this equation is nothing but the universal oper \eqref{eq:univoper} represented on the given module $M$.

The universal Gaudin transfer matrices $\mathtt{t}_\a (z)$ can be written as combinations of the traces of quantum powers of $L(z)$, namely, $\text{Tr}_{\BC^N} L^{[k]} (z)$, $k=2,3,\cdots, N$ \cite{Chervov:2007bb}.\footnote{$\text{Tr}_{\BC^N} L^{[1]} (z) = \text{Tr}_{\BC^N} L(z) = 0$ for $\mathfrak{g} = \fsl(N)$.} The elements of the universal Gaudin algebra $\mathfrak{z}(\widehat{\fsl}(N) )$ are invariant under the action of $\fsl(N) \subset U(\widehat{\fsl}(N)_-)$, due to \eqref{eq:slN} and the cyclicity of the trace $\text{Tr}_{\BC^N}$. Thus, when finding common eigenfunctions of the Gaudin transfer matrices on a $\widehat{\fsl}(N)_-$-module $M$, it is enough to restrict to the space of $\fsl(N)$-invariants $M^{\fsl(N)}$.

\subsubsection{Evaluation map and Gaudin algebra}
The Gaudin algebra is obtained as the image of the universal Gaudin algebra under a given representation, as we now recall. Universal enveloping algebra is a Hopf algebra with the coproduct $\Delta : U(\widehat{\mathfrak{sl}}(N)_{-}) \to U(\widehat{\mathfrak{sl}}(N)_{-}) \otimes U(\widehat{\mathfrak{sl}}(N)_{-})$ given by
\begin{align}
    \Delta (a) = a \otimes 1 + 1 \otimes a.
\end{align}
For a given $z\in \BC$, the evaluation map $\text{ev}_z: U(\widehat{\mathfrak{sl}}(N)_{-}) \to U(\fsl(N))$ is defined by $e_{ab}[-s] \mapsto z^{-s} e_{ab}$. Fix $n$ distinct complex numbers $\mathbf{z} = (z_1,\cdots, z_n) \in \BC^n$, regarding them as the inhomogeneous coordinates for $n$ marked points $S =\{z_1,\cdots, z_n\} \subset \BP^1$. Then consider the map
\begin{align}
    \text{ev}_{S} := (\text{ev}_{z_1} \otimes \cdots \otimes \text{ev}_{z_n}) \circ \Delta^{n-1} : U(\widehat{\mathfrak{sl}}(N)_{-}) \to  U(\fsl(N))^{\otimes n}.
\end{align}
The image of the Lax matrix under this map is computed to be
\begin{align}
          \text{ev}_{S}\left(L(z_0) \right) = - \sum_{i=1} ^{n} \frac{\sum_{a,b=1}^N E_{ab} \otimes e_{ba} ^{(i)} }{z_0-z_i} \in \text{End}(\BC^N) \otimes U(\fsl(N))^{\otimes n} \otimes \BC[(z_0-z_i)^{-1}]_{i=1,\cdots, n},
\end{align}
where $e_{ba} ^{(i)} = 1 \otimes \cdots \otimes 1\otimes \underbrace{e_{ba} }_{i\text{-th}} \otimes 1 \otimes \cdots \otimes 1 $. 

The image of the universal Gaudin algebra $\mathfrak{z} (\widehat{\mathfrak{sl}}_N)$ under the map $\text{ev}_{S}$ is a maximal commutative subalgebra in $U(\fsl(N))^{\otimes n}$, called the Gaudin algebra $\EuScript{Z}_S (\fsl(N))$. By construction, the Gaudin algebra $\EuScript{Z}_S (\fsl(N))$ is generated by the image of the universal Gaudin transfer matrices, $\text{ev}_{S}(\mathtt{t}_\a (z_0))$, called the Gaudin transfer matrices. By taking their Laurent coefficients in $z_0 - z_i$, we obtain explicit expressions for the (higher) Gaudin Hamiltonians and the central elements of $U(\fsl(N))^{\otimes n}$. Note that $\fsl(N) \subset U(\widehat{\mathfrak{sl}}(N)_{-})$ maps to the diagonal $\fsl(N) \subset U(\fsl(N))^{\otimes n}$, and therefore the elements in the Gaudin algebra are invariant under this diagonal $\fsl(N)$-action.

For given $n$ $\fsl(N)$-modules $M_i$, $i=1,\cdots,n$, we can construct a $U({\mathfrak{sl}}(N))^{\otimes n}$-module $M = \bigotimes_{i=1} ^{n} M_i $ by the universality. The Lax matrix is represented as
\begin{align}
         \text{ev}_{S}\left(L(z_0) \right)\vert_{M} =  -  \sum_{i=1} ^{n} \frac{\sum_{a,b=1}^N E_{ab} \otimes e_{ba} ^{(i)} \vert_{M} }{z_0-z_i} \in \text{End}(\BC^N) \otimes \text{End}(M) \otimes \BC[(z_0-z_i)^{-1}]_{i=1,\cdots, n},
\end{align}
where $e_{ba} ^{(i)} \vert_{M} = 1 \otimes \cdots \otimes 1\otimes \underbrace{e_{ba} \vert_{M_i} }_{i\text{-th}} \otimes 1 \otimes \cdots \otimes 1 $. We recognize that the image of the Manin matrix $\p_{z_0} - L(z_0)$ under the map $\text{ev}_{S}$ represented on $M$ is
\begin{align} \label{eq:maninrep}
    \text{ev}_S \left( \p_{z_0} -L(z_0) \right)\vert_M = \frac{\p}{\p z_0} + \sum_{i=1} ^n  \frac{\sum_{a,b=1}^N E_{ab} \otimes e_{ba} ^{(i)} \vert_{M} }{z_0-z_i}.
\end{align}

\subsection{$\ED$-module of coinvariants and universal opers} \label{eq:coinvunivop}
Finally, we connect the algebraic construction of the Gaudin algebra we reviewed so far to the main subject of this work. It is important to note that \eqref{eq:maninrep} is precisely the operator part of the constraints \eqref{eq:slNlax} for the $(n+1)$-point genus-$0$ twisted coinvariants $\boldsymbol\U$, in which $n$ $\widehat{\fsl}(N)$-modules are the ones induced from $(M_i)_{i=1} ^n$ and the $(n+1)$-th $\widehat{\fsl}(N)$-module is the (bi-infinite generalization of) twisted vacuum module. In particular, from \eqref{eq:qoper} and \eqref{eq:slNlax} we deduce
\begin{align} \label{eq:univM}
    0 = \text{ev}_S \left. \left[ \p_{z_0} ^N +  \mathtt{t}_2 (z_0) \p_{z_0} ^{N-2} + \cdots + \mathtt{t}_{N-1} (z_0) \p_{z_0} + \mathtt{t}_{N} (z_0) \right] \right\vert_M \boldsymbol\U,
\end{align}
where the coefficients of the differentials are the Gaudin transfer matrices represented on $M$. As we emphasized in deriving the constraints \eqref{eq:slNlax}, it is indeed important that the operator part of the constraints \eqref{eq:slNlax} does not contain any dependence on the level $k$, since otherwise it would not have been a Manin matrix so that the universal oper could not be defined as its determinant. In this sense, our approach is different from \cite{Chervov:2006xk,Chervov:2009ck}, where a particular value of the level $k+N = 1$ was taken to the usual KZ equation \eqref{eq:KZ} (without insertion of twisted vacuum module or its bi-infinite generalization).\footnote{Note that this special value of the level corresponds to the self-dual limit of the $\O$-background, $k+N = -\frac{\ve_1}{\ve_2} = 1$. On contrary, the geometric Langlands correspondence arises in the limit $\ve_2 \to 0$ in our approach. It has been revealed \cite{Jeong:2020uxz,Nekrasov:2020qcq} (see also \cite{Grassi:2016nnt}) that the $\EN=2$ gauge theories subject to these two limits of the $\O$-background are related by a blowup \cite{Nakajima:2003pg,Nakajima:2003uh}. We also remark that this blowup operation should correspond to the \textit{composition} of junctions in the GL-twisted $\EN=4$ theory and the topological sigma model with boundaries, which was studied in \cite{Frenkel:2018dej}.}

The operator part of the above equation is the universal oper represented on $M$ using the evaluation map, but it is not still quite on the same footing with the oper $\r_{\mathbf{a}}$ \eqref{eq:oper} that we constructed from the canonical surface defect. The problem is that the Gaudin model that we want is defined on the $\fsl(N)$-invariants in $M$, $M^{\fsl(N)} = \left(\bigotimes_{i=1}^n M_i \right)^{\fsl(N)}$, while the $\fsl(N)$-invariance condition obeyed by the $(n+1)$-point coinvariant $\boldsymbol\U$ involves the additional $(n+1)$-th module, i.e., the (bi-infinite generalization of) twisted vacuum module. The problem is cured precisely by taking $k = -N$ and integrating $\boldsymbol\U$ along the $(N-1)$-dimensional cycle $C$ in the space of Hecke modifications $\BP^{N-1}$, which yields the $n$-point coinvariant $\boldsymbol\Psi$ (without the twisted vacuum module) at the critical level acted on by a Hecke operator. We will show that this integral can be taken to the above equation \eqref{eq:univM} to yield twisted differential operators on $\text{Bun}_{PGL(N)} (\BP^1;S)$ as coefficients.

As we have seen in section \ref{subsubsec:hecke} the twisting part $\U_0$ of the $(n+1)$-point coinvariant $\boldsymbol\U$ can be broken into two parts: the twisting $\Psi_0$ for the $n$-point coinvariant and the holomorphic measure on $\BP^{N-1}$, $ (y^{-2} dy)^{\frac{k(N-1)}{2N}} \left( \bigwedge_{a=1} ^{N-1} dx_0 ^a \right)^{-\frac{k}{N}}= (y^{-2} dy)^{\frac{k(N-1)}{2N}} \left( \bigwedge_{a=1} ^{N-1} \frac{d\m_a}{\m_N} \right)^{-\frac{k}{N}} $. Now, we take the critical level $k =-N$. Then we conjugate $\Psi_0$ through the differential operators $\text{ev}_S (\mathtt{t}_k (z_0)) \vert_M$ to the front, while performing the integral on $\boldsymbol\m$ variables along the $(N-1)$-dimensional cycle $C$. The integral precisely yields the Hecke operator acting on the $n$-point coinvariant, leading to
\begin{align} \label{eq:qopereq}
\begin{split}
    0 &=  \left[ \p_{y} ^N +  \hat{t}_2 (y) \p_{y} ^{N-2} + \cdots + \hat{t}_{N-1} (y) \p_{y} + \hat{t}_{N} (y) \right] \left( y^{N-1} dy^{-\frac{N-1}{2}} \right) \oint_C \bigwedge_{a=1} ^{N-1} \frac{d\m_a}{\m_{N}} \lim_{k \to -N}\U(\boldsymbol\g,\boldsymbol\m;y) \\
    & =  \left[ \p_{y} ^N +  \hat{t}_2 (y) \p_{y} ^{N-2} + \cdots + \hat{t}_{N-1} (y) \p_{y} + \hat{t}_{N} (y) \right] H_y \Psi(\boldsymbol\g),
\end{split}
\end{align}
where we have defined the operator-valued meromorphic functions $\hat{t}_k : = \Psi_0 ^{-1} \left( \text{ev}_S (\mathtt{t}_k ) \vert_M \right) \Psi_0$ whose Laurent coefficients are the Gaudin Hamiltonians, given by twisted differential operators on $\text{Bun}_{PGL(N)} (\BP^1;S)$, and the central elements represented on the assigned module $M^{\fsl(N)}$. By construction, the Gaudin algebra $\EZ_S (\fsl(N))$ represented on $M^{\fsl(N)}$ forms a maximal commutative subalgebra of $\text{End}(M^{\fsl(N)})$. The central element represented on $M^{\fsl(N)}$ are simple numbers determined by the weight parameters of $(M_i)_{i=1} ^n$, and the Gaudin Hamiltonians represented on $M^{\fsl(N)}$ commute with each other as twisted differential operators.

At this point, we emphasize that the method that we developed so far is very explicit and constructive for arbitrary $N$. It is only a matter of computation to find the exact forms of the elements of the Gaudin algebra $\EZ_S (\fsl(N))$ $-$ the Gaudin Hamiltonians and the central elements $-$ represented on $M^{\fsl(N)}$. To be very concrete, we will explicitly compute some of the Gaudin Hamiltonians and the central elements, for our example of $\vert S \vert = n =4$ represented on the module $M^{\fsl(N)} = \left( V_{\boldsymbol\z} \otimes H_{\boldsymbol\t-\boldsymbol\z ,\s} \otimes H_{\tilde{\boldsymbol\z}-\boldsymbol\t ,\tilde{\s}}\otimes \tilde{V}_{\tilde{\boldsymbol\z}} \right)^{\fsl(N)}$. 

\paragraph{Quadratic Hamiltonian and Casimirs}
Let us start from the first non-trivial Gaudin transfer matrix $ \text{ev}_S ( \mathtt{t}_2 (z_0) )\vert_M = \text{ev}_S \left.\left( \frac{1}{2} \Tr_{\BC^N}  L ^{[2]} (z_0) \right) \right\vert_M = \text{ev}_S \left.\left( \frac{1}{2} \Tr_{\BC^N}  L ^{2} (z_0) \right) \right\vert_M$. Note that we used $\text{Tr}_{\BC^N} L(z) = 0$. It is straightforward to get
\begin{align} \label{eq:hatt2}
    \hat{t}_2 (y) = \frac{\d_0}{y^2} + \frac{\d_\qe}{(y-\qe)^2} + \frac{\d_1}{(y-1)^2} + \frac{\d_\infty -\d_0 -\d_\qe-\d_1}{y(y-1)} + \frac{\hat{H}_2}{y(y-\qe)(y-1)}.
\end{align}
Here, $\d_i$, $i=0,\qe,1,\infty$ are the quadratic Casimirs represented on the respective modules,
\begin{align}
\begin{split}
    &\d_ 0  = \frac{1}{2}  \Psi_0 ^{-1}  \sum_{a,b=1}^N  \bar{T}_a ^b \vert_{V_{\boldsymbol\z}}  \bar{T}_b ^a \vert_{V_{\boldsymbol\z}}  \Psi_0 \\
    & \d_ \qe  = \frac{1}{2}  \Psi_0 ^{-1} \sum_{a,b=1}^N  \bar{T}_a ^b \vert_{H_{\boldsymbol\t-\boldsymbol\z ,\s}}  \bar{T}_b ^a \vert_{H_{\boldsymbol\t-\boldsymbol\z ,\s}} \Psi_0 \\
    &  \d_ 1  = \frac{1}{2} \Psi_0 ^{-1}\sum_{a,b=1}^N  \bar{T}_a ^b \vert_{H_{\tilde{\boldsymbol\z}-\boldsymbol\t ,\tilde\s}}  \bar{T}_b ^a \vert_{H_{\tilde{\boldsymbol\z}-\boldsymbol\t ,\tilde\s}}   \Psi_0 \\
    & \d_\infty = \frac{1}{2} \Psi_0 ^{-1} \sum_{a,b=1}^N \bar{T}_a ^b \vert_{\tilde{V}_{\tilde{\boldsymbol\z}}}  \bar{T}_b ^a \vert_{\tilde{V}_{\tilde{\boldsymbol\z}}}  \Psi_0,
\end{split}
\end{align}
which are central elements of $U(\fsl(N))^{\otimes 4}$ represented on $M^{\fsl(N)}$. By definition, they are related to the conformal weights of the four primary vertex operators by $\D_i = \frac{\sum_{a,b=1}^N \big( \bar{T}_a ^b \big)\vert_i \big( \bar{T}_b ^a \big)\vert_i  }{2(k+N)} = \frac{\d_i}{k+N}$. By a straightforward computation, we confirm that all the differential operator parts are cancelled as expected, leaving simple combinations of weight parameters,
\begin{align}
\begin{split}
    &\d_0 =  \frac{1}{2}  \sum_{a=1} ^N \left( \sum_{i=a} ^{N-1} \z_i \right)^2 + \frac{1}{2}\sum_{a>b} \sum_{i=b} ^{a-1} \z_i - \frac{1}{2N} \left(\sum_{a=1} ^N \sum_{i=a} ^{N-1} \z_i \right)^2 \\
    &\d_\qe = \frac{(N-1)\s(\s+N)}{2N} \\
    & \d_1 = \frac{(N-1)\tilde\s (\tilde\s+N)}{2N} \\
    &\d_\infty= \frac{1}{2}  \sum_{a=1} ^N \left( \sum_{i=a} ^{N-1} \tilde{\z}_i \right)^2 + \frac{1}{2}\sum_{a>b} \sum_{i=b} ^{a-1} \tilde{\z}_i - \frac{1}{2N} \left(\sum_{a=1} ^N \sum_{i=a} ^{N-1} \tilde{\z}_i \right)^2 .
\end{split}
\end{align}

On the other hand, the last piece $\hat{H}_2$ in $\hat{t}_2$ \eqref{eq:hatt2} is not a simple number but a twisted differential operator on $\text{Bun}_{PGL(N)} (\BP^1;S)$, a non-central element in the Gaudin algebra $\EZ_S (\fsl(N))$ represented on $M^{\fsl(N)}$. It is the quadratic Gaudin Hamiltonian computed as
\begin{align}
    &\hat{H}_2 =   \Psi_0 ^{-1} \sum_{a,b=1} ^N \left( (\qe-1)  \bar{T}_a ^b \vert_{H_{\boldsymbol\t-\boldsymbol\z,\s}} \bar{T}_b ^a \vert_{V_{\boldsymbol\z}}  + \qe \bar{T}_a ^b  \vert_{H_{\boldsymbol\t - \boldsymbol\z ,\s}} \bar{T}_b ^a \vert_{H_{\tilde{\boldsymbol\z} -\boldsymbol\t,\tilde\s}} \right) \Psi_0.
\end{align}
This is exactly the differential operator appearing in the 4-point KZ equation \eqref{eq:redKZ4}, whose explicit form was already obtained as (see \eqref{eq:H-4p})
\begin{align}  \label{eq:quadH}
\begin{split}
    \hat{H}_2 &= (\qe-1) \left[ -\sum_{b>a}^N \frac{\g_b}{\g_a} \left( \g_a \p_{\g_a} + \beta_{a}  \right) \left( (\g_a-\g_b)\p_{\g_b} + \frac{\g_a}{\g_b}\beta_{b} + {\tilde\beta}_{b} \right) \right. \\
    & \quad\quad\quad \left. +\sum_{a=1} ^N  \left( \g_a \p_{\g_a} + \beta_{a}  \right)\left( \sum_{c=a+1}^N \beta_{c} + {\tilde\beta}_{c} + \sum_{i\geq a} \alpha_i \right)  - \frac{1}{N} \left( \sum_{a=1}^N \beta_{a} \right) \sum_{b=1}^N \left( \sum_{c=b+1}^N \beta_{c} + {\tilde\beta}_{c} + \sum_{i\geq b} \alpha_i \right) \right] + \\
    & \quad + \qe \left[  \sum_{a,b=1}^N \frac{\g_b}{\g_a} \left( \g_a \p_{\g_a} + \beta_{a} \right) \left( - \g_b \p_{\g_b} + {\tilde\beta}_{b} \right) - \frac{1}{N} \sum_{a=1} ^N \beta_{a} \sum_{b=1}^N {\tilde\beta}_{b} \right].
\end{split}
\end{align}
By the fact that the quadratic Hamiltonian $\hat{H}_2$ is identical to the differential operator in the KZ equation \eqref{eq:redKZ4}, we confirm that the critical level $k = -N$ (i.e., $\ve_2 \to 0$) limit of the KZ equation yields the spectral equation for $\hat{H}_2$. We do not elaborate more on this here, since we will achieve the spectral equations for all the Gaudin Hamiltonians shortly.

\paragraph{Cubic Hamiltonian and Casimirs} We turn to the next-to-simplest Gaudin transfer matrix, the cubic: $\text{ev}_S ( \mathtt{t}_3 (z_0) )\vert_M = \text{ev}_S \left.\left( \frac{1}{3} \Tr_{\BC^N}  L ^{[3]} (z_0) \right) \right\vert_M $. Note that there is no multi-trace contribution since the Lax matrix is traceless, $\text{Tr}_{\BC^N} L (z_0) = 0$. By a direct computation which is more involved than the previous one for the quadratic Gaudin transfer matrix, we get
\begin{align}
\begin{split}
    \hat{t}_3(y) &= \frac{\l_0}{y^3} + \frac{\l_\qe}{(y-\qe)^3} + \frac{\l_1}{(y-1)^3} - \frac{\l_\infty + \l_0 +\l_\qe +\l_1}{y(y-\qe)(y-1)}  \\
    & + \frac{\hat{H}_2}{ y (y-\mathfrak{q})(y-1)} \left( \frac{\left(\frac{2}{N}-1 \right)\s+1}{y-\mathfrak{q}} + \frac{\left(\frac{2}{N}-1\right)\tilde\s+ 1}{y-1} \right)  + \frac{\hat{H}_3}{y^2 (y-\mathfrak{q})(y-1)}  \\
& + \frac{1-\mathfrak{q}}{y (y-\mathfrak{q})(y-1)^2} \left( \left(\frac{2}{N} -1 \right)\tilde\s +1 \right)  \left( \delta_\infty -\delta_0 - \delta_{\mathfrak{q}} - \delta_1 \right) +\p_y \hat{t}_2  (y). 
\end{split}
\end{align}
Here, $\d_{i}$, $i=0,\qe,1,\infty$ are the quadratic Casimirs which we have already explained above. We observe that the cubic Casimirs, which are higher-order central elements, also enter into the transfer matrix now. Namely, $\l_i$, $i=0,\qe,1,\infty$ are the cubic Casimirs represented on the respective modules, whose expressions are
\begin{align}
\begin{split}
        &\l_ 0  = \frac{1}{3}  \Psi_0 ^{-1}  \sum_{a,b,c=1}^N  \bar{T}_a ^b \vert_{V_{\boldsymbol\z}}  \bar{T}_b ^c \vert_{V_{\boldsymbol\z}} \bar{T}_c ^a \vert_{V_{\boldsymbol\z}}  \Psi_0 \\
    & \l_ \qe  = \frac{1}{3}  \Psi_0 ^{-1} \sum_{a,b,c=1}^N  \bar{T}_a ^b \vert_{H_{\boldsymbol\t-\boldsymbol\z ,\s}}  \bar{T}_b ^c \vert_{H_{\boldsymbol\t-\boldsymbol\z ,\s}}\bar{T}_c ^a \vert_{H_{\boldsymbol\t-\boldsymbol\z ,\s}} \Psi_0 \\
    &  \l_ 1  = \frac{1}{3}  \Psi_0 ^{-1}\sum_{a,b,c=1}^N  \bar{T}_a ^b \vert_{H_{\tilde{\boldsymbol\z}-\boldsymbol\t ,\tilde\s}}  \bar{T}_b ^c \vert_{H_{\tilde{\boldsymbol\z}-\boldsymbol\t ,\tilde\s}}  \bar{T}_c ^a \vert_{H_{\tilde{\boldsymbol\z}-\boldsymbol\t ,\tilde\s}}  \Psi_0 \\
    & \l_\infty = \frac{1}{3}  \Psi_0 ^{-1} \sum_{a,b,c=1}^N \bar{T}_a ^b \vert_{\tilde{V}_{\tilde{\boldsymbol\z}}}  \bar{T}_b ^c \vert_{\tilde{V}_{\tilde{\boldsymbol\z}}} \bar{T}_c ^a \vert_{\tilde{V}_{\tilde{\boldsymbol\z}}}  \Psi_0.
\end{split}
\end{align}
By a direct computation, all the differential operator parts are cancelled as expected, so that the cubic Casimirs are written in the following simple combinations of weight parameters:
\begin{align}
\begin{split}
    &\l_0 = - \frac{1}{3}\sum_{a=1}^N \left( \sum_{i=a} ^{N-1} \z_i \right)^3 + \frac{1}{N}  \sum_{a=1 }^N \left( \sum_{i=a} ^{N-1} \z_i \right)^2  \sum_{b=1} ^N \sum_{j=b} ^{N-1} \z_j + \left(\frac{1}{3N^3}  - \frac{1}{N^2} \right) \left( \sum_{a=1} ^N \sum_{i=a} ^{N-1} \z_i \right)^3  \\
    &  \quad\quad+ \sum_{a=1}^N (2N-3a+1) \left( \sum_{i\geq a}^{N-1} \zeta_i \right)^2  - \sum_{b>a} \left[ \left( \sum_{i\geq a}^{N-1} \zeta_i \right) \left( \sum_{j\geq b}^{N-1} \zeta_{j} \right) + \frac{3}{N} \left( \sum_{c=1}^{N-1} \sum_{j\geq c}^{N-1} \zeta_{j} \right) \left( \sum_{i\geq a}^{b-1} \zeta_i \right) \right] \\
    & \quad\quad - \sum_{b>a} (2N-2b+1) \sum_{i=a}^{b-1} \zeta_i \\
    &\l_\qe = \frac{(N-1) \s (\s+N)((N-2)\s-N) }{3 N^2} \\
    &\l_1 = \frac{(N-1) \tilde\s (\tilde\s+N)((N-2)\tilde\s-N) }{3 N^2}\\
    &\l_\infty =  - \frac{1}{3}\sum_{a=1}^N \left( \sum_{i=a} ^{N-1} {\tilde\z}_i \right)^3 + \frac{1}{N}  \sum_{a=1 }^N \left( \sum_{i=a} ^{N-1} {\tilde\z}_i \right)^2  \sum_{b=1} ^N \sum_{j=b} ^{N-1} {\tilde\z}_j + \left( \frac{1}{3 N^3} -\frac{1}{N^2} \right) \left( \sum_{a=1} ^N \sum_{i=a} ^{N-1} {\tilde\z}_i \right)^3 \\
    & \quad\quad + \sum_{a=1}^N (2N-3a+1) \left( \sum_{i\geq a}^{N-1} \tilde\zeta_i \right)^2  - \sum_{b>a} \left[ \left( \sum_{i\geq a}^{N-1} \tilde\zeta_i \right) \left( \sum_{j\geq b}^{N-1} \tilde\zeta_{j} \right) + \frac{3}{N} \left( \sum_{c=1}^{N-1} \sum_{j\geq c}^{N-1} \tilde\zeta_{j} \right) \left( \sum_{i\geq a}^{b-1} \tilde\zeta_i \right)\right] \\
    & \quad\quad - \sum_{b>a} (2N-2b+1) \sum_{i=a}^{b-1} \tilde\zeta_i.
\end{split}
\end{align}

On the other hand, $\hat{H}_2$ and $\hat{H}_3$ are not simple numbers but twisted differential operators on $\text{Bun}_{PGL(N)} (\BP^1;S)$, being non-central elements in the Gaudin algebra $\EZ_S (\fsl(N))$ represented on $M^{\fsl(N)}$. $\hat{H}_2$ is precisely the quadratic Gaudin Hamiltonian \eqref{eq:quadH} that we already acquired above. The last piece $\hat{H}_3$ is the cubic Gaudin Hamiltonian given by
\begin{align}
    \hat{H}_3 = - \Psi_0 ^{-1} \sum_{a,b,c=1} ^N \bar{T}_a ^b \vert_{V_{\boldsymbol\z}} \bar{T}_b ^c \vert_{V_{\boldsymbol\z}} \left( \bar{T}_c ^a \vert_{H_{\boldsymbol\t- \boldsymbol\z,\s}} +\qe \bar{T}_c ^a \vert_{H_{ \tilde{\boldsymbol\z} - \boldsymbol\t,\tilde\s}} \right)  \Psi_0.
\end{align}
Its explicit form as a twisted differential operator can be obtained as
\begin{align}
\begin{split}
    \hat{H}_3 &= \sum_{a,b,c=1} ^N \left(   -\kq \left( - \g_a \p_{\g_a}  +\tilde\beta_{a} + 1 -\frac{\tilde\s}{N}\d_a ^c \right) - \g_c \p_{\g_a} - \frac{\g_c}{\g_a} \beta_{a} + \frac{\s}{N}  \delta^c_a \right)  \times \\
    & \quad\quad\quad \times \left( \theta_{b>a} \left[ (\g_a - \g_b) \p_{\g_b} + \tilde\beta_{b} +1 +  \frac{\g_a}{\g_b} \beta_{b} \right] + \delta^a_b  \sum_{i=a}^{N-1} \zeta_i  - \frac{1}{N} \delta^a_b  \sum_{d=1}^N \sum_{i=d}^{N-1} \zeta_i  \right) \times \\
    & \quad\quad\quad \times \left( \theta_{c>b} \left[  (\g_b - \g_c) \p_{\g_c} + \tilde\beta_{c} + \frac{\g_b}{\g_c} \beta_{c} \right] + \delta^b_c  \sum_{i=c}^{N-1} \zeta_i  - \frac{1}{N}\delta^b_c  \sum_{d=1}^N\sum_{i=d}^{N-1} \zeta_i  \right) \\
    & \quad - \sum_{a,b=1} ^N \left( (1-\kq) \g_a \p_{\g_a}  + \kq \tilde\beta_{a} + \beta_{a} - \frac{1}{N} \left(  \kq \tilde\s +  \s\right) \right) \left( \sum_{i=a+1}^b \g_i (1-\g_i)\p_{\g_i}  + \zeta_{i-1} \right) \\
    &\quad - \sum_{a=1} ^N \left( (1-\kq) \g_a \p_{\g_a} + \kq \tilde\beta_{a} + \beta_{a}- \frac{1}{N} \left(  \kq \tilde\s +  \s\right)  \right) \left( \sum_{i=a}^{N-1} \zeta_i  \right) \\
    & \quad- \sum_{ b>a} (N-b+1) \left( \qe (\g_a^2 \p_{\g_a})  (1-\g_b)\p_{\g_b}  - (\g_b \p_{\g_a}) \g_a (1-\g_b) \p_{\g_b} \right).
\end{split}
\end{align}

\subsection{Spectral equations from factorization of parallel surface defects}
Finally, we show that the Hecke eigensheaf provides a common eigenfunction of the Gaudin Hamiltonians using our $\EN=2$ gauge theory construction. In section \ref{subsubsec:hecke}, we showed that the vacuum expectation value $\Psi(\mathbf{a};\boldsymbol\g)$ of the regular monodromy surface defect in the limit $\ve_2 \to 0$ gives twisted coinvariants which diagonalize the action of the Hecke operator. Let us remind that the eigensheaf property could be understood as the factorization of the correlation function of the regular monodromy surface defect and the canonical surface defect in the limit $\ve_2 \to 0$.

Now, let us apply the universal oper equation \eqref{eq:qopereq} to the element $\Psi(\mathbf{a};\boldsymbol\g)$ so that we have $H_y \Psi(\mathbf{a};\boldsymbol\g) = \chi(\mathbf{a};y)\Psi(\mathbf{a};\boldsymbol\g)$ by the eigensheaf property. Recall that in the limit $\ve_2\to 0$, the vacuum expectation value of the regular monodromy surface defect is a simple product of the normalized vacuum expectation value and the asymptotics of the $\EN=2$ theory partition function, $\Psi(\mathbf{a}; \boldsymbol\g) = \exp\left( \frac{\widetilde{\EW}(\mathbf{a})}{\ve_2} \right) \psi(\mathbf{a};\boldsymbol\g)$, where $\widetilde{\EW}(\mathbf{a})$ is the effective twisted superpotential that does not depend on the defect parameters $\boldsymbol\g$. Then, the universal oper equation becomes
\begin{align} \label{eq:univopergauge}
     0=\left[ \p_{y} ^N +  \hat{t}_2 (y) \p_{y} ^{N-2} + \cdots + \hat{t}_{N-1} (y) \p_{y} + \hat{t}_{N} (y) \right] \chi_\a (\mathbf{a};y) \psi(\mathbf{a};\boldsymbol\g),
\end{align}
since the effective twisted superpotential part completely factors out from the equation. Here, let us recall that $\chi_\a (\mathbf{a})$ is the normalized vacuum expectation value of the canonical surface defect, which is a solution to the oper $\r_\mathbf{a} \in \text{Op}_{SL(N)} (\BP^1;S)$ \eqref{eq:oper} by our construction in section \ref{subsec:oper}. We remind that this oper equation reads
\begin{align} \label{eq:opergauge}
    0 = \left[ \p_{y} ^N +  {t}_2 (\mathbf{a};y) \p_{y} ^{N-2} + \cdots +{t}_{N-1} (\mathbf{a};y) \p_{y} + {t}_{N} (\mathbf{a};y) \right] \chi_\a (\mathbf{a};y).
\end{align}
In both equations above, $\a=1,2,\cdots, N$ enumerates the choice of the vacuum at infinity for the two-dimensional $\EN=(2,2)$ theory of the canonical surface defect. 

The two equations look almost identical, but in fact they are not because the meromorphic functions appearing as the coefficients of the differentials $\p_y ^{N-k}$, $k=2,3,\cdots, N$, are not exactly the same. The meromorphic functions $(\hat{t}_k (y))_{k=2} ^N$ in the first equation \eqref{eq:univopergauge} are operator-valued in the sense that their coefficients are twisted differential operators on $\text{Bun}_{PGL(N)}(\BP^1;S)$. On contrary, the meromorphic functions $(t_k (\mathbf{a};y) )_{k=2} ^N$ in the second equation \eqref{eq:opergauge}, although they are in exact same form with $(\hat{t}_k(y))_{k=2} ^N$ as Laurent polynomials, have their coefficients given by the combinations of the normalized vacuum expectation values of the local chiral observables in the limit $\ve_2 \to 0$, $\lim_{\ve_2 \to 0} \left\langle \text{Tr}\,\phi^k \right\rangle_{\mathbf{a}}$, which are simply numbers. We denote these $N-1$ independent combinations appearing as the Laurent coefficients of $t_k (\mathbf{a};y)$ by $E_k (\mathbf{a})$, $k=2,3,\cdots, N$.

Now, we can multiply $\psi(\mathbf{a};\boldsymbol\g)$ to the second equation and take the difference of the two equations to get 
\begin{align}
\begin{split}
    0 &= \begin{pmatrix}
        0 & \left( \hat{t}_2 (y) - t_2 (\mathbf{a};y) \right)\psi(\mathbf{a}) & \left( \hat{t}_3 (y) - t_3 (\mathbf{a};y) \right)\psi(\mathbf{a}) & \cdots & \left( \hat{t}_N (y) - t_N (\mathbf{a};y) \right)\psi(\mathbf{a}) \\ 0 & 0 & 0 & \cdots & 0 \\
        \vdots & \vdots & \vdots & \cdots & \vdots  \\
        0 & 0 & 0 & \cdots & 0
    \end{pmatrix} \times \\ 
    & \quad \times \begin{pmatrix}
        \p_y ^{N-1} \chi_1 (\mathbf{a}) & \p_y ^{N-1} \chi_2 (\mathbf{a}) & \cdots & \p_y ^{N-1} \chi_N (\mathbf{a}) \\
        \vdots & \vdots & \cdots & \vdots \\
        \p_y  \chi_1 (\mathbf{a}) & \p_y \chi_2 (\mathbf{a}) & \cdots & \p_y \chi_N (\mathbf{a})  \\
         \chi_1 (\mathbf{a}) &  \chi_2 (\mathbf{a}) & \cdots &  \chi_N (\mathbf{a}) 
    \end{pmatrix}, 
\end{split}
\end{align}
where we organized all the $(N-1)$-jets of the $N$ independent oper solutions into a single $N \times N$ matrix. Due to the mutual independence, this $N\times N$ matrix is of full rank (i.e., $N$) at generic $y \in \BP^1 \setminus \{0,\qe,1,\infty\}$. Thus, we can invert it to arrive at
\begin{align}
    0= \left( \hat{t}_k (y) - t_k (\mathbf{a};y) \right)\psi(\mathbf{a},\boldsymbol\g),\quad\quad k=2,3,\cdots, N.
\end{align}
Since $y$ is generic, all the Laurent coefficients of these equations have to vanish individually. The central elements represented on $M^{\fsl(N)}$ are completely determined by the monodromies of the oper $\r_\mathbf{a}$ at the marked points $S=\{0,\qe,1,\infty\}$. By construction, these monodromies are fixed by the $2N$ hypermultiplet mass parameters $(m^\pm _\a )_{\a=1}^N$ as \eqref{eq:mono}, so that the relevant terms in the Laurent expansion identically vanish. The rest of the Laurent coefficients give
\begin{align} \label{eq:spectralfinal}
    0=\left( \hat{H}_k - E_k (\mathbf{a})\right) \psi(\mathbf{a};\boldsymbol\g), \quad\quad k=2,3,\cdots, N,
\end{align}
where $\hat{H}_k$ is the $k$-th order Gaudin Hamiltonian that we obtained in section \ref{eq:coinvunivop}. Thus, we prove that the sections $\psi(\mathbf{a};\boldsymbol\g)$ of the Hecke eigensheaf, obtained as the normalized vacuum expectation value of the regular monodromy surface defect, are common eigenfunctions of the Gaudin Hamiltonians $\hat{H}_k$. The eigenvalues $E_k(\mathbf{a})$ are the normalized vacuum expectation values of the local chiral observables, which parametrize the space of opers $\text{Op}_{SL(N)} (\BP^1;S)$ by our construction of the oper $\r_{\mathbf{a}}$ and its solutions $\chi_\a (\mathbf{a})$ \eqref{eq:opergauge} from the canonical surface defect. This completes our derivation of the equivalence \eqref{eq:heceigsh} between the Hecke eigensheaf $\D_{\r_{\mathbf{a}}}$ corresponding to the oper $\r_{\mathbf{a}}$ and the quotient $\ED_{\fL}/{\text{ker}}\,\tilde{\r}_{\mathbf{a}} \cdot \ED_{\fL}$ by the ideal generated by the oper $\r_{\mathbf{a}}$.

\section{Discussion} \label{sec:dis}
We have explained the half-BPS surface defects $-$ the regular monodromy surface defect and the canonical surface defect $-$ in the four-dimensional $\EN=2$ gauge theory give rise to the objects appearing in the study of the geometric Langlands correspondence, in agreement with the string duality to the GL-twisted $\EN=4$ theory and the reduction to the topological sigma model with corresponding boundaries and line defects. The vacuum expectation value of the regular monodromy surface defect gives a family of twisted coinvariants, which diagonalizes the action of the Hecke operator realized by the further insertion of the canonical surface defect on top of the regular monodromy surface defect. The \textit{eigenvalue} is shown to be the local system associated to the oper built by the canonical surface defect. Using this construction, we also showed that the distinguished family of twisted coinvariants simultaneously solves the spectral equations for the quantum Hitchin Hamiltonians, establishing the statement of \cite{BD1, BD2} in the $\EN=2$ gauge theory context.

Several future directions deserve more developments. First of all, the mapping of the BPS objects in the $\EN=2$ theory side and the $\EN=4$ theory side should be extended to more general types. Most apparently, non-regular monodromy surface defects in the $\EN=2$ theory and non-regular (deformed) Nahm-pole boundary of the $\EN=4$ theory \cite{Gaiotto:2008sa, Gaiotto:2011nm} are supposed to be dual to each other. The 't Hooft lines in the $\EN=4$ theory labelled by more general dominant integral coweights should correspond to the surface defects in the $\EN=2$ theory engineered by coupling to more general (non-abelian) $\EN=(2,2)$ gauged linear sigma models living on the worldvolume of multiple coincident M2-branes. Moreover, the \textit{composition} operation of junctions in the $\EN=4$ theory, studied in \cite{Frenkel:2018dej}, is expected to be dual to the blowup operation \cite{Nakajima:2003pg,Nakajima:2003uh} with surface defect insertions in the $\EN=2$ theory \cite{Jeong:2020uxz,Nekrasov:2020qcq}.

The string duality between the IIB gauge origami and the twisted M-theory that we outlined in \ref{sec:bbd} should be analyzed in more detail. In the gauge origami setup, we had three independent $\O$-background parameters, $\sum_{i=1} ^4 \ve_i = 0$, and one of them was turned off $\ve_3 = 0$ to perform a T-duality in the simplest fashion, leaving only two independent $\O$-background parameters. It would be desirable to study the $\ve_3 \neq 0$ deformation of the twisted M-theory by properly T-dualizing the IIB background and uplifting to the M-theory. Its implication on the non-commutative five-dimensional Chern-Simons theory \cite{Costello:2016nkh} would be interesting to clarify.

Finally, it would be important to provide a $\EN=2$ gauge theoretical approach to the analytic Langlands correspondence recently developed in \cite{Etingof:2019pni,Etingof:2021eub,Etingof:2021eeu}, where the Hitchin Hamiltonians and the Hecke operators are realized as operators acting on a Hilbert space of states (see also \cite{Teschner:2017djr} for a real form of geometric Langlands correspondence obtained from single-valuedness condition). In the GL-twisted $\EN=4$ theory context, the analytic Langlands correspondence was formulated in \cite{Gaiotto:2021tsq}. It was crucial there to realize the product of two copies of the Hitchin moduli spaces with opposite complex structures, by the folded construction which yields a semi-infinite strip worldsheet where two junctions are glued by a common boundary. It would be nice to establish the corresponding $\EN=2$ theory framework with two fixed points of spacetime isometry corresponding to the two junctions, for which the constructions of this work would provide building blocks.

\appendix
\section{4-point twisted coinvariants and regular monodromy surface defect} \label{app:4-point}
We provide the computational detail of the matching between the KZ equation for the 4-point twisted coinvariants and the constraints on the vacuum expectation value of the regular monodromy surface defect.

\subsection{KZ equation for 4-point twisted coinvariants}
The 4-point genus-$0$ KZ equation written as
\begin{align} \label{eq:app4pointgeo}
    0=\left[ - (k+N)\frac{\p}{\p \qe} + \frac{\hat{\CalH}_0}{\qe} + \frac{\hat{\CalH}_1}{\qe-1} \right]\Psi(\boldsymbol\g;\qe),
\end{align}
where $\hat{\CalH}_0$ and $\hat{\CalH}_1$ are twisted differential operators
\begin{align}
\begin{split}
    \hat{\CalH}_0 &=  \Psi_0 ^{-1} \left[ \sum_{a,b=1} ^N \bar{T}_a ^b \vert_{H_{\boldsymbol\t-\boldsymbol\z,\s}} \bar{T}_b ^a \vert_{V_{\boldsymbol\z}} \right] \Psi_0 + \frac{(N-1)\s(\s+N)}{N}  \\
    &= \Psi_0 ^{-1} \left[ \sum_{a,b=1} ^N \left(-X_2 ^b \frac{\p}{\p X_2 ^a} \right) J^a _b + \frac{1}{N}  \sum_{a=1} ^N X_2 ^a \frac{\p}{\p X_2 ^a}  \sum_{b=1}^N J^b _b  \right] \Psi_0  + \frac{(N-1)\s(\s+N)}{N}, \\
    \hat{\CalH}_1 &=  \Psi_0 ^{-1} \left[ \sum_{a,b=1} ^N \bar{T}_a ^b \vert_{H_{\boldsymbol\t-\boldsymbol\z,\s}} \bar{T}_b ^a \vert_{H_{\tilde{\boldsymbol\z} -\boldsymbol\t,\tilde\s}} \right] \Psi_0 \\
    &= \Psi_0 ^{-1} \left[ \sum_{a,b=1} ^N \left(-X_2 ^b \frac{\p}{\p X_2 ^a} \right) \left(-X_3 ^a \frac{\p}{\p X_3 ^b} \right) - \frac{1}{N}  \sum_{a=1} ^N X_2 ^a \frac{\p}{\p X_2 ^a}  \sum_{b=1} ^N X_3 ^b \frac{\p}{\p X_3 ^b}   \right] \Psi_0   ,
\end{split}
\end{align}
where the twisting factor $\Psi_0$ is given by \eqref{eq:4ptansatz},
\begin{align}
    \Psi_0 =  \left( \frac{z_{2 1} ^2 z_{43}}{z_{41} z_{31} } \right)^{-\D_2} \frac{\prod_{b=1}^N \left( \tilde\pi ^b \left( X_2 \wedge \pi_{b-1} \right) \right)^{\b  _{b}} \left( \tilde\pi ^b \left( X_3 \wedge \pi_{b-1} \right) \right)^{\tilde{\b}  _{b}} \prod_{i=1} ^{N-1} \left( \tilde\pi ^i (\pi_i) \right)^{\a_i} }{z_{14} ^{\D_1 +\D_4 -\D_3} z_{13} ^{\D_1 +\D_3 -\D_4} z_{43} ^{\D_3 + \D_4 -\D_1}} \prod_{i=1} ^4 dz_i ^{\D_i}.
\end{align}
The followings are useful identities:
\begin{align} \label{eq:useful1}
\begin{split}
    &J_a ^b \pi _i = -e_a \wedge \tilde{e}^b \pi_i \\
    &{\tilde\pi} ^c (X_p \wedge J_a ^b \pi_{c-1}) = \th_{c> b} (\d_a ^c X_p ^b - \d_a ^b X_p ^c) \\
    &\tilde\pi^{d} (X_p \wedge J_b ^c J_a ^b \pi_{d-1}) = \th_{d>c} (\d_b ^c + \th_{b\geq d}) \left( \d_a ^c X_p ^d   - \d_a ^d X^c _p  \right) \\
    &\tilde\pi^{e} (X_p \wedge J_c ^d J_b ^c J_a ^b \pi_{e-1}) = -\th_{e>d} (\d_c ^d +\th_{c\geq e})(\d_b^d +\th_{b\geq e})(\d_a ^d X^e _p - \d^e _a X^d _p) \\
    & {\tilde\pi}^i (J_a ^b \pi _i) = -\d_a ^b \th_{i \geq b} \\
    & \tilde{\pi}^i (J_b ^c J_a ^b \pi_i) = \th_{i\geq c} \d_a ^c \left( \d_b ^c+\th_{b > i}  \right)  \\
    &  \tilde{\pi}^i (J_c ^d J_b ^c J_a ^b \pi_i) = -\th_{i\geq d} \d_a ^d (\d_b ^d + \th_{b>i})(\d_c ^d +\th_{c>i}) \\
    &J_a ^b \Psi(\boldsymbol\g;\qe) = \th_{a>b} \left( \frac{X_2 ^b}{X_3 ^a} - \frac{X_2 ^a X_3 ^b}{\left( X_3 ^a\right)^2} \right) \p_{\g_a} \Psi(\boldsymbol\g;\qe)
\end{split}
\end{align}
and
\begin{align} \label{eq:useful2}
\begin{split}
& \frac{\p}{\p X_q ^a} \left( \tilde{\pi}^c (X_p \wedge \pi_{c-1}) )\right) = 
\d_{p,q} \d_a ^c, \\
 &    \frac{\p}{\p X_2 ^a} \Psi(\boldsymbol\g;\qe)= \frac{1}{X_ 3 ^a} \p_{\g_a} \Psi(\boldsymbol\g;\qe), \quad \frac{\p}{\p X_3 ^a}\Psi(\boldsymbol\g;\qe) = - \frac{X_2 ^a}{\left(X_3 ^a \right)^2} \p_{\g_a} \Psi(\boldsymbol\g;\qe).
\end{split}
\end{align}
Using these identities, we obtain the differential operators $\hat{\CalH}_0$ and $\hat{\CalH}_1$ by a straightforward computation as
\begin{align}
    \hat{\CalH}_0 = & - \sum_{b>a} \frac{\g_b}{\g_a} \left( \g_a \p_{\g_a} + \beta_{a}  \right) \left( (\g_a-\g_b)\p_{\g_b} + \frac{\g_a}{\g_b}\beta_{b} + \tilde\beta_{b} \right) + \sum_{a=1}^N \left( \g_a \p_{\g_a} + \beta_{a}  \right)   \left( \sum_{c=a+1}^N \beta_{c} + \tilde\beta_{c} + \sum_{i\geq a} \alpha_i \right)  \nonumber \\
    & - \frac{1}{N} \left( \sum_{a=1}^N \beta_{a} \right) \sum_{b=1}^N \left( \sum_{c=b+1}^N \beta_{c} + \tilde\beta_{c} + \sum_{i\geq b} \alpha_i \right) + \frac{(N-1)\s(\s+N)}{N},
\end{align}
and 
\begin{align}
\begin{split}
    \hat{\CalH}_1 & =  \sum_{a,b=1}^N \frac{\g_b}{\g_a} \left( \g_a \p_{\g_a} + \beta_{a} \right) \left( - \g_b \p_{\g_b} + \tilde\beta_{b} \right) - \frac{1}{N} \sum_{a=1} ^N \beta_{a} \sum_{b=1}^N \tilde\beta_{b}.
\end{split}
\end{align}

\subsection{Constraints on vacuum expectation value of monodromy surface defect}
To derive the constraints on the vacuum expectation value of the regular monodromy surface defect, we consider the $qq$-character in presence of $\G_{34} \times \G_{24}$. It leads to the following gauge origami setup,
\begin{subequations}\label{def:GO-setup-final}
\begin{align}
    \hat{N}_{12} &= \sum_{\o'=0}^{N-1} \left( e^{a_{\o'}} \CalR_0 + e^{m^+_{{\o'}}+\ve_1+\ve_3} \CalR_1 + e^{m^-_{{\o'}}-\ve_1-\ve_3} \CalR_{2}  \right) \otimes \hat{q}_2^{\o'} \fR_{\o'} \nonumber\\
    \hat{N}_{34} & = e^x \hat{q}_2^{\o} \CalR_0 \otimes \fR_{\o}
\end{align}
\end{subequations}
for each choice of $\o\in \{0,1,\cdots, N-1\}$. The fractional $qq$-character raised from such a setup is 
\begin{align}
    \EX_\o(x) = \EY_{\overline{\o+1}}(x+\ve_1+\ve_2 \delta_{\o,N-1}) + \kq_\o \frac{P_{\o+1}^+(x) P_{{\o}}^-(x+\ve_1) }{\EY_\o(x)}
\end{align}
with $P^+_N(x) = P^+_0(x+\ve_2)$. The 1-loop part and and the non-perturbative part of the vacuum expectation value of the regular monodromy surface defects can be obtained as
\begin{align}\label{eq:bulk-final)}
\begin{split}
    & \BE \left[ - \frac{\hat{P}_3 \hat{S}_{12} \hat{S}_{12}^* }{P_1^* \hat{P}_2^*} \right]^{\BZ_3 \times \BZ_N} \\
    & =  \BE \left[ \frac{-SS^* + M^+S^* + S(M^-)^*}{P_{12}^*} \right] \times \BE \left[ \sum_{\o_1<\o_2} \frac{S_{\o_1}S_{\o_2}^*}{P_1^*} - \sum_{\o_1<\o_2} \frac{ M_{\o_1}^+ S_{\o_2}^* + S_{\o_1}(M_{\o_2}^-)^*}{P_1^*} \right],
\end{split}
\end{align}
built from the equivariant Chern characters
\begin{align}
\begin{split}
    & \hat{S}_{12} = \sum_{\o=0}^{N-1} S_\o \hat{q}_2^{\o} \CalR_0 \otimes \fR_{\o} + q_{13} M_\o^+ \hat{q}_2^{\o} \CalR_1 \otimes \fR_{{\o}} + q_{13}^* M_\o^- \hat{q}_2^{\o} \CalR_{2} \otimes \fR_{{\o}} , \\
    & S = \sum_{\o=0}^{N-1} S_\o = N_{12} - P_{12} K_{12,N-1}, \ M^\pm = \sum_{\o=0}^{N-1} M_\o^\pm
\end{split}
\end{align}
satisfies
\begin{align}
    \left[ \frac{\ve_2}{\ve_1}(\qe-1)\kq \frac{\p}{\p \kq} + \hat{ \rm H} \right]\Psi^{\text{1-loop + non-pert}} = 0,
\end{align}
where $\hat{\rm H}$ is the differential operator given by
\begin{align}
\begin{split}
    \hat{\rm H} = \sum_{\omega=0}^{N-1} \frac{\qe-1}{2} \left(  (\nabla^u_\o)^2 - \frac{2 a_\o}{\ve_1} \nabla^u_\o \right) - \kq_\o w_\o \left(\nabla^u_\o + \frac{m_{\overline{\o+1}}^+ -a_\o -\ve_2\d_{\o,N-1}}{\ve_1} \right) \left(\nabla^u_\o+ \frac{m_{{\o}}^- -a_\o}{\ve_1} -1\right),
\end{split}
\end{align}
where we used the notation $\nabla^u_\o \equiv u_\o \p_{u_\o}$ for convenience and defined
\begin{align}
    w_\o = 1 + \kq_{\o+1} + \kq_{\o+1} \kq_{\o+2} + \cdots + \kq_{\o+1}\cdots \kq_{\o+N-1} = \frac{u_{\o+1}+ u_{\o+2} +\cdots + u_{\o+N} }{u_{\o+1}}.
\end{align}

We write the full vacuum expectation value by adding the classical part as 
\begin{align}
\begin{split}
    \Psi(\mathbf{a};\mathbf{u};\qe) &=  \kq^{-\frac{\sum_{\o=0 }^{N-1} a_\o^2}{2\ve_1\ve_2} +  \frac{\ve_1}{\ve_2} \frac{\sigma}{N} \left( - \frac{m^-}{\ve_1} - \frac{N(N-1)}{2} - (N-1)\sigma  \right) } (1-\qe)^{\frac{\ve_1}{\ve_2} \frac{1}{N} \left( \frac{m^--a}{\ve_1} -N \right) \left( \frac{a-m^+}{\ve_1} -N \right)} \\
    & \quad\quad \times \prod_{\o=0}^{N-1} u_\o^{\frac{m_{\o}^--a_\o-\ve_1}{\ve_1}} \prod_{\o=0}^{N-1} \g_{\o}^{-\beta_\o} \; \Psi^{\text{1-loop + non-pert}},
\end{split}
\end{align}
where $\b_\o$ is to be determined with the constraint $\s = \sum_{\o=0} ^{N-1} \b_\o =  \frac{m^--a}{\ve_1} -N$. Here, we are making a change of variables by
\begin{align}
    \g_{\o} = \frac{u_{\o}+ u_{\o+1} + \cdots+ u_{\o+N-1}}{\kq-1}, 
\end{align}
so that $u_\o = \g_{\o+1}-\g_{\o}$ and $\p_{\g_\o} = \p_{u_{\o-1}} - \p_{u_{\o}}$. The change of variables modifies $\kq$-derivative by
\begin{align}
    \p_\kq \to \p_\kq - \frac{\g_{0}}{\kq-1}(\p_{\g_0}+\cdots+\p_{\g_{N-1}}).
\end{align}
Also, the differential operator $\hat{\rm H}$ now takes the form of
\begin{align} \label{eq:Hamil-faf}
\begin{split}
    \hat{\rm H} = 
    & (\kq-1) \sum_{\o,\o'} \g_{\o'} (\g_{\o'}-\g_{\o}) \p_{\g_{\o}} \p_{\g_{\o'}} \theta_{\o'>\o} + \left( \bM^-_{{\o'}} - \bM^+_{\overline{\o'-1}} - 2 \right) \g_{\o'} \p_{\g_{\o}} \theta_{\o'>\o} \\
    & + \d_{\o,\o'} \left( \bM^-_{{\o}} -1 \right) \g_{\o}\p_{\g_{\o}}+ \kq \sum_{\o,\o'} \g_{\o'}\p_{\g_{\o}} \left( - \g_{\o'}\p_{\g_{\o'}} + \bM^-_{{\o'}} - \bM^+_{\overline{\o'-1}} - 2 \right).
\end{split}
\end{align}
Here we denote $\ve_1 \bM^+_\o = m_{\o+1}^+$, $\ve_1 \bM^-_\o = m_{\o}^-$ for $\o=0,\dots,N-1$. Taking account for the contribution of the classical part and the change of variables, the equation satisfied by the full vacuum expectation value $\Psi(\mathbf{a};\boldsymbol\g;\qe)$ becomes
\begin{align} \label{eq:app4pointg}
    \left[ \frac{\ve_2}{\ve_1} \frac{\p}{\p \qe} + \frac{\hat{\CalH}_0}{\kq} + \frac{\hat{\CalH}_1}{\kq-1} \right] \Psi(\mathbf{a};\boldsymbol\g;\qe) = 0,
\end{align}
where the differential operators $\hat{\CalH}_0$ and $\hat{\CalH}_1$ are computed to be
\begin{subequations}
\begin{align}
    \hat{\CalH}_0 = & \sum_{\o,\o'=0}^{N-1} \frac{\g_{\o'}}{\g_{\o}} \left( \g_{\o} \p_{\g_{\o}} + \beta_{\o} \right) \left( (\g_{\o'} - \g_{\o}) \p_{\g_{\o'}} + \beta_{\o'} - \bM^-_{{\o'}} + \bM^+_{\overline{\o'-1}} + 2 - \frac{\g_{\o}}{\g_{\o'}}\beta_{\o'} \right) \theta_{\o'>\o} \nonumber \\
    & + \sum_{\o} \left( - \bM_{{\o}}^- + \o \right) (\g_{\o} \p_{\g_{\o}}+\beta_\o) -  \frac{\sigma}{N} \left( - \frac{m^-}{\ve_1} - \frac{N(N-1)}{2} - (N-1)\sigma  \right)\\
    \hat{\CalH}_1 = & \sum_{\o,\o'=0}^{N-1} \frac{\g_{\o'}}{\g_{\o}} \left( \g_{\o} \p_{\g_{\o}} + \beta_\o \right) \left( -\g_{\o'}\p_{\g_{\o'}} - \beta_{\o'} + \bM^-_{{\o'}} - \bM^+_{\overline{\o'-1}} - 2 - \frac{\ve_2}{\ve_1} \d_{\o',0} \right) \nonumber \\
    &  - \frac{1}{N} \left( \frac{m^--a}{\ve_1} -N \right) \left( \frac{a-m^+}{\ve_1} -N \right)
\end{align}
\end{subequations}
We find the equation \eqref{eq:app4pointg} is identical to the reduced 4-point KZ equation \eqref{eq:app4pointgeo}, provided the mapping of parameters given by
\begin{subequations}
\begin{align}
     &k+N = -\frac{\ve_2}{\ve_1}, \\
    &\beta_{2,\o}  = \beta_\o, \ \o=0,\dots,N-1 \\
    &\beta_{3,\o}  = \bM^-_{{\o}} - \bM^+_{\overline{\o-1}}-2-\frac{\ve_2}{\ve_1} \d_{\o,0} - \beta_{\o},  \ \o=0,\dots,N-1 \\
    &\alpha_{i}  = \bM^+_{i} - \bM^-_{i} + 1, \ i=0,\dots,N-2
\end{align}
\end{subequations}
Note that the equation \eqref{eq:app4pointg} itself does not depend on the Coulomb moduli $\mathbf{a}$. Thus, the vacuum expectation value of the regular monodromy surface defect provides a family of solutions to the 4-point KZ equation enumerated by the Coulomb moduli $\mathbf{a}$.

Let us explicitly write out the mapping between the weight parameters and the gauge theory parameters. We may choose 
\begin{align}
    \beta_\o = \frac{m_\o ^- -a_{\o}-\ve_1}{\ve_1},\quad\quad \o=0,1,\cdots, N-1.
\end{align}
Then we achieve the relation between the weight parameters and the gauge theory parameters as
\begin{subequations}
\begin{align}
    &\zeta_i = \beta_{2,i+1}+\beta_{3,i+1}+\alpha_i = \frac{m_{{i+1}}^-- m_{i}^-}{\ve_1} - 1
    , \quad\quad i = 0,\dots,N-2 \\
    &\tilde\zeta_{i}  = \beta_{2,i}+\beta_{3,i}+\alpha_i = \frac{m^+_{i+1}-m_{{i}}^+  }{\ve_1} -1, \quad\quad i=0,\dots,N-2 \\
    &\tau_{i} = \zeta_{i} - \beta_{2,i+1} + \beta_{2,i}= \frac{a_{i+1} - a_{i} }{\ve_1} - 1, \quad\quad i=0,\dots,N-2 \\
    &\sigma  = \sum_{\o=0}^{N-1} \beta_{2,\o} = \frac{m^--a}{\ve_1} - N, \\
    &\tilde\sigma = \sum_{\o=0}^{N-1} \beta_{3,\o} = \frac{a-m^+}{\ve_1} - N.
\end{align}
\end{subequations}

\section{5-point twisted coinvariants and parallel surface defects} \label{app:5point}
We provide the detail of the matching between the constraints on the 5-point twisted coinvariants with the insertion of the bi-infinite generalization of twisted vacuum module and the constraints on the correlation function of the parallel surface defects.
\subsection{Constraints on 5-point twisted coinvariants}
The constraints on the 5-point twisted coinvariants \eqref{eq:constcoin} are written as
\begin{align} \label{eq:app5pteqgeo}
\begin{split}
    &0= \left[ \frac{\p}{\p y} + \frac{\hat{\CalA}_0}{y} + \frac{\hat{\CalA}_\qe}{y-\qe} + \frac{\hat{\CalA}_1}{y-1} \right] \begin{pmatrix}
        \m_1 \U(\boldsymbol\g,\boldsymbol\m;\qe,y) \\ \m_2 \U(\boldsymbol\g,\boldsymbol\m;\qe,y) \\ \vdots \\ \m_N \U(\boldsymbol\g,\boldsymbol\m;\qe,y)
    \end{pmatrix}  \\
    &0= \left[ -(k+N)\frac{\p}{\p \qe} + \frac{\hat{\CalH}_0}{\qe} + \frac{\hat{\CalH}_y}{\qe-y} + \frac{\hat{\CalH}_1}{\qe-1} \right]\U(\boldsymbol\g,\boldsymbol\m;\qe,y),
\end{split}
\end{align}
where $\left( \hat{\CalA}_i \right)_{ba} = \left( \U_0 ^{(b)} \right) ^{-1} \bar{T}_a ^b \vert_{M_i} \U_0 ^{(a)} $, $i=0,\qe,1$, and $\hat{\CalH}_i =  \left( \U_0 \right) ^{-1} \sum_{a,b=1} ^N \bar{T}_a ^b \vert_{M_\qe} \bar{T}_b ^a  \vert_{M_i} \U_0 $, $i=0,y,1$. Here, we are using the twisting factor given by \eqref{eq:twist5}
\begin{align}
\U_0 = \prod_{p=0,2} \left( \frac{z_{p 1} ^2 z_{43}}{z_{41} z_{31} } \right)^{-\D_p} \frac{\prod_{a=1}^N \prod_{p=0,2,3} \left( \tilde\pi ^a \left( X_p \wedge \pi_{a-1} \right) \right)^{\b  _{p,a}} \prod_{i=1} ^{N-1} \left( \tilde\pi ^i (\pi_i) \right)^{\a_i} }{z_{14} ^{\D_1 +\D_4 -\D_3} z_{13} ^{\D_1 +\D_3 -\D_4} z_{43} ^{\D_3 + \D_4 -\D_1}}\prod_{i=0} ^4 dz_i ^{\D_i},
\end{align}
and also defined $\U^{(a)} _0 := \frac{\tilde\pi^a (X_3 \wedge \pi_{a-1})}{\tilde\pi^N (X_0 \wedge \pi_{N-1})} 
\prod_{i=a} ^{N-1} \left(\tilde\pi^i(\pi_i) \right)^{\frac{2(N-1)}{N-a+1}}  \U_0$, $a=1,2,\cdots, N$, so that $\left( \hat{\CalA}_i \right)_{ba} = \left( \U_0 ^{(b)} \right) ^{-1} \bar{T}_a ^b \vert_{M_i} \U_0 ^{(a)} $.

In addition to the identities \eqref{eq:useful1} and \eqref{eq:useful2}, the following identities are also straightforward for the 5-point case:
\begin{align} \label{eq:useful3}
\begin{split}
    &J_a ^b \U(\boldsymbol\g,\boldsymbol\m) = \th_{a>b} \left[\left( \frac{X_2 ^b}{X_3 ^a} - \frac{X_2 ^a X_3 ^b}{\left( X_3 ^a\right)^2} \right) \p_{\g_a} +\left( \frac{X_0 ^b}{X_3 ^a} - \frac{X_0 ^a X_3 ^b}{\left( X_3 ^a\right)^2} \right) \p_{\m_a} \right]\U(\boldsymbol\g,\boldsymbol\m)
\end{split}
\end{align}
and
\begin{align} \label{eq:useful4}
\begin{split}
& \frac{\p}{\p X_0 ^a} \U(\boldsymbol\g,\boldsymbol\m)= \frac{1}{X_ 3 ^a} \p_{\m_a} \U(\boldsymbol\g,\boldsymbol\m) \\
 &    \frac{\p}{\p X_2 ^a} \U(\boldsymbol\g,\boldsymbol\m)= \frac{1}{X_ 3 ^a} \p_{\g_a} \U(\boldsymbol\g,\boldsymbol\m), \quad \frac{\p}{\p X_3 ^a}\U(\boldsymbol\g,\boldsymbol\m) = - \left(\frac{X_2 ^a}{\left(X_3 ^a \right)^2} \p_{\g_a} + \frac{X_0 ^a}{\left( X_3 ^a \right)^2} \p_{\m_a} \right)\U(\boldsymbol\g,\boldsymbol\m).
\end{split}
\end{align}
Using the identities \eqref{eq:useful1}, \eqref{eq:useful2}, \eqref{eq:useful3}, and \eqref{eq:useful4}, we obtain
\begin{align}
\begin{split}
  \left(\hat{\CalA}_0 \right)_{ba} &=  \left(\U_0 ^{(b)} \right)^{-1} \left( J_a ^b -\frac{1}{N} \d_a ^b \sum_{c=1} ^N J^c _c \right) \U_0 ^{(a)}  - \frac{k(N-1)}{N} \d_a ^b \\
    &=\th_{b>a} \left( \b _{2,b} \frac{\g_a}{\g_b} + \b _{3,b}+1 +(\g_a -\g_b)\p_{\g_b} - \m_b \p_{\m_b} \right) \\
    & \quad + \d^b _a \left( -k  -\sum_{c=a+1} ^N   \sum_{p=2,3} \b _{p,c}  -\sum_{i=a} ^{N-1} \a_i +\sum_{c=a+1} ^N \m_c \p_{\m_c} \right)  \\
    &  \quad +  \frac{1}{N} \d_a ^b \left\{ N+  \sum_{c=1} ^N \left( \sum_{p=0,2,3} \sum_{d=c+1} ^N \b _{p,d} + \sum_{i=c} ^{N-1} \a_i \right) \right\}
\end{split}
\end{align}
Similarly, we compute
\begin{align}
\begin{split}
 \left( \hat{\CalA}_\qe \right)_{ba} &= \left( \U_0 ^{(b)} \right)^{-1} \left( -X_2 ^b \frac{\p}{\p X_2 ^a} + \frac{1}{N} \d^b _a \sum_{c=1} ^N X_2 ^c \frac{\p}{\p X^c _2} \right) \U_0 ^{(a)} \\
     &= - \g_b \p_{\g_a} - \b_{2,a} \frac{\g_b}{\g_a} + \frac{1}{N} \d^b _a \sum_{c=1} ^N \b _{2,c}
\end{split}
\end{align}
and
\begin{align}
\begin{split}
 \left( \hat{\CalA}_1 \right)_{ba} &= \left( \U_0 ^{(b)} \right)^{-1} \left( -X_3 ^b \frac{\p}{\p X_3 ^a} + \frac{1}{N} \d^b _a \sum_{c=1} ^N X^c _3 \frac{\p}{\p X_3 ^c} \right) \U_0 ^{(a)} \\
    &= \g_a \p_{\g_a}+ \m_a \p_{\m_a} -1 -\b_{3,a}  + \frac{1}{N} \d^b _a \sum_{c=1}^N \b _{3,c}.
\end{split}
\end{align}

Next, let us turn to the $\qe$-derivative KZ equation. By a similar computation, we get
\begin{align}
\begin{split}
    \hat{\CalH}_0 &=  \left( \U_0 \right)^{-1} \sum_{a,b=1 }^N \left[ \left( -X_2 ^b \frac{\p}{\p X_2 ^a} \right) J^a _b -\frac{1}{N} \left( -X_2 ^a \frac{\p}{\p X_2 ^a} \right) J^b _b  \right]  \U_0 + \frac{(N-1)\s(\s+N)}{N} \\
    &=   - \sum_{a>b}  \left( \g_a \p_{\g_b} + \b _{2,b} \frac{\g_a}{\g_b}  \right)  \left( \b _{0,a} \frac{\m_b}{\m_a} + \b _{2,a} \frac{\g_b}{\g_a} + \b _{3,a} +  (\g_b-\g_a) \p_{\g_a} + (\m_b -\m_a)\p_{\m_a}  \right) +\frac{(N-1)\s(\s+N)}{N}  \\
    &\quad+ \sum_{a=1}^N \left( \g_a \p_{\g_a} + \beta_{2,a} \right) \left( \sum_{c=a+1}^N \sum_{p=0,2,3} \beta_{p,c} + \sum_{i=a}^{N-1} \alpha_i \right) 
    -\frac{1}{N} \left( \sum_{a=1}^N \b _{2,a} \right) \left( \sum_{b=1}^N \left( \sum_{p=2,3} \sum_{c=b+1} ^N \b _{p,c} + \sum_{i=b} ^{N-1} \a_i \right) \right)
\end{split}
\end{align}

\begin{align}
\begin{split}
    \hat{\CalH}_1 &=  \left( \U_0 \right)^{-1} \sum_{a,b=1}^N \left[ \left(- X_2 ^b \frac{\p}{\p X_2 ^a} \right) \left(- X_3 ^a \frac{\p}{\p X_3 ^b} \right) - \frac{1}{N} \left(- X_2 ^a \frac{\p}{\p X_2 ^a} \right) \left(- X_3 ^b \frac{\p}{\p X_3 ^b} \right) \right] \U_0 \\
    &=\sum_{a,b=1} ^N \frac{\g_b}{\g_a} \left(\g_a \p_{\g_a} + \b_{2,a} \right)\left(-\g_b \p_{\g_b} - \m_b \p_{\m_b} +\b_{3,b} \right) - \frac{1}{N} \sum_{a=1} ^N \b_{2,a} \sum_{b=1} ^N \b _{3,b}
\end{split}
\end{align}

\begin{align}
\begin{split}
    \hat{\CalH}_y &=  \left( \U_0\right)^{-1} \sum_{a,b=1}^N \left[ \left( -X_2 ^b \frac{\p}{\p X_2 ^a} \right) \left( -X_0 ^a \frac{\p}{\p X_0 ^b} \right) -\frac{1}{N} \left( -X_2 ^a \frac{\p}{\p X_2 ^a} \right) \left( -X_0 ^b \frac{\p}{\p X_0 ^b} \right) \right] \U_0\\
    &=\sum_{a,b=1} ^N \left( \g_b \p_{\g_a} + \b _{2,a} \frac{\g_b}{\g_a} \right) \left( \m_a \p_{\m_b} +\b_{0,b} \frac{\m_a}{\m_b}  \right) - \frac{k}{N} \sum_{a=1} ^N \b_{2,a}. 
\end{split}
\end{align}

\subsection{Constraints on correlation function of parallel surface defects}

\subsubsection{$y$-derivative}
We start from the fractional quantum TQ equation \eqref{eq:fracTQ},
\begin{align}\label{eq:TR-final}
    \llangle[\Big] T_\o(\bx) \EQ(\bx) \rrangle[\Big] = P_{\overline{\o+2}}^+(x_{\overline{\o+1}}+\ve_1) \llangle[\Big] \EQ(\bx+\ve_1e_{\overline{\o+1}}) \rrangle[\Big] + \kq_\o P_{{\o}}^-(x_\o+\ve_1) \llangle[\Big] \EQ(\bx-\ve_1 e_\o ) \rrangle[\Big]
\end{align}
The left hand side is a degree 1 polynomial taking the form 
\begin{align}
    T_{\o}(\bx) = x_{\overline{\o+1}} - a_{\overline{\o+1}} + \ve_1 + \ve_1\nabla^u_{\o+1} + \kq_\o (x_\o - m_{\o+1}^+ - m_{{\o}}^- + a_\o + \ve_1 - \ve_1\nabla^u_\o).
\end{align}
with $m_N^+ = m_0^+-\ve_2$. 

Define the Fourier transform of $\EQ(\bx)$ by
\begin{align}\label{def:Upsilon-final}
    \hat{\Upsilon}(y,\boldsymbol\mu) = \qe^{-\frac{\sum_{\o=0}^{N-1} a_\o ^2 }{2\ve_1 \ve_2}}  \prod_{\o=0}^{N-1} u_\o^{\frac{m_{\o}^--a_\o-\ve_1}{\ve_1}} \sum_{\bx\in \bL} \prod_{\o=0}^{N-1}  y_\o ^{-\frac{x_{\o}}{\ve_1}}  \llangle \EQ (\bx) \rrangle, 
\end{align}
Further, we define $\mu_\o$ variables by
\begin{align}\label{def:mu-final}
    y_\o = \frac{\mu_{\o+1}}{\mu_\o}, \ \mu_{\o+N} = y \mu_\o, \ \nabla^\mu_\o = \nabla^y_{\o-1} - \nabla^y_\o. 
\end{align}
so that 
\begin{align}
    \prod_{\o=0}^{N-1} y_\o^{-\frac{x_\o}{\ve_1}} = y^{-\frac{x_{N-1}}{\ve_1}} \prod_{\o=0}^{N-1} \mu_\o^{\frac{x_\o - x_{\overline{\o-1}}}{\ve_1}}
\end{align}
Due to the transformation, the fractional TQ equations are converted into differential equations satisfied by $\hat{\U}$:
\begin{align}\label{eq:TR-F-final}
\begin{split}
    & \left[ -\ve_1 \nabla^y_{\overline{\o+1}} - m_{\overline{\o+1}}^- + 2\ve_1 + \ve_1 \nabla^u_{\o+1} + \kq_\o \left( -\ve_1 \nabla^y_{\o} - m_{\o+1}^+ - \ve_1 \nabla^u_\o \right) \right] \hat\Upsilon(\by) \\
    & = y_{\overline{\o+1}} \left[ -\ve_1\nabla^y_{\overline{\o+1}} - m_{\overline{\o+2}}^+ \right] \hat\Upsilon(\by) + \kq_\o \left[ -\ve_1\nabla^y_\o - m_{\o}^- +\ve_1 \right] \frac{1}{y_\o} \hat\Upsilon(\by) .
\end{split}
\end{align}
The change of variables leads to
\begin{align}
    \nabla^y_\o = \nabla^y + \nabla^\mu_{\o+1} + \cdots + \nabla^\mu_{N-1} = \nabla^y + \hat\nabla^y_\o.
\end{align}
As a result, the above equation becomes
\begin{subequations}
\begin{align}
    & \left[ - (1+\kq_{\o-1}) \nabla^y - \hat\nabla^y_{\o}  + \nabla^u_{\o} - \frac{m_{\o}^-}{\ve_2} + 2  + \kq_{\o-1}\left( -\hat\nabla^y_{\overline{\o-1}} - \nabla^u_{\overline{\o-1}} - \frac{m^+_{\o}}{\ve_1} + 1 \right) \right] \mu_{\o} \hat\Upsilon \nonumber\\
    & = y^{\d_{\o,N-1}} \left[ -\nabla^y - \hat\nabla^y_{\o} - \frac{m^+_{{\o+1}} }{\ve_1} + 1 \right] \mu_{\overline{\o+1}} \hat\Upsilon + \kq_{\o-1} y^{-\d_{\o,0}} \left[ -\nabla^y - \hat\nabla^y_{\overline{\o-1}} - \frac{m_{\o-1}^- }{\ve_1} +2  \right] \mu_{\overline{\o-1}} \hat\Upsilon,
\end{align}
\end{subequations} 
for each $\o=0,1,\cdots, N-1$. Let us define a $N \times 1$ vector
\begin{align}
    \boldsymbol\Upsilon = y^{-2} \begin{pmatrix} \mu_0 \hat\Upsilon \\ 
\mu_1 \hat\Upsilon \\\vdots \\ \mu_{N-1} \hat\Upsilon \end{pmatrix}.
\end{align}
Then, the $N$ differential equations are organized into a single $N \times 1$ matrix equation as
\begin{align}
\begin{split}
    & \left[ \left( \bU +  \bU^{-1} \boldsymbol\kq  - \bI - \bU^{-1} \boldsymbol\kq \bU \right) (\nabla^y) \right. \\
    & \quad + (\bM^+ + \bI) \bU - (\bM^--\bbE) +  \bU^{-1} \boldsymbol\kq (\bM^--\bbE) - \bU^{-1} \boldsymbol\kq (\bM^++\bI) \bU +  \boldsymbol\nabla^z - \bU^{-1}{\boldsymbol\kq} {\boldsymbol\nabla^z} \bU \\
    & \quad \left. +  \hat{\boldsymbol\nabla}^y \bU + \bU^{-1} {\boldsymbol\kq}  \hat{\boldsymbol\nabla}^y \bU - \hat{\boldsymbol\nabla}^y - \bU^{-1} {\boldsymbol\kq} \hat{\boldsymbol\nabla}^y \bU \right] {\boldsymbol\Upsilon} = 0.
\end{split}
\end{align}
Let us explain the matrix notations appearing above. $\bU$ is given by
\begin{align}\label{def:U-matrix-final}
    {\bf U} = \begin{pmatrix}
    0 & 1 & 0 & \cdots & 0 & 0 \\
    0 & 0 & 1 & \cdots & 0 & 0 \\
    & \vdots & & \ddots & & \vdots \\
    0 & 0 & 0 & \cdots & 0 & 1 \\
    y & 0 & 0 & \cdots & 0 & 0
    \end{pmatrix}, \quad 
    {\bf U}^{-1} = \begin{pmatrix}
    0 & 0 & 0 & \cdots & 0 & \frac{1}{y} \\
    1 & 0 & 0 & \cdots & 0 & 0 \\
    0 & 1 & 0 & \cdots & 0 & 0 \\
    & \vdots & & \ddots & & \vdots \\
    0 & 0 & 0 & \cdots & 1 & 0
    \end{pmatrix}.
\end{align}
We defined several diagonal matrices by 
\begin{align}
\begin{split}
    & \ve_1\bM^+ = \text{diag}(m_1^+,\dots,m_{N}^+) \\
    & \ve_1\bM^- = \text{diag}(m^-_0,\dots,m^-_{N-1}) \\
    & \bbE = \text{diag}(1,0,\dots,0) \\
    & \boldsymbol\kq = \text{diag} (\kq_0,\dots,\kq_{N-1}) \\
    & \boldsymbol\nabla^u = \text{diag} (\nabla^u_0,\dots,\nabla^u_{N-1}) \\
    & \hat{\boldsymbol\nabla}^y = \text{diag}(\hat\nabla^y_0,\dots,\hat\nabla^y_{N-1}).
\end{split}
\end{align}
We invert the matrix in front of $\p_y$ by 
\begin{align}
\begin{split}
    & y^{-1} (\bI + \bU\boldsymbol\kq \bU^{-1} - \boldsymbol\kq \bU^{-1} - \bU)^{-1} \\
    & = y^{-1} \left[ (\bI - \bU)(\bI - \boldsymbol \kq \bU^{-1}) \right]^{-1} \\
    & = y^{-1} (\bI - \boldsymbol\kq \bU^{-1})^{-1} (\bI - \bU)^{-1} \\
    & = \frac{1}{(y-\kq)(1-y)}  \sum_{j=0}^{N-1} \left(\boldsymbol\kq \bU^{-1} \right)^j \sum_{j'=0}^{N-1} \bU^{j'}
\end{split}
\end{align}
with 
\begin{subequations}
\begin{align}
    & (\bU - \bI)^{-1}_{\o,\o'} 
    = \frac{1}{1-\bU^{-N}} \left( \sum_{j=1}^N \bU^{-j} \right)_{\o,\o'} = \frac{1}{1-\frac{1}{y}} \left( \frac{1}{y} \right)^{\theta_{\o\leq \o'} } \\
    & (\bI - {\boldsymbol\kq} \bU^{-1})^{-1}_{\o,\o'} = \frac{1}{1-({\boldsymbol\kq} \bU^{-1})^N} \left( \sum_{j=0}^{N-1} ({\boldsymbol\kq} \bU^{-1})^j \right) = \frac{1}{1-\frac{\kq}{y}} \frac{u_{\o+1}}{u_{\o'+1}} \left( \frac{\kq}{y} \right)^{\theta_{{\o<\o'}}} .
\end{align}
\end{subequations}
Consequently, $\boldsymbol\Upsilon$ satisfies 
\begin{align}
   \left[ \p_y + \frac{\bbA_0}{y} + \frac{\bbA_1}{y-1} + \frac{\bbA_\kq}{y-\kq} \right] \boldsymbol\Upsilon = 0,
\end{align}
where 
\begin{subequations}
\begin{align}
    \left( \bbA_0 \right)_{\o,\o'} = & \left[ (\g_{\o}-\g_{\o'})\p_{\g_{\o'}} - \mu_{\o'}\p_{\mu_{\o'}} + \bM^-_{{\o'}} - \bM^+_{\overline{\o'-1}} - 1 \right] \theta_{\o'>\o} \\
    & + \delta_{\o,\o'} (\bM^-_{{\o}} + \hat\nabla^y_{{\o}}) \nonumber\\
    \left( \bbA_1 \right)_{\o,\o'} = & \g_{\o'} \p_{\g_{\o'}} + \mu_{\o'} \p_{\mu_{\o'}} + \bM^+_{\overline{\o'-1}} - \bM^-_{{\o'}} + 1 
    \\
    \left( \bbA_\kq \right)_{\o,\o'} = & -\g_{\o} \p_{\g_{\o'}}
\end{align}
\end{subequations}
Finally, we consider the full correlation function of the parallel surface defects by multiplying the rest of the classical part as
\begin{align} \label{eq:fullparallel}
\begin{split}
\Upsilon(\mathbf{a};\boldsymbol\g,\boldsymbol\m;\qe,y) &= \kq^{ \frac{\ve_1}{\ve_2} \frac{\sigma_2}{N} \left( - \frac{m^-}{\ve_1} - \frac{N(N-1)}{2} - (N-1)\sigma_2  \right) } (1-\qe)^{  \frac{\ve_1}{\ve_2} \frac{1}{N} \frac{m^--a-N\ve_1}{\ve_1} \frac{a-m^+-\ve_2-N\ve_1}{\ve_1}} \times \\
    & \quad\quad \times y^{\frac{m^-}{N\ve_1}-\frac{N+3}{2} }(y-1)^{\frac{m^+-a-\ve_2}{N\ve_1}+1} (y-\kq)^{-\frac{m^--a}{N\ve_1} +1 } \times \\
    &\quad\quad \times  \prod_{\o=0}^{N-1} \mu_\o^{-\eta_\o} \g_{\o}^{-\beta_\o} \; \hat{\Upsilon} \\
    &=  \kq^{ -\frac{\sum_{\o=0}^{N-1} a_\o ^2 }{2\ve_1 \ve_2}+  \frac{\ve_1}{\ve_2} \frac{\sigma_2}{N} \left( - \frac{m^-}{\ve_1} - \frac{N(N-1)}{2} - (N-1)\sigma_2  \right) } (1-\qe)^{  \frac{\ve_1}{\ve_2} \frac{1}{N} \frac{m^--a-N\ve_1}{\ve_1} \frac{a-m^+-\ve_2-N\ve_1}{\ve_1} } \times \\
    & \quad\quad \times y^{\frac{m^-}{N\ve_1}-\frac{N+3}{2} }(y-1)^{\frac{m^+-a-\ve_2}{N\ve_1}+1} (y-\kq)^{-\frac{m^--a}{N\ve_1} +1 } \times \\
    &\quad\quad \times  \prod_{\o=0}^{N-1} \mu_\o^{-\eta_\o} \g_{\o}^{-\beta_\o}   \prod_{\o=0}^{N-1} u_\o^{\frac{m_{\o}^--a_\o-\ve_1}{\ve_1}} \sum_{\bx\in \bL} \prod_{\o=1}^{N}  y_\o ^{-\frac{x_{\overline{\o}}}{\ve_1}}  \llangle \EQ (\bx) \rrangle,
\end{split}
\end{align}
with the constraints
\begin{align}
      \sum_{\o=0}^{N-1} \eta_\o = 0, \quad  \sum_{\o=0}^{N-1} \beta_\o = \s_2 = \frac{m^--a}{\ve_1}-N.
\end{align}
Then, the differential equation satisfied by the full correlation function is
\begin{align} \label{eq:appyder}
0 = \left[ \frac{\p}{\p y} + \frac{\hat{\CalA}_0}{y} +\frac{\hat{\CalA}_\qe}{y-\qe} +\frac{\hat{\CalA}_1}{y-1} \right] \begin{pmatrix}
    \m_0 \U \\ \m_1 \U \\ \vdots \\ \m_{N-1} \U
\end{pmatrix},
\end{align}
where 
\begin{subequations}
\begin{align}
    \left( \hat\CalA_0 \right)_{\o,\o'} = & \ \left[ (\g_{\o}-\g_{\o'})\p_{\g_{\o'}} - \mu_{\o'}\p_{\mu_{\o'}}  + \bM^-_{{\o'}} - \bM^+_{\overline{\o'-1}} - 1 - \eta_{\o'} - \beta_{\o'} \left( 1-\frac{\g_\o}{\g_{\o'}} \right) \right] \theta_{\o'>\o} \\
    & + \delta_{\o,\o'} (\bM^-_{{\o}} + \hat\nabla^y_{{\o}} + \eta_{\o+1} + \cdots + \eta_{N-1}) - \d_{\o,\o'} \left( \frac{m^-}{N\ve_1} - \frac{N-1}{2} \right) \nonumber\\
    \left( \hat\CalA_1 \right)_{\o,\o'} = & \ \g_{\o'} \p_{\g_{\o'}} + \mu_{\o'} \p_{\mu_{\o'}} + \bM^+_{\overline{\o'-1}} - \bM^-_{{\o'}} + 1 + \beta_{\o'} + \eta_{\o'} + \frac{\d_{\o,\o'}}{N} \frac{a-m^+-N\ve_1+\ve_2}{\ve_1}
    \\
    \left( \hat\CalA_\kq \right)_{\o,\o'} = & - \frac{\g_{\o}}{\g_{\o'}} \left( \g_{\o'}\p_{\g_{\o'}} + \beta_{\o'} \right) + \frac{\d_{\o,\o'}}{N} \frac{m^--a-N\ve_1}{\ve_1}.
\end{align}
\end{subequations}

\subsubsection{$\kq$-derivative}
We will consider linear combination of \eqref{eq:TwnN-2}
\begin{align}
    \sum_{\o=0}^{N-1} A_\o w_\o T_{\o}(x=x'_{\overline{\o+1}} - \ve_2\delta_{\o,N-1}) + (1-A_\o) w_\o T_{\o}(x=x'_{\o})
\end{align}
with the coefficient
\begin{align}
    w_\o = 1 + \kq_{\o+1} + \kq_{\o+1} \kq_{\o+2} + \cdots + \kq_{\o+1}\cdots \kq_{\o+N-1} = \frac{u_{\o+1}+ u_{\o+2} +\cdots + u_{\o+N} }{u_{\o+1}}
\end{align}
and $N$ independent parameters $A_\o$ to be chosen later. Taking the Fourier transform \eqref{def:Upsilon-final} of $\EQ(\bx)$ gives
\begin{align}
\begin{split}
    & \left[ \frac{\ve_2}{\ve_1} (\kq-1) \kq \p_{\kq} + \hat{\rm H} \right] \hat\Upsilon (y,\boldsymbol\mu) \\
    &- \left[ \sum_{\o=0}^{N-1} A_{{\o}} u_{{\o}} \left(\nabla^\mu_{\o+1} - \frac{\ve_2}{\ve_1}\d_{\o,N-1} \right) \left( -\nabla^y - \hat\nabla^y_{\o+1} - \bM_{\overline{\o+1}}^- + \nabla^z_{\o+1} \right) \right. \\
    & \qquad \left. + (1-A_\o) \kq_{\o} u_\o \left( - \nabla^\mu_{\o+1} + \frac{\ve_2}{\ve_1} \d_{\o,N-1} \right) \left( - \nabla^y - \hat\nabla^y_\o - \bM_\o^+ -2 - \nabla^z_\o \right) \right] \Upsilon \\
    = 
    & - \left[ \sum_{\o=0}^{N-1} A_\o u_\o  \left( \nabla^\mu_{\o+1} - \frac{\ve_2}{\ve_1} \d_{\o,N-1} \right) \left( -\nabla^y - \hat\nabla^y_{\o+1} - \bM^+_{\overline{\o+1}} -1 \right) {y_{\o+1}} \right. \\
    & \qquad \left. + (1-A_\o) \kq_\o u_\o \left( - \nabla^\mu_{\o+1} + \frac{\ve_2}{\ve_1} \d_{\o,N-1} \right) \left( -\nabla^y - \hat\nabla^y_{\o} -1 - \bM_{{\o}}^- \right) \frac{1}{y_\o} \right] \Upsilon.
\end{split}
\end{align}
The differential operator $\hat{\rm H}$ is exactly the one already obtained in \eqref{eq:Hamil-faf}.
The coefficient of $-\nabla^y$ is
\begin{align}
\begin{split}
    & \sum_{\o=0}^{N-1} A_\o u_\o \left( \nabla^\mu_{\o+1} - \frac{\ve_2}{\ve_1} \d_{\o,N-1} \right) \left( 1 - y_{\o+1} \right) - (1-A_\o) \kq_\o u_\o \left( \nabla^\mu_{\o+1} - \frac{\ve_2}{\ve_1} \d_{\o,N-1} \right) \left( 1 - \frac{1}{y_\o} \right) \\
    = & \sum_{\o=0}^{N-3} A_\o u_\o \nabla^\mu_{\o+1} \frac{\mu_{\o+1}-\mu_{\o+2}}{\mu_{\o+1}} - (1-A_\o) \kq_\o u_\o \nabla^\mu_{\o+1} \frac{\mu_{\o+1} - \mu_{\o}}{\mu_{\o+1}} \\
    & + A_{N-2} u_{N-2} \nabla^\mu_{N-1} \frac{\mu_{N-1}-y\mu_{0}}{\mu_{N-1}} - (1-A_{N-2})\kq_{N-2} u_{N-2} \nabla^\mu_{N-1} \frac{\mu_{N-1} - \mu_{N-2}}{\mu_{N-1}} \\
    & + A_{N-1} u_{N-1} \left( \nabla^\mu_{0} - \frac{\ve_2}{\ve_1} \right) \frac{\mu_{0}-\mu_1}{\mu_{0}} - (1-A_{N-1})\kq_{N-1} u_{N-1} \left( \nabla^\mu_0 - \frac{\ve_2}{\ve_1} \right) \frac{\mu_{0} - y^{-1}\mu_{N-1} }{\mu_{0}}
\end{split}
\end{align}
We will choose $A_0 = \cdots = A_{N-2} = 1$, Using the $y$-derivative constraint that we already derived, we get
\begin{align}
\begin{split}
    & \nabla^y (\mu_{\o}-\mu_{\o+1}) \Upsilon \\
    & = \sum_{\o'=0}^{N-1} \left\{ \left[ (\bbA_0)_{\o+1,\o'}-(\bbA_0)_{\o,\o'} \right] + \frac{y}{y-\kq} \left[ (\bbA_\kq)_{\o+1,\o'}-(\bbA_\kq)_{\o,\o'} \right] \right\} \mu_{\o'} \Upsilon \\
    & = \left[ -\bM^-_\o \mu_\o + (\bM^+_\o+1) \mu_{\o+1} + \hat{\nabla}^y_\o (\mu_{\o+1}-\mu_{\o}) + \sum_{\o'=0}^{N-1} z_{\o} \p_{\g_{\o'}} \left( \theta_{\o'>\o+1} - \frac{y}{y-\kq} \right) \mu_{\o'} \right] \Upsilon
\end{split}
\end{align}
The total contribution is 
\begin{align}
    \sum_{\o=0}^{N-2} (\kq-1) \g_{\o+1} \nabla^\mu_{\o+1} \left[ \p_{z_{\o}} + \frac{1}{\mu_{\o+1}} \sum_{\o'=0}^{N-1} \p_{\g_{\o'}} \mu_{\o'} \left( \frac{y}{y-\kq} - \theta_{\o'> \o+1} \right) \right]
\end{align}
Finally we will choose $A_{N-1}=1$ to get
\begin{align}
\begin{split}
    & \nabla^y (\mu_{0}-\mu_{1}) \Upsilon \\
    & = \sum_{\o'=0}^{N-1} \left\{ \left[ (\bbA_0)_{1,\o'}-(\bbA_0)_{0,\o'} \right] + \frac{y}{y-\kq} \left[ (\bbA_\kq)_{1,\o'}-(\bbA_\kq)_{0,\o'} \right] \right\} \mu_{\o'} \Upsilon \\
    & = - \bM^-_{0} \mu_{0} + (\bM_{0}^+ + 1) \mu_{1}  + \hat\nabla^y_{0} (\mu_{1}-\mu_{0}) + z_{0} \sum_{\o'=0}^{N-1} \p_{\g_{\o'}} \mu_{\o'} \left( \frac{y}{y-\kq}-\theta_{\o'>0} \right).
\end{split}
\end{align}
It gives
\begin{align}
    (\kq-1) \g_{0} \left( \nabla^\mu_{0} - \frac{\ve_2}{\ve_1} \right) \left[ \p_{z_{N-1}} + \frac{1}{\mu_{0}} \sum_{\o'=0}^{N-1} \p_{\g_{\o'}} \mu_{\o'} \left( \frac{y}{y-\kq} - \theta_{\o'>0} \right) \right].
\end{align}
We obtain the $\kq$-derivative constraint
\begin{align}
\begin{split}
    & \left\{ \frac{\ve_2}{\ve_1} \p_\kq + \frac{\ve_2}{\ve_1} \frac{\g_{N-1}}{(1-\kq)\kq} (\p_{\g_{0}}+\cdots+\p_{\g_{N-1}}) + \frac{\hat{\rm H}}{\qe(\qe-1)} \right. \\
    & \quad \left. -\frac{1}{\kq} \sum_{\o=0}^{N-1} \left( \nabla^\mu_\o - \frac{\ve_2}{\ve_1}\d_{\o,N-1} \right) \g_{\o} \left[ \frac{\kq}{\kq-1}(\p_{{\g_{\o+1}}}+\cdots+\p_{\g_{\o+N}}) + \sum_{\o'=0}^{N-1} \frac{\mu_{\o'}}{\mu_\o} \p_{\g_{\o'}} \left( \frac{y}{y-\kq} - \theta_{\o'>\o} \right) \right] \right\} \Upsilon = 0
\end{split}
\end{align}
It can be organized into 
\begin{align}
    \left[ \frac{\ve_2}{\ve_1} \frac{\p}{\p\kq} + \frac{\BH_0}{\kq} +\frac{\BH_y}{\kq-y}+ \frac{\BH_1}{\kq-1}  \right] \hat\Upsilon=0
\end{align}
with the Laurent coefficients given by
\begin{subequations}
\begin{align}
    \BH_0 = &\sum_{\o,\o'} \g_{\o'} \p_{\g_{\o}} \left[ (\g_{\o'}-\g_{\o}) \p_{\g_{\o'}} - (\bM^-_{{\o'}}-\bM^+_{\overline{\o'-1}}-2) \right] \theta_{\o'>\o} \\
    & + \sum_{\o,\o'} \left[ \nabla^\mu_{\o'} - \frac{\mu_{\o}}{\mu_{\o'}} \left( \nabla^\mu_{\o'} - \frac{\ve_2}{\ve_1} \d_{\o',0} - 1 \right) \right] \g_{\o'} \p_{\g_{\o}} \theta_{\o'>\o} \nonumber\\
    & + \sum_{\o} \left(-\bM^-_{{\o}} + \o  \right) \g_{\o} \p_{\g_{\o}} \nonumber\\
    \BH_y = & \sum_{\o,\o'=0}^{N-1} \frac{\mu_{\o'}}{\mu_{\o}} \left( \nabla^\mu_\o -1 - \frac{\ve_2}{\ve_1}\d_{\o,0} \right) \g_{\o} \p_{\g_{\o'}}\\
    \BH_1 = &\sum_{\o,\o'=0}^{N-1} \g_{\o'} \p_{\g_{\o}} \left( - \g_{\o'}\p_{\g_{\o'}} - \mu_{\o'}\p_{\mu_{\o'}} + \bM^-_{\o'} - \bM^+_{\overline{\o-1}}-2 \right) .
\end{align}
\end{subequations}
Finally, we consider the full correlation function of parallel surface defects \eqref{eq:fullparallel} by multiplying all the classical part by
\begin{align} 
\begin{split}
\Upsilon(\mathbf{a};\boldsymbol\g,\boldsymbol\m;\qe,y) &= \kq^{ \frac{\ve_1}{\ve_2} \frac{\sigma_2}{N} \left( - \frac{m^-}{\ve_1} - \frac{N(N-1)}{2} - (N-1)\sigma_2  \right) } (1-\qe)^{  \frac{\ve_1}{\ve_2} \frac{1}{N} \frac{m^--a-N\ve_1}{\ve_1} \frac{a-m^+-\ve_2-N\ve_1}{\ve_1}} \times \\
    & \quad\quad \times y^{\frac{m^-}{N\ve_1}-\frac{N+3}{2} }(y-1)^{\frac{m^+-a-\ve_2}{N\ve_1}+1} (y-\kq)^{-\frac{m^--a}{N\ve_1} +1 } \times \\
    &\quad\quad \times  \prod_{\o=0}^{N-1} \mu_\o^{-\eta_\o} \g_{\o}^{-\beta_\o} \; \hat{\Upsilon} \\
    &=  \kq^{ -\frac{\sum_{\o=0}^{N-1} a_\o ^2 }{2\ve_1 \ve_2}+  \frac{\ve_1}{\ve_2} \frac{\sigma_2}{N} \left( - \frac{m^-}{\ve_1} - \frac{N(N-1)}{2} - (N-1)\sigma_2  \right) } (1-\qe)^{  \frac{\ve_1}{\ve_2} \frac{1}{N} \frac{m^--a-N\ve_1}{\ve_1} \frac{a-m^+-\ve_2-N\ve_1}{\ve_1} } \times \\
    & \quad\quad \times y^{\frac{m^-}{N\ve_1}-\frac{N+3}{2} }(y-1)^{\frac{m^+-a-\ve_2}{N\ve_1}+1} (y-\kq)^{-\frac{m^--a}{N\ve_1} +1 } \times \\
    &\quad\quad \times  \prod_{\o=0}^{N-1} \mu_\o^{-\eta_\o} \g_{\o}^{-\beta_\o}   \prod_{\o=0}^{N-1} u_\o^{\frac{m_{\o}^--a_\o-\ve_1}{\ve_1}} \sum_{\bx\in \bL} \prod_{\o=1}^{N}  y_\o ^{-\frac{x_{\overline{\o}}}{\ve_1}}  \llangle \EQ (\bx) \rrangle,
\end{split}
\end{align}
with 
\begin{align}
    \sum_{\o=0}^{N-1} \eta_\o = 0, \ \sum_{\o=0}^{N-1} \beta_\o = \s_2 = \frac{m^--a}{\ve_1}-N.
\end{align}
The $\qe$-derivative constraint for the full correlation function becomes
\begin{align} \label{eq:app5pteqg}
\begin{split}
    &\left[ \frac{\ve_2}{\ve_1} \frac{\p}{\p\qe} + \frac{\hat\CalH_0}{\kq} + \frac{\hat\CalH_y}{\kq-y} + \frac{\hat\CalH_1}{\kq-1} \right] \Upsilon(\mathbf{a};\boldsymbol\g,\boldsymbol\m;\qe,y) = 0,
\end{split}
\end{align}
where the Laurent coefficients are given by
\begin{subequations}
\begin{align}
    \hat\CalH_0 = & \sum_{\o,\o'} \frac{\g_{\o'}}{\g_{\o}} \left( \g_\o \p_{\g_{\o}} + \beta_\o \right) \left[ (\g_{\o'}-\g_{\o}) \p_{\g_{\o'}} - (\bM^-_{{\o'}}-\bM^+_{\overline{\o'-1}}-2) + \beta_{\o'} \left( 1 - \frac{\g_{\o}}{\g_{\o'}} \right) \right] \theta_{\o'>\o} \\
    & + \sum_{\o,\o'} \left[ \nabla^\mu_{\o'} + \eta_{\o'} - \frac{\mu_{\o}}{\mu_{\o'}} \left( \nabla^\mu_{\o'} + \eta_{\o'} - \frac{\ve_2}{\ve_1} \d_{\o',0} - 1 \right) \right] \g_{\o'} \p_{\g_{\o}} \theta_{\o'>\o} \nonumber\\
    & + \hat\sum_{\o} \left(-\bM^-_{{\o}} + \o  \right) \left( \g_{\o} \p_{\g_{\o}} + \beta_\o \right) -  \frac{\sigma_2}{N} \left( - \frac{m^-}{\ve_1} - \frac{N(N-1)}{2} - (N-1)\sigma_2  \right) \nonumber\\
    \hat\CalH_y = & \sum_{\o,\o'=0}^{N-1} \frac{\mu_{\o'}}{\mu_{\o}} \left( \nabla^\mu_\o -1 - \frac{\ve_2}{\ve_1}\d_{\o,0} + \eta_{\o} \right) \frac{\g_{\o}}{\g_{\o'}} \left( \g_{\o'}\p_{\g_{\o'}} + \beta_{\o'} \right) - \frac{k}{N} \frac{m^--a-N\ve_1}{\ve_1} \\
    \hat\CalH_1 = &  \sum_{\o,\o'=0}^{N-1} \frac{\g_{\o'}}{\g_{\o}} \left( \g_{\o} \p_{\g_{\o}} + \beta_{\o} \right) \left( - \g_{\o'}\p_{\g_{\o'}} - \mu_{\o'}\p_{\mu_{\o'}} + \bM^-_{\o'} - \bM^+_{\overline{\o-1}}-2 - \beta_{\o'} - \eta_{\o'} \right) \\
    & -\frac{1}{N} \frac{m^--a-N\ve_1}{\ve_1} \frac{a-m^+-\ve_2-N\ve_1}{\ve_1} \nonumber.
\end{align}
\end{subequations}

We see the exact matching between the constraints on the 5-point twisted coinvariants with the quotient of the bi-infinite module \eqref{eq:app5pteqgeo} and the constraints on the vacuum expectation values of the parallel surface defects \eqref{eq:appyder}, \eqref{eq:app5pteqg}, provided the following mapping of the parameters:
\begin{subequations}
\begin{align}
    &k+N = -\frac{\ve_2}{\ve_1} \\
    &\beta_{0,\o}  = -\frac{\ve_2}{\ve_1} \d_{\o,0} -1 + \eta_{\o}, \quad \o=0,\dots,N-1, \\
    &\beta_{2,\o}  = \beta_\o, \quad \o=0,\dots,N-1, \\
    &\beta_{3,\o}  = -\bM^+_{\overline{\o-1}} + \bM^-_{\o} - 2 - \beta_{\o} - \eta_{\o}, \quad \o=0,\dots,N-1, \\
    &\alpha_{i}  =  \bM^+_i - \bM^-_i + 1, \quad i =0,\dots,N-2 .
\end{align}
\end{subequations}
We may further choose 
\begin{align}
   \beta_\o = \frac{m^-_{\o}-a_{ \o}-\ve_1}{\ve_1}. 
\end{align}
As a consequence, we find the mapping between the weight parameters and the gauge theory parameters given by
\begin{subequations}
\begin{align}
    \zeta_{i} &= \beta_{0,i+1}+\beta_{2,i+1}+\beta_{3,i+1} + \alpha_i = \frac{m_{{i+1}}^-- m_{i}^-}{\ve_1} - 1
    ,\quad\quad i=0,\dots,N-2 \\
   \tilde\zeta_i &= \beta_{0,i}+\beta_{2,i}+\beta_{3,i} + \alpha_i = \frac{m^+_{i+1}-m_{{i}}^+ }{\ve_1} -1 , \quad\quad i=0,\dots,N-2 \\
   \tau_{i} &= \zeta_{i} - \beta_{2,i+1} + \beta_{2,i} = \frac{a^-_{{i+1}}-a_{{i}}^- }{\ve_1} - 1, \quad\quad i=0,\dots,N-2 \\
    \sigma_0 & = \sum_{\o=0}^{N-1} \beta_{0,\o} = k \\
    \sigma_2 & = \sum_{\o=0}^{N-1} \beta_{2,\o} = \frac{m^--a}{\ve_1} - N \\
    \sigma_3 & = \sum_{\o=0}^{N-1} \beta_{3,\o} = \frac{a-m^++\ve_2}{\ve_1} -N.
\end{align}
\end{subequations}
As noted earlier, $\b_{0,\o}$ are not free parameters, determined in terms of other weight parameters by the constraints \eqref{eq:constsln}. In particular, in the open subset where we have the twisted vacuum module $\b_{0,\o} = k \d_{\o,N-1}$, we get
\begin{align}
    \eta_\o = -\frac{\ve_2}{\ve_1} (\d_{\o,N-1} - \d_{\o,0})+1 -N \d_{\o,N-1}.
\end{align}

\section{Bi-infinite generalization of twisted vacuum module} \label{app:gentwvac}
Here, we verify our claim that the bi-infinite $\widehat{\fsl}(N)$-module $\mathbb{H}^k _{\boldsymbol\t,\s}$ contains a proper submodule if $\s=k$. We start by constructing the $\fsl(N)$-module by
\begin{align}
    (J^b_a)_0 \ket{\bn} = (h_b+n_b) \ket{\bn + \d_a - \d_b}
\end{align}
with $\s = \sum_{a=1}^N h_a+n_a$. 
A state $(J^b_a)_{-1}\ket{\bn}$ on degree 1 has weight 
\begin{align}
\begin{split}
    & (J^c_c)_0 (J^b_a)_{-1} \ket{\bn} \\ 
    & = \left[ (J^b_a)_{-1} (J^c_c)_0 + \d^c_a (J^b_c)_{-1} - \d^b_c (J^c_a)_{-1} \right] \ket{\bn} \\
    & = (n_c+h_c+\d^c_a - \d^b_c) (J^b_a)_{-1} \ket{\bn}
\end{split}
\end{align}
The following $N^2$ rank 1 states share the same weights $(n_1,\dots,n_N)$: 
\begin{align}
    (J^b_a)_{-1} \ket{\bn+\d_b-\d_a}
\end{align}
The $N^2$ rank 1 states can be organized into 4 groups
\begin{itemize}
    \item $(J^N_N)_{-1} \ket{\bn}$,
    \item $(J^b_N)_{-1} \ket{\bn+\d_b}$,
    \item $(J^N_a)_{-1} \ket{\bn-\d_a}$, 
    \item $(J^b_a)_{-1} \ket{\bn+\d_b-\d_a}$.
\end{itemize}
We search for the null vector at degree 1 such that it is annihilated by all $(J^d_c)_1$, whose action on the degree 1 states are 
\begin{align}
\begin{split}
    & (J^d_c)_1 (J^b_a)_{-1} \ket{\bn + \d_b-\d_a} \\
    & = [ \d^d_a (J^b_c)_0 - \d^b_c (J^d_a)_0 + k \d^b_c \d^d_a ] \ket{\bn+\d_b-\d_a} \\
    & = \d^d_a (n_b+h_b+1-\d^b_a) \ket{\bn-\d_a+\d_c} - \d^b_c (n_d+h_d+\d^{bd}-\d^d_a) \ket{\bn+\d_b-\d_d} \\
    & \quad  + k \d^b_c\d^d_a \ket{\bn+\d_b-\d_a}
\end{split} 
\end{align}
with $\d_N = 0$. 
We will consider the linear combination with $b=1,\dots,N-1$
\begin{align}
\begin{split}
    0 = & (J^d_c)_1 \left[ C^N_N (J^N_N)_{-1} \ket{\bn} + C^b_N (J^b_N)_{-1} \ket{\bn+\d_b} + \sum_{a=1}^{N-1} C^N_a (J^N_a)_{-1} \ket{\bn-\d_a} + C^b_a (J^b_a)_{-1} \ket{\bn-\d_a+\d_b} \right] \\
    = & \ C^N_N \left( \d^d_N (n_N+h_N) \ket{\bn+\d_c} - \d^N_c (n_d+h_d) \ket{\bn-\d_d} + k \d^N_c \d^d_N \ket{\bn} \right) \\
    & + C^b_N \left( \d^d_N (n_b+h_b+1) \ket{\bn+\d_c} - \d^b_c (n_d+h_d+\d^{bd}-\d^d_N) \ket{\bn+\d_b-\d_d} + k \d^d_N \d^b_c \ket{\bn+\d_b} \right) \\
    & + \sum_{a=1}^{N-1} C^N_a \left( \d^d_a (n_N+h_N+1)\ket{\bn-\d_a+\d_c} - \d^N_c(n_d+h_d+\d^{dN}-\d^d_a) \ket{\bn-\d_d} + k \d^N_c \d^d_a \ket{\bn-\d_a} \right) \\
    & + \sum_{a=1}^{N-1} C^b_a \left( \d^d_a (n_b+h_b+1-\d^b_a) \ket{\bn-\d_a+\d_c} - \d^b_c (n_d+h_d+\d^{bd}-\d^d_a) \ket{\bn+\d_b-\d_d} + k \d^b_c\d^d_a \ket{\bn+\d_b-\d_a} \right)
\end{split}
\end{align}
\begin{itemize}
\item \textbf{Case 1. $c=d=N$:} 
\begin{align}
\begin{split}
    \left( C^N_N k + C^b_N (n_b+h_b+1) - \sum_{a=1}^{N-1} C^N_a (n_N+h_N+1) \right) \ket{\bn} = 0
\end{split}
\end{align}
\item \textbf{Case 2. $d=N$, $c=1,\dots,N-1$:}
\begin{align}
\begin{split}
    & C^N_N (n_N+h_N) \ket{\bn+\d_c} + C^b_N (n_b+h_b+1) \ket{\bn+\d_c} - C^b_N \d^b_c (n_N+h_N-1+k) \ket{\bn+\d_b} \\
    & - \d^b_c \sum_{a=1}^{N-1}C^b_a (n_N+h_N) \ket{\bn+\d_b} = 0
\end{split}
\end{align}
\item \textbf{Case 3. $ c=N$, $d=1,\dots,N-1$:} 
\begin{align}
\begin{split}
    & -C^N_N (n_d+h_d) \ket{\bn-\d_d} + \sum_{a=1}^{N-1} C^b_a \d^d_a (n_b+h_b+1-\d^b_a) \ket{\bn-\d_a} \\
    & + \sum_{a=1}^{N-1} C^N_a \left( \d^d_a (n_N+h_N+1+k) \ket{\bn-\d_a} - (n_d+h_d-\d^d_a) \ket{\bn-\d_d} \right) = 0
\end{split}
\end{align}
\item \textbf{Case 4. $c,d=1,\dots,N-1$:}
\begin{align}
\begin{split}
    & - C^b_N \d^b_c (n_d+h_d+\d^{bd}) \ket{\bn+\d_b-\d_d} + \sum_{a=1}^{N-1} C^N_a \d^d_a (n_N+h_N+1) \ket{\bn-\d_a+\d_c} \\
    & + C^b_d (n_b+h_b+1-\d^b_d) \ket{\bn-\d_d+\d_c} + \d^b_c \sum_{a=1}^{N-1} C^b_a(n_d+h_d+\d^{bd}-\d^d_a+k\d^{d}_a) \ket{\bn+\d_b-\d_d} = 0
\end{split}
\end{align}
\end{itemize}
The solution is obtained as
\begin{align}
\begin{split}
    &\s=k,\\
    & C^b_N = -\frac{n_N+h_N}{n_b+h_b+1} C^N_N, \ C^N_a = \frac{n_a+h_a}{n_N+h_N+1} C^N_N, \ C^b_a = - \frac{n_a+h_a}{n_b+h_b+1-\d^b_a} C^N_N.
\end{split}
\end{align}
In particular, note that the solution exists only when $\s=k$. The corresponding degree 1 null-states are (up to an overall scaling of $C^N_N$)
\begin{align}
\begin{split}
    v^b_{\bn}[1] = & \ (J^N_N)_{-1} \ket{\bn} - \frac{n_N+h_N}{n_b+h_b+1} (J^b_N)_{-1} \ket{\bn+\d_b} \\
    & + \sum_{a=1}^{N-1} \frac{n_a+h_a}{n_N+h_N+1} (J^N_a)_{-1} \ket{\bn-\d_a} - \frac{n_a+h_a}{n_b+h_b+1-\d^b_a} (J^b_a)_{-1} \ket{\bn+\d_a-\d_b}.
\end{split}
\end{align}
We can define a generating function 
\begin{align}
    \ket{\Sigma(\bx)} \equiv \sum_{\bn \in \BZ^{N-1}} \frac{\bx^\bn}{\Gamma\left( \sigma+1 - \sum_{c=1}^{N-1} n_c+h_c \right)\prod_{c=1}^{N-1} \Gamma(n_c+h_c+1) } \ket{\bn}, \quad \bx^\bn = \prod_{c=1}^{N-1} (x^c)^{n_c} ,
\end{align}
so that all the null-states, for each $a=1,2,\cdots, N-1$, are organized into
\begin{align}
\begin{split}
    &  \left[(J^a_N)_{-1} - x^a \sum_{b=1}^{N} x^b (J^N_b)_{-1} + \sum_{b=1}^{N-1} x^b (J^a_b)_{-1} \right] \ket{\Sigma(\bx)} \\
    & = -x^a \sum_{\bn \in \BZ^{N-1}} \frac{\bx^\bn}{\Gamma\left( \sigma+1 - \sum_{c=1}^{N-1} n_c+h_c \right)\prod_{c=1}^{N-1} \Gamma(n_c+h_c+1) } v^a_{\bn}[1],
\end{split}
\end{align}
where we are using the notation $x^N =1$.

\bibliographystyle{utphys}
\bibliography{reference}

\end{document}